# Instability and Information

Doctoral dissertation by Felix Patzelt

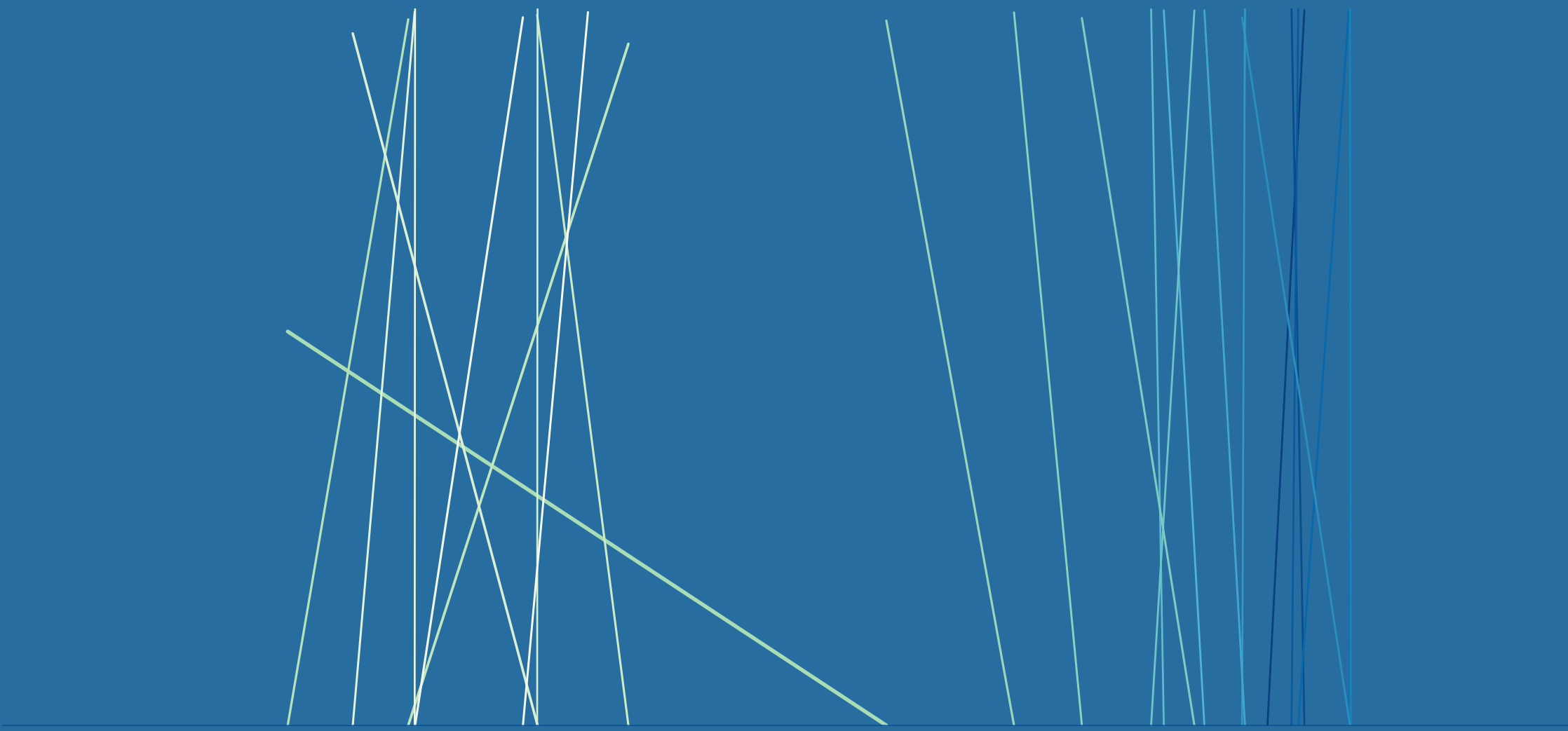

First referee and supervisor: Prof. Dr. Klaus Pawelzik
Second referee: Prof. Dr. Stefan Bornholdt

Department of Physics
University of Bremen

Defence: June 4, 2014

# Instability and Information

Doctoral dissertation by Felix Patzelt

| | |
|---|---|
| First referee and supervisor: | Prof. Dr. Klaus Pawelzik |
| Second referee: | Prof. Dr. Stefan Bornholdt |

Department of Physics
University of Bremen

# Instability and Information

Vom Fachbereich für Physik und Elektrotechnik
der Universität Bremen

zur Erlangung des akademischen Grades eines

Doktor der Naturwissenschaften (Dr. rer. nat.)

genehmigte Dissertation

von

Dipl. Phys. Felix Patzelt

aus Bremen

| | |
|---|---|
| Eingereicht am: | 30. April 2014 |
| Datum des Kolloquiums: | 4. Juni 2014 |
| | |
| Erster Gutachter: | Prof. Dr. Klaus Pawelzik |
| Zweiter Gutachter: | Prof. Dr. Stefan Bornholdt |
| Erster Prüfer: | Prof. Dr. Monika Fritz |
| Zweiter Prüfer: | Prof. Dr. Dr. Ulrich Krause (FB 03) |

Of markets and men,
seesaws and sticks,
and extreme events

# Preface

> “ I can calculate the motions of the heavenly bodies,
> but not the madness of people. ”
>
> Isaac Newton

Human behaviour exhibits extremes of many different kinds. An extreme is only extreme in relation to a set of expectations. Behaviour can only be studied in relation to an environment; it is always an interaction. This work examines extreme events that emerge as global consequences of local interactions contrary to intuitive expectations.

The first research topic covers hand-eye coordination during tasks such as balancing a stick on a finger tip. The second topic covers speculative trading. These two paradigms may appear completely unrelated. They share, however, certain properties, which allow them to be studied using similar methods. In both cases, subjects have an incentive to use observable information about the world to their advantage. Large, unexpected fluctuations in the behavioural dynamics, quantified by balancing errors and logarithmic price changes, pose presumably undesirable risks. Nevertheless, extreme fluctuations that are orders of magnitude larger than the typically observed ones occur repeatedly, even though the probability that they occur at all would be negligible if they were Gaussian (i.e. normally) distributed. Their statistics resemble natural catastrophes.

A well-known example for an extreme economic event is the South Sea Bubble. Over the course of 1719, the price per share of the South Sea Company stock rose from about £100 to £1,000 and fell back to almost where it started [Gar90]. Motivated by a successful trade, Isaac Newton bought a larger number of shares near the peak of the bubble and lost £20,000. After the loss, Newton made the statement quoted above and never wanted to hear about the South Seas again [PA05]. Many others were also financially devastated by the bubble burst. But was it all madness? After all, Hoare's Bank, which suspected that the stock was overvalued, was able to trade highly profitably throughout the bubble [TV04].

Rather than despair, like Newton, we here investigate whether the apparent madness of people can be better understood with the help of modern scientific tools adopted from nonlinear dynamics, statistical physics, neuroscience, control- and information theory. We pay particular attention to the distributions and correlations of extreme events. The significance of such statistics in general is two-fold. First, they are important for risk assessment, especially since costs are often dominated by the most extreme events [Sor04]. Second, since the latter in many cases represent crises or catastrophes, it may be tempting to create credible-sounding post-hoc narratives around them [BP95]. Thereby, rare and extreme observations are perceived as disconnected from more typical ones even though they may share common causes. The statistics of extreme events reveal common underlying structures that might be difficult to observe directly and only become apparent when they lose their balance.

An important related concept is power-law scaling, which reflects a scale invariant (e.g. self-similar) structure of the respective quantity. Different scales, therefore, cannot be understood in isolation. Examples for power-law distributed event magnitudes include earthquake energies, forest fire areas, and several types of avalanches including those in neuronal activity [Sor04, BP03]. Power laws for the probabilities of large events and for long-range spatial or temporal correlations are commonly associated with the so-called critical phenomena. They typically stem from the divergence of a characteristic length or of the susceptibility to perturbations, for example in systems at the boundary of order and disorder. There are, however, also other mechanisms, including certain multiplicative noise processes, that can lead to similar effects [Sor04].

To identify the specific mechanisms in any particular system, it is therefore necessary to consider meaningful model structures in the respective context. For this reason, considerable parts of this work are devoted to the existing literature on the two main topics: human motor control and financial economics. Yet the new contributions begin with the introduction of minimalistic models, where fundamental effects emerge generically, without detailed parameter tuning or even completely independent of parameters. Features are then added or varied carefully step by step, to account for empirical findings while still maintaining a detailed understanding of the model dynamics. Furthermore, behavioural experiments were performed to motivate model assumptions and to test predictions.

Nevertheless, the mechanisms for extreme events presented throughout this work all turn out to follow an overreaching principle, which fits surprisingly well with many existing findings in the respective fields–some of which were previously thought to contradict each other. It is found, seemingly paradoxically, that locally minimising fluctuations can increase a dynamical system's sensitivity to unpredictable perturbations and thereby facilitate global catastrophes. Just like the methods used throughout this work were adopted from many different fields of research, this mechanism for extreme events may be relevant in fields of research outside of the present scope. Concrete examples for this conjecture are presented in the final discussion.

## Thesis structure

The following chapters are grouped into four parts. The structure was planned such that this work can–but doesn't have to–be read from cover to cover. Readers interested only in certain chapters should nevertheless also read the final conclusions in part IV.

In Part I, complex systems, critical phenomena, and important statistical methods are introduced very briefly. The next two parts constitute the majority of this work. Balancing tasks are discussed in Part II and speculative markets in part III. Each of these two main parts features an introduction to the existing literature in the field, and more detailed investigations of preexisting models that are most closely related to this work. After the respective main results, which are motivated and previewed in slightly more detail below, each of these two parts includes a detailed discussion from the perspective of the respective field. Finally, in part IV, the main results are summarised more concisely and discussed from a more general point of view.

We try to explain results as generally understandably as possible, particularly in the introduction and discussion chapters, and in the introductions to each chapter. For the detailed results, however, we assume basic knowledge of higher mathematics, such as probabilities. Some sections also require more advanced knowledge in special fields. In part I, the most important methods for data analysis in the later parts are explained. Furthermore, complexity, criticality, extreme events, self-similarity in time and space, as well as processes generating such phenomena are introduced briefly.

### Balancing tasks

In our everyday lives, we may experience that it is currently impossible to build machines which come even close to human high-level skills like reason. However, even abilities that are often taken for granted can pose unsolved engineering problems. For example, many human (and animal) motor skills are still very difficult or even impossible to replicate in robots. Yet, this remarkable performance is achieved despite unreliable sensors, slow nerve conduction, and muscles which are outperformed by artificial actuators [Hun92]. Therefore, one may argue that the Central Nervous System (CNS) employs motor control strategies which are vastly superior to current engineered solutions. Like in the famous quote “skate where the puck’s going, not where it's been"(Wayne Gretzky), the CNS's success often lies in the ability to predict impending dynamics. This presumably involves the use of internal models for body and environment (see e.g. [FW11], and chapter 5)–a concept which also aligns well with the sensation of stable percept of the world. Yet humans sometimes appear to perceive and memorise only minimal information about their environment (see sec. 9.6.4). A better understanding of the mental representations of the world can be expected to eventually contribute to solving many problems in philosophy, medicine, and engineering.

Since the actual internal representations in a living brain cannot be read out directly[1], simple movement- and object manipulation tasks are important paradigms to investigate motor control behaviour. Movements typically exhibit highly stereotyped patterns across repetitions and across subjects [WG00]. These regularities are commonly attributed to a high degree of optimisation in the aforementioned internal representations. However, movements also show a high degree of variability even between repeated trials where conditions are kept as constant as possible [FSW08]. In the past, this variability has been attributed to mostly physiological noise sources that affect movement execution, but this stance has been questioned in a few more recent publications (e.g. [CAS06], see chapter 5 for more details).

In Part II, we investigate movement variability in the a classic–and perhaps the most fundamental–control problem: the inverted pendulum. Several solutions exist in control theory, but humans exhibit surprising and hitherto unexplained dynamics, including extreme events that follow spatio-temporal scaling laws. The deviations from perfect stability when balancing a stick on a fingertip or performing similar virtual tasks indicate amounts of multiplicative noise that vastly

[1] Current electrophysiological- and imaging techniques can only measure small (and distinct) fractions of the neuronal activity at any point in time. Furthermore, despite much progress over the last century, the neuronal code remains mostly undeciphered as of this writing.

exceed plausible execution noise. We here focus on virtual balancing tasks.

It was argued before that the CNS may add parametric (i.e. multiplicative) noise to prolong escape times in inherently unstable balancing scenarios [CM02]. This model, however, has several shortcomings, especially with respect to reproducing the fluctuations during virtual balancing with a low task difficulty. It is also inconsistent with the more mainstream literature on motor control in other experimental paradigms (see sec. 7.1 for a summary and chap. 6 for more details).

In chapter 7, a model will be introduced where parametric noise arises from the rapid online estimation of parameters which are used in the controller's internal representation of the task. This parsimonious model can–in contrast to previous ones–reproduce the experimental findings in much detail. The models makes concrete predictions, which will be tested in chapter 8. Furthermore, the results demonstrate for the first time that balancing behaviour in the same task changes qualitatively after training depending on how subjects are rewarded.

The findings imply that the CNS uses a highly adaptive representation of the control problem and efficiently extracts local trends from observations only as needed. This is consistent with much of the literature outlined above, but also adds substantial new insights.

### Speculative Markets

Markets play an important role in the life of most humans. Well-functioning markets facilitate trade and enable the decentralised, self-organised distribution and allocation of resources in a society. In mainstream economics and many other economic theories, markets are characterised as structures that transform information about the traded goods into prices. The latter serve as signals for the coordination of the economy.

Financial markets are particularly well suited for scientific study for several reasons, including many participants and a relatively high degree of transparency. This is also reflected in the great amount of

available–though not always free–data, which facilitates the comparison of theory and reality. The difficulty to perform experiments in financial markets notwithstanding, the amount and quality of empirical data on financial markets has attracted many researchers also from outside of economics, including the natural sciences.[2] An introduction to the field is found in chapter 10.

Financial markets are often described as "informationally efficient" such that predictable price changes are eliminated by traders exploiting them, leaving only residual unpredictable fluctuations. This classical view of markets operating close to an equilibrium is challenged by extreme price fluctuations which occur far more frequently than can be accounted for by external news. Criticism has been raised from several directions, which will be discussed in more detail in chapter 10. It was supposed, for example, that excessive price fluctuations emerge because markets are inefficient and irrational, which is often blamed on market psychology. Proponents of market efficiency, however, stress its theoretical virtues and empirical evidence for some of its aspects or implications. So far, no consensus has been reached. A major problem in this discussion is that efficiency cannot be directly tested empirically, but only in joint hypotheses.

Physicists have been particularly interested in the statistics of price changes (log returns), which exhibit scaling relations resembling those for critical phenomena (see above). These so-called "stylised facts" are remarkably stable over time and across very different markets (sec. 10.3). How such large-scale collective dynamics can emerge from the interactions within a market is commonly studied using multi-agent models. It was argued, that the "stylised facts" emerge from the complexity of many interacting elements, possibly at a phase transition. We will discuss several prominent multi-agent models, all of which exhibit specific market failures that give rise to large fluctuations, in section 10.7.

---

[2] Note, however, that bidirectional exchange between economics and the natural sciences has existed for centuries. Some examples given in [FSS05, TML08].

In the following chapters, it will be shown that information efficiency itself can be a force that drives markets towards states of extreme susceptibility. We will thereby disclose rigorous links between the "stylised facts" and market efficiency with respect to self-generated information. We will further clarify several aspects of different means of adaptation and information generation in multi-agent models.

In chapter 11, a parsimonious trading model will be introduced where collective information efficiency emerges via self-organisation. The perfect balance of trading strategies, however, becomes extremely susceptible to perturbations. The model quantitatively reproduces the aforementioned "stylised facts" of real log returns.

In chapter 12, we will investigate group experiments with real subjects. The "seesaw game" will be introduced, which maps many results of the preceding chapters in a mathematically precise way to a particularly simple and very illustrative game. Furthermore, the elimination of predictable short-term price changes is shown to induce bubbles. The experimental results can be captured in a simple and analytically tractable model. If one allows for a nonlinear pricing rule that matches demand and supply, the prices become highly volatile during bubbles.

In chapter 13, an extended seesaw game is introduced which exhibits realistic price dynamics even for moderately sized groups. It can, as of this writing, be played online at seesaw.neuro.uni-bremen.de. Thanks to coverage in newspapers, on the radio, and online, results were obtained for a large number of subjects.

Taken together, the findings show how the consequences of information efficiency can change completely when the self-interaction of the market is taken into account.

## Abbreviations

| | |
|---|---|
| CCAPM | Consumption Capital Asset Pricing Model |
| CCDF | Complementary Cumulative Distribution Function |
| CNS | Central Nervous System |
| EMH | Efficient Market Hypothesis |
| FX | Foreign Exchange |
| GARCH | Generalised Autoregressive Conditional Heteroskedasticity |
| GCMG | Grand Canonical Minority Game |
| IID | Independent and Identically Distributed |
| IAI | Information Annihilation Instability |
| KS | Kolmogorov-Smirnov |
| MG | Minority Game |
| MSE | Mean Squared Error |
| OFC | Optimal Feedback Control |
| OMG | Original Minority Game |
| OOI | On-Off Intermittency |
| PDF | Probability Density Function |
| PSD | Power Spectral Density |
| SOC | Self-Organised Criticality |
| VSB | Virtual Stick Balancing |

# Contents

# Part I.

# Introduction to noise and scaling in complex and critical systems

# 1. What is complexity?

> “If you try and take a cat apart to see how it works,
> the first thing you have on your hands
> is a nonworking cat.”
>
> Douglas Adams

Many of the greatest breakthroughs in science were made possible by identifying phenomena on different scales which have little influence on each other and, therefore, can be treated independently. This is especially apparent in systems that can be completely understood by dividing them into smaller parts and understanding of each part. The kinetic theory of gases, for example, reduces the absolute temperature of a gas to the average kinetic energy of the molecules of which the gas is composed. The basic interaction of two adjacent water molecules is the same whether the molecules are in an ocean or a bucket.

There are, however, systems whose properties emerge from the interaction of their parts, often over many different scales, in a way which cannot be understood from the properties of the isolated parts alone. A wide range of systems from diverse disciplines that fit under this definition–or similar ones–are called “complex systems”. A system whose dynamical behavior changes depending on external influences is often called called a “complex adaptive system”. These systems “compute in the broadest sense by transforming information received from the environment into actions on it” [Sch01]. The most prominent example is the brain. Complex systems often exhibit metastable states where a small change in conditions can cause a major change in the systems dynamics.

One of the difficulties in understanding complex systems lies in the interaction between processes happening on many different levels. For example, a DNA molecule is a carrier of meaningful biological information only in the environment of a cell where the genes are expressed. The differentiation of a cell into a specialised type, defined by its pattern of gene expression, depends on the cell's environment within an organism. The whole organism, however, consists of cells. Therefore, the properties of the parts of this system are not independent of the whole. Similarly, a species' success depends on an ecosystem of other species and further factors that may be affected by them. Individual behaviour in socioeconomic systems depends on the collective behavior of others, for example on states and markets. Complex feedback loops between different scales are also found in the climate system.

A well-understood complex phenomenon in physics, which has been applied also to systems in many other fields, is criticality. A critical state is characterised by a cascade of correlations across all scales in the system. It is extremely susceptible to disturbances: even small perturbations can have large consequences throughout the whole system. Critical systems typically exhibit self-similar structures. Self-organised criticality is considered to be one of the mechanisms for complexity.

A striking property of complex and critical systems is large-scale collective behaviors. They may provide insights into the underlying structures that may be difficult to observe directly. In the following, we will discuss the statistics of extreme events, which often defy the more commonly known properties of Gaussian distributions. Since the following chapters are concerned with the analysis of time series, measures of temporal correlations are also introduced. Then, critical phenomena are explained briefly, as well as intermittent systems that exhibit complex dynamics from low-dimensional mechanisms. A comprehensive introduction to these topics can be found in [Sor04].

# 2. Extreme events

What is the probability that someone is twice your height? Unless you are a small child or exhibit an extreme case of dwarfism, that probability is essentially zero. In contrast, there is a good chance that someone is twice, or even a hundred times richer than you unless you happen to be among the few richest people in the world. In the following we will introduce quantitative measures of such extreme distributions.

## 2.1. General measures of distribution shapes

The shape of any distribution, for example the probability distribution of a random variable $x$, can be characterised by its moments $\mathrm{E}(x^n), n = 1, 2, \ldots$, or by the following closely related named measures. The mean is the (raw) first moment (i.e. $n = 1$). It is the most well-known measure of central tendency. The variance $\mathrm{Var}(x) = \mathrm{E}\big((x - \mathrm{E}(x))^2\big)$ is the central second moment and measures deviations from the mean. A Gaussian distribution is completely characterised by its mean and its width, quantified by its standard deviation $\mathrm{Std}(x) = \sqrt{\mathrm{Var}(x)}$.

Higher moments are usually standardised, yielding dimensionless measures which are invariant under any linear change of scale. The third standardised moment is the skewness, which measures asymmetry. The kurtosis is the fourth standardised moment

$$\mathrm{Kurt}(x) = \frac{\mathrm{E}\left(\left(x - \mathrm{E}(x)\right)^4\right)}{\mathrm{E}\left(\left(x - \mathrm{E}(x)\right)^2\right)^2}. \tag{2.1}$$

Sometimes, the same symbol is used for the excess kurtosis, which equals to $\mathrm{Kurt}(x) - 3$ in the above notation. The kurtosis measures the mass of the distribution tail, in other words, how much influence rare, large events have on the variance of distribution. It is therefore going to be an important measure in the following chapters. The kurtosis of a Gaussian distribution is three. In other words, a kurtosis above three, that is, a positive excess kurtosis, characterises distributions that are more heavy-tailed (leptokurtic) than a Gaussian one. The exponential distribution, for example, has a kurtosis of nine. The lowest kurtosis, one, is that of the Bernoulli distribution with $p = 1/2$. The latter describes the distribution of outcomes when flipping a fair coin.

For some distributions, in particular certain power laws (see below), higher moments may diverge. Even if a moment is finite, it is often very sensitive to extreme events. Therefore, elementary statistic courses often advise to discard outliers. This leads to highly deceptive results if the true underlying distribution is heavy-tailed.

A more robust measure of central tendency than the mean is the median $m$ where $P(x \leq m) \geq 1/2$ and $P(x \geq m) \leq 1/2$. In other words, the median separates the higher half of a sample or probability distribution from the lower half. For a Gaussian distribution, mean and median are identical. The concept can be generalised to quantiles. For example, a sample can be divided into four quartiles.

## 2.2. The central limit theorem

The central limit theorem states that the sum of $N$ Independent and Identically Distributed (IID) random variates (numbers), normalised by $1/\sqrt{N}$, with zero mean and finite variance $\sigma^2$, converges to a Gaussian distribution with variance $\sigma^2$. The mean has to be zero for the limit to exist and the sum to not diverge. The normalisation by $1/\sqrt{N}$ ensures that the characteristic scale of the distribution is constant. For finite, but large $N$, the central limit theorem is well approximated and the scale of the distribution will be finite even without normalisation. The theorem also holds for weak correlations and for random variates from different distributions with similar variances [Sor04]. Accordingly, the sum of many random forces acting on a particle or the collective activity of many independent elements should generally converge towards a Gaussian distribution.

There are, however, some exceptions and complications to this result. First of all, the Lévy distributions (see also sec. 2.3) , which have an infinite variance, are stable under summation. Second, even processes with a finite variance but long-range correlations may converge arbitrarily slow. This includes higher order correlations, for example due to a variance that slowly varies over a wide range of scales. Hence, understanding why any real system deviates from the central limit theorem requires an explanation why the underlying factors exhibit such a remarkable behaviour. See section B.3 for an example of slow convergence involving a market model.

## 2.3. Power-law distributed events

In the following chapters, Probability Density Functions (PDFs) with power-law tails $p(y) \propto |y|^{-\delta}$ for large event magnitudes $|y|$ play a significant role. Note that $\delta > 1$ and that the power law behaviour can only be valid above a threshold or asymptotically for large $|y|$. Otherwise the PDF could not be normalised. Figure 2.1 shows a comparison of a time series generated from Gaussian distributed random variates and one from power-law-tailed random variates. Both processes have the same variance. The Gaussian distribution is very localised. After a few dozen samples, the range of likely fluctuations is clearly visible. The power-law-tailed process, however, exhibits extreme outliers that effectively never occur for a Gaussian. Since these events are rare, the bins in the tail of the empirical probability density are undersampled. This problem can be mitigated by using logarithmically spaced bins, but then the largest bin becomes extremely wide. There is, however, a way to show all data points and the full distribution at the same time.

Figure 2.2 shows the Complementary Cumulative Distribution Functions (CCDFs) in double logarithmic coordinates for the same

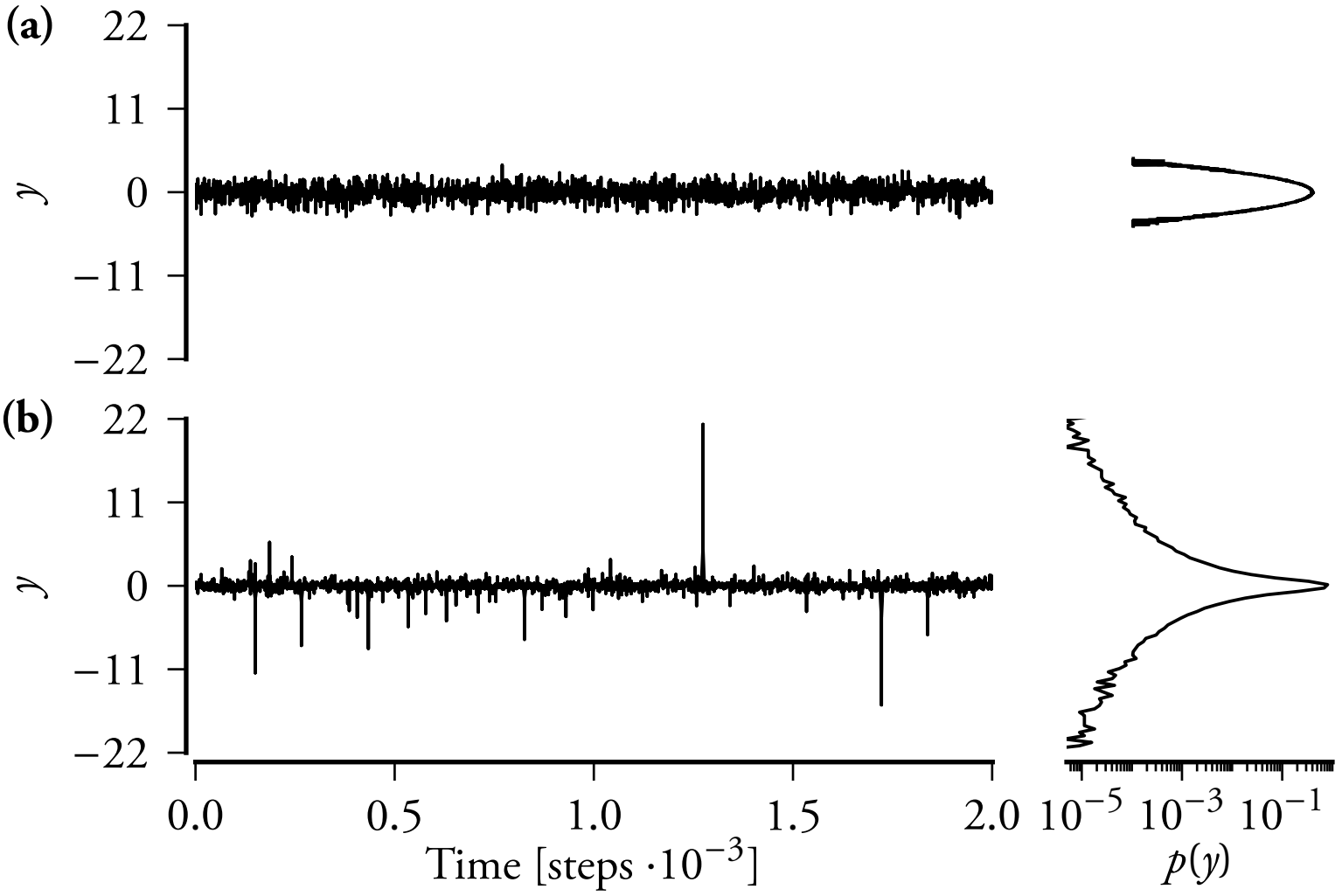

Figure 2.1.: Comparison of two white noise processes (left column) with unit variance but differently shaped distributions. The empirical probability densities (right column) were calculated from $10^6$ time steps and binned linearly. **(a)**: Gaussian. **(b)**: power-law-tailed with PDF tail exponent $\delta = 3$, that is, CCDF tail exponent $\xi = 2$. More precisely, the latter distribution is equivalent to a Gaussian below a threshold and then follows a power law. At the threshold, it has a continuous derivative. Independent random variates were generated using inverse transform sampling.

processes also shown in figure 2.1. The value of the CCDF (i.e. the position on the y-axis) for the magnitude of each event (i.e. the position on the x-axis) gives the probability to observe an even larger one. Since this probability is equal to the area under the PDF for all larger event magnitudes, a power law in the PDF with exponent $\delta$ is equivalent to a power law in the CCDF,

$$P(|y| > x) \propto |y|^{-\xi}, \tag{2.2}$$

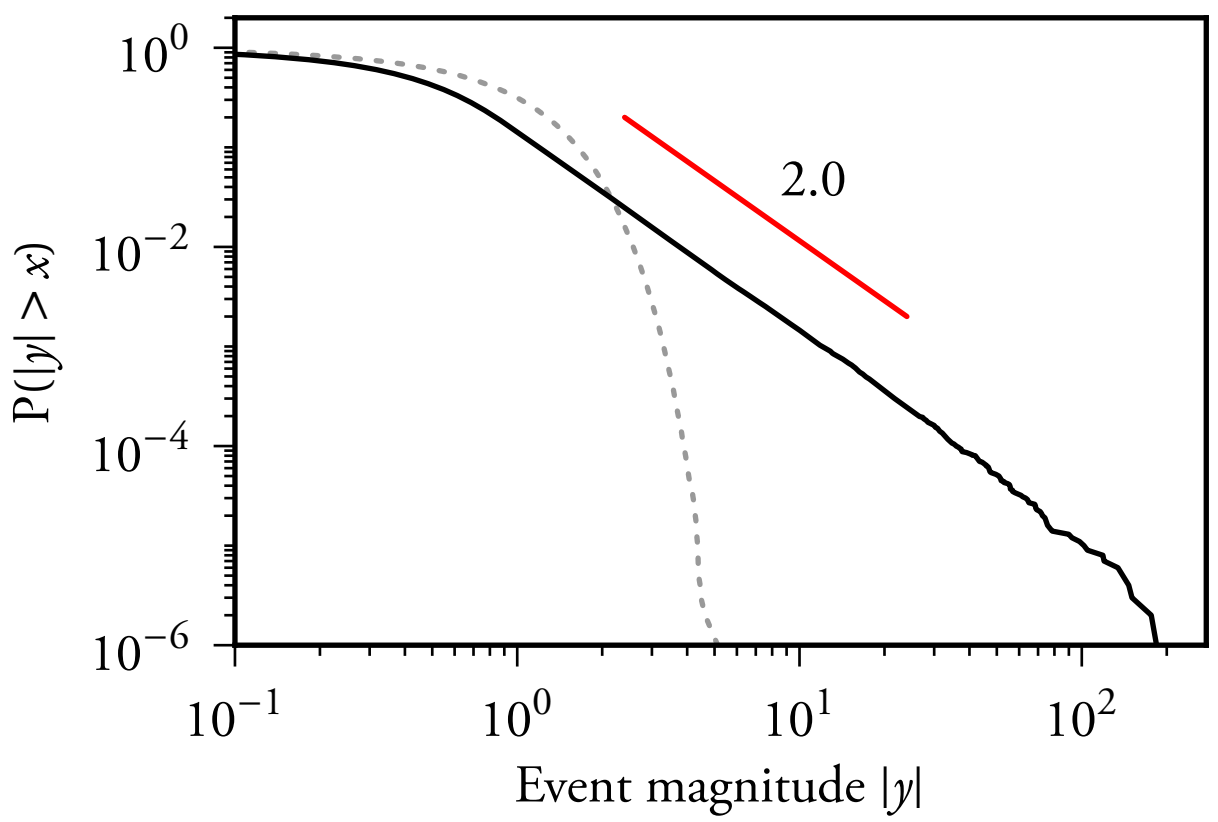

Figure 2.2.: CCDFs for the same processes shown in fig. 2.1. Grey dotted line: Gaussian. Solid black line: power-law-tailed. Short red line: the tail-exponent of the analytical distribution from which the random variates were drawn.

with exponent $\xi = \delta - 1$.[1] The CCDF for a given sample is calculated by sorting all observed events by their magnitudes in descending order. The values of the CCDF then follow by dividing each rank by the sample size. The CCDF, therefore, shows all events. The random deviations from the underlying true distribution increase in the end of the tail. When the sample size is increased, the tail for a true power law extends towards increasingly rare and large events. This effect highlights a particular property of power-law-distributed events: if one waits long enough, there will always come an event which surpasses all previous ones by a significant margin.

[1] The Complementary Cumulative Distribution Function (CCDF) is most commonly used instead of the cumulative distribution function to depict power law distributions. A likely reason is that the CCDF visually resembles the PDF in the sense that large events are shown on the right-hand side of the x-axis. Note that also the empirical CCDF is commonly called just "CCDF". We here follow this convention unless the difference in between the analytical ideal and the empirical realisation is to be explicitely emphasised. Note, furthermore, that we will not always distinguish between the notation for random process and a particular realisation of that process, as long as the meaning is clear from the context.

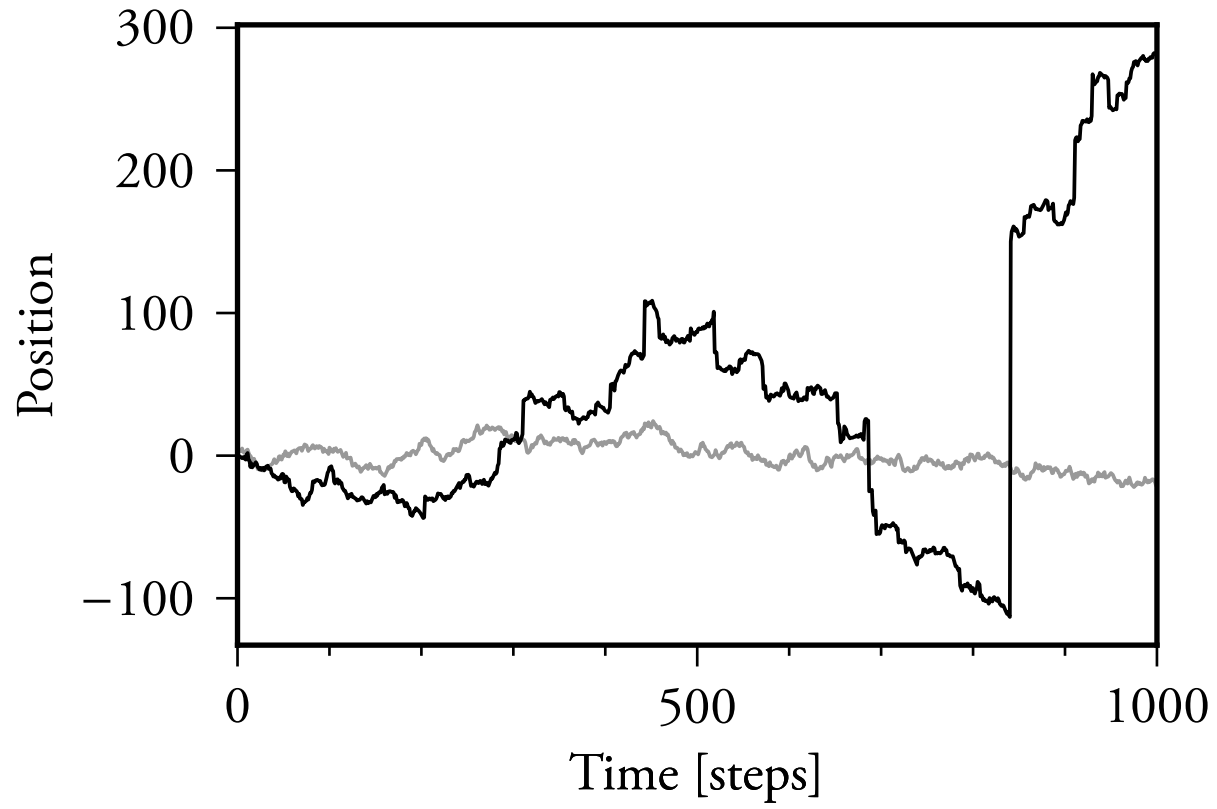

Figure 2.3.: Random walks with differently distributed increments, generated by forming the cumulative sum over similar processes as shown in fig. 2.1. Grey line: Gaussian distributed increments. Black line: a so-called Lévy flight with power law distributed increments with CCDF tail exponent $\xi = 1.5$. Because this distribution has a vanishing variance, the increments for both walks were normalised to unit median magnitude.

The $n$-th moment of a power-law distribution only exist if $n < \xi$ and diverges otherwise. As mentioned above, Lévy distributions are stable under summation and thereby beat the central limit theorem. Their cumulative distributions exhibit tail exponents $0 < \xi < 2$.

IID random variates are commonly used to represent the increments of random walks or diffusion processes. Figure 2.3 shows two random walks with the same typical increment size, but very different distribution tails. The walk with Gaussian increments is closely related to the Wiener process.[2] The latter is well known, for example, as a model for the Brownian motion of small particles in a fluid, resulting from their random collisions with atoms or molecules. Clearly, the changes in position from one time step to the next are all of the same scale. In contrast, the walk with power-law-distributed increments becomes dominated by rare, extreme jumps.

[2] More precisely, it is used to numerically simulate the Wiener process, which is the continuous-time limit of the random walk.

## 2.4. Self-similarity and power laws

The laws of physics do not change if one walks from one side of a room to the other. The invariance of physical laws with respect to symmetry transformations are very important in physics, since they reflect the fundamental structure of nature. In the following, we briefly discuss the symmetry of scale invariance.

The most well-known form of scale invariance is presumably the self-similarity of fractals, where geometric features are repeated on each scale. A simple fractal, the Koch curve, is shown in figure 2.4. There are many more small structures than large ones: at the $n$-th iteration, the curve consists of $4^n$ segments of size $1/3^n$. After many iterations, the curve looks exactly the same at any scale of observation. Physical objects are only self-similar over a finite range of scales, and often only approximately. Many spatially extended objects, including the distribution of galaxies, geological structures, or blood vessels appear to be self-similar fractal structures. The most strikingly self-similar vegetable is the Romanesco. A specimen is shown in figure 2.5.

Mathematically, a scale invariant function (curve, observable) $f(x)$ at two different scales depends only on the ratio of the two scales:

$$\frac{f(\lambda x)}{f(x)} = \lambda^\alpha \Leftrightarrow \tag{2.3}$$

$$f(\lambda x) = \lambda^\alpha f(x) \tag{2.4}$$

for some exponent $\alpha$ and for all $\lambda$. Therefore, f is a homogeneous function of degree $\alpha$. A solution to (2.4) is a power law

$$f(x) \propto x^\alpha, \tag{2.5}$$

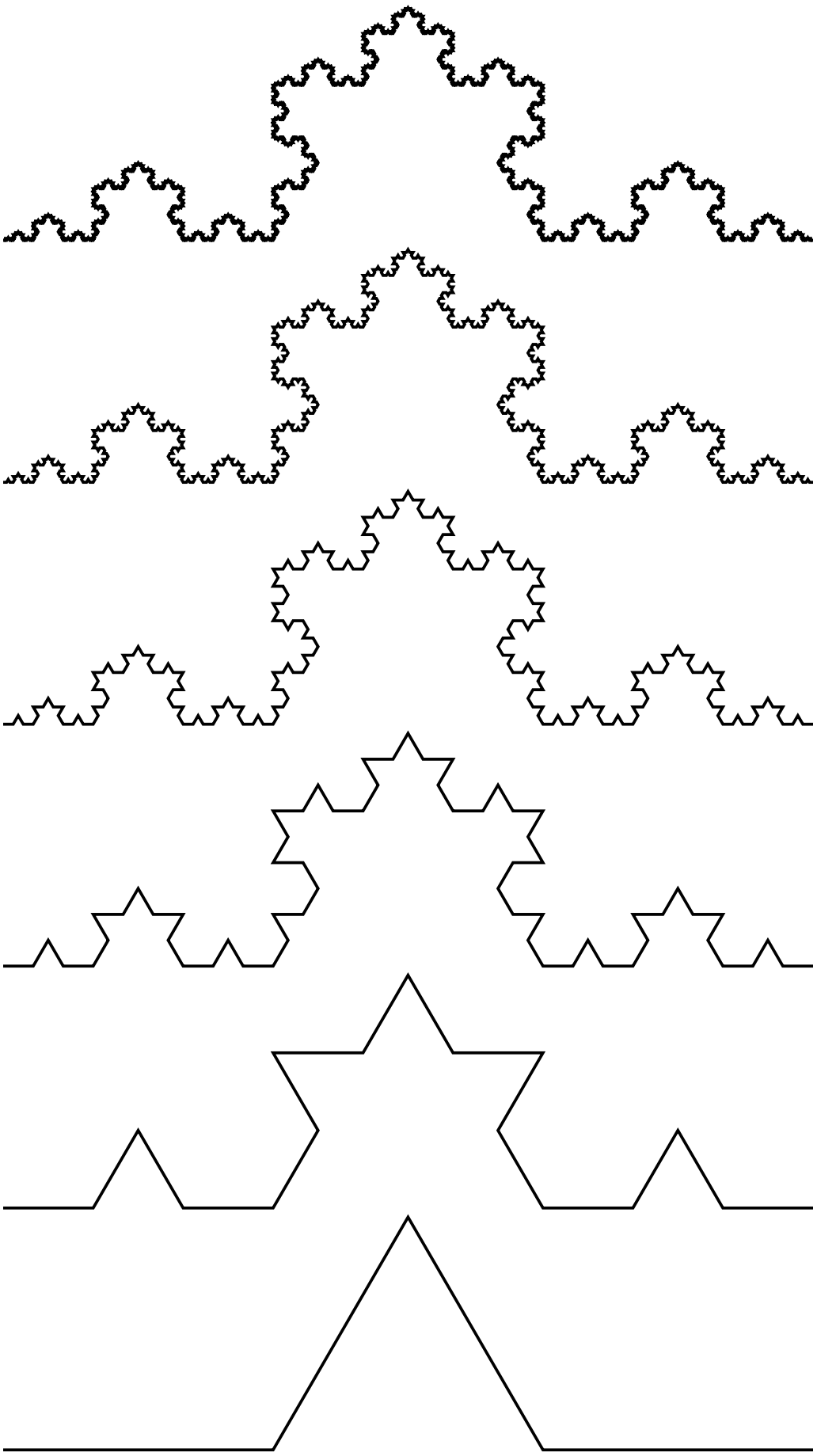

Figure 2.4.: The famous Koch curve is constructed by starting with the unit interval as the initial segment. In each subsequent iteration, each segment is dividing into three smaller segments of equal length. The central one is replaced by an equilateral triangle. The base of the triangle is removed. These three steps are repeated for an infinite number of iterations, forming a self-similar fractal. The first six iterations are are shown above.

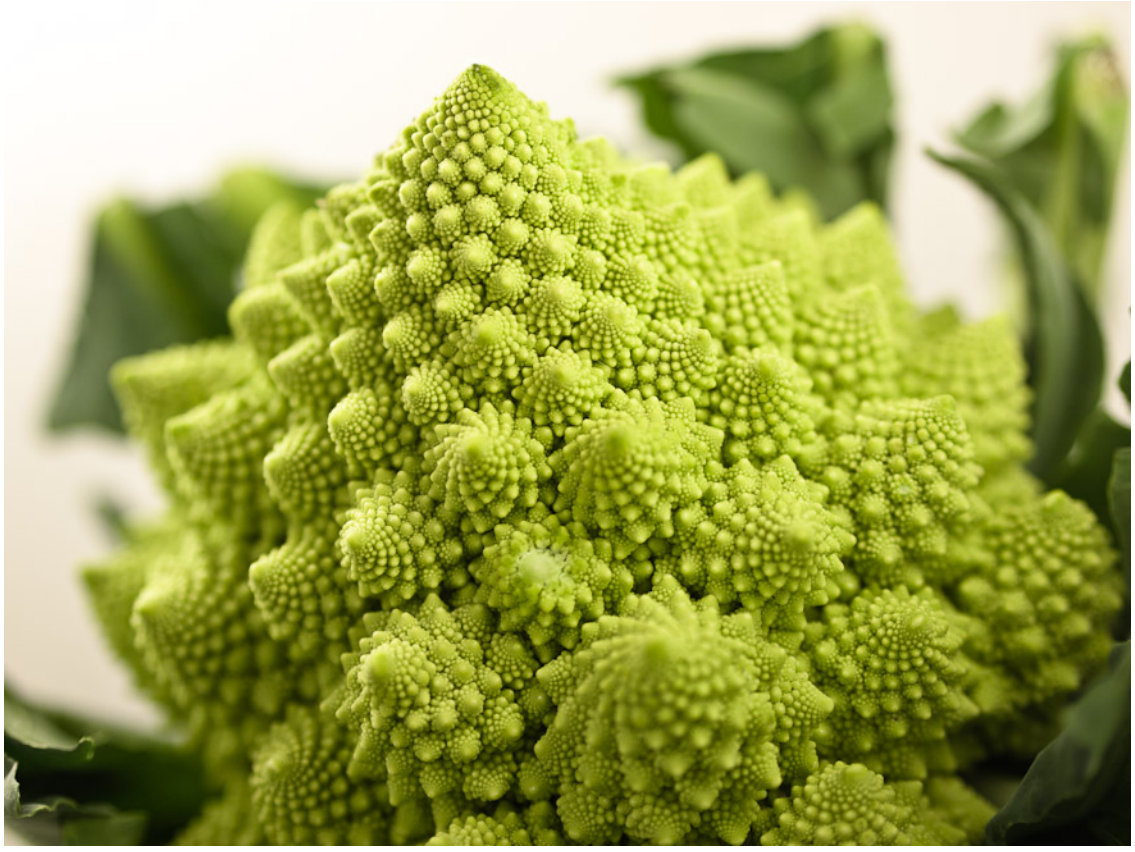

Figure 2.5.: A Romanesco.

which can be verified by insertion.

Power-law distributed events exhibit the same statistics at all scales of observation above the lower limit (cutoff) of the scaling regime, as shown in figure 2.6. In practise, the largest scale will be determined either by the length of the time series or by a second upper cutoff to the power law due to, for example, the finite size of the system. Particularly for steeper power-laws, extremely large sample sizes are required to obtain events over several orders of magnitude. Then, the limiting factor is most likely the sample size.[3]

[3] This note clarifies the notion of self-similarity in random processes, and what is considered an (extreme) event in the following chapters.

A Wiener process, that is, the integral of a continuous-time Gaussian white noise process, is self-similar. Its increments, however, are Gaussian distributed. The Wiener process, therefore, only exhibits velocities within a narrow range. Nevertheless, the process may, after a sufficient amount of time, wander arbitrary far from where it started. If there is a mean reverting drift towards the origin, however, the process is not self similar on longer time scales.

Control errors, for example deviations of a balanced stick from the upright position, are quickly mean reverting. Otherwise, the system would not be stable. As shown in sec. 6.2 ff., however, control errors during human balancing tasks are power-law distributed and, therefore, self-similar from a few millimeters up to the possible movement range in the task. In practise, the upper limit for the observed range in all experiments was the sample size. We consider these control errors relevant events since large errors bear the risk that the subject loses control.

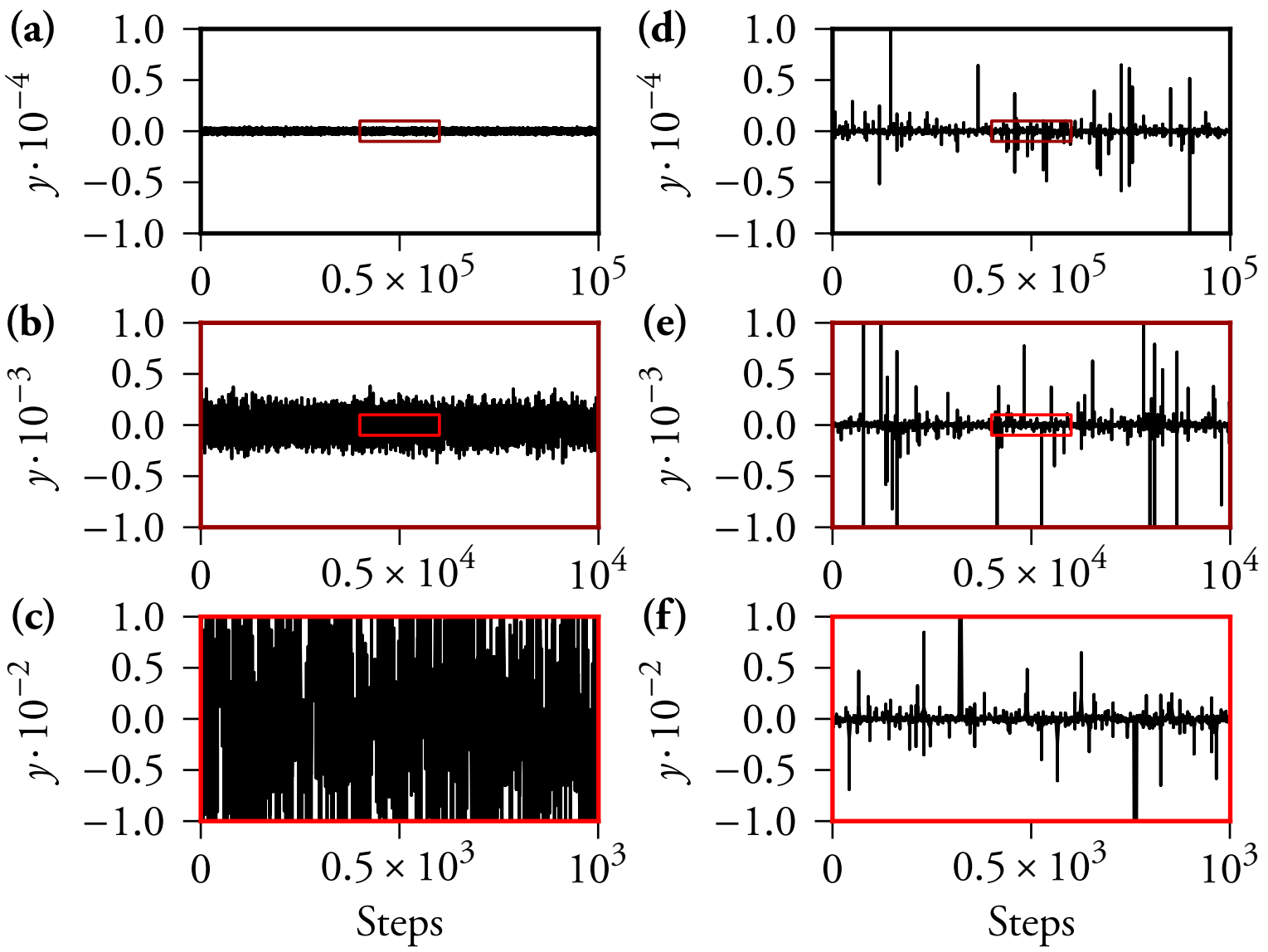

Figure 2.6.: Two time series at different magnifications. **(a, b, c)**: Gaussian white noise. **(d, e, f)**: control errors in the minimal balancing model described in sec. 7.3, which exhibits a power-law CCDF-tail with exponent $\xi = 1$ in the analytical limit for large fluctuations.

As another example, logarithmic price movements in financial markets are often modelled as a Wiener process, but this severely underestimates the probability of large price changes. As shown in sec. 10.3, the increments of price changes actually follow a power law. A large price change may, for example, be a market crash–an extreme event. There is possibly no hard upper limit for price changes, but exchanges may suspend trading on major events.

## 2.5. Fitting power-law distributed events

Estimating parameters of distributions of rare, extreme events generally requires large data sets and great care. Binning creates several problems which can be avoided by using a maximum likelihood estimator [Soro4]. Power-law tail exponents can be obtained from the Hill estimator after rank-ordering the event magnitudes, which is also done when calculating the CCDF (see above). More precisely, the magnitudes $x = |y|$ of the events $y$ are sorted in descending order, such that

$$x_1 \leq x_2 \leq x_3 \leq \ldots. \tag{2.6}$$

The self-similarity of fluctuations with a power law CCDF with exponent $\xi$ is expressed in the ratio of the probabilities for observing two event magnitudes $x_i$ and $x_j$,

$$\frac{p(x_i)}{p(x_j)} = \left(\frac{x_i}{x_j}\right)^{-\xi}, \tag{2.7}$$

which is also a power law.

The Hill estimator uses the first $r$ ranks, which correspond to the largest events up to a cutoff. Smaller events, which may belong to a different regime, are discarded. The maximum likelihood estimator for $\xi$, based on the ratio of the $x_i$, where $i = 1, 2, \ldots, r$, and the event magnitude at the cutoff $x_r$, is [Soro4]

$$\xi_r = \left(\frac{1}{r}\sum_{i=1}^{r} \ln\left(\frac{x_i}{x_r}\right)\right)^{-1}. \tag{2.8}$$

The Hill estimator is sensitive to the cutoff, the optimal value of which is generally unknown. It furthermore changes for different data sets or when changing the parameters of a model. To improve the robustness of the estimator while minimising manual interaction, the following method was developed. The exponents $\xi_r$ for 100 different logarithmically spaced cutoff ranks $r$ are calculated. For each cutoff, the exponent is calculated using the Hill estimator equation (2.8). For each of these power law fits, the Kolmogorov-Smirnov (KS)-statistic [MTW03, NIS14] follows as

$$D_r = \max_i \left( F_i - \frac{i-1}{r}, \frac{i}{r} - F_i \right), \quad \text{with} \tag{2.9}$$

$$F_i = \left( \frac{x_i}{x_r} \right)^{-\xi_r}. \tag{2.10}$$

It is calculated as if only the fitted tail would constitute the whole distribution. This particular normalisation ensures that the expected $D_r$ decreases with increasing $r$ unless the distribution deviates from the power law. The optimal fit is

$$\xi_{\text{opt}} = \xi_{r_{\text{opt}}}, \quad \text{with} \tag{2.11}$$

$$r_{\text{opt}} = \operatorname*{argmin}_r \; D_r. \tag{2.12}$$

The range of cutoffs is chosen to be as wide as possible while avoiding pathological cases which may occur in some particular situations. Alternative methods to determine the cutoff were tested, but the above one proved to be the most robust. In the following, we generally call the best fit for the exponent $\xi$, dropping the index.

Since the Hill estimator tests the tail of the distribution, it is a highly appropriate method to quantify the asymptotic behaviour for large events. Methods that make more assumptions about the full underlying distribution may lead to deceptive results. For example, a method based on the assumption that the distribution is within the Lévy regime always yields fitted exponents $\xi < 2$. Such a method based on the return probability to the origin was found to yield erroneous fits for the tails of logarithmic price returns [GPA+99, GPL+00, Lux06]. Regardless of this problem, the same method was used for stick balancing in several studies (see sec. 6.1).

Testing whether the underlying distributions in balancing time series truly follow power laws is discussed in sec. A.1.

# 3. Temporal correlations

Events in a time series are characterised not only their distribution, but also by their temporal structure, which manifests as correlations. Two measures of temporal correlations will be introduced below: the power spectral density and the autocorrelation. The prior is more common in motor control and useful to analyse the balancing tasks in part II, while the autocorrelation is more common in quantitative economics and therefore used in part III. Other measures are discussed briefly in sec. A.5.

The Power Spectral Densities (PSDs) shown in the following were estimated using Welch's method with a Hanning window. This method sacrifices some frequency resolution to reduce random variability. In addition, because the following PSD-figures will all use double logarithmic axes, the periodograms were binned logarithmically.[1] Therefore, the periodograms exhibit less random variability at higher frequencies.

Figure 3.1 shows a comparison of different types of noise. White noise contains all frequencies with equal power. This is equivalent to a signal that is completely uncorrelated in time. All processes shown in figures 2.1 and 2.6 are types of white noise. Noise with temporal correlations, in particular with a PSD that decays with increasing frequencies as $S(f) = f^{-\lambda}$, is called coloured noise.

A Wiener process is defined as the integral of white noise. This introduces a factor $1/f$ in the Fourier transform, which becomes $1/f^2$

[1] Low frequencies with at most one data point per bin were not binned. The algorithm was tested with synthetic data and did not introduce significant biases. For some uses other than plotting, more aggressive binning was required, which is described in sec. 8.7.1.

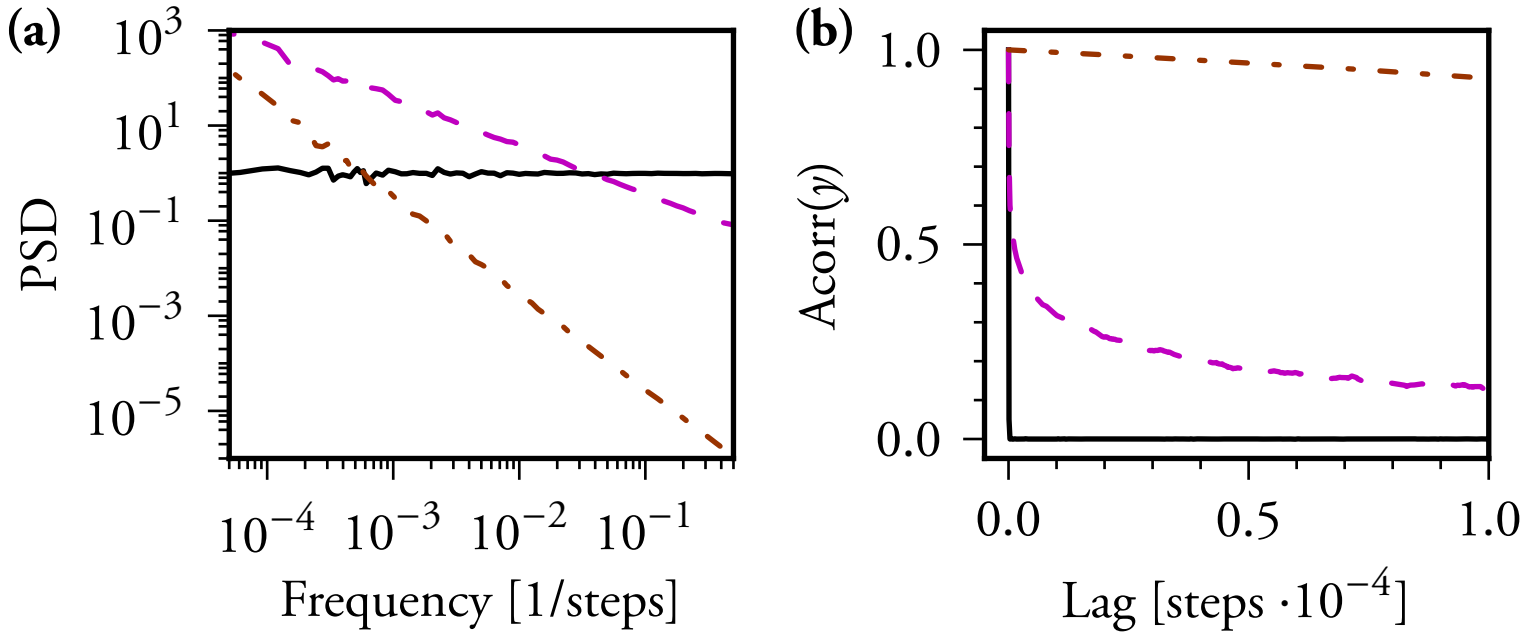

Figure 3.1.: Comparison of different types of noise. Solid black lines: white noise. Dashed magenta lines: Pink noise, also called flicker- or $1/f$ noise. Dash-dotted brown lines: Brown noise, also called Brownian- or red noise. Colored noise was generated using the method from [TK95]. All random variates were Gaussian distributed, but the results are completely independent of the respective distributions and only reflect the temporal structure. **(a)** Power Spectral Density (PSD). **(b)**: Autocorrelation.

in the power spectrum: the PSD decays with $\lambda = 2$ (fig. 3.1 (a)). In other words, the power density per frequency falls 20dB per decade or 6dB per octave. This so-called "brown noise" is observed for Brownian motion or other random walks with independent increments, for example those shown in figure 2.3. Because the state (position) of an unconstrained walk at each time is the integral or sum over all previous states, the time series is extremely long-range correlated (fig. 3.1 (b)). Yet, the process is memoryless in the sense that its increments are independent: the Wiener process at each point in time depends only on its current state and its infinitesimal change. The discrete-time random walk at each time step $t$ depends only on the position at $t-1$ and on the corresponding increment.

Pink noise has a PSD which decays with $\lambda \approx 1$, that is, 10dB per decade or 3dB per octave (fig. 3.1 (a)). It is, therefore, often called $1/f$-noise. The total energy per octave is constant. The time series is also long-range correlated (fig. 3.1 (b)), but correlations decay faster than brown noise because the relative power of low frequencies is smaller. $1/f$-noise is observed in many systems, particularly at low frequencies–at high frequencies it may be masked by other noise sources. Examples include astronomic signals, electronic devices ("flicker noise"), the flow of rivers and oceans, traffic, communication, biological systems, and many more. $1/f$-noise can be seen as the half-integral of white noise, an operation that is non-local in time. Correspondingly, fractional derivatives may be used to describe dynamical systems with a long memory. There is, however, no universal explanation why $1/f$ noise appears to be so ubiquitous. See [Pre78, Sor04] and also sec. 4.2 for more in-depth discussions.

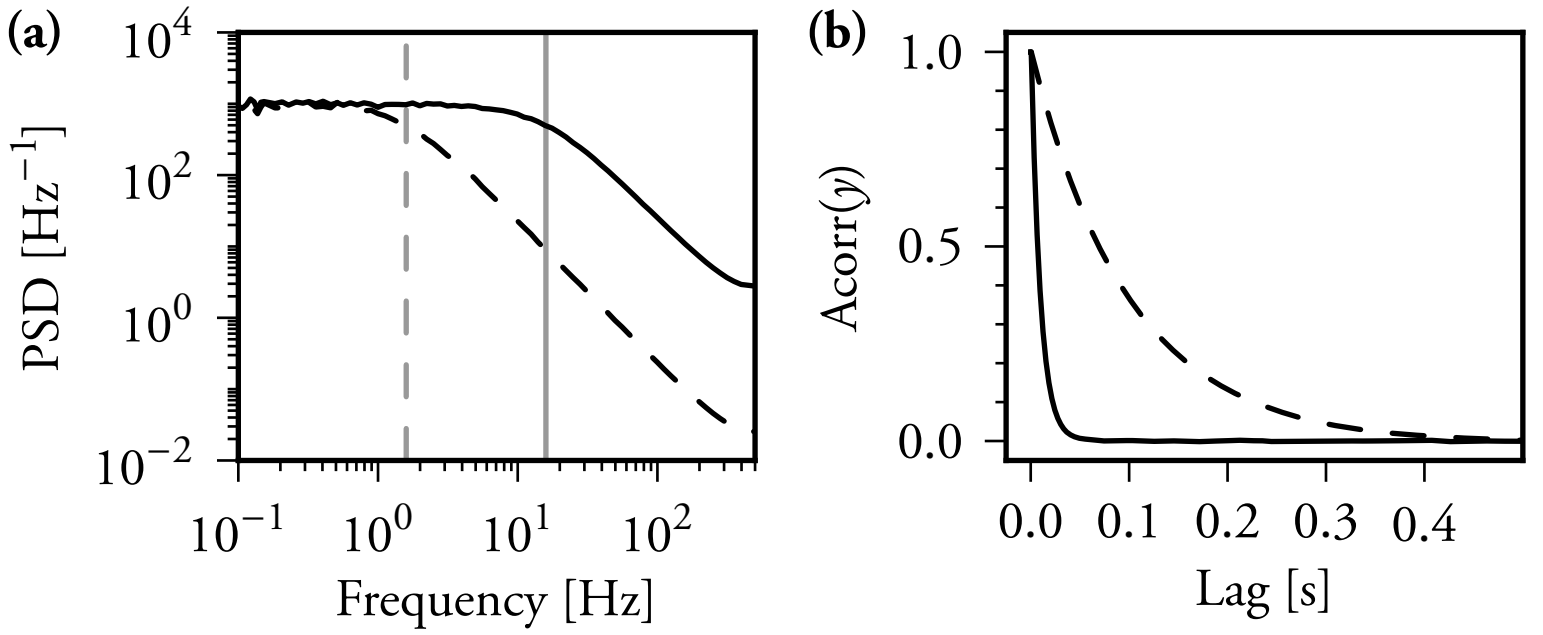

Figure 3.2.: Lowpass filtered continuous-time white noise, simulated with 1ms time steps. Solid lines: $\tau = 10\,\text{ms}$. Dashed lines: $\tau = 100\,\text{ms}$. **(a)** Black lines: Power Spectral Density (PSD). Grey lines: the cutoff frequencies. **(b)**: Autocorrelation.

Simple dynamical systems or random walks within bounds exhibit short-range correlations. Figure 3.2 shows results for a passive lowpass filter

$$\dot{y}(t) = -\frac{1}{\tau}\,y(t) + \beta(t) \tag{3.1}$$

with Gaussian white noise input $\beta(t)$. Fluctuations decay exponentially with time constant $\tau$. This behaviour is observed for a simple RC-circuit or for the velocities in a diffusive process with friction. The PSD is constant for low frequencies up to a cutoff $f_c = 1/(2\pi\tau)$. The system effectively integrates all higher frequencies: the PSD decays with $1/f^2$ like brown noise. The autocorrelation decays exponentially with time constant $\tau$, just like the fluctuations in $y$.

Higher-order filters have an even faster decaying high-frequency response with even integer exponents (not shown). For example, the position of a diffusing particle with friction is described by the integral of the above first-order system equation (3.1).

# 4. Random processes and collective phenomena

In the following, several fundamental mechanisms for collective behaviours and power laws, which are related to the mechanisms discussed in later chapters, will be introduced briefly. We begin with high-dimensional systems by introducing phase transitions and then Self-Organised Criticality. Then, we will discuss intermittency and multiplicative noise in simple, time-discrete maps.

## 4.1. Critical points in thermodynamic phase transitions

The laws of nature exhibit many symmetries with respect to transformations such as translation, rotation, and reflection. The actually realised states, however are typically less symmetric. The structure of a solid state (crystal), for example, is only invariant against discrete translations and special rotations. A thermodynamic system at a phase transition abruptly changes between a less ordered but more symmetric state, and more ordered and less symmetric one, depending on external influences. The spontaneous breaking of symmetries is an important principle for the description of, for example, the differentiation of the fundamental forces and particles in the universe [oS14] or for the onset of cell differentiation in an embryo [LB10]. It is at the critical point, however, where the compromise between order and disorder allows for complex behaviours and new symmetries like scale invariance to emerge.

There are different types of phase transitions, most of which can be classified as first or second order. Introductions to the field are found, for example, in [Sor04, Sch04].

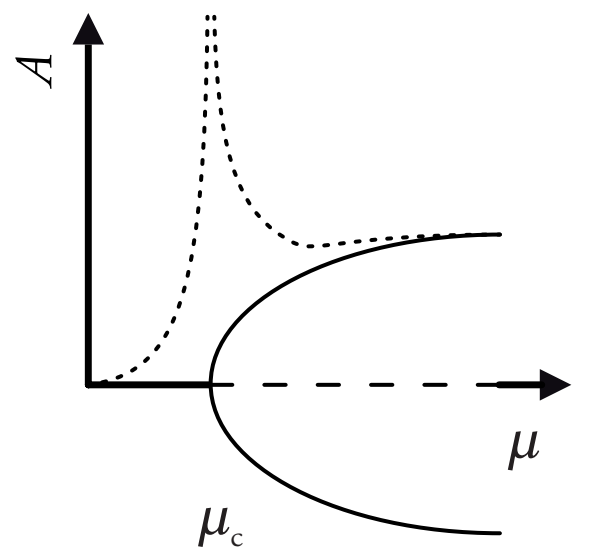

Figure 4.1.: Bifurcation Diagram of a supercritical bifurcation. At the critical value $\mu_c$ of the control parameter $\mu$, the equilibrium of the order parameter $A$ at $\mathrm{E}(A) = 0$ becomes unstable and new solutions emerge. Close to $\mu_c$, fluctuations in $A$ increase dramatically.

In mathematical terms, a sudden qualitative change in the behaviour of a system depending on a small, smooth parameter change is called a bifurcation. Close to such a transition, many complex and dynamical systems can be simplified in a way allowing to characterise important aspects of their behaviour by a single order parameter $A$. The behaviour of the system is determined by a control parameter $\mu$. At a critical value $\mu_c$, an equilibrium (fixed point), represented by $A$, becomes unstable.[1]

Figure 4.1 shows a supercritical bifurcation, which corresponds to a second order (also called "critical") phase transition. As an example for such a phase transition, consider a ferromagnet. The macroscopic magnetisation is the result of a competition between order and disorder. The alignment of spins in the material through local interactions lowers the system's energy, but thermal fluctuations destroy these correlations. The temperature $T$ takes the role of the control parameter and the macroscopic magnetisation that of the order parameter. At high temperatures, the system is paramagnetic, that is, globally disordered unless it is exposed to a strong external magnetic field. The systems susceptibility is small since alignment cannot propagate over longer scales. At low temperatures, the alignment of spins dominates and the system exhibits spontaneous macroscopic magnetisation without an external field. The rotational symmetry of the spins is broken: the system must be in one of several distinguishable macroscopic states

[1] There are more general formalisations which we will not discuss.

(e.g. magnetised in one of two possible directions in the Ising model). This order is robust against small perturbations. Exactly at the Curie temperature $T_C$, however, the whole system forms one fractal cluster [Con89]. Because of this self-similar structure, even small perturbations can propagate over all scales. Several quantities, including the susceptibility and the correlation length, diverge according to power laws $|T - T_C|^{-\gamma}$.[2] In practise, however, the largest possible fluctuations close to a critical point may be limited by saturation [Sor04].

Because a system close to criticality is dominated by collective behaviours, global properties such as dimensionality and symmetry are often more important than microscopic details. Therefore, very different systems may exhibit the same universal exponents $\gamma$.

Not all bifurcations are phase transitions and not all bifurcations or phase transitions are critical. A subcritical bifurcation, which corresponds to a first order phase transition, does not exhibit the divergences described above. Here, the order parameter discontinuously jumps to a finite value. When the system moves through the critical region, the old phase becomes metastable and the new one stable. Examples include the phase transitions between liquid and solid, as well as gas and liquid phases.

Today, many phenomena in fields as diverse as depression, epileptic seizures, opinion formation in groups, climate change, and ecosystems are described as bifurcations. Many critical systems do not exhibit the precise universal exponents $\gamma$ found in thermodynamics. There are, however, sometimes other "universal" features over broad classes of systems such as early warning signs close to phase transitions or other tipping points [SBB+09].

## 4.2. Self-Organised Criticality (SOC)

Following the above considerations, the widespread occurrence of persistent critical phenomena in nature – $1/f$-noise (sec. 3), fractal structures (sec. 2.4), and power-law distributed events (sec. 2.3) – still cannot be explained sufficiently. What is missing in the classical paradigm is at least a plausible explanation why the control parameters for so many systems should be tuned exactly to a critical point.

[2] The different quantities generally have different exponents.

The first possible explanation for the robust emergence of critical phenomena without fine tuning of parameters was given by SOC [BTW87, BP95]: certain classes of dynamical systems have a critical point as an attractor. SOC is typically observed in slowly driven (i.e. non-equilibrium) systems with extended degrees of freedom and non-linear interactions.

The canonical example is the formation of a sandpile. The system is externally driven by adding grains of sand one after another. The state with the lowest energy is a completely flat surface. When a grain is added, however, it will roll away from where it landed only until it reaches a locally stable position. The grains therefore pile up. If the slope of the surface is too large, the pile will collapse. Because the grains are added incrementally, however, the the average slope reaches a critical value where the pile is marginally stable: A single grain can start an avalanche of any size. The dynamical response of the sandpile to small random perturbations is characterised by $1/f$-noise and the avalanches are power-law-distributed.

For many SOC-models, the distribution of fluctuations can be mapped to the first return time of a random walk [Sor04]. These distributions are characterised by the scaling exponent $3/2$. This is a hint on simpler mechanisms that can generate critical behaviour. There are, however, also many very different models that have been considered SOC. There is no general understanding what precisely the conditions for SOC are, no unifying framework, and no universality in the strict sense found in thermal critical phase transitions [Sor04]. This limits the transfer of results for one system to another one, but not the applicability of specific mechanisms for SOC in any particular case. What is so far missing for the application in this work, however, is a connection to behaviour, adaptation, and information processing.

Figure 4.2.: Caricature of the construction of an intermittend bursting system in phase space. The system spends long times at small distances y from the the origin (laminar phase). During bursts large distances y and velocities v occur, but the trajectories are reinjected into the origins vicinity.

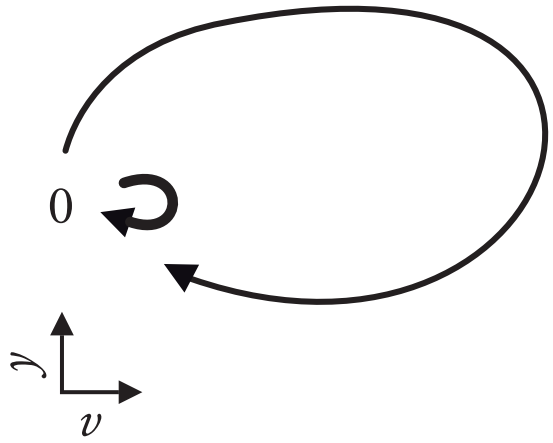

## 4.3. Intermittency

Systems that switch between qualitatively different kinds of oscillations in an appearently random way are called intermittent. Such systems can be constructed around (quasi-) invariant objects in a system's phase space near which the system will tend to spend long times. One example is an invariant unstable object combined with a reinjection mechanism driving the system back to the instability when it is far away from it. The mechanism is depicted in figure 4.2.

### 4.3.1. On-Off Intermittency (OOI)

A special type of intermittency arises from of the repeated forcing of a control parameter through a bifurcation point [PST93, HPH94]. As an example, consider a logistic map

$$y(k+1) = a\,x(k)\; y(k)\bigl(1 - y(k)\bigr) \tag{4.1}$$

with a random parameter $x(t)$ drawn independently at each time from the uniform distribution $\mathcal{U}(0,1)$. The map passes through a bifurcation at $y = 0$ when $ax = 1$. For small $ax$, the system is stable. When $a$ is above an intermittency threshold, the system randomly switches between quiet (laminar) and bursting phases, as shown in figure 4.3 for $a = 2.75$. OOI exhibits some universal properties, for example of the distribution of laminar phases (see also sec. A.10). It has been suggested as a mechanism in many systems with extreme events like stick-balancing and stock markets. However, as shown in figure 4.3 (b), OOI generally does not exhibit power-law distributed fluctuations because of the system's nonlinear saturation, which is necessary because the fixed point becomes unstable during bursting phases [Sor98].

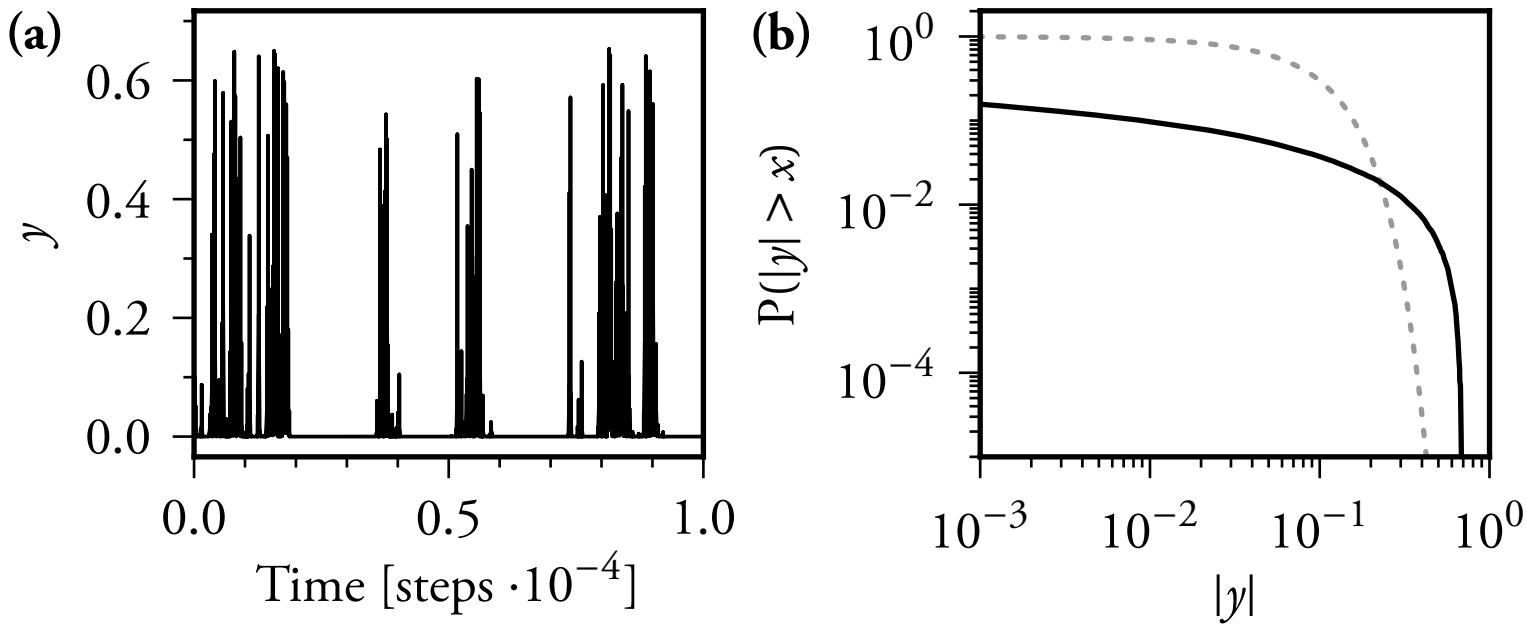

Figure 4.3.: On-Off Intermittency in a logistic map driven by multiplicative uniform noise drawn from $[0, 2.75]$.**(a)**: time series **(b)**: CCDF for the logistic map (black line) and a Gaussian distribution with the same variance (dashed grey line).

## 4.4. Multiplicative (parametric) noise

Some multiplicative noise processes different from the one discussed above can produce a special kind of intermittency with power-law-distributed fluctuations. These processes can be completely linear and produce critical behaviour under relatively mild conditions for the driving noise. Consider the random map

$$y(t+1) = \alpha(t)y(t) \tag{4.2}$$

where $\alpha$ is a stochastic variable with probability distribution function $P(\alpha)$. The logarithm of $y(t)$ is the sum over $t$ IID variables. As long as the central limit theorem is valid for $P(\alpha)$, this sum converges to a Gaussian distribution. This means that $y$ is distributed according to the log-normal distribution. To turn $P(y)$ into a power law, we need to make sure that (4.2) contracts, on average, and is repelled from the origin. While keeping the random map linear, this is fulfilled for a negative logarithmic growth rate $\mathrm{E}(\ln|\alpha|) < 0$ by introducing an additive noise term $\beta(t)$. The resulting process

$$y(t+1) = \alpha(t)y(t) + \beta(t) \tag{4.3}$$

is is called the Kesten process. The essential results can be summarised as:

- If $\alpha(t)$ and $\beta(t)$ are IID real-valued random variables and if $\mathrm{E}\big(\ln|\alpha(t)|\big) < 0$, then $y(t)$ converges in distribution and has a unique limiting distribution $P(y)$.
- If, additionally, $\beta(t)/\big(1-\alpha(t)\big)$ is nondegenerate and if there exists a $\mu > 0$ with
  1. $0 < \mathrm{E}\big(|\beta(t)|^{\xi}\big) < +\infty$,
  2. $\mathrm{E}\big(|\alpha(t)|^{\xi}\big) = 1$ and
  3. $\mathrm{E}\big(|\alpha(t)^{\xi} \ln^{+}|\alpha(t)|\big) < +\infty$

then the limiting distribution for $y(t)$ is for large $y(t)$ asymptotic to $P(y) \propto y^{-\delta}$ as $y \to \infty$ with $\delta = \xi + 1$ [Kes73, Sor04].

While the multiplicative term causes the process to contract, on average, fluctuations across the stability boundary $\ln|\alpha(t)| = 0$ cause intermittent amplification. Figure 4.4 shows the time series and distribution. While perfect tuning onto the stability boundary is not required to yield observable power laws in the distribution of $y$, the logarithmic growth rate still has to be adjusted to be negative and close to 0. The exponent $\xi$ is also not universal, but parameter dependent. The Kesten process can also be used as a model for long-range correlations [Sor04].

The Kesten process will serve as a conceptual starting point for control models with power law distributed fluctuations discussed in part II. The market models in part III involve state-dependent noise as well. The main purpose of these models, however, is to understand how such dynamics arise from fundamental principles in the specific contexts.

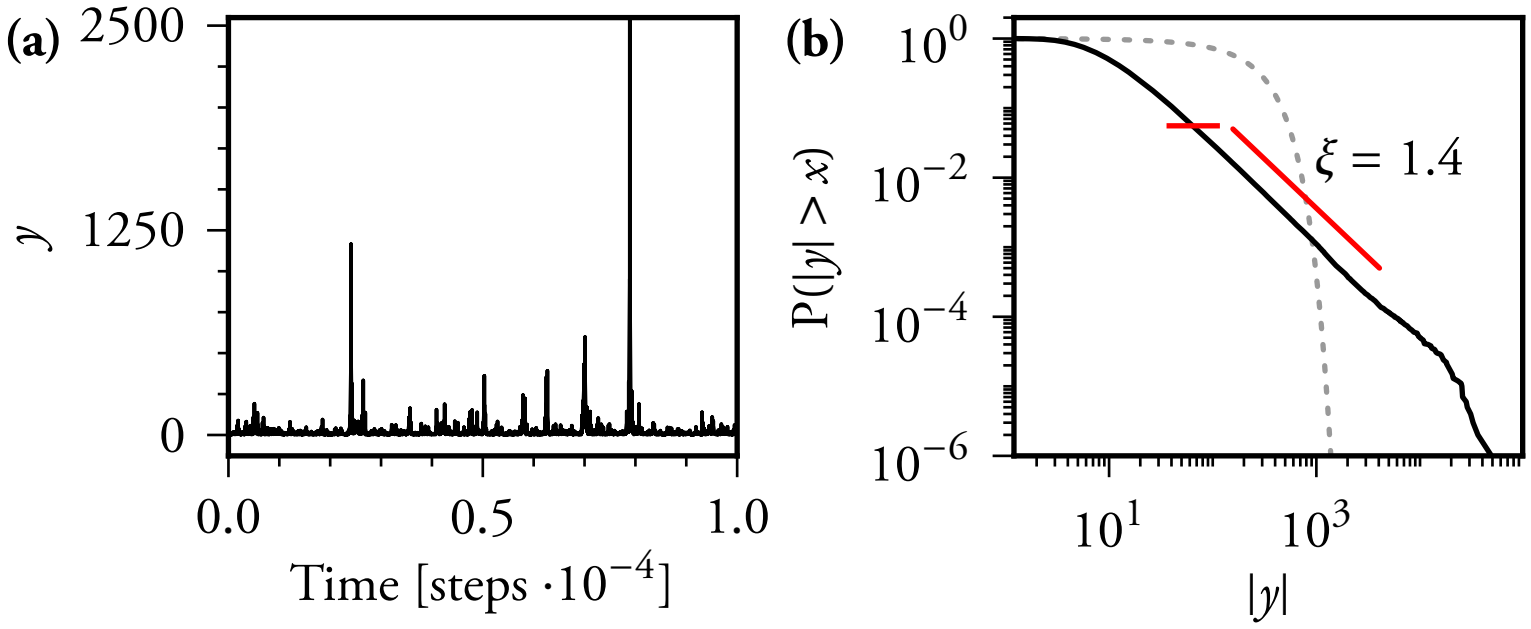

Figure 4.4.: The Kesten process driven by uniform noise. Additive noise was drawn from $\beta(t) \sim \mathcal{U}(0,1)$, multiplicative from $\beta(t) \sim \mathcal{U}(0.48, 1.48)$. **(a)**: Time series. **(b)**: CCDF. Black line: the Kesten process. Short red line: power law fit (see sec. 2.5). Dashed grey line: a Gaussian distribution with the same variance as the black line.

# Part II.

# Balancing

# 5. Introduction to human motor control

Many aspects of human motor control remain unknown. Nevertheless, computational principles have emerged that provide a theoretical framework for movement neuroscience [WG00, FS01, Sch02, FW11, WDF11]. This introduction briefly summarises the main aspects of this framework that will be relevant in later chapters.

As an illustration of how computational principles can be inferred from observed behaviour, consider the following common situation: Someone wants to get a carton of milk from the refrigerator. The carton is grasped with one hand and then lifted faster and higher than necessary since, unexpectedly, it is nearly empty. A fraction of a second later the subject slows down the movement, realises the problem, and then decides to put the box back into the refrigerator.

This simple example[1] suggests that humans are able to anticipate the outcome of certain actions. This requires some prior assumptions about the dynamics of their own bodies and of the objects which are being manipulated. If the estimates of the parameters or even the structure of a dynamical system are incorrect, they can be rapidly, but not instantaneously, corrected online. This action-perception cycle will be discussed in the next section. Then the role of noise will be discussed in more detail, followed by control strategies and cost functions. In the next chapter, human stick balancing tasks are introduced. This paradigm is central to the following chapters, where we investigate the mechanisms behind rare, extreme missteps in adaptive, predictive motor control.

[1] A similar task was investigated in [JW88]: When lifting an object, subjects scale the lifting force applied by the fingertips in anticipation of the object weight.

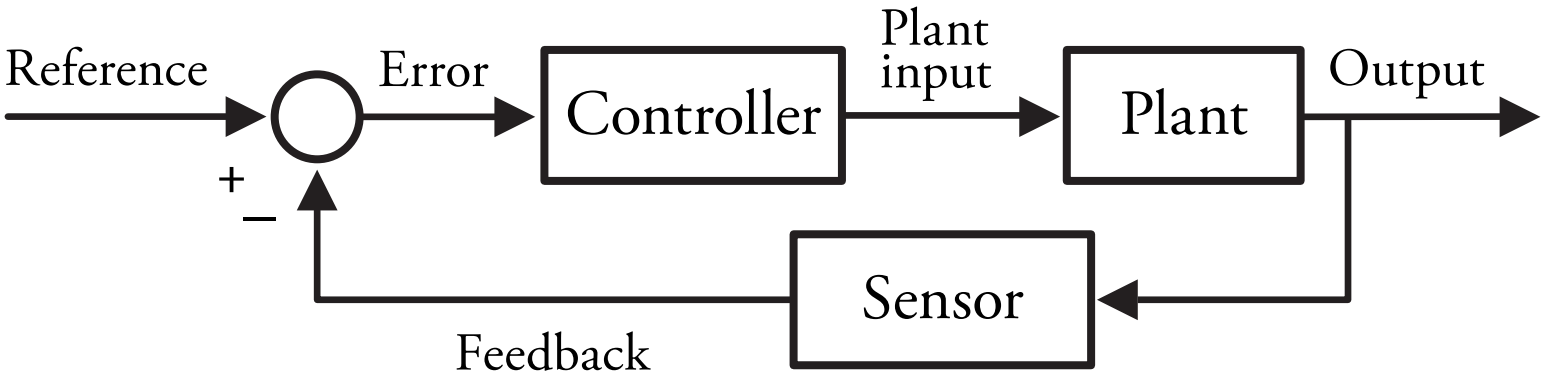

Figure 5.1.: Control diagram of a closed-loop controller.

## 5.1. The action-perception cycle

The Central Nervous System (CNS), with the brain at its highest level, regulates virtually all human activity. Many of its functions can be formalised using control theory, which deals with the behaviour of dynamical systems with inputs. An external input to the control system is called the reference. It can provide information about the desired state of the system, which is called the output. The controller manipulates one or more system variables to make the actual output follow the reference. The subsystem to be controlled is called the plant.

Figure 5.1 shows the classic closed-loop controller. Here, the output is measured using sensors. The difference of the measured output and the reference is the control error. This error is fed back as an input to the controller. This kind of negative feedback loop can be used to stabilise a system's dynamics. For example, cruise control stabilises a car's measured velocity $v$ by accelerating if $v$ is too low, and decelerating if $v$ is too high.

For human motor control, the situation becomes a lot more complicated. Here, movement is created by skeletal muscles applying forces to bones and joints. The output state may be, for example, the set of muscle activations or the position and velocity of the hand. This state changes continuously during movement. Other parameters, like physical properties of the body or the identity of a manipulated object, may change on other timescales or discretely. External disturbances may influence the state in unexpected ways. Furthermore, sensory feedback as well as the planning and execution of motor commands involves substantial delays and uncertainties that have to be incorporated into the control strategy.

For more than a hundred years, simple reaction times for visual stimuli have been reported to be around 190ms on average [JMY+85]. That is, the time required for an observer to respond to the presence of a stimulus by e.g. pressing a button. This is slower than reaction times for auditory or haptic stimuli. Reaction times also increase with task complexity. [Don68]

In many tasks, predictive control strategies move delays out of the feedback loop. For example, when tracking a target on a touch screen with a finger, humans predict the movement of the target to reduce control errors [ES00]. For unpredictable changes in the movement of the target, it still takes around 250ms to the onset of corrective finger movements and several hundred milliseconds more to reacquire the target.

How the CNS preforms predictive control is commonly explained by distinguishing three stages in the sensorimotor loop [WG00] as shown in figure 5.2. A motor command is generated based on information about the task, context and output state. To do this, an inverse internal model maps the desired consequences of a control action to an appropriate command. Execution of this motor command changes the output state. The loop gets closed when the actual changed output state causes new sensory feedback. Internally, the consequences of the execution of the motor command are predicted by a forward dynamic model. Based on the predicted change of the output state, a forward sensory model predicts the expected new feedback. Therefore, further actions can be planned predictively before the sensory feedback arrives. Several studies support the use of forward models in the sensorimotor system, and provide evidence that estimates of the body state use both sensory feedback and a model of the world which can be adaptively reconfigured [FW11, WDF11].

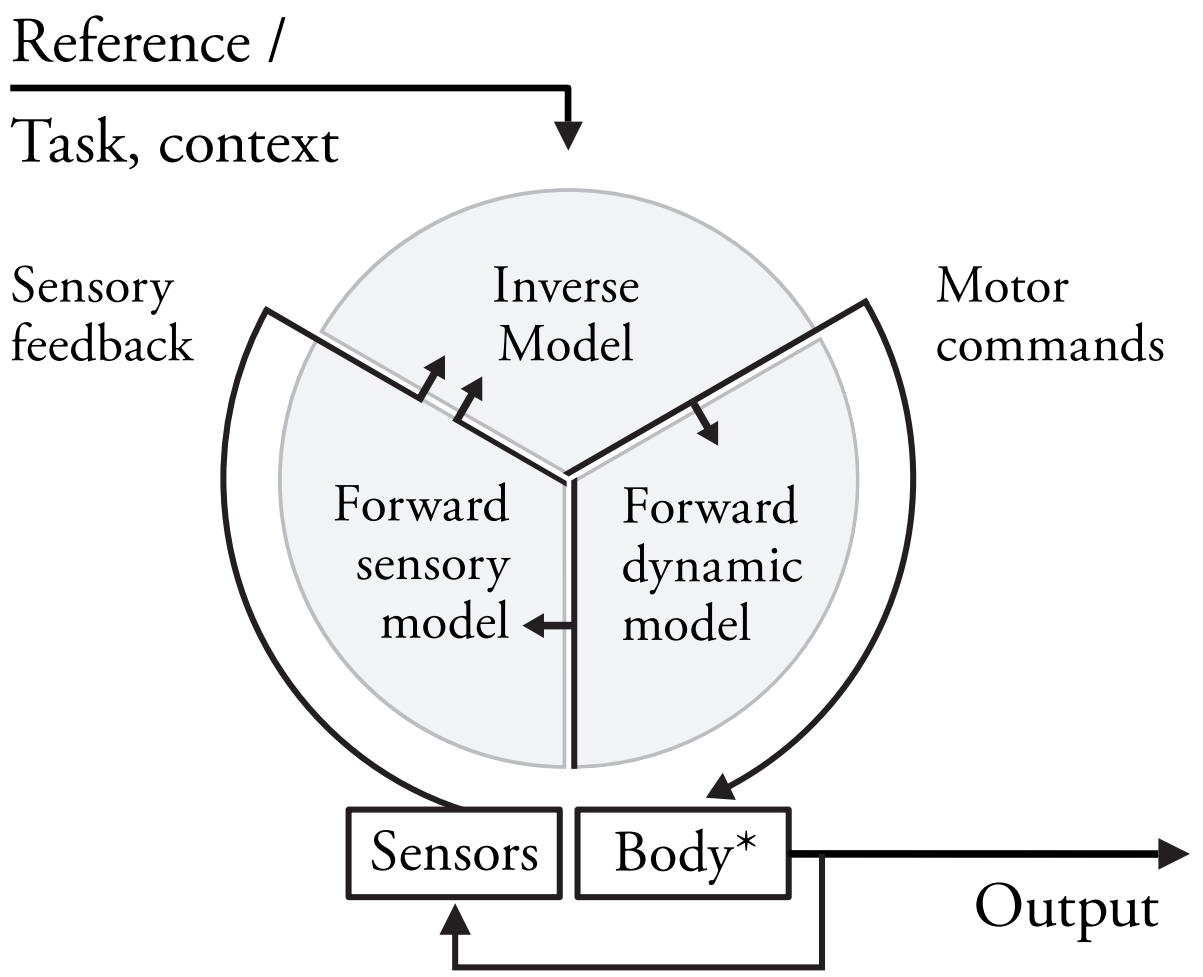

Figure 5.2.: A concept of the human sensorimotor loop with three stages: motor command generation, state transition, and feedback generation. Motor commands to the muscles initiate mechanical movement of parts of the body and thereby potentially of manipulated objects. That is, in control theoretic terminology, the plant (*). Sensors report the changed output state. The expected dynamics of the plant and of the consequent sensory feedback are predicted by the CNS using internal models.

Error based learning helps to adapt to changing contexts, to calibrate behaviours, and to correct for systematic biases [WDF11]. Lack of feedback for prolonged times can lead to erroneous behaviour like walking in circles [SFSE09]. Trial based findings support adaptation on different timescales coexisting in parallel [LS09]. Further, memory decay rates depend on how quickly the learning environment changes [HS09]. Behavioural flexibility is also present on shorter time scales. Recently, trial to trial adaptation to force fields which were present only during a part of each trial was investigated. It was found that adaptation depends on hand movements within each trial before the force field is switched on, but this memory decays quickly and vanishes after approximately 600ms [HFW12]. In another study, changes of mind during movements suggest that motor strategies can even be adjusted to information that was still in the processing pipeline when a movement was initiated [RKWS09].

Visuomotor transformations are an integral part of the action-perception cycle [FS01, CHM11]. Since both different sensors and effectors in the body have many different frames of reference, the CNS has to perform coordinate transformations. In addition, the use of higher-level coordinate systems can solve the following problem: Consider the approximately 600 muscles in the human body as being either contracted or relaxed. Even in this simplified situation, there are $2^{600}$ possible motor activations. A look-up-table from motor activations to sensory feedback alone would need more entries than a list of all the atoms in the observable universe. Hence, information about the system state that is necessary for a given task needs to be represented in a far lower dimensional space. For example, egocentric spatial coordinates are likely used for working memory, eye movements, and for ongoing or intended arm movements [FS01]. These coordinates are often eye- or head-centred, but coexist with allocentric representations (i.e. spatial coordinates relative to an external cue) [CHM11]. Such high-level representations not only simplify movement planning, but also facilitate imitation: There is evidence that watching another person perform an action engages high-level sensorimotor representations of the observed action [WDF11]. This includes mirror neurons which fire when performing and when observing actions (ibid.).

Transformations to motor commands are likely performed by an inverse model after response selection [FS01, CHM11]. This is possibly done in several steps, first calculating joint coordinates, then the required joint torques, and at last the motor commands [FS01]. On the lowest level, reflex loops connect muscles and neurons in the spinal cord. They help to stabilise joint angles by adjusting muscle contractions much faster than signals can be send to and received from the brain.

## 5.2. Noise and uncertainty

The CNS has to deal with uncertainties due to incomplete knowledge about the world. This is partly the consequence of incomplete observations of new, complex, or unpredictable environments. Ambiguity also arises since sensors like eyes only capture a lower dimensional projection of the world. However, there are also many intrinsic noise sources in the CNS [FSW08]. This includes the biochemical transduction of signals on the receptor level. Fluctuations of ion channels in cell membranes and the stochastic transmission at synapses add to cellular noise. Noise can also emerge at the network level, for example due to interference.

Given the aforementioned noise sources, it should come to little surprise that a significant amount of variability is observed between repeated movements. Yet, several studies put little emphasis on planning noise and focus on execution noise instead. It was suggested that the latter accounts for at least a large proportion of movement variability [vBHW04]. Of particular interest is the finding that movements show significant signs of signal-dependent noise. Forces generated by voluntary muscle contractions were found to fluctuate with a standard deviation of $2-3\%$ [JdCHW02]. This constant signal-to-noise ratio corresponds to a linear amplification of both mean force and fluctuations. There is evidence that this signal-dependent noise is the consequence of the physiological organisation of the motor-unit pool [JdCHW02]. That is, of the motor neurons that innervate a single muscle. These findings are consistent with the ubiquity of Fitt's Law [Fit54] which describes a trade-off where faster and longer movements are less precise.

However, the situation for some behavioural tasks is more complicated. Some findings indicate that in addition to movement speed and path, stiffness is a separate dimension which influences movement variability. When pointing to targets, higher co-contraction of antagonist muscles can even reduce endpoint variances. Therefore, it was suggested that increased motor-command noise during high-contraction movements can be compensated by less susceptibility to noise due to an increased impedance of the arm, and possibly changed feedback [OKI+04]. It is further possible that the impact of observation- and planning noise increases in some situations. For example, recorded preparatory activity of cortical neurons in monkeys suggest that at least half of the observed movement variability likely had its source during motor preparation even during a highly practised reach task [CAS06].

There are several ways for the CNS to combat these problems. The most basic method is averaging over redundant information sources such as sensors in close proximity or neuronal populations with common input. Averaging over time can take place at the cellular level because of the temporal-integration properties of the neuronal cell membrane. The behavioural relevance of temporal averaging is supported, for example, by electrophysiological animal studies [FSW08]. Additionally, the CNS uses prior knowledge about the world to deal with ambiguous and noisy observations [WG00, FS01, FSW08, FW11]. This includes the combination of sensory data from different modalities which is often well described by Bayesian estimators. It has further been suggested that temporal averaging and prior knowledge are combined in an optimal way which resembles the famous Kalman Filter. The latter can also be considered a recursive Bayesian filter: An estimator is repeatedly updated using new observations and prior knowledge about a systems structure such that the Mean Squared Error (MSE) is minimised. Optimisation principles can also be used to minimise the impact of execution noise; this topic is discussed in the next section. Finally, there are also situations where noise may be beneficial to the functioning of the CNS (e.g. stochastic resonance, avoiding local minima in associative learning, ...).

## 5.3. Costs and strategies

Movements are usually performed in order to achieve some goal. Achieving this goal is connected to some kind of reward or utility.

For instance, grasping a piece of chocolate and moving it into one's mouth may lead to reward signals in the brain. Failure to perform an action due to uncertainties, for example concerning the location and amount of chocolate, likely leads to less or no reward.

Maximising the utility of movement outcomes in the face of uncertainty may be considered a decision problem [WL12]. In the framework of statistical decision theory, this can be formalised as minimising a loss function. This function can include not just the movement error: Choices can be optimised given uncertainties, prior knowledge, as well as costs and benefits of any outcomes that may occur.

For instance, unpredictable external forces may be compensated with a high co-contraction of agonist and antagonist muscles increasing the stiffness of joints. However, this impedance control is associated with high movement costs due to faster exhaustion. Therefore, humans try to minimise both the variability of motor output and the required effort [OBD09]. For example, stiffness decreases when subjects learn to act more predictively [WDF11]. However, predictable forces are not the only perturbations which can be optimised for.

Movements can also be optimised such that the effect of execution noise and other unpredictable perturbations with known statistics is minimised. For a linear scaling of noise with the control signal, this leads to smooth movements with minimal change of acceleration, a model which is consistent with experimental findings [HW98, FS01]. Even with a trade-off between effort and accuracy in more complex object manipulation tasks, smooth trajectories were still found to be optimal [NBW09]. However, there are special situations where humans may apply forces in a highly nonlinear or discontinuous way. Especially during ballistic movements high, accelerations are produced over short times to minimise the time required for the movement. Examples include throwing or punching movements. Whether such strategies may be relevant for balancing is discussed in chapter 9.

A different type of cost is constituted by the use of CNS resources, in particular memory. For example, a tradeoff between memory and motor effort was found in a visual search task [KK11]. Change blindness (sec. 9.6.4) in object manipulation tasks even suggests that for certain computations, information extracted from the fixation point is used only when immediately needed to solve the current goal [TBHS03]. Despite these findings, many models of motor control treat perception and control strategy as separate problems. Typically, an observer model and a planning model are used which are optimised separately [WG00]. One notable exception is Optimal Feedback Control (OFC) (see sec. 9.7).

The utility of a reward may also be modulated by factors which are not direct costs. Time can be such a factor through temporal discounting. That is, a reward is treated as less valuable the longer it takes to obtain. For example, treating the execution of an eye movement as a reward and assuming sub-second temporal discounting enables prediction of dynamic features of saccadic eye movements [SOdXXWS12]. Furthermore, several recent studies suggest that subjects do not always maximise expected reward in sensorimotor tasks. Instead, they are sensitive to risk as well [WL12].

Since concepts like utility and risk aversion originate in economics, a direct comparison of decision making in motor control and economic tasks suggests itself. It was found that subjects tend to be more risk-seeking in motor tasks than in classical economic tasks. One study concluded that the same subjects distort probability, but not value, differently when making identical decisions in motor and classical form [WDM09]. This is consistent with several studies on movement planning. For example, in a pea shooter task it was found that subjects act according to a loss function that punishes large errors less than predicted by a quadratic loss function. For small errors, however, the function is very well approximated by the MSE [KW04]. In another study, it was suggested that the nervous system estimates the likelihood that an error has been caused by either the motor system or the external world and uses this estimate to adapt optimally. In one Bayesian model, this leads to an adaptation which reacts strongly to small errors and ignores large ones [WK09].

## 5.4. Postural sway: continuous control and scaling

Many models in sensorimotor control only account for endpoint errors. This is intuitive, for example, in pointing tasks; but some tasks have a different structure. Consider one of the most basic human motor skills: while quiet standing may appear to be static, it is a complex dynamical control task. The human upright posture is unstable and requires ongoing active balancing, leading to body sway. Numerous studies have investigated the movement of the center of pressure under the feet of humans standing on force platforms. Body movements have also been measured using ultrasonic or infrared reflectors [WvHR88].

By analysing the scaling of mean displacements over different time scales, the fluctuations in human postural sway have been found to exhibit three regimes. Postural sway can be modelled as a correlated random walk for short timescales. At a correlation time of approximately one second, a crossover to an anti-correlated random walk occurs [CDL94]. For very long timescales, the time-series become uncorrelated [CC95]. To explain these features, postural sway was modelled as an overdamped inverted pendulum with simple feedback control [CDL93, Pet00]. The actual control strategy employed by the CNS, however, is more complex but still unknown [KZJ11].

Upright standing was previously compared to other tasks that involve balancing of an inverted pendulum. Foremost, balancing a stick on a fingertip or comparable virtual settings were investigated. These tasks are in the focus of the following chapters. In particular, noise sources and cost functions in adaptive, predictive control, are investigated with respect to their consequences for the statistics of rare extreme control errors. The analogy between stick balancing and upright standing is discussed further in section 9.6.1.

# 6. Balancing tasks

Stabilising an inverted pendulum is probably the most elementary control problem. However, although numerous solutions exist, it is still not completely understood how humans perform this task. It can be realised by balancing a stick on a finger tip, which is discussed in the next section. Observed fluctuations of displacement angles of the stick show intermittent bursts of fluctuations.

Similar fluctuations where observed in Virtual Stick Balancing (VSB) where an unstable target on a computer screen is stabilised. The analysis of a VSB data set establishes the main features of human balancing behaviour which this thesis part seeks to explain. First, control errors are power-law distributed. Therefore, much more extreme magnitudes are observed than what would be expected from a Gaussian distribution. Second, Power Spectral Densities (PSDs) show a characteristic knee with two distinct scaling exponents.

In the final section of this chapter, the strengths and shortcomings of an existing balancing model are discussed. This model suggested that reaction delays and strong multiplicative noise are key ingredients for the observed fluctuations in stick balancing.

A more in-depth analysis of VSB data, including learning of balancing skills under different objectives, is found in chapter 8. There, a larger data set is presented that was recorded in an optimised experimental paradigm based on the findings to be presented in chapter 7.

## 6.1. Stick balancing on the fingertip

Several studies investigated humans balancing sticks on their fingertips, usually while sitting on a chair [CM02, CBE+04, CLM06, CM12].

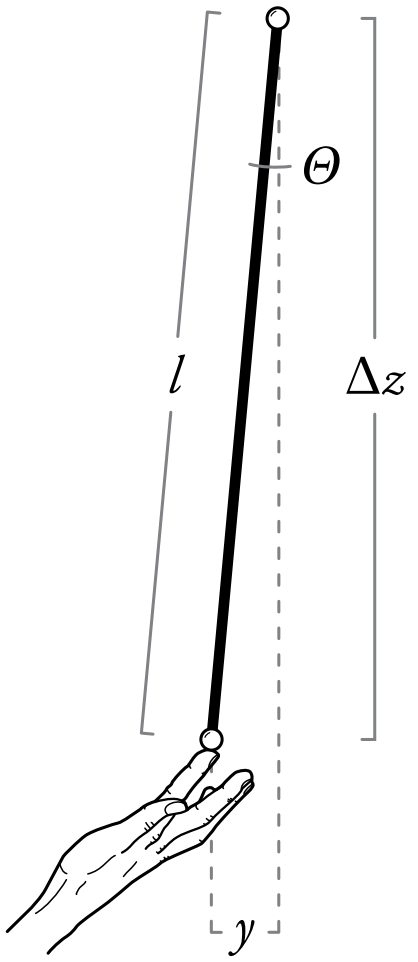

Figure 6.1.: Stick balancing on the fingertip is characterised by the stick's length $l$ and the vertical displacement $\Delta z$. These quantities are related to the displacement angle by $\cos(\theta) = \Delta z / l$. The projection of the endpoints onto the horizontal plane $y = l\,(1 - \cos(\theta))$ becomes relevant for VSB in section 6.2.

The movements of the sticks' endpoints in three dimensions were tracked using motion capturing cameras. Priority was given to the analysis of the dynamics of the relative vertical displacement $\Delta z / l$ illustrated in figure 6.1.

The most prominent reported features of these dynamics are intermittent fluctuations [CM02]. Here intermittency denotes the random alternation between phases with extremely low movement amplitudes and phases with high movement amplitudes. Initially, analyses focused on scaling properties of velocity increments. Later, heavy tailed distributions of $\Delta z / l$ were also found. It was suggested that they follow power-laws, but even more extreme fluctuations might occur when the stick is about to fall [CM12].

Power spectra exhibit two distinct scaling regimes [CM02] similar to the ones shown for VSB in figure 6.5 (section 6.2). For $0.1 - 1$Hz, a scaling exponent close to $1/2$ was reported. A pronounced knee is found above one Herz. For higher frequencies, the scaling exponent was reported close to 2.5.

These effects have been attributed to the interplay of multiplicative noise and reaction delays. As an explanation, it was proposed that the system is tuned very close to a stability boundary, with parametric noise causing fluctuations across this boundary [CM02]. A model for this theory is discussed in section 6.3. For the rest of this section, we discuss a variety of results that play ancillary roles in the motivation of the next chapters and in the discussion of their respective results in the context of the research field.

Eventually, the stick always falls down. Survival times have been reported to be fit well by the Weibull distribution [CM12]. For a 56cm stick, average survival times were 14s for novice subjects and more than 100s on the third day of practise. For a 25cm stick, survival times were dramatically shorter.

The effects of practise are also reflected in the distribution of velocity increments, which are more heavy-tailed for skilled subjects [CBE+04]. It was claimed that these changes can be quantified as changing Lévy flights (see sec. 2.3), but the fits were produced using a method which is very susceptible to problems (see section 2.5): it severely overestimates scaling exponents if the process at hand is not actually a Lévy flight. The figures shown in [CBE+04, CB09] clearly show this problem: the fits are far more heavy tailed than the data sets. Still, despite the problematic method used for quantification, the differences between high and low skill levels are substantial.

It was argued that the intermittency observed in stick balancing is of the on-off type [CM02]. As discussed in section 4.3, On-Off Intermittency (OOI) arises due to parametric noise which drives a control parameter across a stability boundary. There are two arguments for this conclusion. First, OOI is associated with a power-law scaling in the power spectrum with exponent 1/2. Reported power-spectra for stick balancing scale are close to 1/2 in the 0.1 – 1Hz range, but not over all frequencies. Second, the laminar phases (sec. A.10) in both OOI and stick balancing scale with exponent 3/2. Such scaling, however, is not exclusive to OOI. A different example is found in section A.10. Also, OOI is not necessarily associated with power-law probability distributions (see section 4.3). Therefore, it is possible that the mechanism underlying the fluctuations in stick balancing is very similar, but not necessarily identical to what is typically considered to be OOI.

The distributions of laminar phases further lead to another observation: most of the waiting times between crossings of a small threshold from above (i.e. from large to small displacement angles) are shorter than the reaction time [CM02]. This was interpreted as evidence that parametric noise helps to stabilise the system faster than reactive control. While the proposed model shows that parametric noise can indeed contribute to a more stable system (section 6.3), this effect requires some parameter settings that are not easy to justify. A similar distribution of laminar phases can also be found for extremely simple stable systems with additive noise (section A.10). Therefore it is unclear whether these threshold passings can be interpreted as a type of intentional corrective movements which emerge from parametric noise, as was claimed in [CM02].

On a final note, stick balancing is predominantly driven by visual input since, in contrast to e.g. upright standing, it is not possible to balance a light stick with closed eyes. It was further reported that when subjects closed their eyes and moved their hands to mimic the movements during stick balancing, power spectra did not contain a significant region with slope 1/2 [CM02].

## 6.2. Virtual Stick Balancing (VSB)

VSB facilitates data acquisition and setup manipulations compared to real stick balancing [BCME04]. It typically involves only visual feedback about the target, forgoing haptic feedback which may play an ancillary role in real stick balancing. Further, movements in real balancing are limited, for example, by the length and inertia of the subject's arm while in VSB additional limitations include the size of the display, the movement range of the input device, or the latency of the setup. The most common input device is a computer mouse which is restricted to two dimensions. Many setups use linearised dynamics of first or second order. Nevertheless, VSB time series exhibit some statistical features which are very similar to real stick balancing.

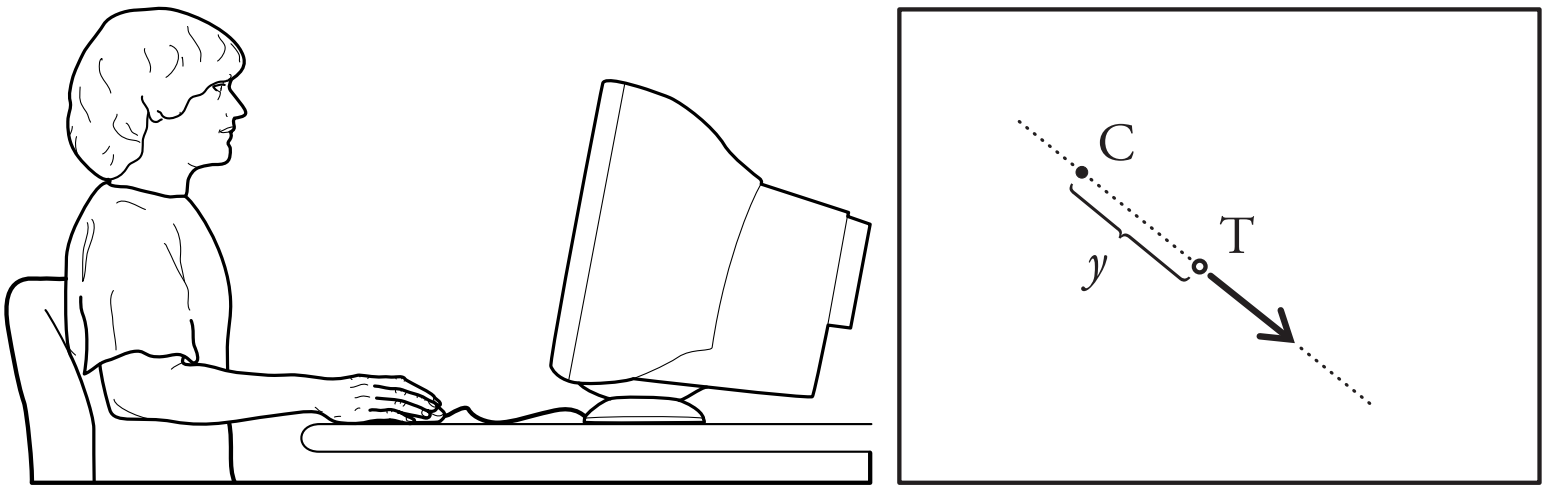

Figure 6.2.: A typical virtual balancing setup. Subjects try to catch a virtual target T with cursor C which they control using a computer mouse. Both C and T are displayed on a computer screen (rectangle). T moves according to an unstable dynamics. Therefore, the distance $|y|$ grows if C is not moved.

### 6.2.1. Experimental setup

Here main features of an experimental data set recorded by Marcus Riegel are presented. The experimental setup was prepared by Udo Ernst. Representative results for one subject are shown and the ranges of the findings for all subjects are stated. The complete data set also serves as a guideline for the modelling done in chapter 7. We first published an analysis of this data in [PREP07]. The setup is depicted in figure 6.2. In this task, a cursor C and a target T are displayed on a computer screen. C is controlled by the subject's hand and moves linearly proportional to the position of a computer mouse. T is moved by the computer according to

$$\vec{T}_{t+1} = \vec{T}_t + \frac{\Delta_t}{\tau}\,(\vec{T}_t - \vec{C}_t), \tag{6.1}$$

where $\Delta_t$ is the sampling interval, and $\tau$ is a time constant which determines how fast the distance $|y|$ between C and T grows when C is not moved. Subjects were told to keep C and T as close together as possible without one of them running out of the screen. This situation can be thought of as a highly stylised form of balancing a light stick: after linearisation and disregarding inertia, C and T correspond to the projections of the two ends of a stick onto the horizontal plane (see figure 6.1).

The horizontal and vertical positions of mouse and target were recorded with 85Hz, the same frequency as the refresh rate of the screen. At least 10 trials per day were recorded for at least four days for each subject. A trial was discarded if either T of C left the screen. The experiments were authorised by the ethics committee of the University of Bremen.

Seven subjects participated in the experiments. Subjects 1, 2, and 3 performed the tasks with $\tau = \text{const}$. According to their skills and training, trials with $\tau = \frac{1}{3}\,\text{s}$, and $\tau = \frac{1}{4}\,\text{s}$ were recorded. For the others, $\tau$ was randomly switched every second to a value in $\{\frac{1}{3}\,\text{s}, \frac{1}{4}\,\text{s}, \frac{1}{5}\,\text{s}, \frac{1}{6}\,\text{s}\}$.

### 6.2.2. Results

Since the axes of the screen represent an artificial coordinate system which is unlikely to be of any distinct meaning to the brain and since the task is point symmetric, we exclusively analyse on the radial distance $|y|$ between mouse and target. $|y|$ can be considered the (absolute) control error. We further normalise $|y|$ such that magnitudes are in units of the standard deviation. Normalisation has no effect on scaling in the Complementary Cumulative Distribution Function (CCDF) and PSD, and allows for an easier comparison between different distributions. A time series is shown in figure 6.3. Similar results were obtained for all subjects (see sec. 7.5). Note the frequency of events which are more than one order of magnitude larger than the standard deviation.

The corresponding CCDF is shown in figure 6.4. Control error distributions for all subjects strongly deviate from Gaussians, exhibiting power-law tails with CCDF scaling exponents $\xi$ in the range of two to four. The apparent cutoff of the power law for large $|y|$ is caused by temporal correlations: The few highest events all belong the same

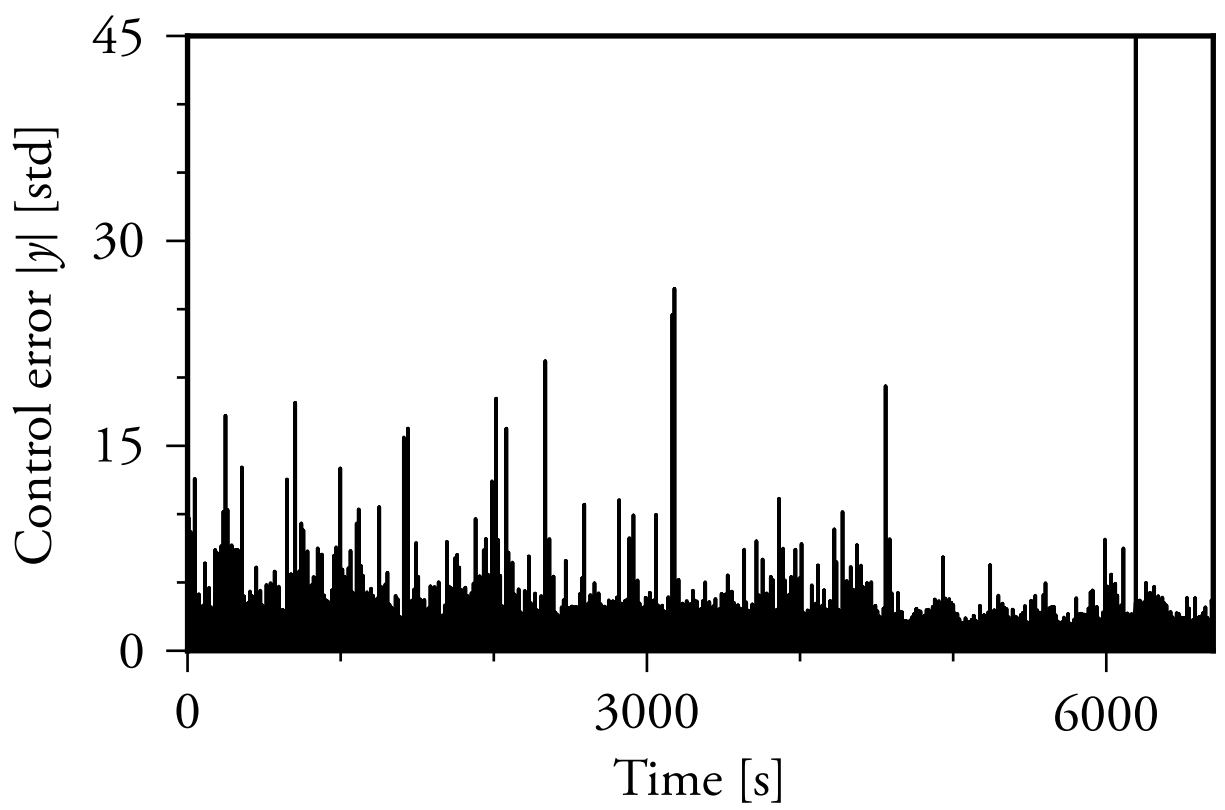

Figure 6.3.: Time series of normalised controller-target distances $|y|$ (in units of Std($y$)) of the combined trials of four days for subject 3. $\tau = \frac{1}{4}$ s. Trial length was 3min.

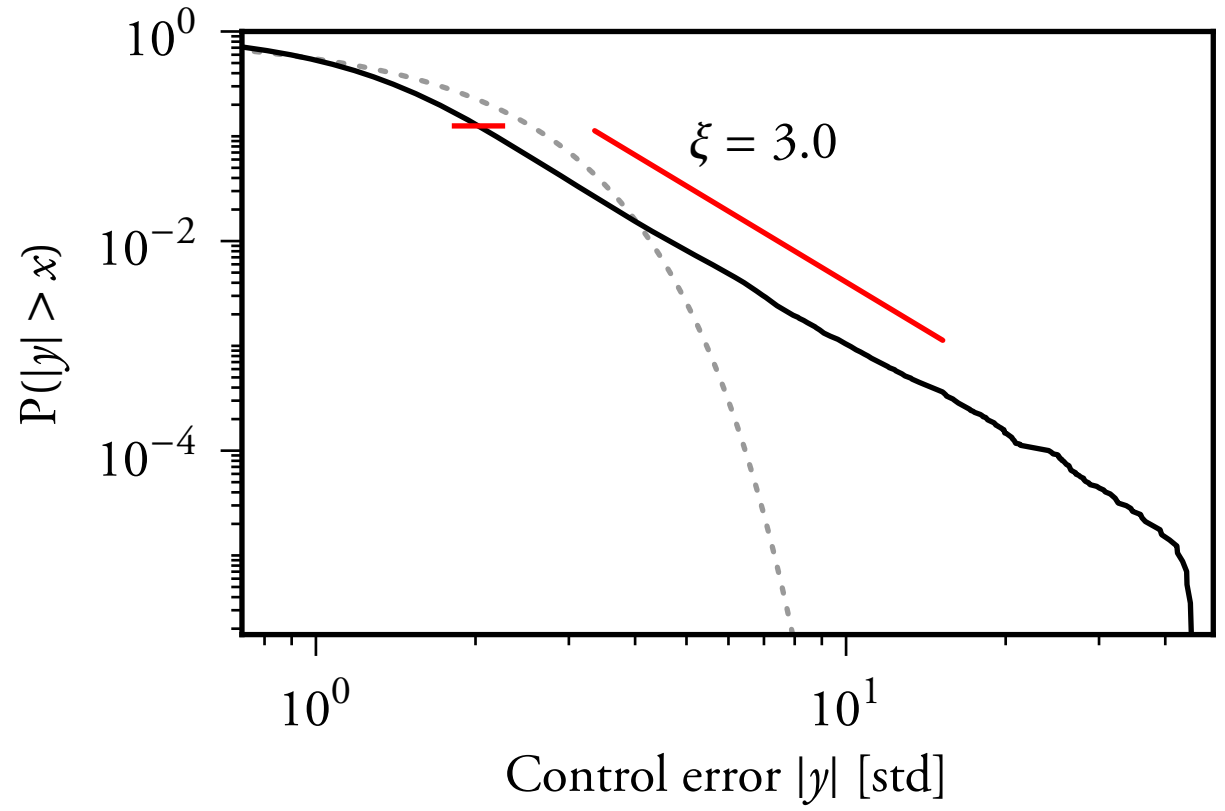

Figure 6.4.: Complementary cumulative distribution $F_c$ of the normalised control errors shown in figure 6.3. Red diagonal line: power-law fit (see section 2.5). Short red horizontal line: estimator cutoff. Grey dotted line: Gaussian with unit variance[1].

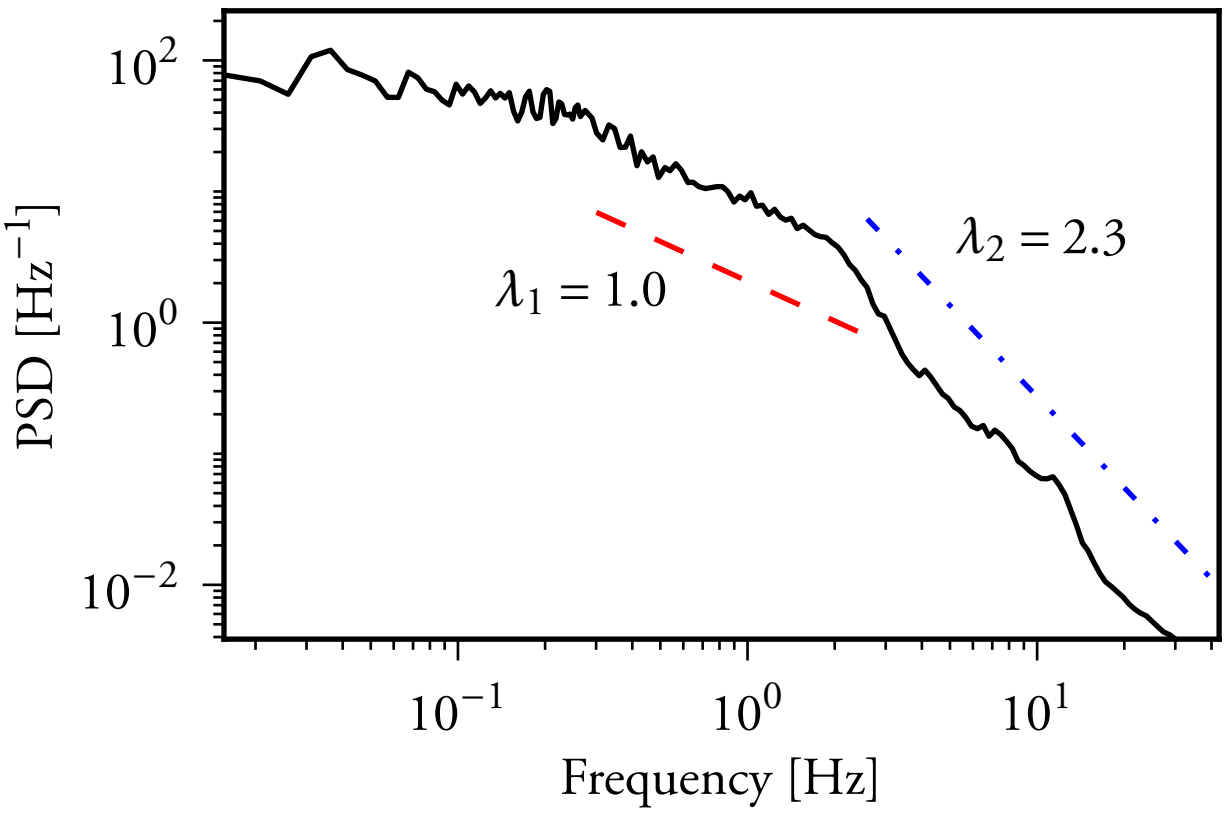

Figure 6.5.: Power Spectral Density (PSD) for the same time series shown in figure 6.3. Densities were calculated as described in sec. 3. The scaling exponents $\lambda_{1,2}$ have been estimated using linear regression. The two cutoffs were obtained by minimising the MSE for a fit of the whole PSD with three linear regimes. The fits for the low (red dashed) and high (blue dash-dotted) regimes are depicted with a vertical offset only, so as not to obstruct the data.

peak which occurs close to 5000s in figure 6.3. A similar effect is found for all single- and combined trials for all subjects. The position of this apparent cutoff depends on the length of the analysed time series. By using appropriate surrogate data, it is possible to show that experimental CCDF tails are not significantly different from power laws if temporal correlations are correctly accounted for (see secs. 7.7, and 7.7)

Figure 6.5 shows the PSD for $|y|$. Spectra are constant for low frequencies and above 0.1 Hz approximate broken power-laws for all subjects. The first scaling exponents $\lambda_1$ are above 1/2 and below 2 and

[1] Since we here compare only positive events, a half-normal distribution is used. For simplicity, albeit slightly imprecise, we use the term “Gaussian” for both normal and half-normal distributions throughout this thesis.

the second ones $\lambda_2$ above two and below four. A knee is observed in between one and five Hertz.[2]

The scaling features in the CCDF and PSD are stable over time and can be found even in single trials. We found neither systematic trends over time across subjects nor differences between the two conditions (constant and variable $\tau$) that reach significance[3] [PREP07]. However, especially for the CCDF there is a strong trial-to-trial variability within each day. A more in-depth discussion is found in chapter 7 after introducing an appropriate model to which the data is compared.

### 6.2.3. Other VSB experiments

The effects discussed above were reproduced in [MFC+11] in a similar setup, but with an unstable second order dynamics. It was further found that in more difficult tasks (faster acceleration of the target) average control errors increase, but their distribution becomes less heavy tailed. This effect is discussed in section A.6.

A different setup using a manipulandum as the input device was studied in [MS02]. It was found, that subjects could tolerate blackouts with no feedback for up to 600ms. A comparison of the data with different generic controller types lead to the rejection of most possibilities. The authors concluded that subjects used a forward model in the sensory preprocessing stage of the control loop, allowing them to predict movements during the blackouts.

---

[2] Note that the exponents for, for example, a damped harmonic oscillator would be even integers. The finding that $\lambda_1 \approx 1$ hints at a process that depends on states at different points in time (see sec. 3). The cutoffs, however, imply that this temporal non-locality is restricted to a relatively short timescale.

[3] A re-analysis of the data for the first four days of each subject using the methods from chapter 8 confirmed these results. However, the group with constant $\tau$ is either very small or contains trials with different values of $\tau$. This limits the power of the study to detect small differences between trials grouped by condition or day. For example, some subjects could only perform in more difficult trials with smaller $\tau$ after some training. Therefore, training did have some effect on some subjects.

## 6.3. The Cabrera & Milton model

The fluctuations observed during balancing lead to the hypothesis, that the system is tuned very closely to a stability boundary [CM02, CLM06, CM12]. It was argued that parametric noise causes fluctuations across this boundary like in OOI (see section 6.1). This behaviour was modelled as an overdamped inverted pendulum with delayed feedback and multiplicative noise:

$$\begin{aligned} \ddot{\theta}(t) &= -\Gamma\dot{\theta}(t) + q\sin\theta(t) - F\big(t, \theta(t-t_r)\big) \\ F\big(t, \theta(t-t_r)\big) &= \big(R_0 + \eta(t)\big)\theta(t-t_r), \end{aligned} \tag{6.2}$$

where $\theta$ is the displacement angle (see figure 6.1). The influence of friction is scaled by $\Gamma = \gamma/m$ where $\gamma$ is the damping coefficient and $m$ is the mass of the stick. $q = g/l$ determines the acceleration due to gravity. $l$ is the length of the stick. $g$ is the acceleration of gravity. The feedback $F$ is delayed by the reaction time $t_r$. It exerts a restoring force scaled by a parameter $R_0$ that is perturbed by Gaussian white noise $\eta$ with variance $\sigma^2$.

For carefully adjusted parameters, the model shows intermittent fluctuations which appear similar to the observations for stick balancing. A time-series is shown in figure 6.6 (a). Comparison of the simulations with and without noise reveals a main aspect of this model: Multiplicative noise can prolong escape times. That is, the average time until the stick tilts beyond the horizontal position and inevitably falls down (detailed statistics not shown). In other words, if it's not possible to keep control using targeted movements, one might as well wildly wiggle the stick around to get closer to the upright position by chance. Occasionally, the model may even end up so close to the origin that it stays stable (not shown).

Figure 6.6 (b) shows the PSD for the example time-series. This author was not able to reproduce a broken power-law with a shallow scaling regime close to 1/2 or even close to 1 for any parameter combination. While it was claimed that the model can reproduce such PSDs, the only published evidence for this in the literature known to the author are three data points at the low-frequency end of the power spectrum shown in figure 3 in [CM02]. Therefore, it might be possible that there the low-frequency cutoff of the system's response was mistaken for another scaling regime. Another explanation is, that very sensitive parameter-tuning is required in order to see the claimed effect which makes reproducing it extremely difficult.[4] However, whether or not it is possible to find a parameter set where experimental power spectra are reproduced is not central for the arguments pursued in the following.

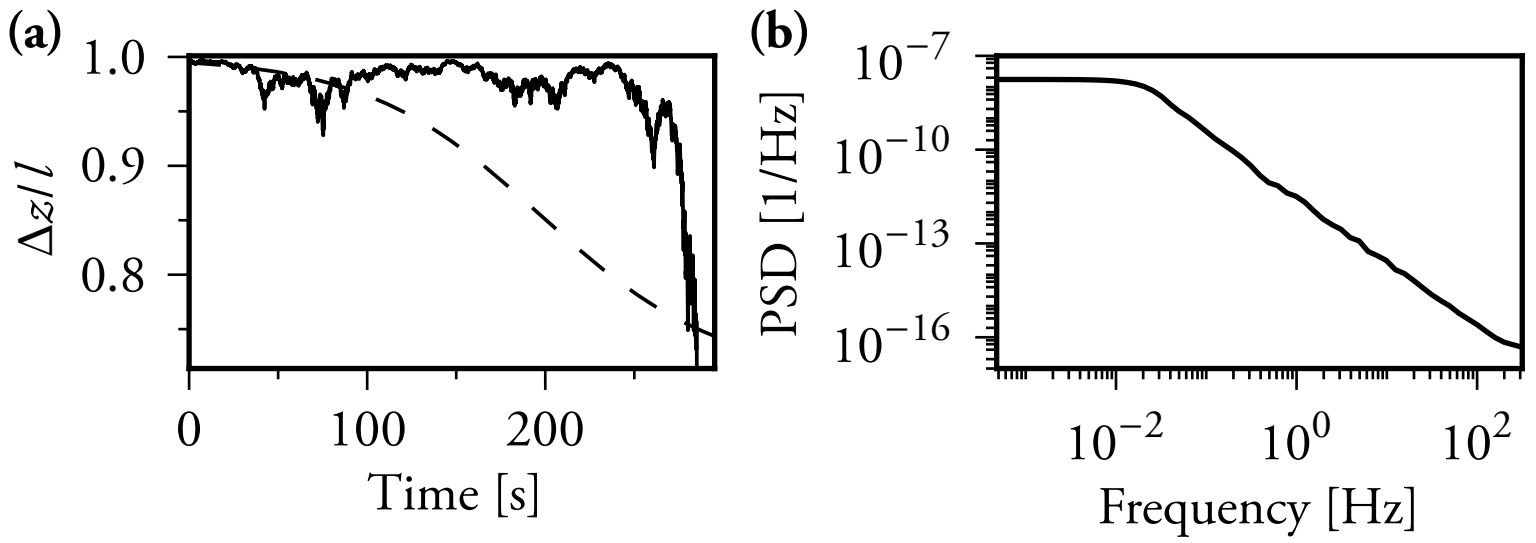

Figure 6.6.: (a): Time-series for the relative vertical displacement of a balanced stick simulated according to (6.3). Parameters: $q = 10\,\mathrm{s}^{-2}$, $R_0 = 9\,\mathrm{s}^{-2}$, $t_r = 0.2\,\mathrm{s}$, $\Gamma = 100\,\mathrm{s}^{-1}$. Discretisation step $h = 0.01\,\mathrm{s}$. Solid line: $\sigma = 10\,\mathrm{s}^{-3/2}$ Dashed line: $\sigma = 0$. (b): Power Spectral Density for the same time series as the solid line in (a).

Note that in figure 6.6, parameters are different than in previous publications, which assumed reaction times of 70ms. Such short reaction times are inconsistent with the vast majority of the literature on reaction times with visual stimuli (see section 5.1 or for a balancing example [MS02]), as well as with experimental findings presented in this thesis. Therefore, a more realistic feedback delay of 200ms is used.

[4] The publications on this model tend to include only incomplete information about the parameters used to generate the figures. During personal communications, one of the authors suggested that parameter tuning in this model is difficult.

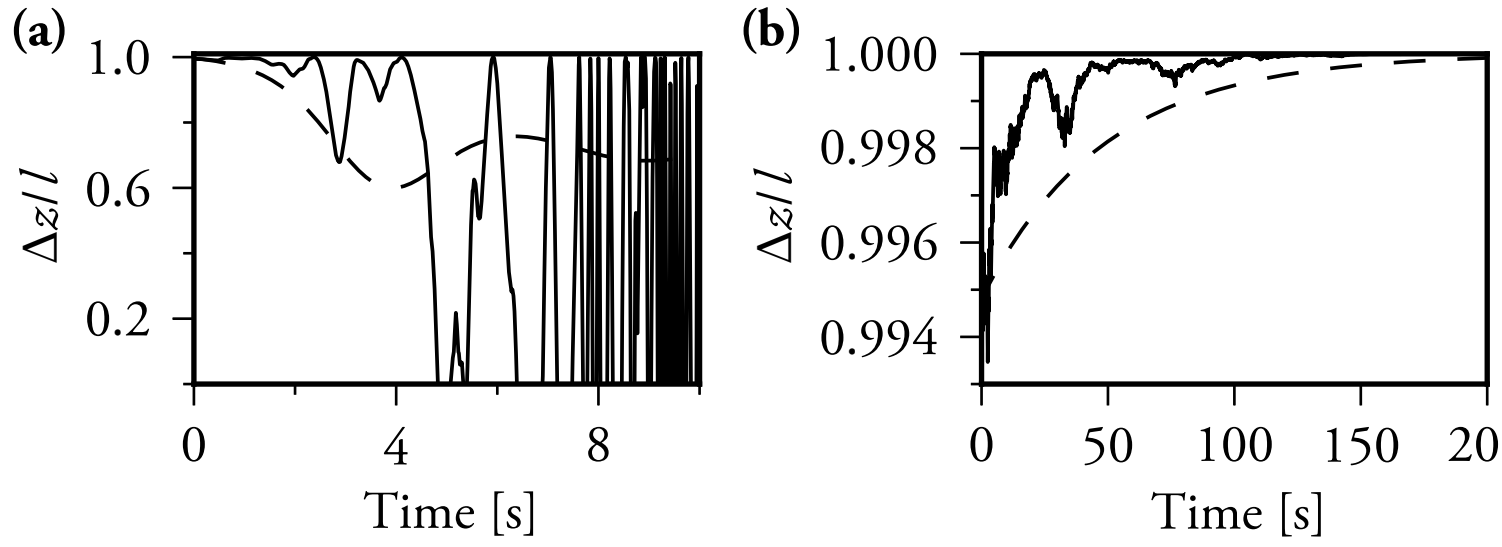

Figure 6.7.: Model time-series with (a): with low damping ($q = 10$, $R_0 = 9$, $t_r = 0.2$, $\gamma = 2.5$, $\sigma = 10$, $h = .01$), (b): slight overdamping ($q = 8$, $R_0 = 9$, $t_r = 0.2$, $\gamma = 100$, $\sigma = 10$, $h = .01$).

However, the exact value of the delay does not change the qualitative findings for this model. Furthermore, the value of $q$ corresponds to a stick which is almost a meter long, but very similar results can be obtained for a wide range of stick lengths.

The choice of the right parameter combination, however, is very important. If the amount of multiplicative noise is too low, the system just fluctuates around the position where it ends up in the noiseless case. If the noise is too high, the stick escapes immediately. If the friction is too low, the system oscillates. Noise then feeds energy towards these oscillations, causing them to grow over time as shown in figure 6.7 (a). This stick quickly falls, followed by wild fluctuations which are not expected in the real world – the stick would turn upside down and back up. This can be understood by considering the structure of (6.3): even if the control feedback were not delayed, a frictionless system could at best oscillate, but never be stabilised. Friction is necessary to remove kinetic energy from the system. Therefore, very strong friction is a prerequisite for the multiplicative noise to have a stabilising effect.

Further, a just barely overdamped stick will quickly either reach the upright position or fall down. As shown in figure 6.7 (b), the model becomes stationary once the stick reaches the upright position. This is due to the lack of a re-injection mechanism like the additive noise in the Kesten process (section 4.4). The system is always either stable or unstable and all observable dynamics are only transient. Therefore, ongoing activity over more than few seconds requires both, high friction and precise tuning to the stability boundary. No mechanism has been proposed for the self-tuning of the system.

Despite the importance of the high damping in this model, its origin has–to this author's knowledge–not been discussed in detail before. This circumstance seems quite peculiar since the stick by itself is most certainly not highly damped. In other situations such as upright standing, muscle reflex loops can act as dampers on shorter timescales than the visuomotor control loop. A stick on the fingertip, or even on a Ping Pong racket as used in some task variations [CM12], only loosely sits on the support point with one end while the other end moves freely. In section 7.8 it is shown how this problem can be solved for the modelling approach presented in the next chapter.

A similar model to the one presented above is found in [BCME04], but that model can per definition only be unstable or oscillate. It also uses highly unrealistic parameters and is therefore not discussed any further.

Taken together, the above model yielded some insights in the dynamics of marginally stable or unstable systems with noise. However, difficulties to reproduce the statistical features of section 6.2, open questions regarding parameter choices, and few connections to the motor control literature reviewed in chapter 5, remain. Therefore, in the following chapters, we take a slightly different approach which we motivate in detail from the totality of the insights presented so far.

# 7. Modelling critically adaptive control

Here we start over with a new perspective on balancing. Observing a stick as it starts to fall allows for predicting where it is going to fall. A controller can then stabilise the stick by moving its suspension point. However, bringing the stick back to the upright position removes all predictable dynamics from the system. Then, it is only possible to observe residual unpredictable noise like hand jitter or wind. In this situation an adaptive controller becomes very susceptible to missteps because the deterministic movement of the stick cannot be clearly discerned from the noise.

We can formulate this idea as a quite general physical principle [PP11]: Stabilisation of a dynamical system annihilates observable information about its structure. This mechanism can induce critical points as attractors in locally adaptive control.

In the following sections, we will first explain why this new approach to balancing models is called for. Then, the underlying mechanism is characterised in detail. Next, a complete balancing model is introduced. It involves a reaction time, an unbiased short-term parameter estimator, and an optimal forward model based on this prediction. In short, subjects move their hand in the direction where they, based on their most recent observations, predict the target to move next. This model can quantitatively reproduce the features of human balancing behaviour discussed in the previous chapter. The model further suggests that power-law error distributions arise because subjects minimise mean control errors by eliminating random local trends at the cost of rare, extreme errors [PP11]. This hypothesis is tested in chapter 8. This chapter will conclude with a more detailed investigation of the power law fits and their significance, as well as a higher order model. The latter, however, primarily serves to justify the simpler model mentioned above.

## 7.1. Motivation

Most motor control models add multiplicative noise in the movement execution stage. Some studies even suggested that movement variability arises mostly due to the organisation of the motor unit pool, but for many tasks the situation is more complicated (see sec. 5.2). In stick balancing (chapter 6), erratic bursts of fluctuations and high movement variability hint at amounts of multiplicative noise that seem difficult to justify with the few percent of signal dependent noise measured during force production (e.g. in [JdCHW02]). Furthermore, if the observed noise were an immutable trait of movement generation, fluctuations like those observed during balancing should occur when similar movements are performed without actually balancing a stick. This is apparently not the case (see sec. 6.1), although the author is not aware of any rigorous comparisons published as of yet. Studies on other tasks like tracking (e.g. [ES00]) did not report similar extreme bursting fluctuations. In chapter 8, it is confirmed that the signal-dependent noise is indeed task dependent.

A different hypothesis than motor unit noise was proposed in the existing literature on stick balancing (sec. 6.1). Parametric noise can, in specific conditions, prolong escape times for unstable systems. It has therefore been argued that the system is tuned closely to the boundary of a stability domain across which it fluctuates due to parametric noise. This implies that subjects add multiplicative noise to increase the chance of catching the stick outside the basin of attraction where otherwise it would be certainly lost. However, if this intention were indeed the only reason for the strong multiplicative noise, one would expect the latter to vanish in scenarios where subjects can easily maintain control using delayed, but planned movements. This is not the case: even in VSB tasks where subjects can easily maintain control for minutes, we still observe signs of strong multiplicative noise (sec. 6.2).

It has even been reported that error distributions are more heavily tailed for easy tasks than for difficult ones (sec. 6.2.3). This raises the question of why extreme fluctuations, which pose the risk of losing control, are not avoided.

Furthermore, the corresponding model by Cabrera and Milton (sec. 6.3) cannot explain how the system stays close to the stability boundary: if the simulated stick is brought in the upright position once, it stops moving (figure 6.7 (b)). This is not a big problem in the tasks for which this model was originally devised: difficult tasks where control can only be maintained over short periods of time. Yet the model cannot explain why subjects in easier tasks should act as if they were barely able to maintain control. Most importantly, however, this model cannot reproduce the scaling relations discussed in section 6.2 and requires some hard-to-justify parameter choices (see sec. 6.3). Another more technical problem is that since the intermittent behaviour in this model is only transient, a stationary distribution of fluctuations does not exist. In conclusion, previous studies on human stick balancing found that multiplicative noise and reaction delays play a central role, but could not provide a complete and convincing model based on these findings.

In the following we present a new class of models where the main source for parametric noise is the online estimation of parameters which are then used for movement planning. Justification for this approach includes evidence that human motor control is predictive, and that humans are often capable of adapting to quickly changing environments by rapidly extracting information from their environment as needed (see [MS02, PREP07], and chap. 5). The basic structure is illustrated in figure 7.1: Subjects remove the predicted control errors based on an online estimation of the systems dynamics.

In the next section, it is shown that such a system can self-organise towards a critical state. In section 7.4, we show that the simplest realistic model of this class can solve all the problems stated above and explain VSB data in much detail.

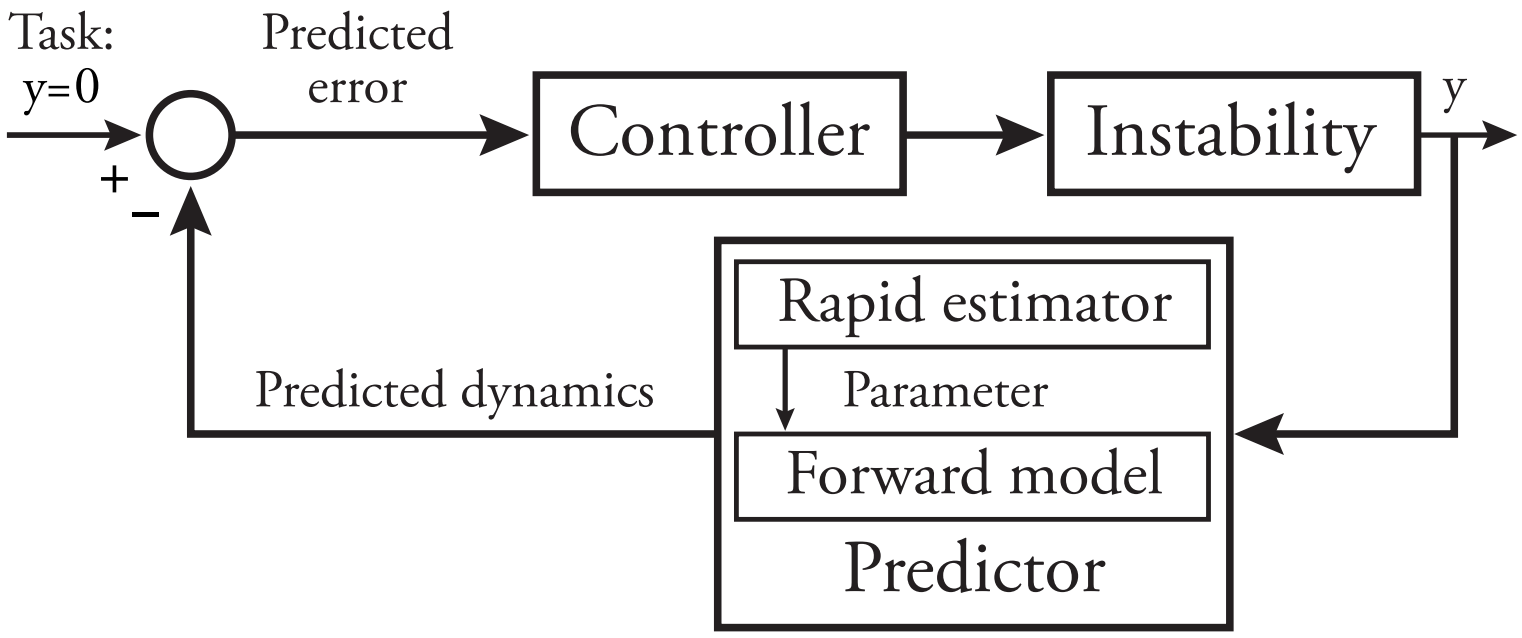

Figure 7.1.: Standard models of the action-perception cycle (sec. 5.1) include forward models allowing subjects to predict the future dynamics of their environment. In the following it is demonstrated that including rapid online estimations of parameters can explain the strong multiplicative noise observed in balancing tasks.

## 7.2. Information Annihilation Instability (part one)

A predictive and highly adaptive controller, as described above, can run into an instability because successful control annihilates observable information about the controlled system: it is only possible to predict where a stick is going to fall once it starts falling. When the stick is brought into the upright position, all predictable dynamics are removed, leaving only random fluctuations to observe. This principle maps small control errors to large uncertainties about future dynamics and vice versa, thereby inducing self-similar fluctuations. It can be illustrated in a simple linearised example (published in [PP11]). Consider a system with expected dynamics without control

$$\dot{y}(t) = \vartheta y(t). \tag{7.1}$$

$y(t)$ denotes the system's deviation from some target value (e.g. a stick's upright position) and $\vartheta$ is a hidden parameter[1]. Assume that the system is observed at a given location $y$. The observer has access to noisy observations of $y$ and $\dot{y}$ with respective probability distributions $p(\dot{y})$ and $p(y)$. The noise may be either inherent to the system or to the measurement process. The likelihood function of $\vartheta$ given an observation at the location $y$ is

$$\mathscr{L}(\vartheta|y,\dot{y}) = p(y,\dot{y}|\vartheta) \quad (7.2)$$
$$= p(\dot{y}|y,\vartheta)\,p(y|\vartheta) \quad (7.3)$$
$$= p(\dot{y}|y,\vartheta)\,p(y). \quad (7.4)$$

[1] The following considerations can be transferred also to higher-order systems if the derivative at the fixed point is zero. This is demonstrated in section 7.8

Further assume, that $\dot{y}$ at the observed location is Gaussian distributed:

$$p(\dot{y}\,|\,y,\vartheta) \propto \mathcal{N}(\vartheta y, \sigma_{\dot{y}}). \quad (7.5)$$

Maximizing the log-likelihood with respect to $\vartheta$ gives the unbiased estimator[2]

$$\tilde{\vartheta} = \frac{\dot{y}}{y}\,. \quad (7.6)$$

The expected amount of information about $\vartheta$ that an observation contains may be expressed using Fisher information:

$$\mathscr{I} = -\mathrm{E}\left(\frac{\partial^2}{\partial\vartheta^2}\ln\mathscr{L}\middle|\vartheta\right) = \frac{\mathrm{E}(y^2)}{\sigma_{\dot{y}}^2}. \quad (7.7)$$

The MSE of the estimator is given by the Cramer-Rao bound

$$\mathrm{Var}(\tilde{\vartheta}) \geq 1/\mathscr{I} \quad (7.8)$$

which represents an uncertainty principle [FS95]. When observing the system at the origin such that $\mathrm{E}(y^2) \to 0$, the susceptibility of the estimator to random fluctuations diverges. Hence, stabilizing control added to equation (7.1) evolves the system towards a critical point.

[2] The proof is found in sec A.11. (7.6) also follows from equation (7.13) or alternatively if the independent variable is $\dot{y}$. Equation (7.5) implies, that $y$ will be controlled for.

Consequent control errors could be reduced using additional independent observations. However, to do this optimally the controller has to know a priori the exact form of a possible state- or time dependency in $\vartheta$. In the following, we settle for the minimal assumption that $\vartheta$ can be considered constant over a small set of subsequent observations.

## 7.3. A minimal model

> “With four parameters I can fit an elephant, and with five I can make him wiggle his trunk.”
>
> John von Neumann

To underline the generality of Information Annihilation Instability (IAI) and to understand how it can be realised in an actual dynamics, we here study the most simple example. Consider a random map where dynamics, control and observation of the system (7.1) take place only at discrete times $t \in \{1,2,3,\ldots\}$. Then,

$$y_{t+1} = \alpha y_t + \beta_t, \tag{7.9}$$

with parameter $\alpha$ and Gaussian distributed independent random variables $\beta_t \sim \mathcal{N}(0,\sigma)$. Control by removing a prediction of $y_{t+1}$ in time-step $t+1$ gives

$$y_{t+1} = (\alpha - \tilde{\alpha}_{t+1}) y_t + \beta_t. \tag{7.10}$$

The estimation $\tilde{\alpha}_{t+1}$ that minimises the expected error given the two most recent observations $\mathrm{E}(y_{t+1}^2 | y_t, y_{t-1})$ is

$$\tilde{\alpha}_{t+1} = \frac{y_t}{y_{t-1}} + \tilde{\alpha}_t. \tag{7.11}$$

We here assumed, that $\alpha$ can be considered constant over the span of two observations, but that nothing else is known about its time- or state dependency. Equations (7.10) and (7.11) represent a minimal adaptive control system with a restricted memory. It can also be written as

$$y_{t+1} = -\frac{y_t}{y_{t-1}}\beta_{t-1} + \beta_t \tag{7.12}$$

revealing that the dynamics are independent of $\alpha$, and that $\beta$ only determines the absolute scale of the fluctuations, not the relative relation of subsequent values of $y$. A time-series is shown in figure 7.2 (a). The corresponding CCDF is shown in figure 7.2 (b). It was shown analytically that the tail of the distribution obeys a power-law $P(|y| > x) \propto |y|^{-\xi}$ with exponent $\xi = 1$ independently of $\alpha$ and $\beta$ [PREP07]. Using $m \geq 2$ time-steps for the estimator numerically yields $\xi = m - 1$ (ibid.). The PSD is mostly white, except for a small dip at very high frequencies since the magnitudes of two subsequent $y$ are correlated (see eq. (7.12)).

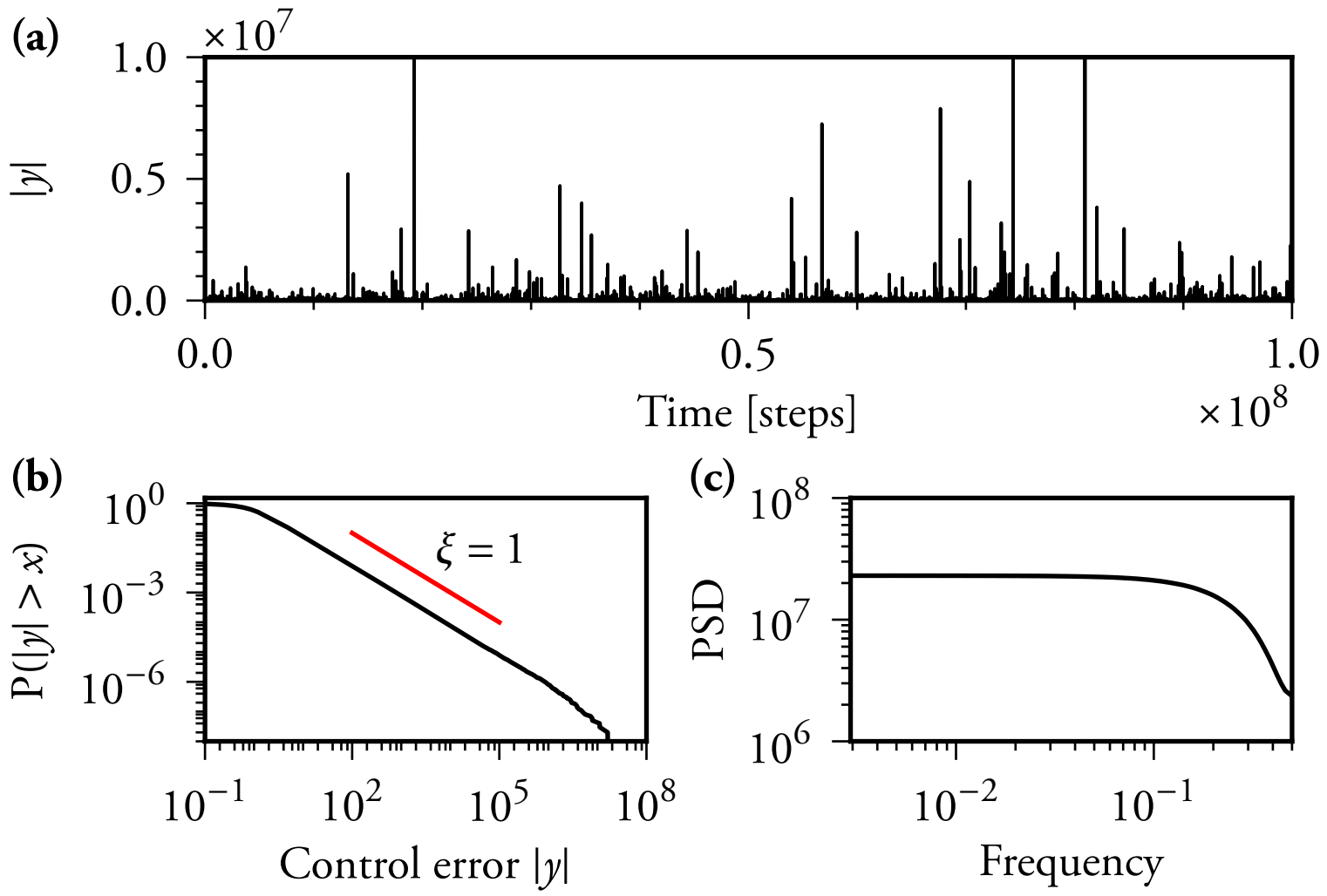

Figure 7.2.: Analysis of the minimal adaptive control model exhibiting IAI. Var($\beta$) = 1. (**a**): time-series. (**b**): CCDF and analytical exponent. (**c**): PSD. Note that the variance for this model diverges (see sec. 2.3). Therefore, also the power per frequency increases with the length of the time-series.

This simple control mechanism yields critical behavior without the need of parameter tuning [EP05]. It further demonstrates that multiplicative noise can be the result of locally optimal parameter estimation. These findings foreshadow a solution to the problems specified in section 7.1. However, the model is still very abstract, the CCDF exponent is too low, and the power spectrum lacks the characteristic shape found experimentally. These problems are solved in the next section.

## 7.4. A realistic model

Many real control systems including human motor control are not time-discrete and movements are subject to limitations like maximum forces. Therefore, we now consider the continuous control problem posed by the stochastic differential equation

$$\dot{y}(t) = \frac{1}{\tau}\, y(t) + \beta(t) \tag{7.13}$$

where fluctuations grow exponentially with time constant $\tau$. $\beta(t)$ is Gaussian white noise, i.e. $\mathrm{E}(\beta(t)\beta(t')) = \sigma^2\delta(t-t')$. A real controller has a finite reaction time making stabilization non-trivial. It has to remove a prediction $\tilde{y}(t)$ of $y(t)$ based on observations only up to some earlier time $t - t_r$. Furthermore, a controller cannot remove $\tilde{y}(t)$ from equation (7.13) completely and instantly without reaching infinite velocities. Instead, it may continuously remove a term proportional to $\tilde{y}(t)$. To stabilize the system, the proportionality factor has to be bigger than $1/\tau$. Thus, we get

$$\dot{y}(t) = \frac{1}{\tau} y(t) - \gamma\, \tilde{\vartheta}(t)\; \tilde{y}(t) + \beta(t) \tag{7.14}$$

with $\tilde{\vartheta}(t)$ as estimator for $1/\tau$ and a gain factor $\gamma > 1$. Since the controller has already determined its own actions for all times $t' < t$, the probability density $p(y(t') \mid y(t-t_r))$ is a Gaussian whose mean evolves according to equation (7.14), dropping $\beta(t)$. Solving for $y(t)$ with known actions for $\{t' \mid t-t_r \le t' < t\}$ and initial condition $y(t-t_r)$ yields the prediction

$$\tilde{y}(t) = \mathrm{E}\big(y(t) \mid y(t-t_r),\ \gamma,\ \{\tilde{\vartheta}(t')\}, \{\tilde{y}(t')\}\big) \qquad (7.15)$$

$$= e^{\tilde{\vartheta}(t)\,t_r}\Big(-\gamma \int_{t-tr}^{t-0} e^{\tilde{\vartheta}(t)\,(t-t_r-t')}\tilde{\vartheta}(t')\tilde{y}(t')\ \mathrm{d}t' + y(t-t_r)\Big).$$

We now focus on an estimator for the hidden parameter $1/\tau$. The exact continuous record log-likelihood function [PY09] for equation (7.14) can be derived analytically:

$$\begin{aligned}\ln \mathscr{L}(1/\tau) = \int_{t_0}^{t} \frac{y(t')}{\tau\sigma^2}\Big(\dot{y}(t') + \gamma\tilde{\vartheta}(t')\,\tilde{y}(t')\Big)\mathrm{d}t' \\ -\frac{1}{2}\int_{t_0}^{t} \frac{y(t')^2}{\tau^2\sigma^2}\mathrm{d}t'.\end{aligned} \qquad (7.16)$$

Since we are interested in the drift without control, the bracket in the first term contains the observed velocity minus the controller's contribution. Maximising equation (7.16) with respect to $1/\tau$ yields the estimator:

$$\tilde{\vartheta}(t+t_r) = \frac{\int_{t_0}^{t} y(t')\left(\dot{y}(t') + \gamma\tilde{\vartheta}(t')\,\tilde{y}(t')\right)\mathrm{d}t'}{\int_{t_0}^{t} y(t')^2\ \mathrm{d}t'.}. \qquad (7.17)$$

equations (7.14), (7.15), (7.17) define a delayed predictive continuous control system. By setting $t_0$ to $t - t_m$ in equation (7.17), we can restrict the integration window to an interval of fixed length $t_m$. While this constraint is sufficient to induce criticality [PP11], we here introduce exponential forgetting to better approximate real forgetting curves (see also subsec. 9.6.3). Keeping $t_0$ fixed e.g. at $-\infty$, exponentially decaying factors $\exp(t'/\tau_m)$ with time constant $\tau_m$ under both integrals in equation (7.17) create a smooth shifting integration window. The numerator and denominator can then be expressed in differential form:

$$\begin{aligned}\dot{A}(t) &= -A(t)/\tau_m + \big(\dot{y}(t) + \gamma\, \tilde{\vartheta}(t)\; \tilde{y}(t)\big)\, y(t) \\ \dot{B}(t) &= -B(t)/\tau_m + y(t)^2 \\ \tilde{\vartheta}(t+t_r) &= \frac{A(t)}{B(t)}. \end{aligned} \tag{7.18}$$

This form of the estimator is essentially the quotient of two lowpass filters. It is both easier to implement numerically, and biologically more plausible than equation (7.17). Note, that $B$ will always be positive if $y \not\equiv 0$.

The model investigated in the following is defined by equations (7.14), (7.15), and (7.18). It is the continuous-time equivalent to the minimal control model (eq. (7.12)). This can be shown by means of a limiting case using a stroboscopic mapping (sec. A.12). The variance of the driving noise $\sigma$ still only determines the absolute overall scale of the dynamics and therefore is only relevant for absolute errors, not for scaling relations. Yet the introduction of a reaction delay breaks the invariance with respect to the task difficulty (determined by $\tau$) which was found for the minimal model. Formally, one could renormalise the model such that all times and time constants are defined relative to $\tau$ to recover the invariance. However, since we are interested in scenarios where the experimenter sets $\tau$ and the subjects CNS determines the other parameters–some of which may be biologically predetermined–the model in the following is discussed as a controller with three parameters facing a task determined by one (external) parameter.

## 7.5. Comparison with experimental data

A time-series for the model described in the last section is shown in figure 7.3. The model generates a stable distribution of extreme events of similar magnitudes as the experimental data shown in figure 6.5. The model time-series further quantitatively reproduces the main scaling features found in the VSB data-set.

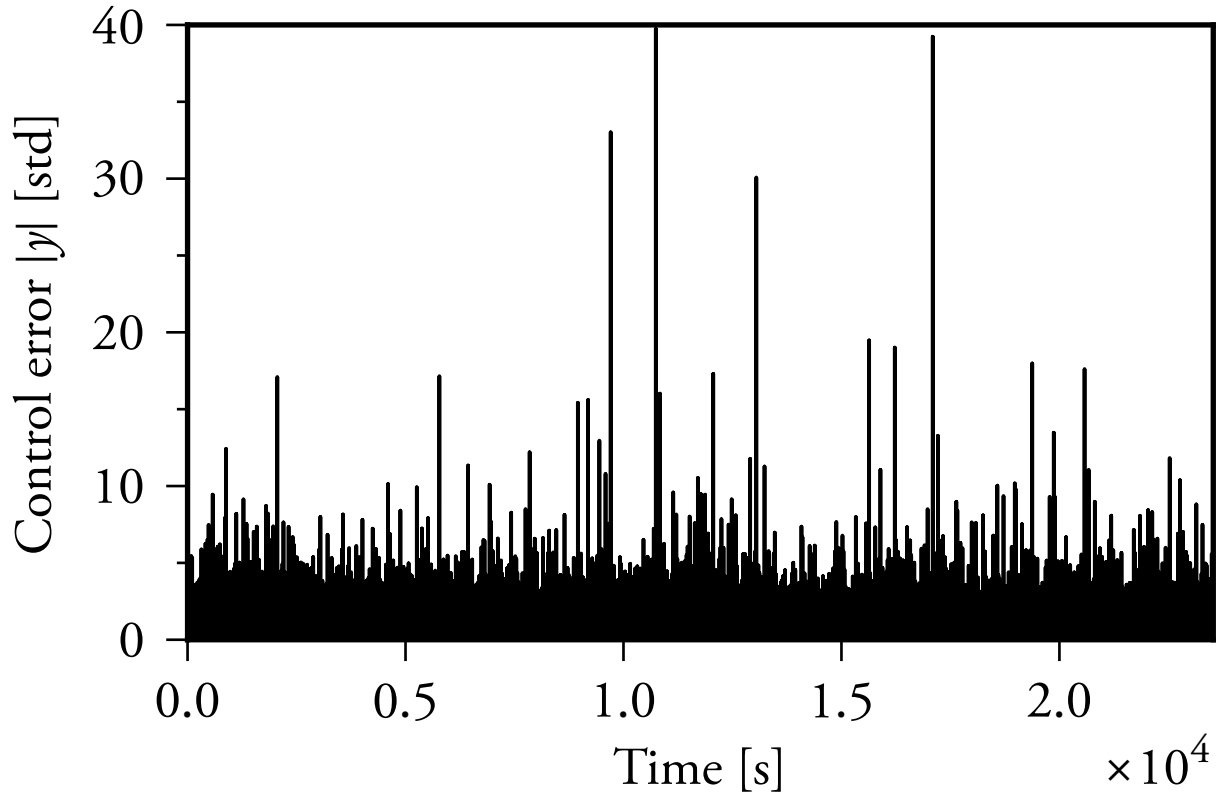

Figure 7.3.: time-series of normalised control errors for the continuous model. The simulation was performed for $\tau = 250\,\mathrm{ms}$ which is close to the average over all experimental conditions described in sec. 6.2. Controller parameters were $t_r = 180\,\mathrm{ms}$, $\gamma = 1.1$, $\tau_m = 120\,\mathrm{ms}$ ($\sigma$ is irrelevant for normalised errors). Time discretization: 85Hz.

Figure 7.4 (a) shows the CCDF. Small fluctuations are dominated by the additive baseline noise while apparent deviations from the power law for large fluctuations are caused by temporal correlations only (see sections 6.2 and A.1). Figure 7.4 (b) shows the PSD. The model exhibits a characteristic broken power law which is also found for the VSB experiments.

The combination of CCDF and PSD scaling allows to narrow down the plausible parameter range for the model. This facilitates a deeper understanding of the experimental observations: The model predicts reaction times comparable to simple visual tasks (see sec. 5.1), an even slightly faster adaptation to observed trends, and cautious movements that avoid overshooting.

In this parameter range, the exponent $\xi$ is to a large extent determined by the controller's memory $\tau_m$. The faster the controller adapts, the more heavy-tailed the distribution of control errors becomes. In addition, there is some interaction with other parameters: Longer reaction times also decrease $\xi$, as do higher gains $\gamma$. However, the latter dependence is very weak unless $\gamma$ gets close to 2. For much larger $\gamma$, the system becomes uncontrollable.

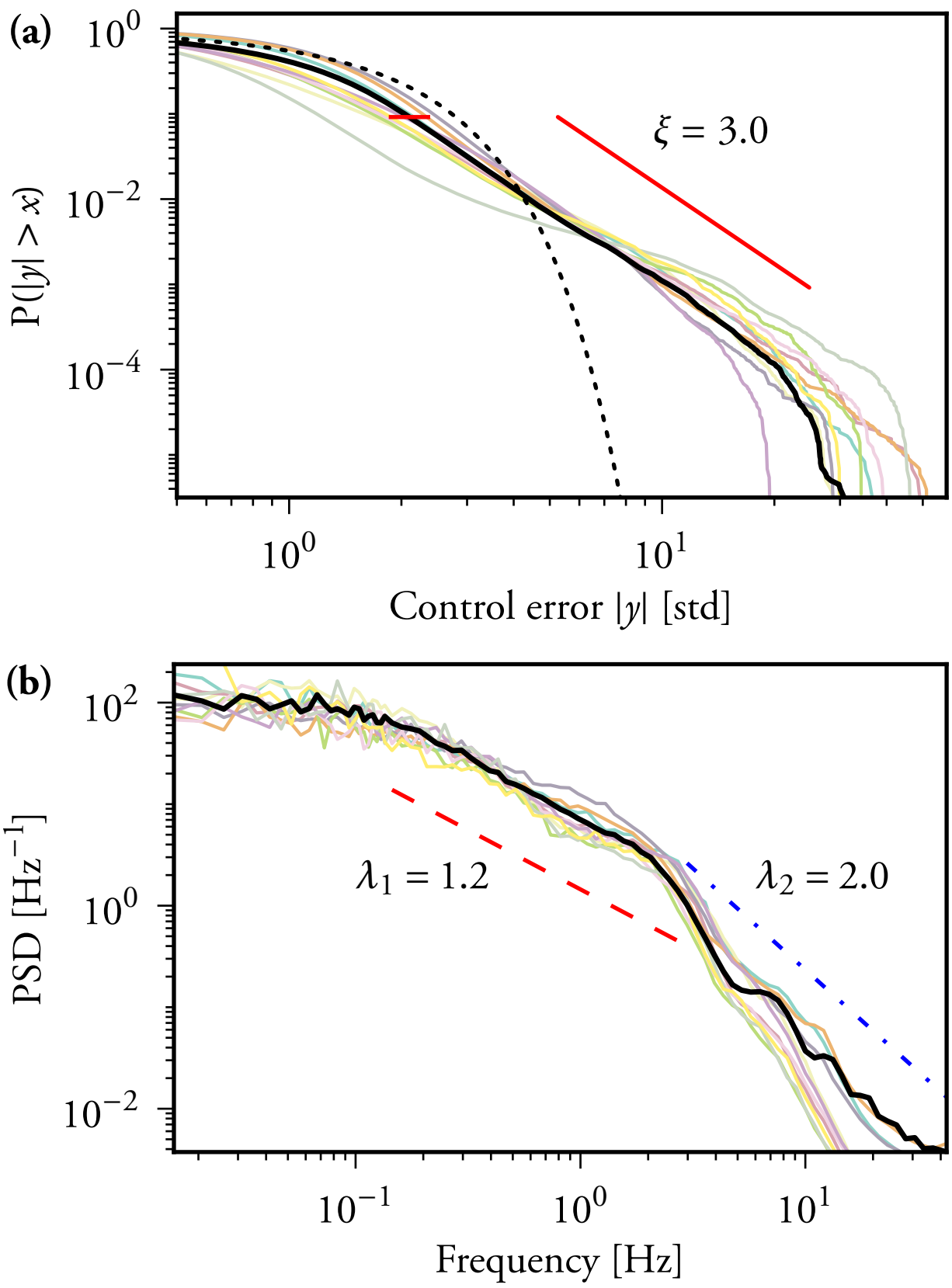

Figure 7.4.: Comparison of normalised control errors for the model (thick black lines) with the VSB time-series for several subjects (light colored lines), for each of which combined trials of several days totalling in several hours of data are shown. Trials with constant but different $\tau$ for the same subjects are kept separate. The simulation is the same time-series shown in figure 7.3. **(a)**: CCDF. Diagonal line: power-law fitted to the simulation (see section 2.5). Dotted line: Gaussian. **(b)**: Power Spectral Density and scaling exponents $\lambda_{1,2}$ were estimated like in figure 6.5.

Realistic exponents $\lambda_1$ are found for a combination of controlling cautiously with $\gamma$ just above one and adapting fast with $\tau_m \leq t_r$. This strongly reduces correlation strengths on time scales less than a second, but leaves small correlations that persist over a few seconds. The transition from the constant super-low frequency response to low frequency scaling with exponent $\lambda_1$ depends mostly on $\gamma$: Lower gains lead to longer persisting correlations and therefore shift the scaling onset towards lower frequencies. There is, however, some interaction with $\tau_m$ which becomes relevant later in this thesis.

The knee position depends on the reaction time. $t_r$ in between 170 and 200 ms yield good fits.

The high frequency scaling regime represents frequencies above the controller's active response, with $\lambda_2$ being mostly parameter independent. Only the onset of this regime close to the knee depends on parameters, which can slightly influence the fitted exponents[3]. $\lambda_2$ is also the only measured quantity where the model as it is discussed here inevitably lies at the very edge of the experimentally observed range. However, passive damping increases $\lambda_2$. It can be included using either of the model extensions discussed in section A.4 and section 7.8. High frequency ripples are found for several subjects in combined and single-trial time series. These signatures are characteristic and subject-dependent. In the model, the ripples are more pronounced. However, if model time-series with slightly different parameters are combined, the ripples are greatly reduced [PP11]. In addition, also the aforementioned model extensions can reproduce a high frequency regime with less ripples.

[3] Fitting only very high frequencies that are truly unaffected is not possible for the experimental data. For the model, an example is shown in section A.4

The effects presented above are robust to parameter changes including $t_r \to 0$. $\tau$ may also be time dependent. Introducing a time dependent $\sigma$ creates additional higher order temporal correlations. State dependencies reduce the increase of $\xi$ with $\tau_m$ and cause additional clustering of control errors.

Based on CCDFs and PSDs, we can rule out other parameter regimes, and several model modifications. Examples include multiplicative execution noise without adaptation (sec. A.3), and control with a high gain (sec. A.2).

## 7.6. Why do subjects adapt rapidly?

The estimated values of two of the three parameters of the controller model can be well justified from the existing literature (chap. 5). If subjects increased their gain, one would expect that all planning errors and the signal dependent noise in muscle activation should be scaled up linearly. A small gain further minimises the risk of overshooting or oscillation buildup in delayed controllers. All of this is consistent with established findings like the well known Fitts' Law [Fit54] which states that faster movements become less precise.

Second, even though so far we could only indirectly measure reaction times, findings are at the lower limit of biological restrictions. Therefore, it is unlikely that subjects could react even faster. All else equal, there are no obvious benefits associated with reacting slower.

This leaves us with explaining the memory time constant $\tau_m$. Why should subjects adapt to trends on such short time scales if the task parameter is constant? This problem is resolved by a minimum in the variance of $y$ found for memory time constants $\tau_m$ close to the reaction time $t_r$. Figure 7.5 (a) shows the variance in dependence of the ratio between reaction time and memory length for different reaction times. For small $\tau_m$, the variance diverges. The minimum's exact position is parameter-dependent: the optimal $\tau_m$ increases slightly slower than $t_r$.

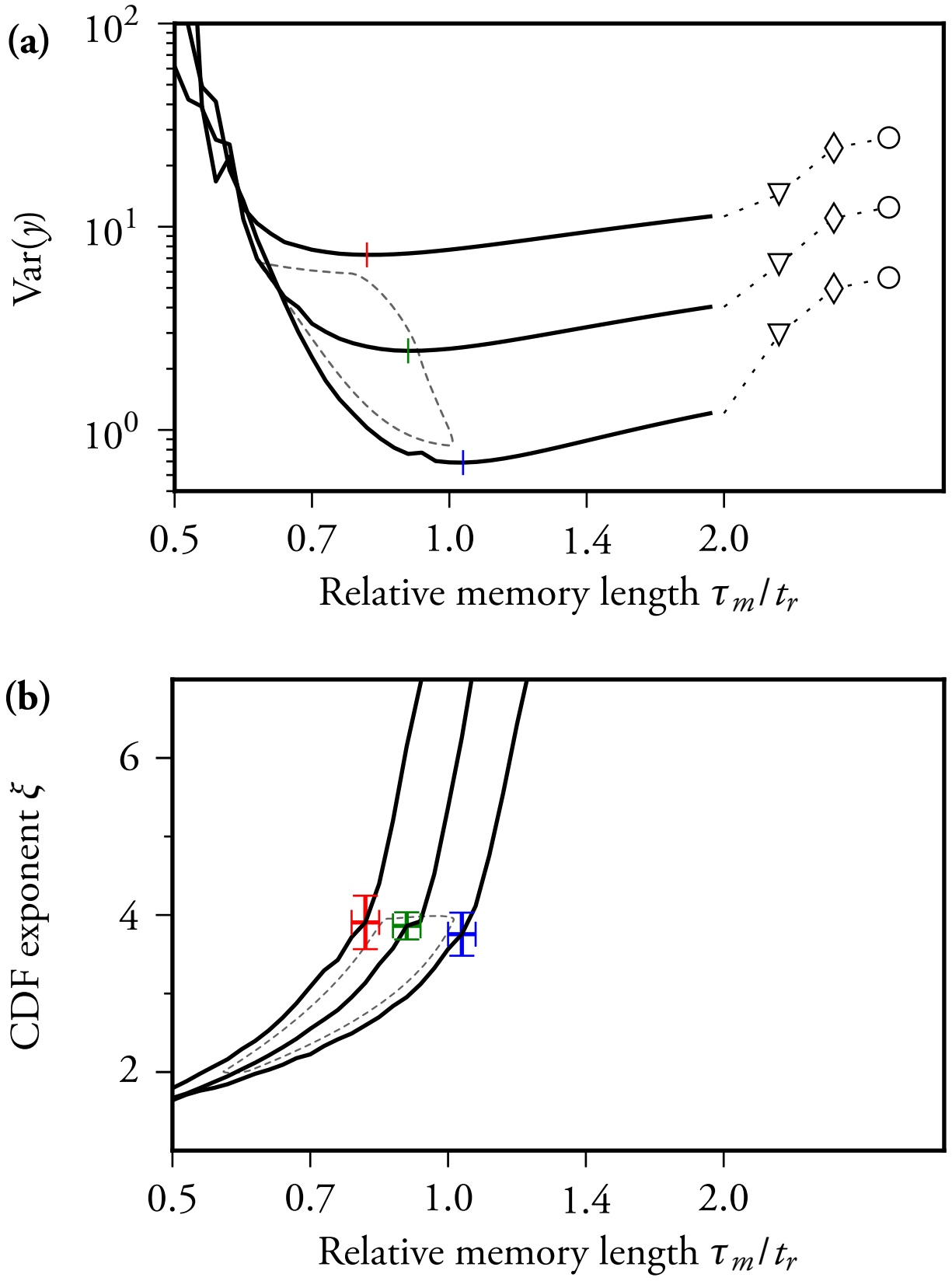

Figure 7.5.: Statistical properties depending on the ratio between the memory time constant $\tau_m$ and the reaction time $t_r = 300$, 200 and 100 ms (different curves top to bottom). $\tau = 250$ ms, $\sigma = 1\,\mathrm{s}^{-0.5}$, $g = 1.05$. Curves are averages from 50 simulations of length $2 \cdot 10^7$ s with discretisation step 10 ms. Dashed contours: possible range for VSB subjects. **(a)**: Variances correspond to mean squared control errors. Vertical lines: minima. Symbols: $\tau_m = 1$ s (triangle), 10 s (diamond), a controller using the true $\tau$ instead of equation (7.18) (circle). **(b)**: Corresponding exponents $\xi$. Error bars: positions of minimal variances.

Figure 7.5 (b) shows the corresponding CCDF tail exponents $\xi$. Positions where variances are minimal are marked with error bars. Here, $\xi$ is just below four for all conditions. However, because correlation lengths increase with $t_r$ and $t_m$, distributions may appear different for limited data set sizes.

Summarising the above two paragraphs, fast adaptation minimises mean balancing errors by tolerating rare, large errors in favour of the removal of random trends (figure 7.5). Corresponding CCDF scaling exponents lie in the highest range of those observed experimentally. Therefore, subjects perform quite close to being optimal in the sense of minimising the mean distance between target and controller given the inevitable presence of a reaction delay and some baseline noise. This hypothesis leads to several predictions which are investigated experimentally in chapter 8. Until the end of this chapter, however, we will first investigate two more aspects of critically adaptive control.

## 7.7. Significance and variation of power-law distributions

A commonly asked question for heavy-tailed empirical distributions is whether they truly follow a power-law or just look approximately linear on a log-log plot. This scrutiny stems from the difficulty to discern power-laws from other heavy-tailed distributions in smaller data sets. In short, we need surrogate data to which we can compare the data that is to be tested. This surrogate data is generated such that it follows a true power-law. It further shares the shape of the cutoff, the steepness, and also temporal correlations with the data to be tested[4] [PP11]. There is no standard way to generate such surrogate data, but luckily we have a model with closely reproduces these features for the experimental VSB time series. The model is rigorously tested in a two-step procedure in section A.1 and found to follow a true power law. Therefore, testing whether power-law fits to the experimental time-series perform worse in the Kolmogorov-Smirnov (KS)-test than fits for the model tells us if deviations of the experimental CCDFs from an analytical power law could have happened by chance. As shown in figure 7.6, we cannot reject this null hypothesis for all but one sample. The latter consists of the combined trials from 5 days for one subject. Analysing the trials for each day separately does not lead to a rejection of the null hypothesis. Therefore, we can conclude that VSB control errors generally cannot be distinguished from a truly power-law-tailed distribution. More details, including the analysis of single trials, are found in the supplement to [PP11], where the data sets were grouped slightly differently.

[4]A similar method is used in [CSN09] but no time-series were analysed. Nevertheless, the authors claim broad applicability of their methods, despite ignoring the possibility of correlated measurements. To this authors knowledge, the methods presented here represent the only attempt in the literature to rigorously test for power laws in time-series and other data sets where measurements are not perfectly independent.

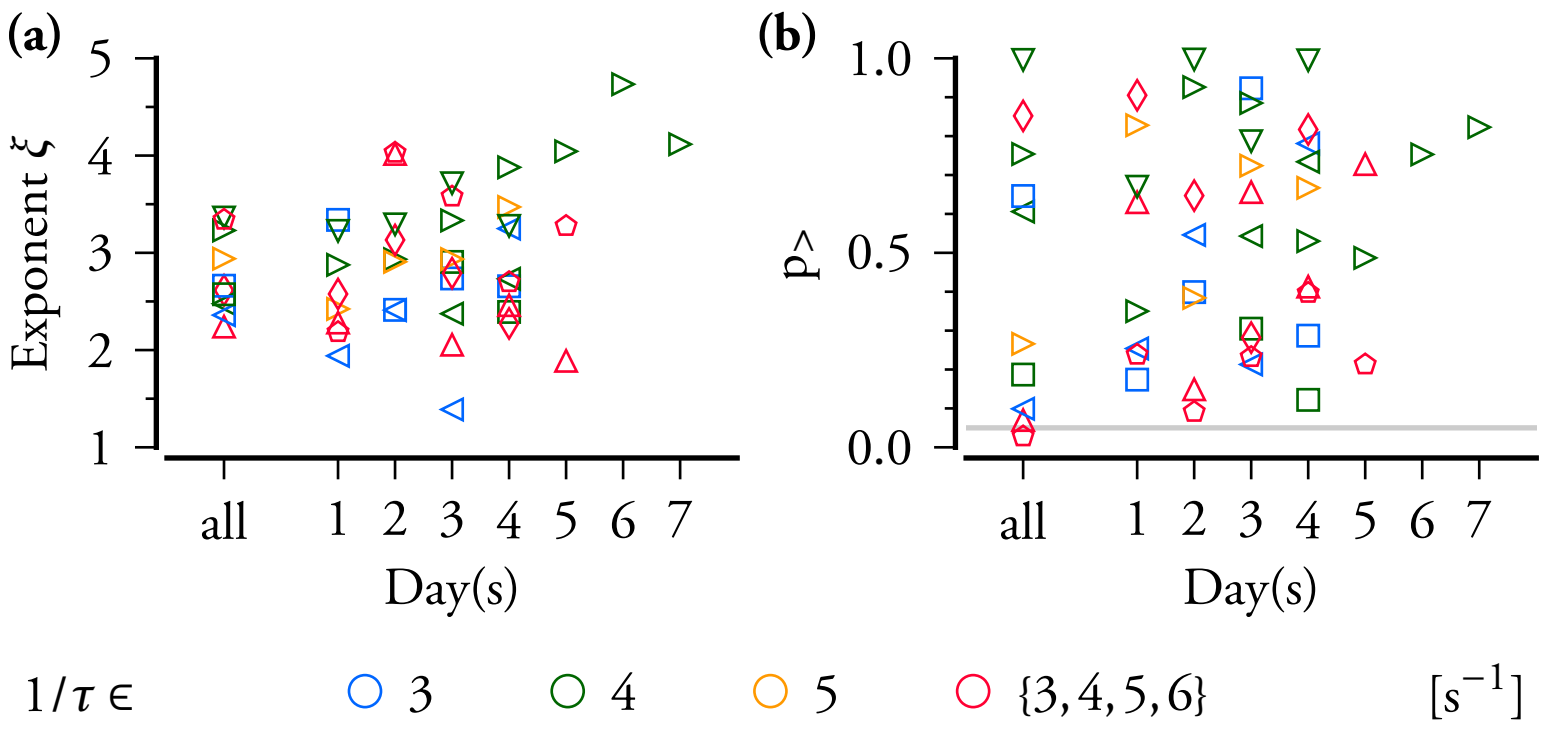

Figure 7.6.: CCDF fits for the combined trials of each subject for all days, and for each day separately. Each subject is marked by a different symbol, and each condition by a different color as indicated above. Some subjects recorded sets of trials with different fixed $\tau$ on the same day. **(a)**: CCDF exponents. Only one subject (triangle pointing to the right) shows a clear trend, all others don't. **(b)**: Probabilities, that the KS-statistics for power-law fits for the model (with parameters as in fig. 7.4) are worse than for experimental time series. As expected if control errors for model and experimental data were distributed equally, goodness of fits scatter around 0.5. Grey line: significance level $p_{>} = 0.05$.

On a note related to figure 7.6 (a), the coloring of the trials highlights a possible reason as to why no significant trends over time could be detected: Subjects with constant $\tau$ during each trial performed at different difficulties and sometimes in two different tasks on the same day. To achieve a higher discriminability, experiments with more subjects under identical conditions are required (see chap. 8).

## 7.8. Second order control

So far we only studied the simplest possible balancing task and the simplest possible model for movement generation. That is, equation (7.14) describes a first order dynamical system where predicted control errors decay exponentially. This matches the VSB experiments shown in section 7.5 since they featured an equivalent instability. Mechanical systems, however, follow second-order dynamics. Therefore, real hand movements eventually require the generation of forces. Furthermore, very similar results have been reported for other VSB experiments that did feature second order dynamics, as well as for real stick balancing tasks (see chap. 6). Hence, a demonstration of IAI in a second order system seems warranted, as does a justification for as to why a first order description may suffice in certain situations.

To investigate the effect of second-order dynamics, consider control of an inverted pendulum linearised for small angles[5] and perturbed by Gaussian white noise $\beta$:

$$\ddot{y}(t) = \vartheta y(t) + c(t) + \beta(t) \tag{7.19}$$

$$c(t) = -\gamma_a \tilde{\vartheta}(t)\tilde{y}(t) - \gamma_v \tilde{\dot{y}}(t), \tag{7.20}$$

[5] This simplification well-known from the "mathematical pendulum" should be unproblematic since successful balancing of a real stick presumably involves small angles. However, introducing a nonlinearity is straight forward.

where again the system state is predicted based on observations that are delayed by the reaction time, and on all planned control actions $c(t')$ for the intermediate times $\{t' \mid t - t_r \leq t' < t\}$:

$$\begin{aligned} \tilde{y}(t) &= \mathrm{E}\big(y(t) \mid y(t-t_r), \dot{y}(t-t_r), \{c(t')\}\big) && (7.21) \\ \tilde{\dot{y}}(t) &= \mathrm{E}\big(\dot{y}(t) \mid y(t-t_r), \dot{y}(t-t_r), \{c(t')\}\big). && (7.22) \end{aligned}$$

The system parameter is again estimated using a maximum likelihood estimator with exponentially decaying memory

$$\begin{aligned} \dot{A}(t) &= -A(t)/\tau_m + \big(\ddot{y}(t) - c\big)\, y(t) \\ \dot{B}(t) &= -B(t)/\tau_m + y(t)^2 \\ \tilde{\vartheta}(t+t_r) &= \frac{A(t)}{B(t)}. && (7.23) \end{aligned}$$

Equation (7.20) features one notable difference with respect to the first order case equation (7.14): two different control terms with separate gain factors. This is necessary for two reasons. First, imagine that the controller would only counteract the acceleration due to the instability in equation (7.19). Then, given finite initial $y$ or $\dot{y}$, the system would oscillate and never be able to stop. Second, the term proportional to the velocity acts like friction. It removes kinetic energy from the system and can bring it to halt, but it cannot move the system back towards the origin. Hence, successful stabilisation requires both counteracting the instability and the removal of kinetic energy.

Figure 7.7 shows distributions and spectra for different control regimes. Power-law CCDFs are found for a large parameter range (figure 7.7 a, c, e). However, the tail exponent here depends on the gain parameters in a more complex way than in the first order model.

For underdamped control, the system shows strong oscillations (b). These oscillations are suppressed when $\gamma_v$ is increased (d). Further increasing $\gamma_c$ slows down the compensation of control errors, which creates more long-ranged correlations. This effect is demonstrated by a shift of the onset of the low-pass filtering in (f) towards lower frequencies than in (d). The second order model for many parameter sets exhibits a steeper high frequency scaling than the first order one–even steeper than those observed for most subjects in VSB experiments. For the combination of both high $\gamma_a$ and $\gamma_v$, however, the second-order model can exhibit $\lambda_2$ as low as two (not shown). Nevertheless, as discussed above, the second-order model does not exactly match the task for the experimental VSB data presented in this thesis. A matched set of experiments would be required in order to determine the parameters that fit human behaviour.

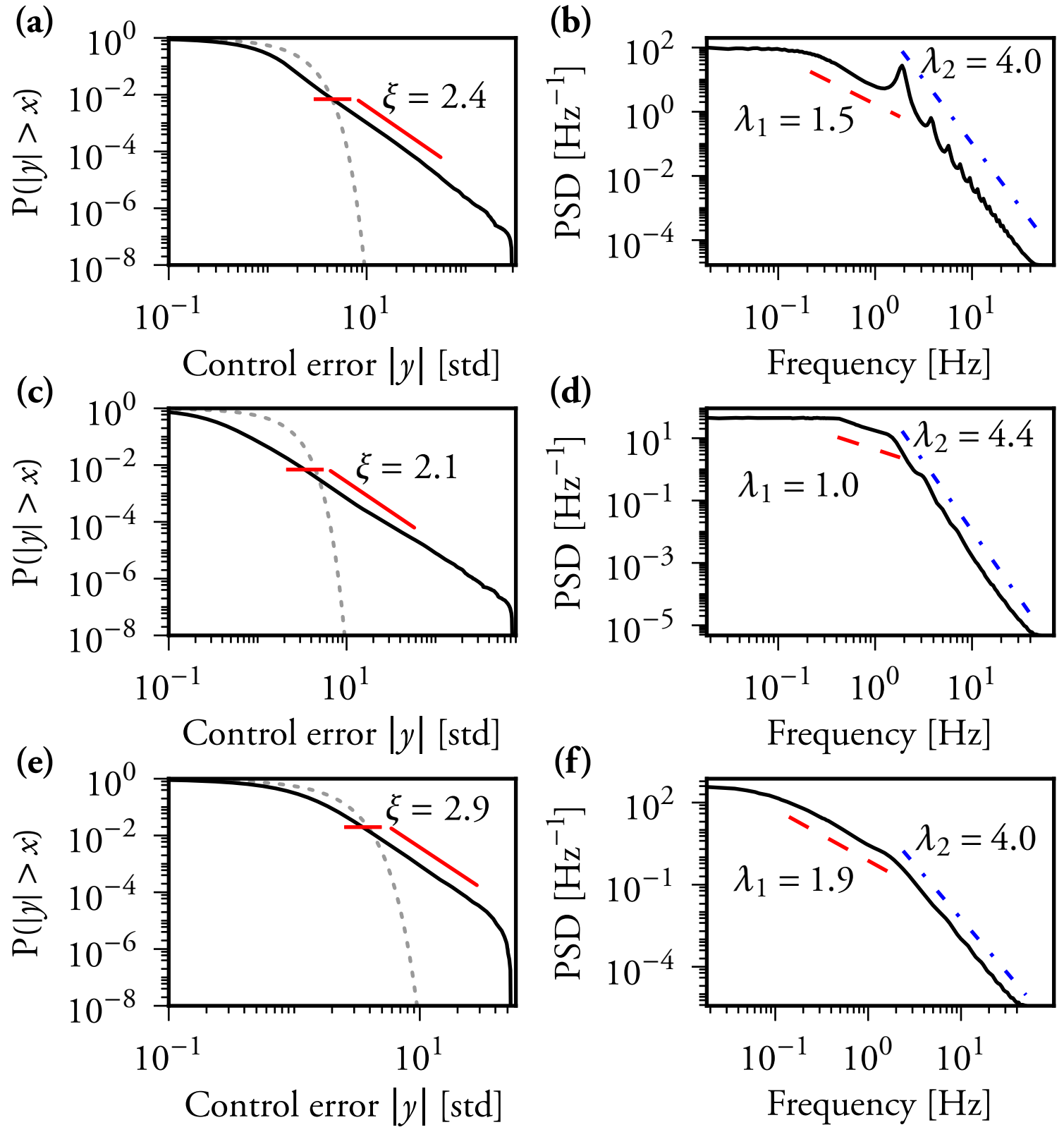

Figure 7.7.: Comparison of CCDFs (left column) and PSDs (right column) for simulations of the model defined by equations (7.19) ff. in different dynamical regimes. **(a), (b)**: underdamped. $\gamma_a = 5$, $\gamma_v = 1$, $\tau_m = 140\,\text{ms}$. **(c), (d)**: nearly critically damped. $\gamma_a = 5$, $\gamma_v = 6$, $\tau_m = 150\,\text{ms}$. **(e), (f)**: overdamped. $\gamma_a = 1.2$, $\gamma_v = 10$, $\tau_m = 100\,\text{ms}$. All simulations: $\vartheta = 10$, $t_r = 200\,\text{ms}$, discretisation step 10ms. Analysis like in figure 7.4.

To summarise: realistic time-series with power-law CCDFs and broken power-law spectra are obtained when the first-order control term in equation (7.20) dominates the system's tendency to oscillate. This finding explains why the first-order model approximation to human balancing control described by equation (7.14) is an excellent quantitative fit to the experimentally observed scaling–except for the highest frequencies, which are more strongly damped for some subjects. Further, for light sticks and VSB, it is plausible that humans plan where to move their hand as opposed to planning which torques to apply directly. The latter may then be calculated further down in the control hierarchy (see chap. 5).

In any case, findings so far indicate that the scaling properties in human balancing behaviour emerge during trajectory planning and are–at least qualitatively–largely independent of many details of movement force generation (see also sec. A.8).

As a final remark, the second-order model also offers a new perspective on the balancing model discussed in section 6.3. The latter includes a substantial amount of friction which cannot be justified for a stick on a fingertip. Here, in contradistinction, the controller causes the system to behave as if it was overdamped. The responsible term in equation (7.20) is proportional not directly to the relative velocity of controller and target (or the displacement angle of the stick), but to the controllers prediction of said velocity. This influence, in contrast to ordinary kinetic friction, is possible by only moving the suspension point of the stick. Hence, an effective overdamping of a balanced stick's dynamics likely stems from the subjects desire to suppress oscillations and to stop the stick from moving.

# 8. Training and task objectives (VSB II)

In chapter 7, a critically adaptive balancing model was presented where heavy-tailed error distributions resembling those observed in human balancing tasks arise from a rapid online estimation of parameters. The model suggests that subjects are highly adaptive even in stationary task conditions, due to a trade-off between eliminating random trends that accumulate during movements and a susceptibility to missteps. This implies that if subjects are able to also employ different strategies, the distribution of errors should depend on the task objective.

Here we test this hypothesis by imposing two different objective functions on subjects in virtual balancing tasks: (S) minimising average errors, and (K) minimising rare and atypically large errors. Experimental results confirm the expected trade-off between minimal average errors and heavy-tailed fluctuations. Further model predictions, and the effect of training will be discussed as well. Considering the results of this study, we conclude it to be highly unlikely that the other mechanisms for multiplicative noise found in the literature are the dominating factors in this task.

In the following, after establishing the new experimental setup, we will show that the scaling features discussed in section 6.2 are observed in condition (S). Next, we will qualitatively investigate the changes of the error distributions over time with respect to the two task objectives. Then we will quantify their statistics rigorously and test further model predictions. Finally, we will compare model fits for naive and trained subjects.

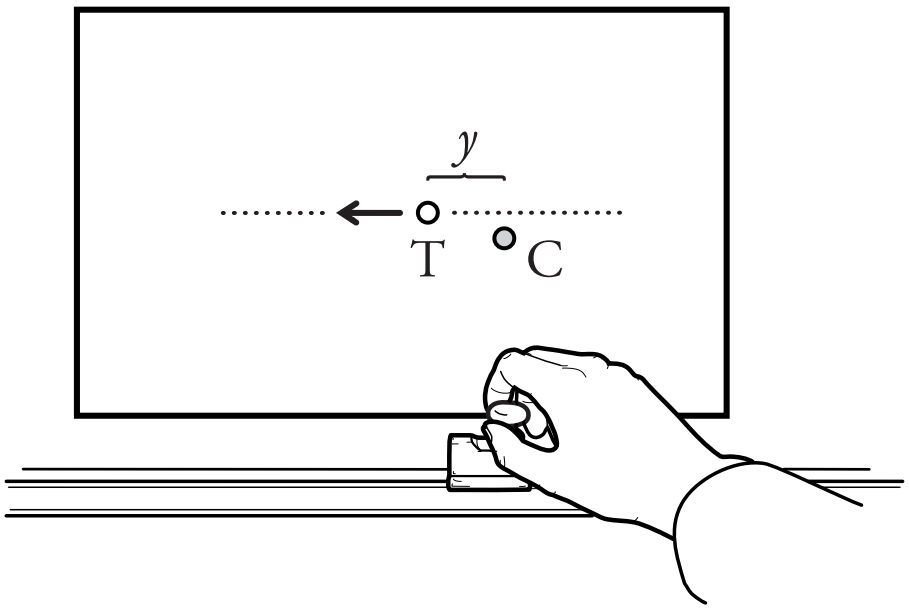

Figure 8.1.: In one-dimensional VSB, subjects control the cursor C using a position encoder on a rail. The unstable target T is displayed with a fixed vertical offset, but also moves horizontally only. $y$ is the horizontal displacement.

## 8.1. A new paradigm in VSB

Previously, we explained many features of control error statistics in two-dimensional human balancing tasks (chap. 6) using a model where movements occur only in one dimension (chap. 7). Therefore, we here introduce a simplified VSB task corresponding directly to this model. The setup is shown in figure 8.1.

This task corresponds to an inverted pendulum on a cart, a well known control problem. Subjects move a light low-friction slider along a rail. On a computer screen behind the rail, a cursor C is displayed at the same horizontal position as the hand-grip of the slider. A target T is moved by the computer according to

$$\dot{T}(t) = \frac{1}{\tau}\big(T(t) - C(t)\big), \tag{8.1}$$

where $T(t)$ and $C(t)$ are the positions of T and C, respectively, and at each time $t$. $\tau = 250\,\text{ms}$. Hence, the distance $|y(t)|$ grows exponentially if C is not moved. Here, the horizontal displacement is given by

$$y(t) = T(t) - C(t) \tag{8.2}$$

### 8.1.1. Experimental protocol in brief

Subjects were randomly assigned to one of two groups with different instructions: (S) minimising the standard deviation of $y$, and (K)

minimising the kurtosis of $y$ (see sec. 2.1). (S) was explained to the subjects as minimising average errors. (K) was explained as minimising the influence of rare, extreme errors. That is, avoiding rare errors that are so much larger than the average ones that these rare events dominate the average over all errors. Each trial lasted 3 minutes. After each trial, subjects received feedback on their performance: The trial score (i.e. the standard deviation or kurtosis) as well as an individual top 10 high score table was displayed.

To ensure that all subjects performed under comparable conditions, a strict schedule was arranged. Each subject recorded trials on four days in a row. Trials were aborted and discarded if T left the screen[1]. Each subject tried to complete 10 trials every day. If at least one trial was aborted, one more trial was recorded on that day. Therefore, in order to keep the amount of training comparable among subjects, some of them completed less then 10 trials on some days. Trial completion rates are found in section A.6.1. New trials were started by the subjects after a 1min pause enforced by the computer. Subjects were instructed to relax and focus on more distant objects in between trials.

For each subject and on each day, the target position was disturbed by random fluctuations during two randomly selected trials. These noise trials served a dual purpose: measuring reaction times directly, and testing how error distributions depend on absolute movement amplitudes. We did not record more noise trials so as to not interfere with the training process too much.

10 subjects participated in each condition. The results of one subject in (S) and two subjects in (K) had to be discarded.

### 8.1.2. Detailed Protocol

The input device was a magnetoresistive position encoder (WayCon MAB, linear with 0.02% error). The analog output of the position encoder was digitised at 5kHz and 16bit. While this sensor measures the position of a magnet without requiring mechanical contact, we added a slider that can only be moved in one dimension. A small round grip was firmly attached to the slider allowing for comfortable and precise usage. The slider could be moved with low friction, comparable to a computer mouse[2].

[1] The rail in this setup is long enough such that C can be moved outside of the screen.

Stimuli were presented on a $(365 \pm 1)$mm wide CRT-screen (Samsung SyncMaster 950p) at a horizontal resolution of 1600px. The screen was updated at $(100.002 \pm 0.001)$Hz. The stimuli consisted of two solid discs with a radius of 5px and an additional 0.5px wide smooth edge. They were rendered with 8× anti-aliasing[3]. The controller was rendered in bright green and the target in white. The background was black.

To ensure optimal timings, we used the Psychotoolbox [Bra97, Pel97] to generate the stimuli and to record the data. The target dynamics were simulated in double precision floating point format at the data sampling frequency of 5kHz with $\tau = 250$ms. For every screen refresh, the then current state of the simulated stimuli was rendered to a frame. The time needed to get new data from the AD-converter, update the simulation, and draw a new frame was shorter than the time between frames. We also recorded the analog VGA vertical synchronisation pulse in a separate channel an found that the system never failed to draw a new frame in time.

Recorded target positions exhibit high frequency noise with a maximum amplitude below 3 quantisation steps. This corresponds to jumps of the target position significantly below one pixel. Subjectively, no jitter in the displayed target position could be perceived.

[2] Increasing downwards pressure on the slider slightly increased friction. We had no means of measuring this force during trials. Nevertheless, we instructed the subjects to use little pressure and suspect that they tried to minimise the effort required to move the slider. In any case, it is highly unlikely that friction between slider and rail impaired the subjects' ability to perform either precise or fast movements

[3] Therefore, movements were displayed very smoothly and even sub-pixel movements were visible. Data from some subjects reveals that they actually tried to compensate sub-pixel distances between controller and target.

For the noise trials, we had to use slightly low-pass filtered noise because otherwise, erratic jumps in the target position were too distracting. Therefore, we used Gaussian white noise and applied a finite impulse response low-pass filter with a time constant of 100ms. This filter is linear phase, that is, it adds a frequency-invariant delay which we subtracted to calculate linear responses. If a noise trial was not completed and there were less than 11 trials already recorded on the same day, the number of noise trials was set to 3 and the total number of trials on that day to 11. See also section A.7.

Subjects sat on an chair with individually adjusted height, back-, and armrests. The screen viewing distance was approximately 65cm. The room was lit evenly and indirectly from behind the screen.

We recruited healthy subjects aged 18 to 35 using a bulletin on the university campus and website. Subjects stayed for approximately one hour on each day. Payment was 8 euros per day. On the first day, subjects read and signed an information sheet. The experimenter filled out a data sheet. The blank documents can be found in section A.14. They were developed with advice from (past) members of the centre for cognitive sciences (ZKW) of the University of Bremen, including Marc Shipper and Cathleen Grimsen. Subjects were assigned random codes in order to collect both the personal data and measured trials pseudonymously. All subjects declared to be free of all of the following properties:

- known neurological, ophthalmological, and cardiac diseases
- medication that may influence driving ability
- heavy use of caffeine, alcohol, or nicotine
- regular use of other drugs.

Ametropia was required to be fully corrected by either glasses or contact lenses. All subjects reported to be able to clearly see the experi-mental stimuli. Handedness was tested according to the Edinburgh Handedness Inventory. All subjects except one were right-handed. The left-handed subject did not perform significantly different from the right handed ones. All subjects performed the task with their dominant hand.

On the first day, subjects completed a training program to ensure that they could perform the task. This training consisted of three trials à 30s with increasing difficulty: $\tau$ decreased from 500ms to 250ms. If a subject failed to complete any of the three trials, the training program was repeated once. Training trials were included neither in the high scores nor in the results of this study. We tried to have each subject perform at the same time of day on each of the four subsequent recording days. Only a few exceptions had to be made. On each day before starting their trials, subjects had to rate their subjective fitness and the amount of sleep in the previous night.

High-score tables were separate for all subjects to ensure equal motivation. However, the individual high scores were initially filled with dummy entries that were identical for all subjects within each task condition respectively. These entries were chosen such that some were easy to beat and others were moderately challenging.

One subject's results were discarded due to failure to participate on four days in a row. Two subjects' results were discarded because they had severe difficulties in concentrating on the task and performing fast and precise movements–both were computer science students.

The computer was prepared by David Rotermund, who also wrote a custom low latency driver. The Psychotoolbox setup, including the framework for timing and stimulus updates, was prepared by Udo Ernst. He further built the analog breakout box for the AD-converter. This box and the cable connections were then slightly modified by the author to include a balanced analog signal transmission and optimised shielding. The author further wrote the experiment-specific code, planned and conducted the experiments.

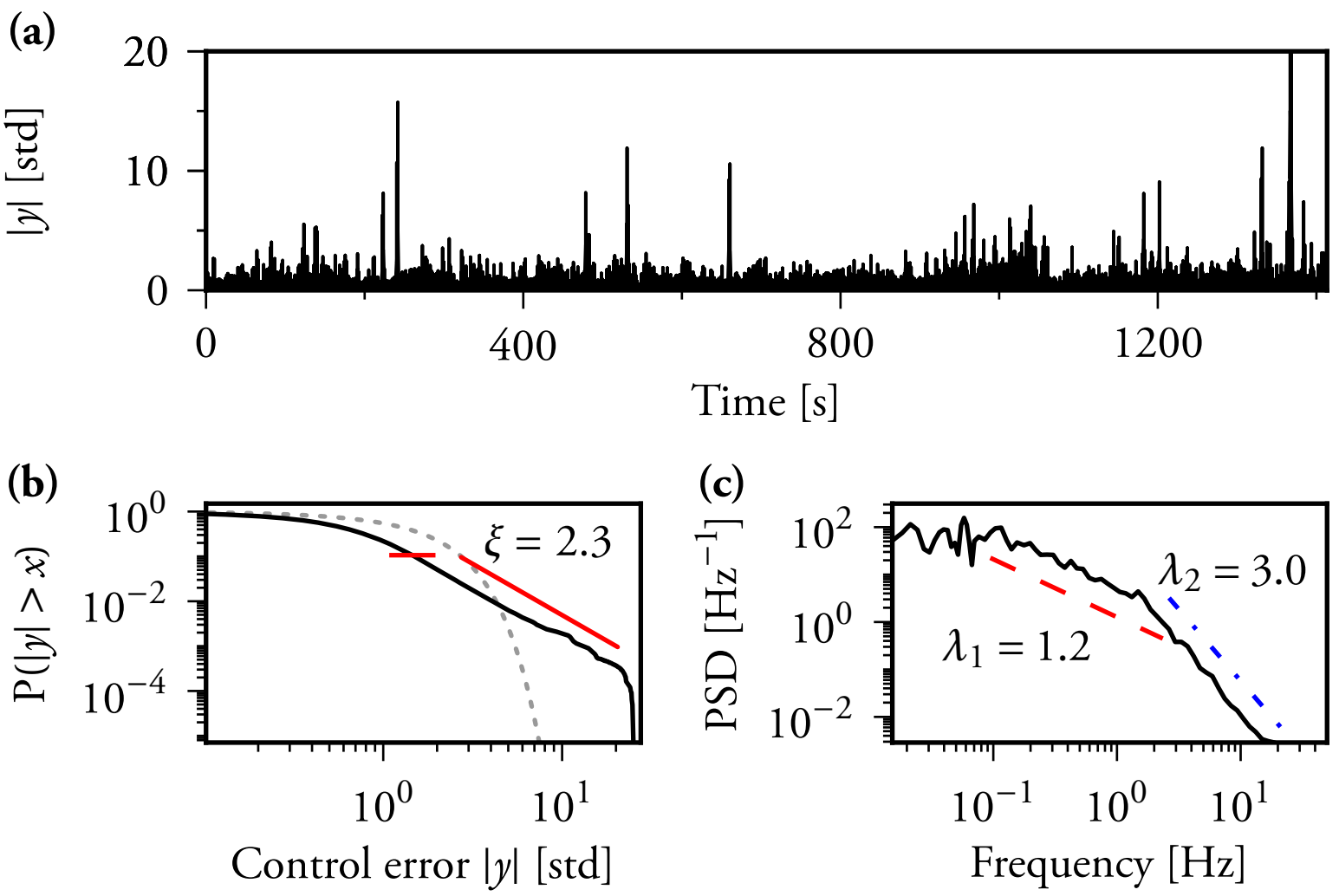

Figure 8.2.: Normalised control errors for the combined trials of day one of subject one who was in (S). **(a)**: Time series. **(b)**: Complementary Cumulative Distribution Function (CCDF) (solid black), normal distribution (dashed grey), and power law (red diagonal line) fitted using the hill estimator with a cutoff (short red line) that optimises the KS-statistics. **(c)**: Power Spectral Density (PSD) and fitted scaling exponents $\lambda_{1,2}$ minimising MSEs. High frequencies eventually level out. Frequencies above his point where excluded from the analysis. Methods: see figs. 6.4 and 6.5

The experiments were approved by the ethics committee of the University of Bremen and conducted according to the data privacy law of the state of Bremen.

## 8.2. Scaling in one-dimensional VSB

Figure 8.2 (a) shows the control errors $|y|$ for the combined trials of day one for one subject in (S). The time series was analysed using the same methods as in previous chapters. Results closely resemble those for the mouse-based VSB experiments: a power-law tail in the CCDF (fig. 8.2 (b)) and a broken power law in the PSD. This result confirms a non-trivial model prediction: the most prominent features of human balancing error statistics can be modeled by considering only movement amplitudes; the process of planning the movement direction within the two-dimensional plane can be neglected.

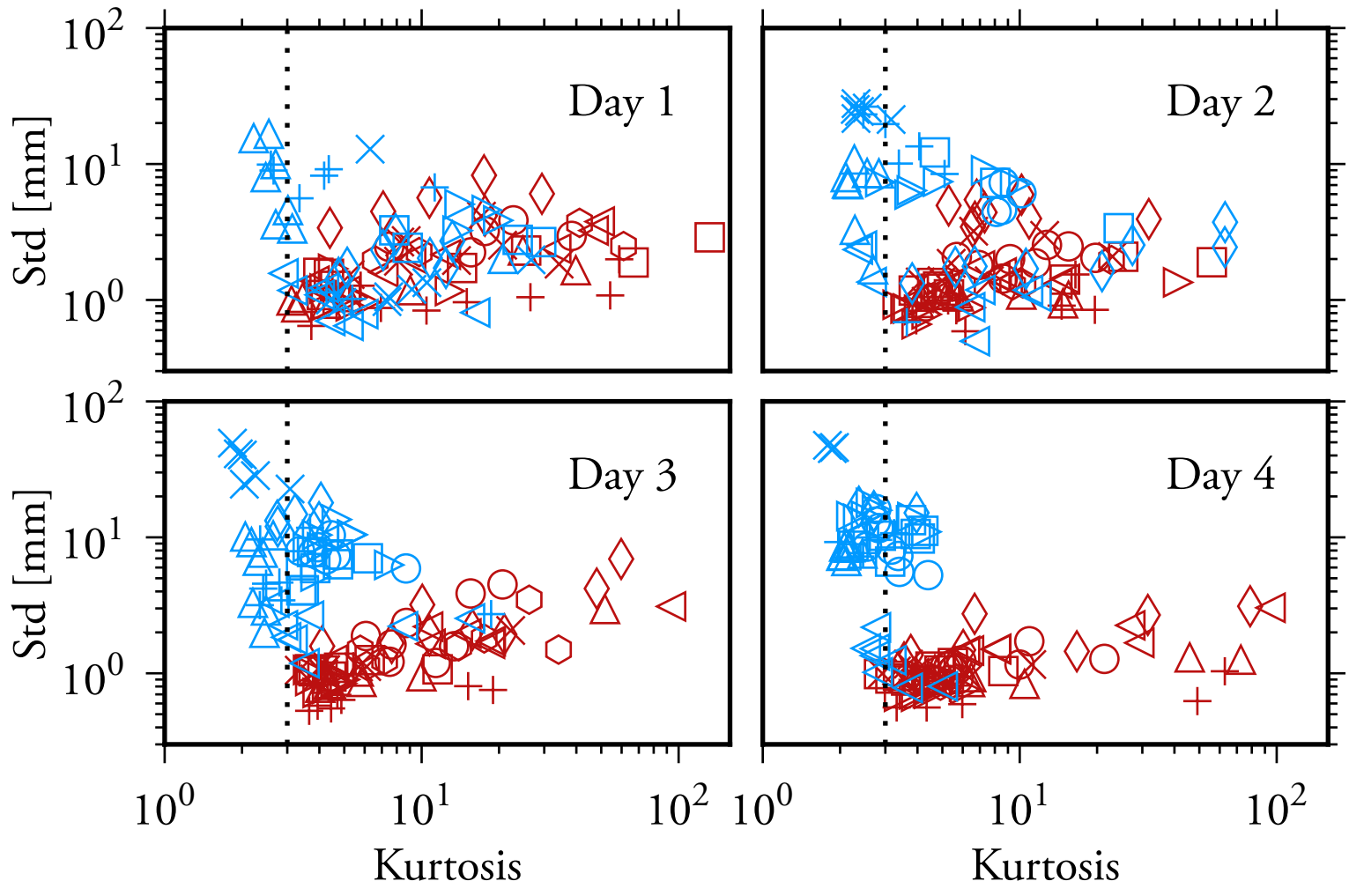

Figure 8.3.: Over four days, subjects either had to minimize the standard deviation (red) or the kurtosis (blue) of the horizontal controller-target displacement $y$. Each marker represents both statistics for one completed trial. Each combination of symbol and color corresponds to a different subject. Dashed line: The kurtosis of a Gaussian is 3.

## 8.3. Standard deviation versus kurtosis

Having established that the new protocol yields the expected results for naive subjects, we focus on the main question of this study: What is the effect of intense training with different task objectives? Figure 8.3 shows standard deviations and kurtoses of the controller-target displacement $y$ for all completed trials of all subjects on each day.

Over time, the two conditions separate into distinct clusters. On day four, almost all trials in condition (S) are heavy tailed. In contrast, kurtoses for most trials in (K) are close to those for a Gaussian or even below. All except one subject (symbolised by a triangle pointing to the right) from (K) exhibit much higher standard deviations than those in (S). However, even for this subject lower standard deviations coincide with higher kurtoses and vice versa. For each group, there appears to be a higher degree of variability within the other statistic than the respective score. More statistical differences between the two groups are investigated in detail in section 8.5. First, however, we investigate how the differences between the two groups arise.

## 8.4. Different control strategies

To understand the behavioural differences between trained subjects in the two score conditions, consider the 30s time-series parts shown in figure 8.4. All four examples were recorded on day four. For each example, controller and target positions as well as their displacements are shown.

For (S), even though there is some variation between subjects and even between trials for the same subject, figure 8.4(a) is quite representative (despite being one of the more extreme examples). In this condition, all subjects managed to keep C and T closely aligned most of the time. This is reflected by the low standard deviations close to a millimetre (fig. 8.3). Nevertheless, both C and T are constantly moving. From time to time, subjects fail to perfectly keep track of the target's movements. This manifests itself in under- or overcorrection of small fluctuations. Sometimes, like in the depicted example at 9s, even brief movements in the wrong direction are observed. Such missteps can lead to errors that are many times larger than the standard deviation. Nevertheless, as soon as subjects notice the target's sudden escape, they are usually able to correct their mistake. Precise corrective movements re-stabilise the target with little to no oscillations. More examples are shown in section A.8.

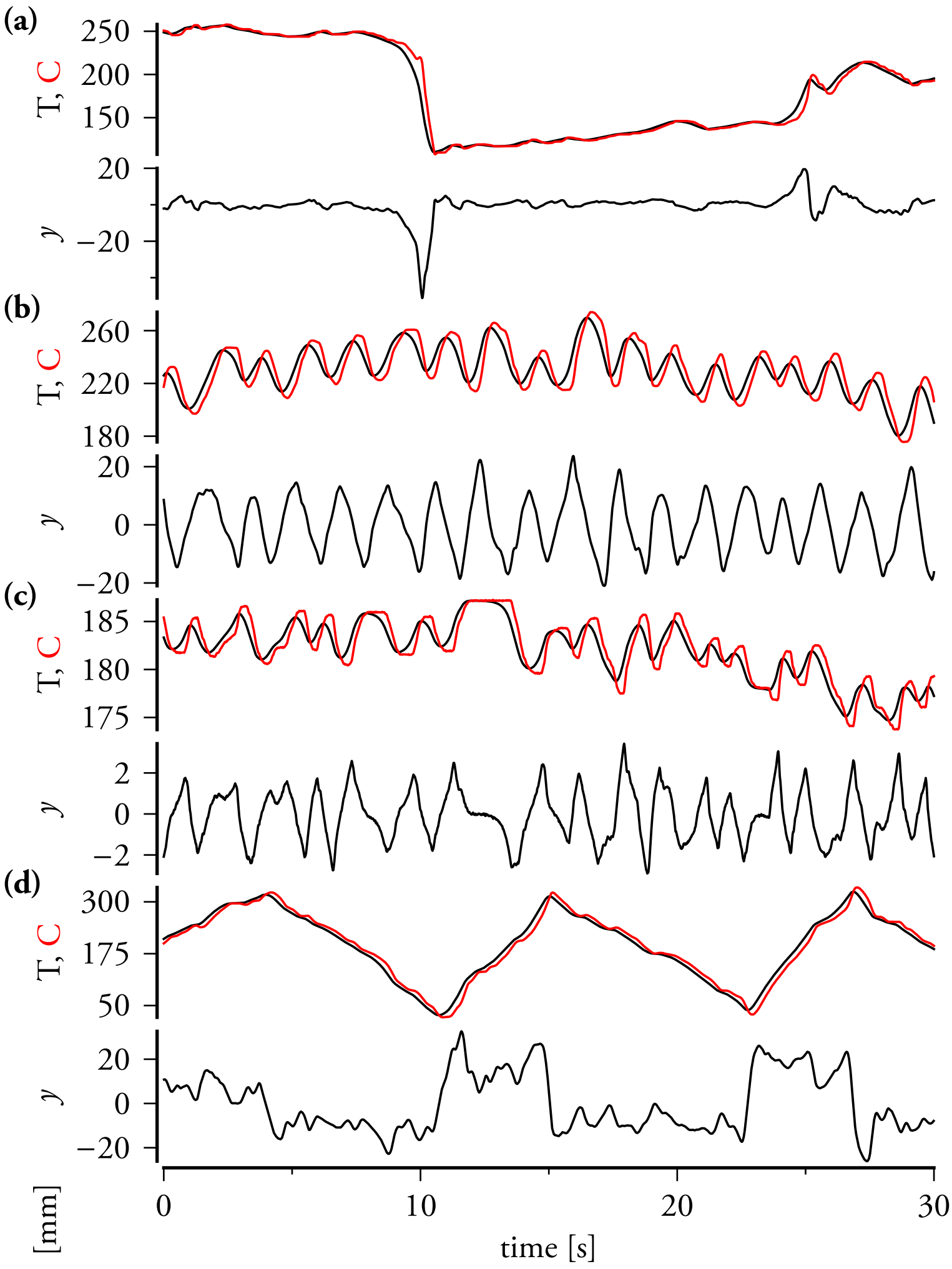

Figure 8.4.: Target and Controller positions and the corresponding horizontal displacements $y$ for different subjects. All time-series are parts of trials recorded on day 4. The y-axes have different scales to improve the visibility of small movements. The allowed range for T extended from 0 – 365mm. The given task was to minimise **(a)**: the standard deviation, or **(b-d)**: the kurtosis. Subjects from top to bottom: 8, 2, 10, 16.

For (K), subject strategies are much more diverse. However, all of them involve avoiding control movements while lingering near $y = 0$. Here we discuss three trials with sub-Gaussian kurtoses (i.e. $< 3$). In figure 8.4(b), the subject performs regular oscillations while trying to restrict T to a small part of the screen. In contrast, the subject in figure 8.4(c) tries to overshoot as little as possible and stops if $y$ gets too small. Yet another strategy is shown in figure 8.4(d): this subject keeps T drifting slowly between both edges of the screen. The movements are performed in a slightly intermittent manner: C repeatedly closes up T and slows down again.

Naturally, avoiding close co-movements of C and T leads to much larger Std($y$). Whether larger movement amplitudes per se have any side effects is investigated in section 8.6.

## 8.5. Statistics of day to day changes

Here we will quantify the changes in numerous parameters between conditions and days. There are inter-subject differences even within groups, but we do not discuss them in detail.

### 8.5.1. Median statistics

As is apparent from figure 8.3, the trial score distributions are highly non-Gaussian. We therefore apply robust, non-parametric tests. Standard box plots are used to visualise quartiles and more extreme deviations that are not outliers[4]. For an example, see figure 8.5. To test for significant changes between samples, we apply two different tests.

[4] For each sample, 25% of the data points lie below the lower edge of the box and another 25% above the upper edge. Whiskers (vertical lines above and below each box) indicate variability outside of this range. The upper whisker extends to the largest data point that is within the upper quartile plus 1.5 times the interquartile range. The lower whisker is calculated analogously. Extreme outliers are not shown to leave space for labels.

First, the median test is a nonparametric test for the null hypothesis that the medians of the populations from which two samples are drawn are identical. However, it has a relatively low statistical power.

Second, the rank-sum test (also called e.g. the Mann-Whitney-Wilcoxon test) has a much higher power. For non-normal distributions it surpasses the t-test in efficiency. On normal distributions it is nearly as efficient. However, it tests whether the probability of an observation from one population exceeding an observation from the other one is not equal to 0.5 [FP10]. This equals a shift in median only if the distributions of both populations have the same shape (except for the shift).

We follow the common practise to use stars to indicate significance. One star: $p < 0.05$. Two stars: $p < 0.01$. Three stars: $p < 0.001$. However, we use a compact visualisation for two tests at once. Outlines are drawn for stars corresponding to the rank-sum test. The median test is indicated by filling the smaller area within each star. For example, two stars only one of which is filled correspond to $p < 0.01$ for the rank-sum test, but only $p < 0.05$ for the median test. If only outlines are drawn, only the rank-sum test found significant differences. This indicates that samples from one populations are more likely to be greater than samples from the other one, but we could not detect a significant shift in the median. Such a finding may be caused by the higher sensitivity of the rank-sum test. However, it is also possible that the populations have differently skewed distributions with identical medians. In none of the following figures did the median test indicate a lower p-value than the rank-sum test (this would be drawn as a small filled star due to the missing outline).

### 8.5.2. Score differences

Box plots for the standard deviations and kurtoses for the trials in each score condition and for each day are shown in figure 8.5. As expected from figure 8.3, distributions for both conditions evolve differently over time. On day four, the distributions are highly separated with (S) exhibiting lower standard deviations and higher kurtoses than (K). In addition, some more subtle changes are found. For (S), both standard deviation and kurtosis distributions have slightly yet significantly lower means after training.

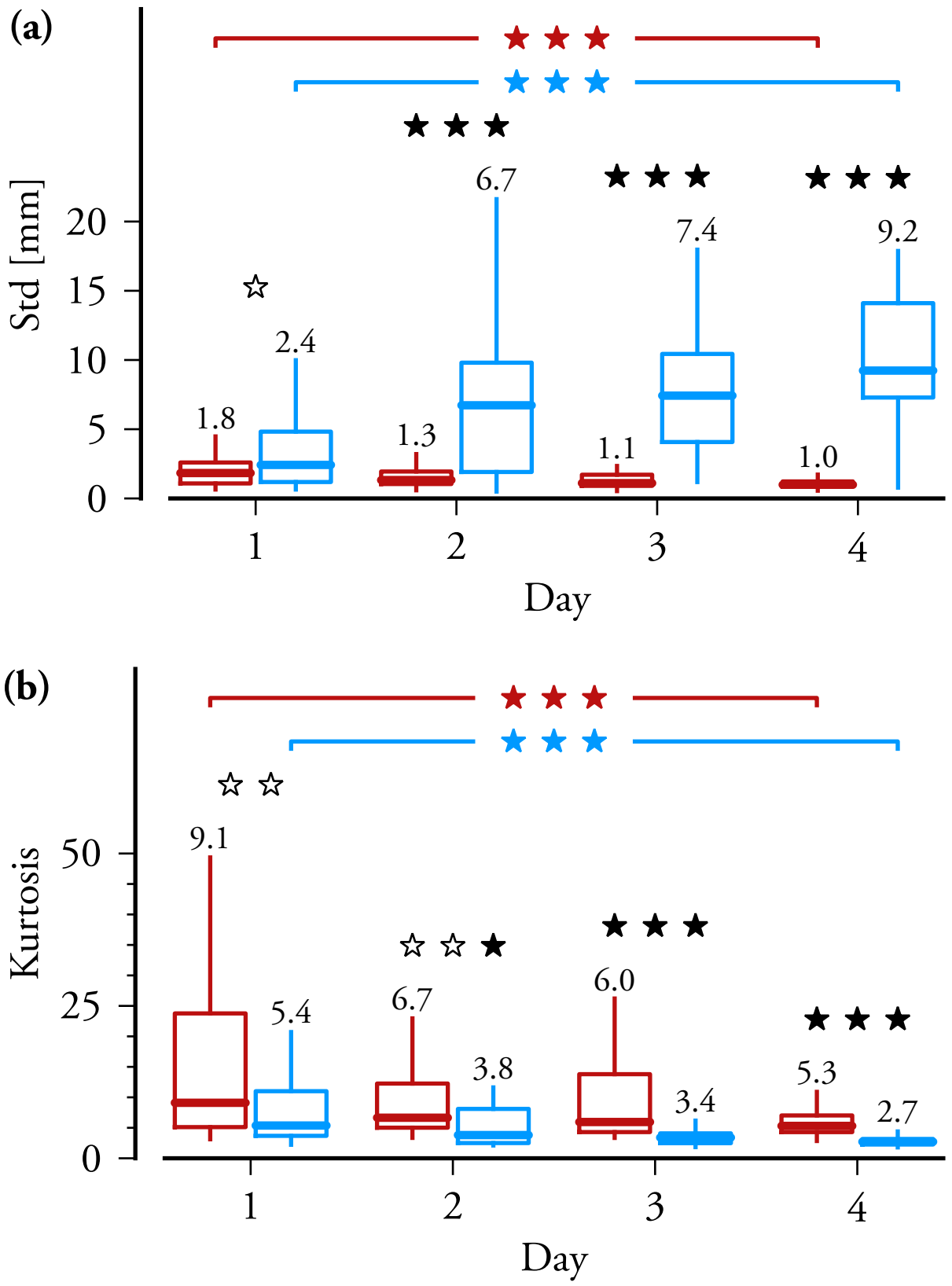

Figure 8.5.: Box plots of **(a)**: standard deviations, and **(b)**: kurtoses for the trials also shown in figure 8.3. For each day, (S) (red, left) and (K) (light blue, right) are analysed separately. Medians (thick horizontal lines across each box) are also stated as numbers above the upper whiskers. P-values are indicated by star outlines for the rank-sum test and star fillings for the median test. E.g. on day 2, kurtoses distributions for the two conditions are found to be different with $p < 0.001$ (three stars) for the rank-sum test, but only $p < 0.05$ (one star) for the median test. See also section 8.5.1, and footnote 4.

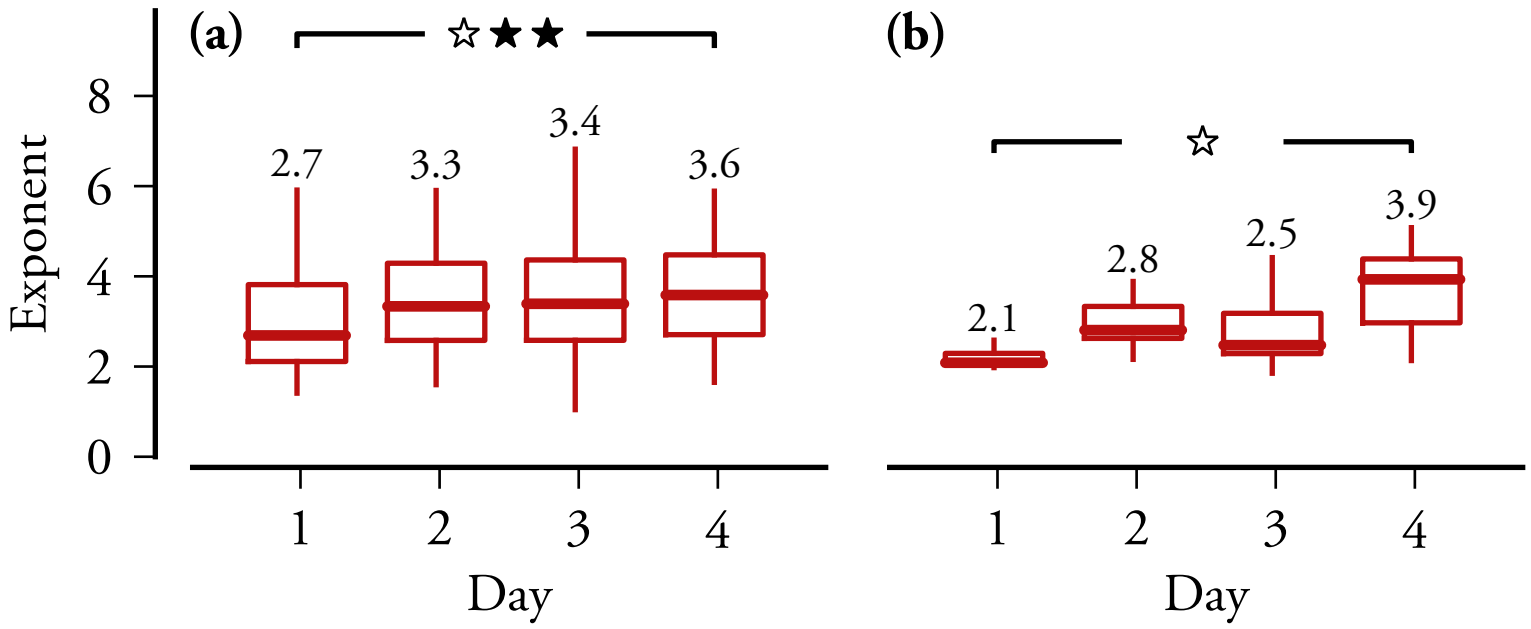

Figure 8.6.: CCDF-tail exponents $\xi$ on each day for subjects in (S). **(a)**: single trials. **(b)** combined trials for each subject. Exponents were fitted like in fig. 8.2.

### 8.5.3. Scaling in (S)

Figure 8.5(a) shows the evolution of tail exponents fitted to the cumulative distributions of the trials in (S). A highly significant increase in the median is found. For the combined trials of each subject on each day, the change appears even more extreme (fig. 8.5(b)). On day four, the measured median exponent is very close to $\xi = 4$, the optimum predicted in section 7.6.

However, while combining trials into longer time series tends to give better fits (which we verified using surrogate data), we also obtain less fits and therefore lose statistical power. For the combined trials, we only find a significant difference when using the more sensitive rank-sum test. Therefore, we can say that (in the combined trials) we tend to find higher exponents on day four, but not that the median increased. There is one more possible effect. The inter-subject variation is very low on day one and much higher on day four. This may indicate that training does not affect all subjects to the same extent. This point is discussed again in section 8.7.3. The finding that the variation among the kurtosis trials is reduced over time is consistent with an increase in the tail exponent for surrogate data (not shown).

We cannot reject the hypothesis that the distribution tails follow true power laws for any of the combined trials in the error condition[5].

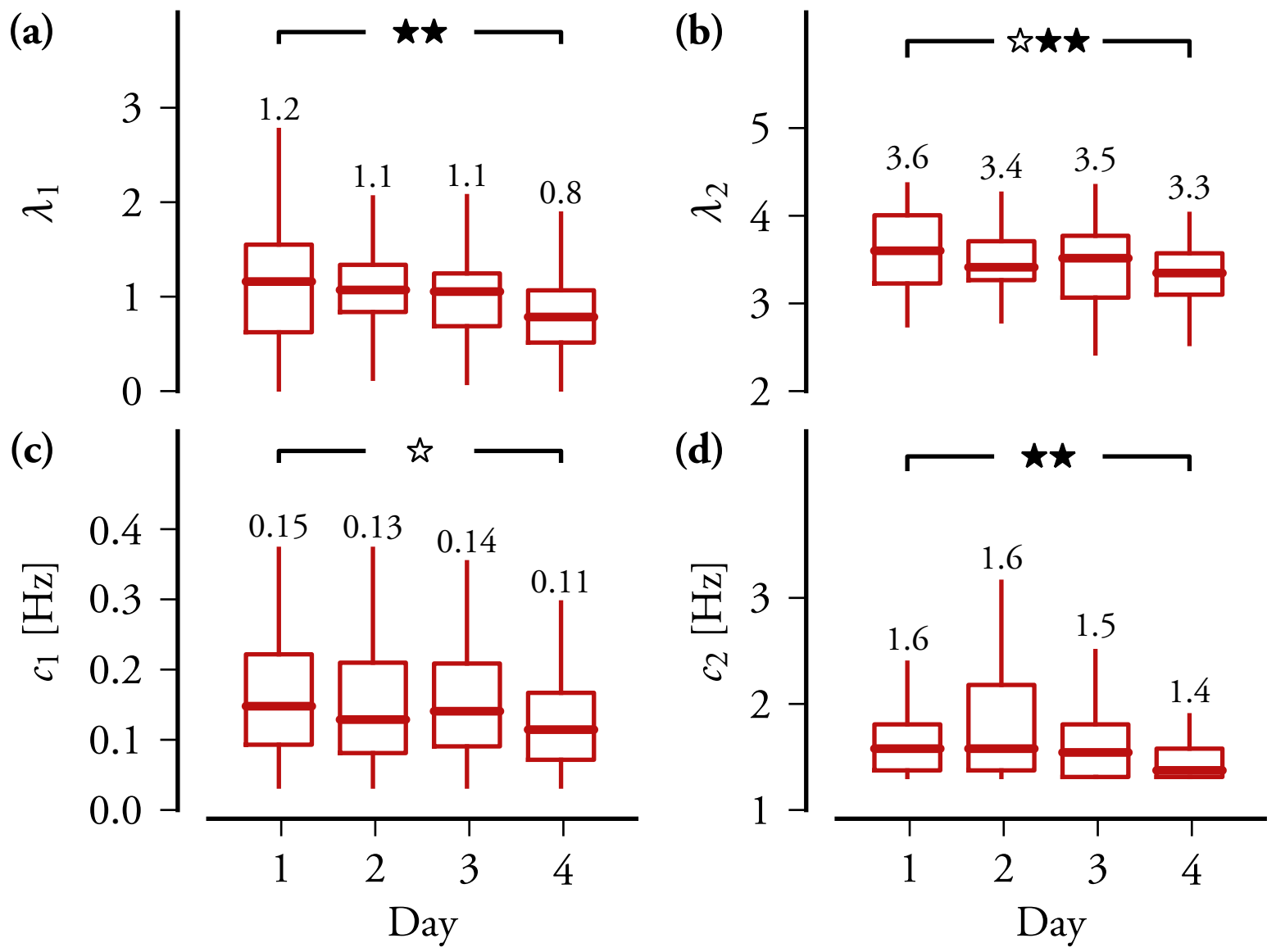

Figure 8.7.: For combined trials, similar values are found, but the changes from day one to four are not significant.

Figure 8.7 shows statistics for different parameters fitted to the PSDs using the same methods as in figure 8.2. Median low frequency scaling exponents $\lambda_1$ are 1.2 on day one and significantly decrease to 0.8 on day four (fig. 8.7(a)). Median high frequency scaling exponents $\lambda_2$ are 3.6 on day one and significantly decrease to 3.3 on day four (fig. 8.7(b)). In other words: both scaling regimes become shallower over time, but the change is more pronounced for the low frequency regime.

[5] $p_> = 0.44 \pm 0.08$ for day one, and $p_> = 0.48 \pm 0.07$ for day four; 0.5 would be expected if model and data were perfectly identically distributed. See also sec. 7.7. Here, we used the median model parameters shown in fig. 8.10. For single trials we obtained $p_> = 0.62 \pm 0.04$ and $p_> = 0.65 \pm 0.03$, respectively.

The position $c_1$ of the low-frequency scaling onset tends to occur at slightly lower frequencies on day four than on day one (fig. 8.7(c)). However, this might be an artefact of the automated fitting caused by the smaller difference between the flat, ultra-low frequencies and the low-frequency scaling regime which is shallower on day four than on day one.

An even smaller yet highly significant change is found in the position $c_2$ of the knee. Here, median values decrease from 1.6Hz to 1.4Hz. While one might again suspect that this is an artefact of the fitting method, there is independent evidence that this is not the case (presented in the following sections).

Findings for the spectra of the combined trials are consistent with those presented above, but the trends in this case fail to reach significance due to the then low number of samples (not shown). Fits for $\lambda_1$ tend to be slightly higher on average for combined trials than for single trials (not shown); this effect is also found for fits to model time series of corresponding lengths (fig. 8.12).

## 8.6. Noise trials

Figure 8.8 shows a comparison of the trial scores for normal trials and noise trials. The random perturbations of the target caused significantly higher movement errors. This effect was more pronounced in (S) where median standard deviations increased approximately by a factor of 6. Despite this large effect of the artificial noise, we found no significant change in median kurtoses. If there were a small effect that failed to reach significance due to the relatively small number of noise trials, it would be a very small increase in kurtosis for the noise trials. Therefore, we can rule out the possibility that an increase in standard deviation per se causes smaller kurtoses. This is consistent with the hypothesis that in order to avoid the IAI, the deterministic movements of the target have to be large enough to be clearly distinguishable from the noise.

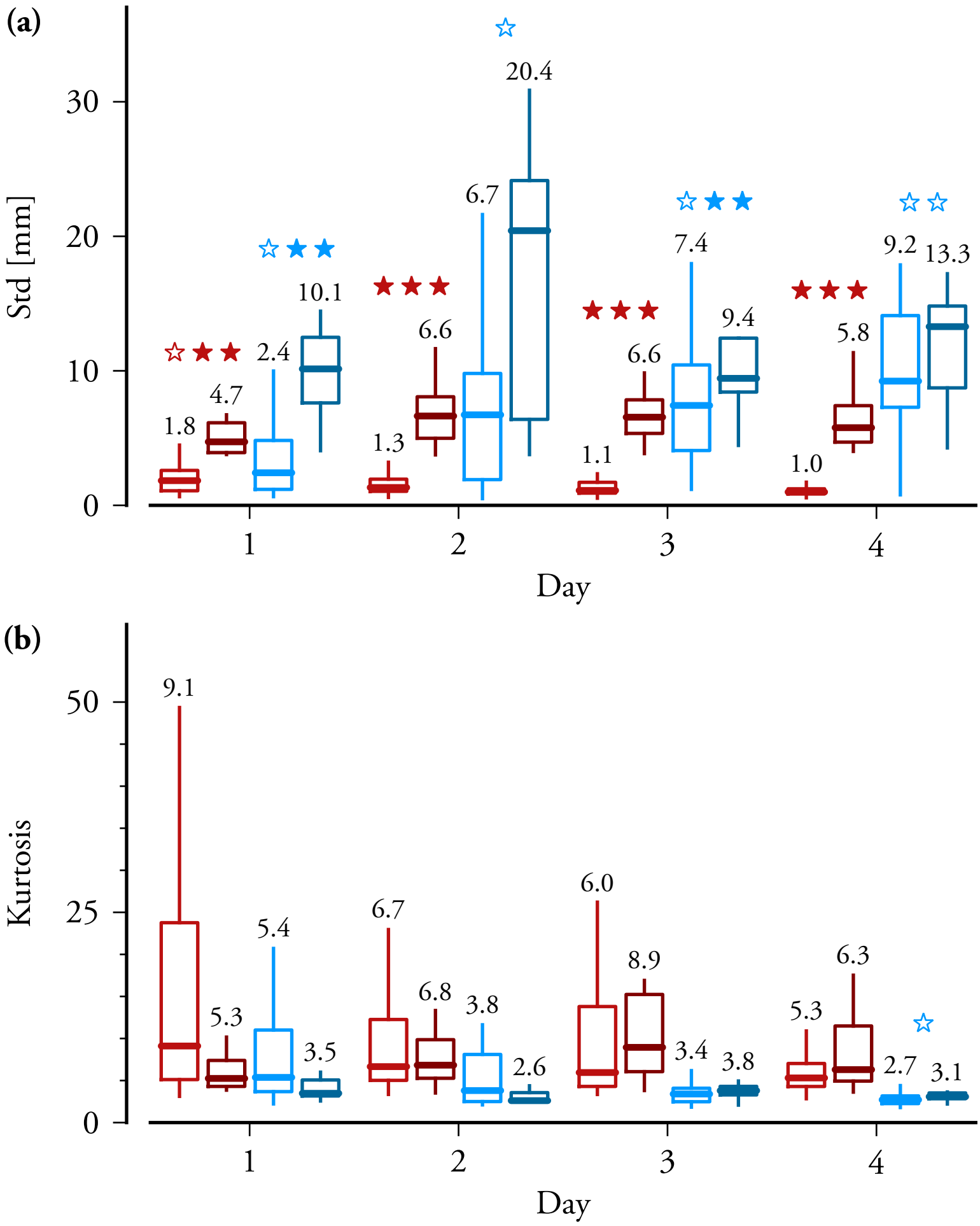

Figure 8.8.: Comparison of **(a)** standard deviation and **(b)** kurtoses for trials with and without artificial noise. The sample groups for each day are arranged as follows: First (S) (reds), then (K) (blues). Within each condition, first normal trials (lighter colours) and then noise trials (darker colours). That is, conditions left to right are {(S), normal}, {(S), noise}, {(K), normal}, {(K), noise}. Stars indicate significant differences between the trials with and without noise for each score condition. To avoid confusion, no tests for differences between days are shown. Tests are explained in section 8.5.1.

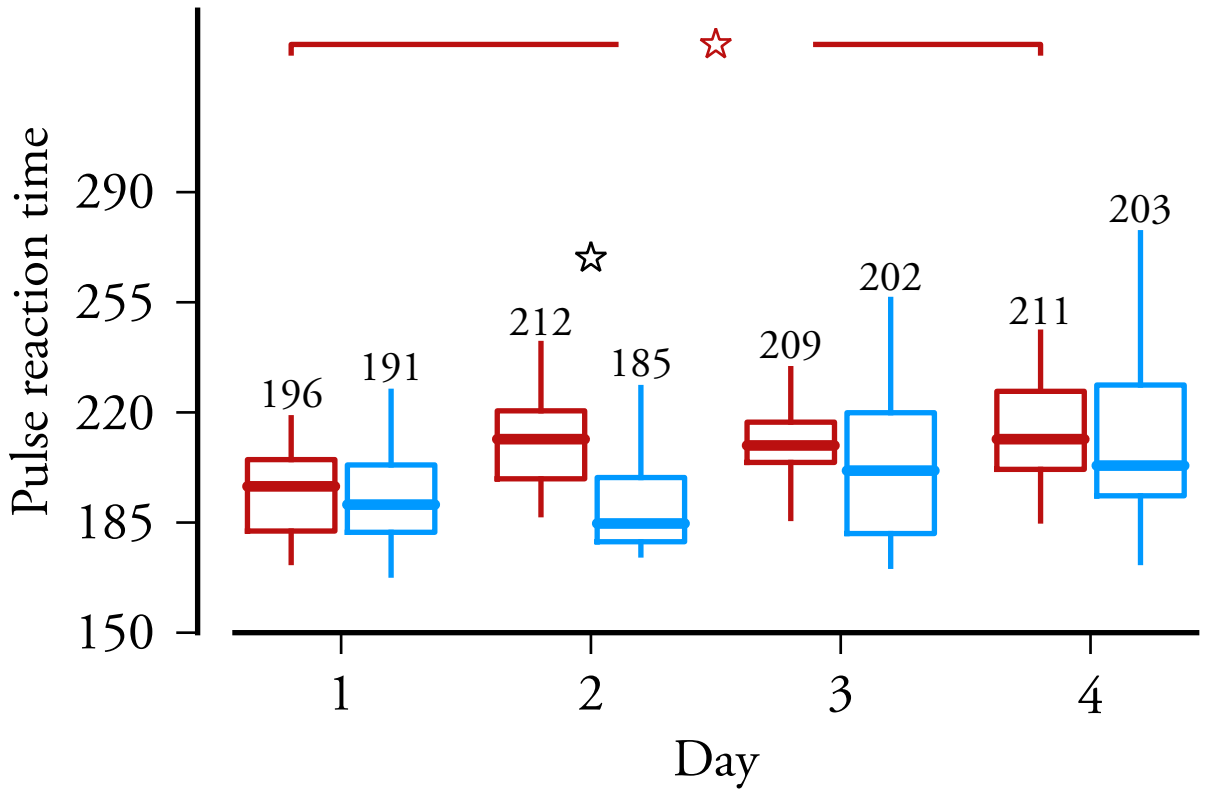

Figure 8.9.: Reaction times calculated from the linear responses measured during noise trials. That is, the average response time of each subject to a delta pulse.

In the PSDs, the only robust effect is an increase in the high-frequency-scaling exponent (not shown). This is likely caused by the artificial noise dominating all other noise sources.

The noise trials can be used to directly measure the subjects' reaction times. This is done by calculating linear responses from the recorded movements and the known noise tracks. An example is shown in section A.7. Figure 8.9 shows how the results are distributed for each day and condition. Reaction times are found to be close to 200ms. Further, reaction times in (S) tend to be longer on day four than on day one. While we only have a quite limited number of noise trials and therefore a low statistical power in this test, the result is consistent with the hypothesis that the knee in the PSD correlates with the reaction time. In figure 8.7 we found lower median frequencies for the fitted knee positions on day four than on day one. It appears quite unlikely that two completely different measurements would produce spuriously significant effects in the same direction. A third method of detecting the same effect results discussed in the next section.

## 8.7. Fitting changes during score condition (S)

How and how well can the model fit the above results? To answer this question, the first step is to find appropriate model parameters.

### 8.7.1. Obtaining model parameters

In order to track the changes in the distributions and spectra over the days, we use automated fitting methods. Since the model has only three free parameters to fit the trials ($t_r$, $\gamma$, and $\tau_m$), they can be arranged in a three-dimensional grid. We considered 14208 parameter combinations, each of which forms a node in the aforementioned cubic grid. The parameter ranges were chosen such that no fitted parameter sets lie on the surface of the cube. To avoid potential biases that may arise from comparing parametric fits which themselves have to be automatically calculated, we took a different approach.

For each node, we simulated 50 time series of the same length as the experimental trials. For each simulation, we calculated the CCDF and the PSD. To be able to average and store these results for all nodes and to quickly compare them to the experimental trials, we first binned each CCDF along the probability axis. These averages were stored for all parameter combinations creating a hypercube. Later, we processed each trial in exactly the same way. A modified KS-test suitable for the binned data was used to determine for each trial the dissimilarity to each average simulated CCDF. For the PSDs, we binned along the frequency axis and used MSE to determine the dissimilarities.

We then calculated a combined dissimilarity measure by normalising the KS-statistics and the MSEs by their respective medians over all parameter combinations and then added the two results. For each trial, the simulation average with the smallest dissimilarity was then determined to correspond to the most fitting parameter set.

There were several requirements for the binning process which slightly increased its complexity. For both CCDFs and Probability Density Functions (PDFs) we calculated 100 logarithmically spaced bins using the same method. These bins were overlapping and all data points were weighted with a Blackman window, and according to their non-uniform spacing within each bin. Hence, each binned data set is smoother than the original data set, yet overlaps it perfectly. The point of this exercise is that logarithmic binning without properly considering that each bin has more data points in one half than it does in the other half can cause biases. Furthermore, the lower-frequency data points in the PSDs are quite far apart. Empty bins are avoided since the bins are overlapping and centre-weighted.

Since we know that the model's ability to fit the PSD's high frequency regime for individual subjects is limited, we only considered frequencies up to 6Hz to determine the best model parameters. This allows us to stick to the most simple continuous-time balancing model which otherwise fits the data extremely well. The alternatives (secs. 7.8 and A.4) would mean introducing unknown new parameters like a correlation length for the driving noise or the inertia of the arm.

### 8.7.2. Model parameters for each day

Figure 8.10 shows how the model parameters fitted to the trials of each subject in (S) evolved over time. First, median reaction times $t_r$ slightly yet significantly increased from 198ms on day one to 214ms on day four. This is consistent with the PSD-knee positions and reaction times from linear responses reported above. Second, gains $\gamma \approx 1.1$ were found for all four days. Third, median memory time constants $\tau_r$ were found to significantly increase from 115ms on day one to 145ms on day four.

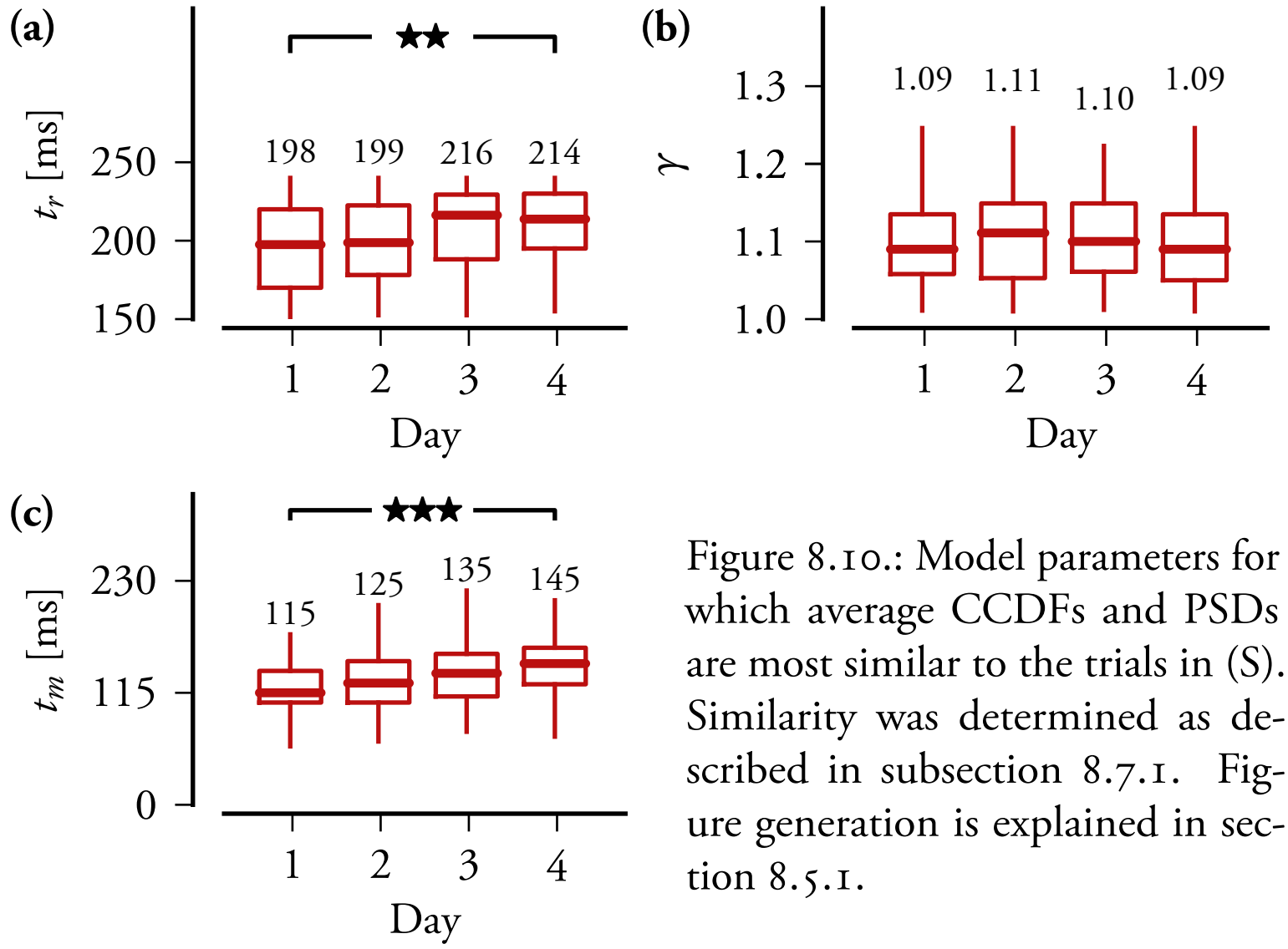

Figure 8.10.: Model parameters for which average CCDFs and PSDs are most similar to the trials in (S). Similarity was determined as described in subsection 8.7.1. Figure generation is explained in section 8.5.1.

### 8.7.3. Changes in distributions and Spectra

To better comprehend how the statistics of naive and trained subjects differ, we repeated the analysis performed for the mouse-based VSB time-series (section 7.5) for the new data set, but for days one and four separately. Figure 8.11(a) shows the CCDFs for the combined trials of day one for each subject in (S), and for a model simulation with parameters according to figure 8.10 for the same day. Figure 8.11 (a) shows the same analysis for day four. In both cases, the model reproduces the data very well despite using parameters that were fitted to single trials. Consistent with figure 8.8, there appears to be a higher inter-subject variation among the exponents on day four than on day one.

Figure 8.12 shows the PSDs for the same time series as in Figure 8.11. As expected, the high frequency scaling for the model is not steep enough to perfectly fit the experimental data. However, this discrepancy is greatly reduced on day four (consistent with fig. 8.7). Experimental spectra level much earlier than the Nyquist frequency. This is presumably related to either perception or movement thresholds since it is not observed for larger movement amplitudes. Nevertheless, low frequencies as well as the position and shape of the knee are fitted very well.

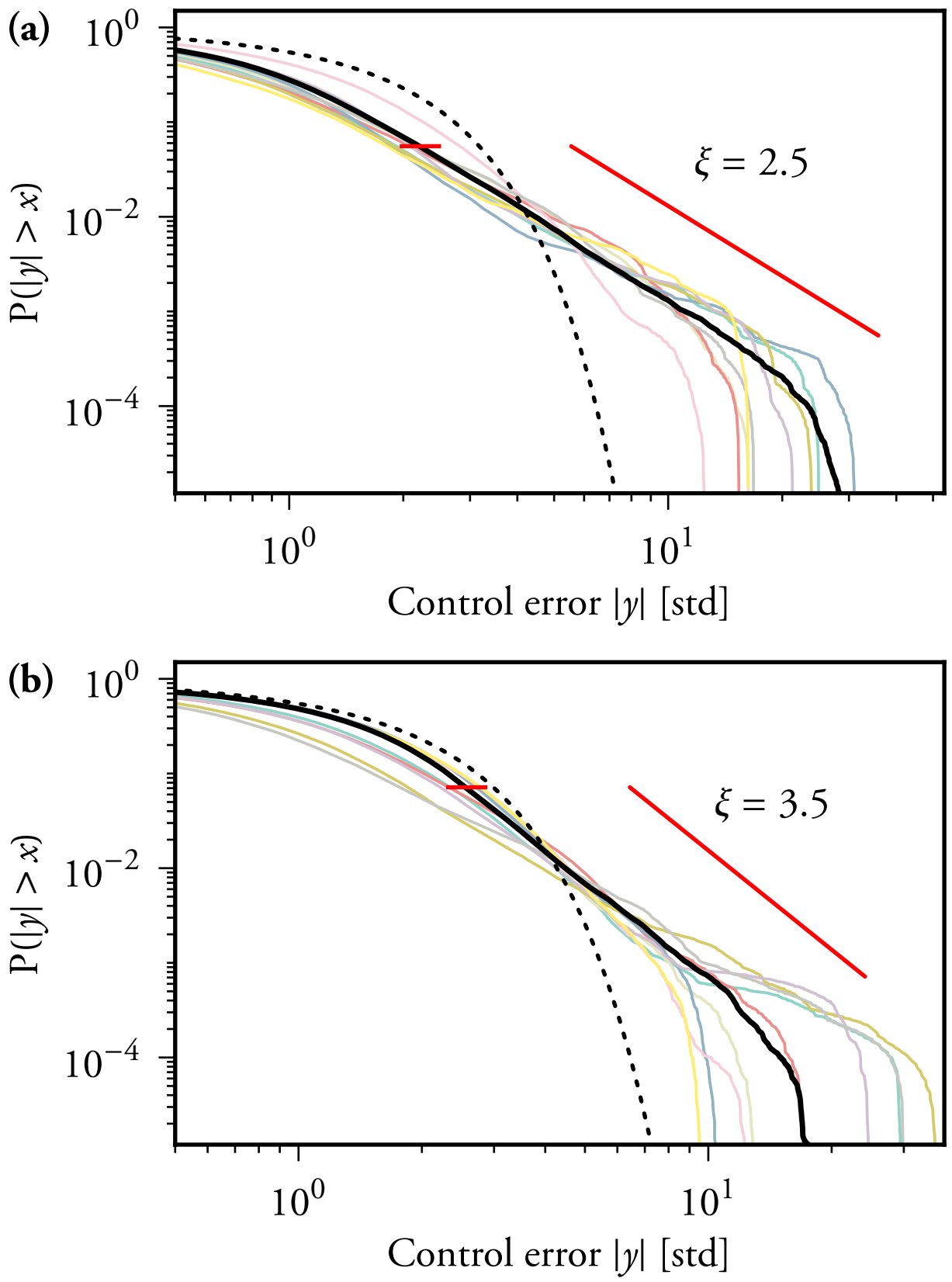

Figure 8.11.: Comparison of the CCDFs for the normalised control errors for the model (thick black lines) with the VSB time-series (thin red lines) for the subjects on **(a)**: day one, and **(b)**: day four. For each day, the trials for each subject were combined into one time series. For each day, model parameters correspond to the median best fits to the single trials shown in figure 8.10.

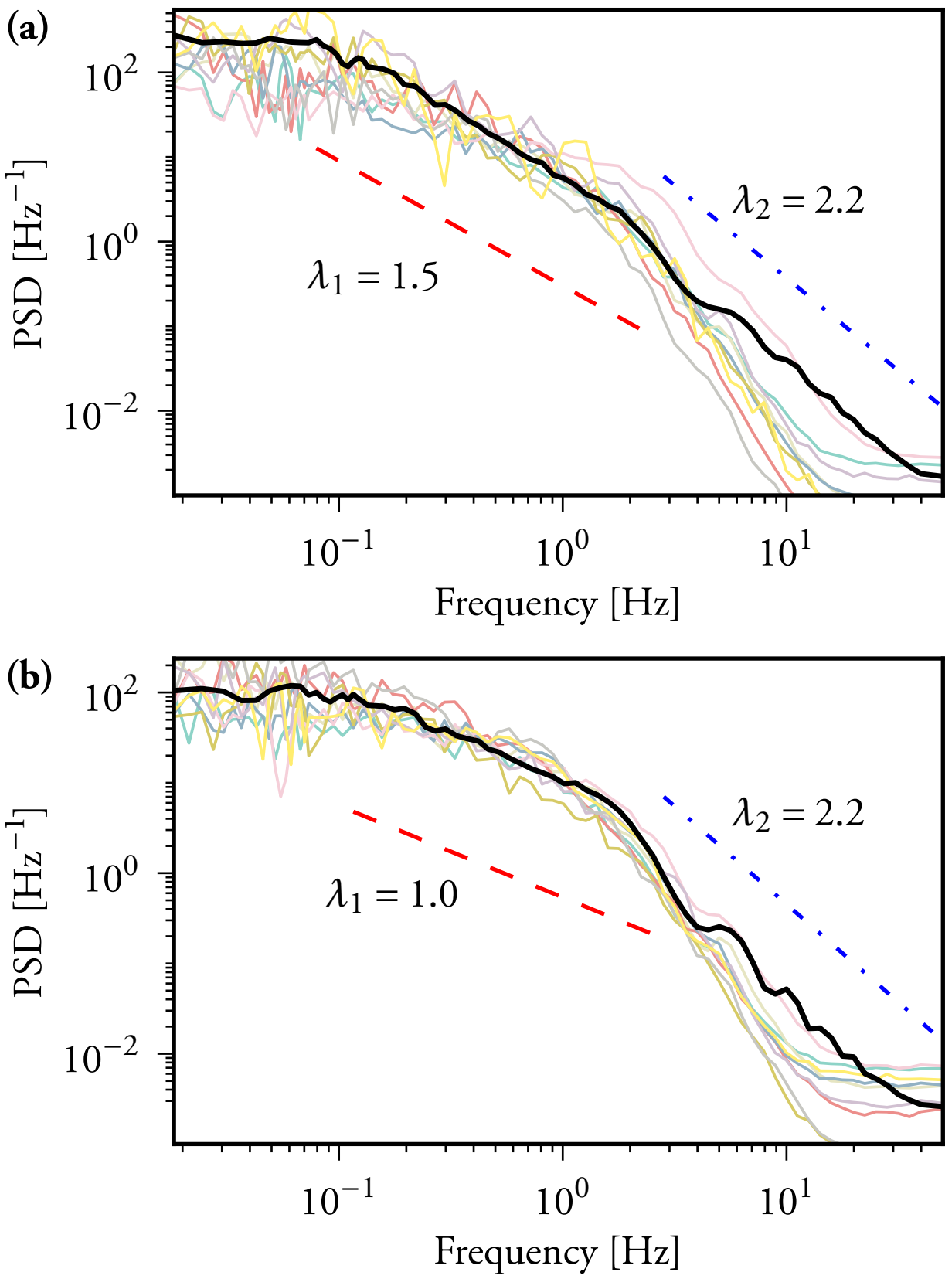

Figure 8.12.: Comparison of Power Spectral Densities for the model (thick black lines) with the VSB control-error time-series (thin red lines) for the subjects on **(a)**: day one, and **(b)**: day four. For each day, model parameters correspond to the median best fits to the single trials shown in figure 8.10.

## 8.8. Error distributions in score condition (K)

Since there is an enormous variability among the subjects' strategies in (K) (section 8.4), we will not consider individual fits here. Instead, we focus more on the qualitative differences between the error distributions for the two score conditions.

Figure 8.13(a), and (b) show the PDFs for the combined trials of all subjects on day four in condition E and K, respectively. Here we did not use CCDFs in order to show that the error distributions for trained subjects in (K) are often bimodal. Note the differences on the $y$-axes in figure 8.13 (a), and (b). In the latter case, the distributions are much less broad and some are even shallower than the normal distribution. As a qualitative model, we extend our standard balancing model (section 7.4) to include a threshold $\epsilon$ below which no control actions take place. A comparison of the model with and without threshold is shown in figure 8.13 (c). The threshold greatly reduces the weight of the distribution tails. For sufficiently large $\epsilon$, $p(y)$ becomes bimodal. A short time series for the model with threshold is shown in figure 8.13 (d).

In contrast to condition (S), the behaviour of (at least several) subjects in (K) cannot be reproduced well by the model. Therefore, testing for the power-law hypothesis in (K) using the same method as in (S) is not well justified. If the test is performed anyway, the high variability among the subjects in this condition shows again. The hypothesis can be rejected only for 3 subjects on day four. However, mean fitted exponents are very high at $6.8 \pm 0.8$ (note that the algorithm always returns an exponent, even if the fitted distribution is not a power law). Visually inspecting the distributions reveals that some indeed appear to follow very steep power-laws (not shown). In any case, the distributions are so steep that only a very small range of values is observed in finite time series (fig. 8.13 (b)). Therefore, one can hardly speak of scaling for any of the trials in (K). Also, average $p_>$-values are $0.19 \pm 0.07$, respectively. This is much lower than in the error condition (sec. 8.5.3).

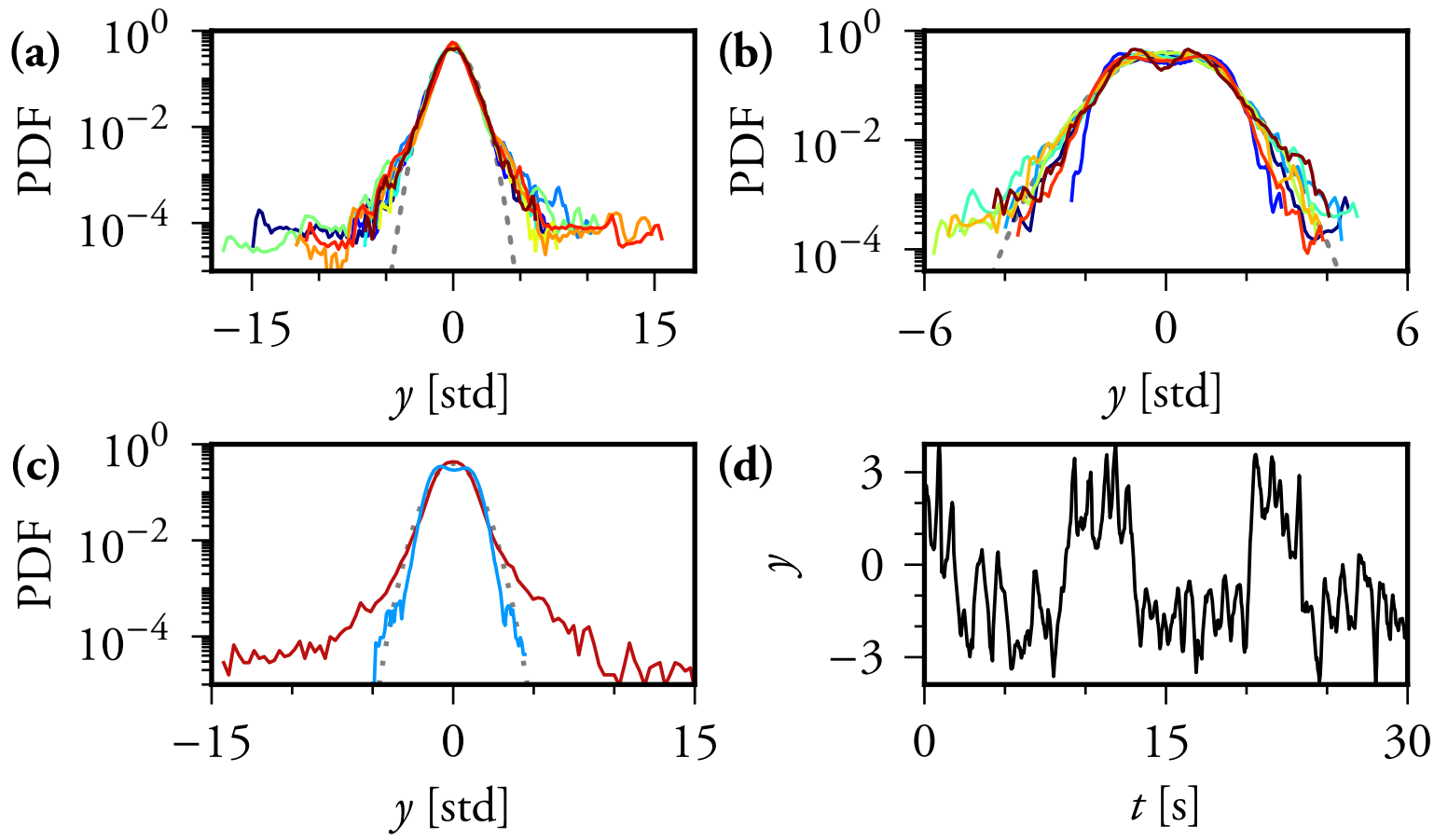

Figure 8.13.: Comparison of error distributions in the two score conditions on day four, and appropriate model variations. In the upper row, PDFs are shown for the combined trials of each subject in **(a)**: (S), **(b)**: (K). In **(c)**, and PDFs for two models are shown. Red line: $\tau = 250\,\text{ms}$, $t_r = 210\,\text{ms}$, $g = 1.1$, $\tau_m = 145\,\text{ms}$, $\sigma = 1\,\text{s}^{-2}$. Blue line: A model with the same parameters and an additional threshold $\epsilon = 1$. For $|y| < \epsilon$, no control actions are performed. In all PDFs, normal distributions are drawn as dotted lines. **(d)**: A part of the time series for the model with a threshold.

Power Spectral Densities vary even more then the PSDs. As one may expect from figure 8.4, some of them show peaks at very different frequencies. Hence, no superimposed spectra are shown as the result would be very confusing. Please turn to the next chapter for a discussion of the results of this part of the thesis.

# 9. Summary and Discussion

Human balancing dynamics are characterised by intermittent bursts of control errors and distinct regimes of temporal correlations. These features were ascribed to the interplay of reaction delays and parametric (i.e. multiplicative) noise [CM02]. However, existing explanations for multiplicative noise focus predominantly on execution noise and are inconsistent with several experimental observations (section 7.1). In contrast, a new model introduced in chapter 7 explains these empirical findings in great detail. In this model, parametric noise arises from the adaptive online estimation of parameters.

The following review and discussion of the results presented in the previous chapters concludes our investigation of human balancing behaviour. First, the main experimental and theoretical results are summarised. Subsequently, we discuss why the new adaptive model is consistent with experimental findings, while previous explanations fail. Next, the structure and parameters of the new adaptive model are discussed in more detail. We will then broaden the scope of the discussion to consider other behavioural tasks and possible implications of this work for our understanding of internal models in a more general context, that is, mental models of the world that humans may use in order to plan their behaviour. Finally, we briefly discuss possible future theoretical and experimental opportunities that are directly connected with this thesis. A much broader discussion including the relevance of the theoretical results for completely different fields of research is found in part IV.

## 9.1. The critically adaptive model in brief

The balancing model (chapter 7), which is central to the previous chapters, makes only few assumptions. First, human reaction times close to $t_r = 200\,\mathrm{ms}$ are compensated by using an internal model. This is consistent with standard models of human motor control (chap. 5). The major innovation in this model is the use of a forward model based on recently observed trends. In simpler terms, a stick is balanced by predicting where it is going to fall and moving the suspension point towards the predicted position slightly faster. The model controller has three parameters: A reaction time $t_r$, a time constant for the exponentially decaying weights of past observations $\tau_m$, and a gain $\gamma$ which determines how fast predicted controller-target displacements $y$ are compensated. The task difficulty $\tau = 250\,\mathrm{ms}$ in the model is chosen to match the experimental conditions. Finally, the model is driven by additive baseline fluctuations: Gaussian white noise which represents the sum of all additive noise sources in the control system. The amplitude of the noise $\sigma$ only determines the absolute scale of the fluctuations in the model; all other features like scaling exponents are invariant. We therefore focus on normalised time-series and thereby eliminate the influence of $\sigma$ on the results.

Successful control (bringing the stick into the upright position) removes predictable dynamics from the system. This creates a self-similar instability where small $|y|$ lead to more uncertainty about future dynamics, and vice versa. This Information Annihilation Instability (sec. 7.2) also plays a major role in part III of this thesis.

## 9.2. VSB experiments

Our analysis in the previous chapters involves two VSB experiments. That is, tasks where subjects control a cursor C to stabilise an unstable target T on a computer screen. The adaptive control model can reproduce statistical features of the measured control errors $|y|$ quantitatively in great detail. First, large $|y|$ are distributed according to a power law. Deviations from this description can be explained by chance (sec 7.7). Second, the PSDs exhibit a broken power law with a pronounced knee.

In the first experiment (sec. 6.2), the control problem is two-dimensional and subjects interacted with the computer using a mouse. Fitting the model to the experimental trials requires cautious, smooth movements with gains just above one, and fast adaptation with $\tau_m$ just below $t_r$. The model further predicts that the position of the knee in the PSD is determined by $t_r$.

These findings raise the question of why subjects adapt on such short timescales even in stationary experimental conditions. As an answer, the model predicts that this control strategy reflects a nearly optimal compromise between the elimination of random local trends and rare large errors. The optimum is characterised by CCDF distributions with a power law tail with slope $\xi = 4$ while experimental trials are found in between 2, and 4. Therefore, we predicted that intense training while minimising $\mathrm{Var}(y)$ should lead to exponents closer to 4, but not to a Gaussian distribution ([PP11], sec. 7.6). Training with a different objective function, on the other hand, may lead to completely different error statistics.

To test these predictions, a new task was developed. Since the model only includes control errors in one dimension, in the new task a slider is moved on a rail to control the cursor (fig. 8.1). After each trial, a high score is shown. Subjects were randomly assigned to one of two score conditions. (S): minimisation of the standard deviation $\mathrm{Std}(y) = \sqrt{\mathrm{Var}(y)}$, and (K): minimisation of the kurtosis $\mathrm{Kurt}(y)$ (i.e. avoiding heavy-tailed distributions). All subjects recorded trials on four subsequent days according to a strict training schedule.

On day four, differences between the error distributions for the two conditions are highly significant. In (S), subjects were able to closely follow the target with median standard deviations of just 1mm. Median kurtoses were measured at 5.3, distinctively higher than the kurtosis for a Gaussian distribution which is 3 (fig. 8.5). Over time, median tail exponents increased from 2.7 to 3.6 for single trials, and from 2.1 to 3.9 for the combined trials of one day of each subject (fig. 8.8). Therefore, subjects in this condition moved closer to the optimum predicted by the model. In (K), however, standard deviations were 9.2mm, and kurtoses only 2.7. These subjects further used a number of different strategies that avoid closely aligning C and T, and thereby avoid the IAI (sec. 8.4).

At this point, one may ask whether the excess kurtosis in (S) is due to the very small distance between C and T. For example, there may be a lower threshold below which errors are not detectable by the subjects, or additive, heavy tailed motor execution noise which dominates during very small corrective movements. Therefore, during two trials per day for each subject, we disturbed the movements of T with additive Gaussian distributed noise. Thereby, the median Std($y$) in (S) is increased sixfold. This as high as for (K) for days two and three, for both of which there is a highly significant difference between the measured kurtoses for the two score conditions. However, in (S), the kurtoses during normal trials and noise trials are not significantly different. Therefore, we can rule out the possibility that error amplitudes per se (and thereby also the velocities at which T and C move, see sec. A.9) have a significant effect on the shape of the error distributions. This confirms one more prediction of the adaptive control model.

For (K) Std($y$) is increased in noise trials, too. A small increase in Kurt($y$) relative to the normal trials is found on day four (but not on the other days), which is only just significant in one of two tests. Therefore, it might be a chance result. Nevertheless, it is quite remarkable that high and very noticeable amounts of artificial noise (which also had to be slightly non-white to be less confusing) did not have any more pronounced effects. If the increase in kurtosis on day four were real, it would still go in the opposite direction of the effect found during normal trials (low standard deviations leading to high kurtoses). Therefore, the differences in kurtoses between the two score conditions cannot be attributed to different standard deviations per se. As predicted by the IAI, rare extreme missteps are avoided only if larger movements lift $|y|$ above the unpredictable baseline noise.

The adaptive control model can fit the measured distributions for both naive and trained subjects in (S) (see figs 8.10, 8.11, and 8.12). Automatic fitting using the CCDFs and PSDs yields values for both $\tau_m$, and $t_r$ that significantly increase with training. Fitted gains $\gamma$, however remain constant over time. The increment in reaction time is very small, but it is confirmed by directly measuring linear responses using the known perturbations during noise trials. This change in reaction time also coincides with a significant shift in the position of the knee in the PSD, as predicted by the model. Further, the fitted parameters are very close to those measured during mouse-based tasks.

In contrast, subjects in (K) can only be approximated by a modified model since they employ a variety of different strategies, all of which avoid proximity to the unstable fixed point. By introducing a threshold to the model below which control errors are not corrected, normalised error distributions become quantitatively very similar to the experimental ones in (K) (fig. 8.13).

## 9.3. Existing explanations in the literature

Most motor control models include involuntary execution noise to explain movement variability (sec. 5.2). However, this mechanism only explains noise amplitudes up to few percent of the muscle force independently of the task. The extreme bursts of fluctuations observed in balancing tasks, however, indicate much higher amounts of multiplicative noise[1].

Furthermore, a task dependency is found in chapter 8. In two task conditions, which only differ in the objective function that is conveyed by a high score displayed after each trial, error distributions follow completely different statistics (see above). In particular, subjects in condition (S) exhibit lower standard deviations and heavier distribution tails (kurtoses) than subjects in (K). The higher errors and lower kurtoses in (K) further coincide with higher velocities and accelerations of the subjects' hands (fig. A.12). We can therefore rule out constant multiplicative noise in force generation generation (motor unit noise), as well as classical speed-accuracy tradeoffs (e.g. Fitts' law, see sec. 5.2) as explanations of the intermittent error bursts: Neither mechanism can explain a decrease of the bursting fluctuations for increased movement amplitudes, velocities, and accelerations.

[1] In [JdCHW02], 3% of motor unit noise were reported. E.g. in figs. 6.6 and 6.7, the noise surpasses the average applied force. See also sec. 4.4, as well as figs. A.4, and A.11 (d).

A prevalent explanation for the bursts of fluctuations in balancing tasks is that high amounts of multiplicative noise drive the system across a stability boundary. These fluctuations were further presumed to be added by the CNS for stabilisation. (see secs. 6.1 and 6.3). This mechanism may conceivably be task dependent and can explain higher amounts of multiplicative noise than motor unit noise. A conceptual similarity to the results presented in this thesis is that the parametric noise is considered to be related to a useful control strategy as opposed to limitations or even defects of the movement apparatus. However, if there were a task dependency, and if the noise were added for its allegedly stabilising effects, it should prevail in difficult tasks and for large or fast movements. Neither is the case: the intermittent bursting occurs in the scenario which is the least demanding for the movement generation apparatus. That is, a situation where subjects successfully minimise fluctuations and retain control for minutes. Further, since subjects are able to retain control in the face of either artificial perturbations or while generating large voluntary movements themselves, the control task is far away from any stability boundary. Therefore, we have to reject both assertions of this mechanism. This result is confirmed by the recent finding that easy tasks lead to more heavily-tailed error distributions than difficult tasks [MFC+11]–another observation that is reproduced only by the adaptive balancing model[2]. In addition, bursts of control errors in the experiment described in chapter 8 occur suddenly and without the slow buildup which is typical for processes with multiplicative execution noise (sec. A.8). Furthermore, the standard model for multiplicative execution noise during balancing exhibits several problems (sec. 6.3). These include the inability of the model to reproduce the scaling relations found in VSB tasks, as well as unrealistic parameters[3]

On a final note, a semi-predictive controller was suggested for balancing tasks with delay in one purely theoretical study [IMS13]: A proportional-derivative-acceleration controller corresponds to a predictive controller without including the controller's own contributions in the prediction. It was shown–unsurprisingly–to perform better than a proportional-derivative controller without acceleration feedback. No argument was given as to why this controller should be biologically more plausible than an actually predictive one. The only missing component would be an efference copy which is commonly assumed to a be available in the motor cortex ([WDF11]). The study does not include a comparison with experimental fluctuations. Therefore, it does not provide a notable contribution to the present discussion.

## 9.4. Rail- vs mouse-based VSB

During one-dimensional VSB, error distributions for subjects in (S) closely resemble those in the earlier, mouse-based experiment. This finding is a non-trivial test of a model prediction: the dominating effects only rely on the absolute distance between C and T.

[2] See sec. A.6. This finding is highly non-trivial. First, due to the well known speed-accuracy tradeoffs, testing the system at its absolute limits should increase the chance to measure multiplicative execution noise. Second, purposeful addition of multiplicative noise which hypothetically could have a stabilising effect should be more prevalent in difficult tasks.

[3] In particular, the model assumes high amounts of friction that are not present neither physically while balancing a freely moving stick on the fingertip, nor in the VSB tasks discussed in this thesis. The main problem here is that the model by Cabrera and Milton is not predictive. A second order variant of our adaptive (and predictive) control model allows for a deeper understanding as to why balancing dynamics can appear to be overdamped. See sec. 7.8

One may ask, however, why no significant signs of an improved performance after training were found in the mouse-based study. There are several possible answers to this question. In the study described in chapter 6.2, the number of subjects was much lower than in the rail-based one (chap. 8). Subjects in the mouse-based study further had an irregular training schedule: For some subjects, there were days without any trials in between the days were the experiments took place. There were also many different task conditions and some subjects recorded trials in different conditions on the same day (sec. 7.7). Finally, subjects in the mouse-based experiments had no objective feedback on their performance. We therefore expect that more controlled experiments using a mouse would lead to training effects that are very similar to those reported for the rail-based setup.

## 9.5. Structure and parameters of the adaptive model

As discussed above, many experimental findings are reproduced only by the adaptive control model (chap. 7). Another important feature of the model is its biological plausibility. First, the model is consistent with standard models of motor control and the action-perception cycle. The crucial change with respect to these established findings is the introduction of online parameter estimation as a dominant source of parametric noise (fig. 7.1).

Second, the model is well-suited for a neuronal implementation: In addition to leaky integration, our model only requires normalisation which may, too, be an ubiquitous neuronal mechanism in the brain [CH12]. This is a desirable feature, since the low reaction times of balancing subjects (close to simple reaction times for visuomotor tasks) might be interpreted as an indication of a control strategy which is effectively automated at the neuronal level.

Since the model includes only three free parameters for the controller, automated fits to single trials are highly feasible (see above).

These fits suggest that well trained subjects in (S) adapt slower and also react slightly slower than untrained ones, while the gain of their corrective movements remains unchanged (see above). This finding corresponds to moving closer to the optimum in figure 7.5 (a). It may therefore be seen as an indication that subjects successfully optimise their trade-off in between average and freak errors within their biological limits.

The fitted reaction times were validated using direct measurements (figs. 8.9, and 8.10). Since there is no evidence that subjects could possibly react much faster (sec. 5.1), reaction times require no further justification. However, the combination of rapid estimation and smooth movement execution warrants some further discussion beyond the finding, that only these model parameters are consistent with the experimental data (see e.g. sec. A.2, and A.3).

### 9.5.1. Memory

The controller's memory decay time constant $\tau_m$ determines the time scale on which it estimates the time constant $\tau$ of the control problem. As it turns out, values of $\tau_m$ just below $t_r$ minimise Var($y$). A controller that adapts rapidly to observed trends leads to smaller average errors than a controller that uses the true $\tau$. These exploitable trends during which the system deviates from its average dynamics emerge from random perturbations of the smooth control movements. Since eliminating these trends increases the system's susceptibility to the IAI, there is an optimal tradeoff in between the two effects.

However, we found that the behaviour of naive subjects is best explained with a model that adapts slightly faster than optimal in the above sense. Over time, trained subjects appear to move closer to the predicted optimum[4]. There are several possible interpretations for this result. First, differences in movement execution between the model and real subjects may add a bias to the estimated parameters or the predicted optimum. Yet, the effects of training in stationary conditions indicate that the result is at least qualitatively correct.

[4] When comparing the prediction (figure 7.5 (b)) and the subjects' estimated exponents (figure 8.8), subjects appear to be closer to being optimal than the automated model parameter fits (figure 8.10 (c)) suggest. However, the distributions of these estimates do overlap and also show the same trend over time.

A shorter than optimal memory on naive subjects may have a variety of reasons. At first, subjects have little knowledge about the system details. Therefore, they might construct an ad-hoc forward model by extracting trends. Adapting on very short timescales may indicate that subjects initially expect the system to be non-stationary. Furthermore, since we also found an increase in reaction times $t_r$ with training, there might be a neuronal tradeoff in between slower adaptation and $t_r$. Alternatively, there may be neuronal noise which accumulates during a more prolonged integration of observations. In this case, the CNS would have to optimise structural details that are not included in the model.

However, even naive subjects are most likely within in the parameter range where the dependence of Std($y$) on $\tau_m$ is rather shallow (fig. 7.5 (b)). Therefore, subjects may optimise for additional factors. They might minimise the use of memory resources, underestimate the impact of rare extreme events, interpret small errors as correctable motor errors and large errors as singular externally triggered events, or be risk-seeking. All of these factors have been suggested to be relevant in other motor tasks (see sec. 5.3). In addition, if naive subjects used imperfect internal models, fast adaptation would be useful to realign the internal model of the world with observations.

---

Note, that exponents for experimental time series have to be estimated from very limited sample sizes. Further, even on a single day, real time-series are potentially not perfectly stationary. We did not quantify the subjects' distances from the predicted optimum more rigorously, since it presumably cannot be reliably distinguished how much such a measure would depend on both, subject parameters and model abstractions.

### 9.5.2. Movement execution

A real controller cannot remove predicted errors immediately. Instead, it has to continuously approach the desired state. While the adaptive model does not include physical movement limitations like finite muscle forces in detail, they are still approximated. In reality, the planning and execution of movements is a complex, high-dimensional problem which may include nonlinearities and non-stationary parameters. A very prominent feature of real hand trajectories are bell shaped velocity profiles [Sch02]. They might be considered one of the most obvious goals of a model extension.

Luckily, however, the linear first-order approximation to continuous movement execution has proven to suffice for the model to reproduce many experimental findings even for single subjects in much detail. These results range from error distributions and spectra to linear response functions. Only some stereotypical features of the shapes of single error bursts (or peaks) are reproduced merely approximately (sec. A.8). Nevertheless, the underlying mechanism that initiates these bursts is based on a very general principle (sec 7.2). Hence, highly similar results were found for a second order model (sec. 7.8). Therefore, in order to avoid unnecessary model complexity and overfitting, the first order model can be considered appropriate for many purposes.

Statistical features of experimental time-series are best reproduced by the model for low gains $0.5 < \gamma < 1.2$. No correlation was found between $\gamma$ and the amount of training of the subjects. This may indicate that cautious movements are optimal given biological costs or constraints (for an overview see sec. 5.3). Subjects might, for example, minimise effort, smoothness, or optimise a speed-accuracy tradeoff. This result is neutral with respect to different strategies for muscle contractions, including even intermittent activations, as long as the resulting trajectories are smooth. Note, however, that the model parameter fits represent averages over whole trials. Therefore, subjects might act cautiously most of the time, yet occasionally perform ballistic movements.

Subjects in (K), that is, minimising kurtoses, generated regular oscillations which increased Std($y$) much more than necessary according to the model. One may speculate that these strategies are easier to perform than constantly stopping and accelerating again. Since small Std($y$) were not directly rewarded in (K), subjects may have chosen to create substantial amounts of observable and predictable movements. Note, however, that T moves faster for larger $|y|$. Therefore, maintaining easier detectable deterministic movements is in conflict with minimising effort. Faster and larger movements may also potentially reduce accuracy (sec. 5.2), although we found no evidence that this is was a problem (sec. 8.6). Still, the finding that most subjects in (K) seem to prefer similar standard deviations (fig. 8.3) might indicate an optimal tradeoff.

## 9.6. Some similar and dissimilar tasks in the literature

### 9.6.1. Postural sway

Maintaining an upright posture has previously been compared to the stabilisation of an inverted pendulum (sec. 5.4). Compared to stick balancing, however, upright standing is a much more complex task featuring proprioceptive, visual, and vestibular feedback [LSB77, FTM92]. Ankles stiffness and local feedback loops act dampening, and are even sufficient to stand. Yet, the different sensory modalities are integrated dynamically [vdKP11], and possibly used predictively (e.g. [WPP+98]).

Despite this increase in complexity, postural sway time-series still exhibit three distinct scaling regimes that are very similar to those found for balancing errors (sec. A.5). However, the natural frequency of quiet standing is on the order of one second [WPP+98]. Accordingly, the transition from superdiffusion for short timescales (high frequencies) to subdiffusion also occurs closer to 1s than for balancing (compare fig. A.6 (b) to [CDL94, CC95]. Possible causes for this shift may include the higher inertia of the full body compared hand movements, or additional processing time required for the integration of the additional sensory modalities. There are, however, even more possibilities. For instance, it has been reported that maintaining balance with minimal effort is most likely prioritised over minimal sway [KZJ11].

Whether IAI plays a substantial role during upright standing is unknown and most likely context dependent. It could, however, provide an interesting new explanation (in addition to minimising effort) for as to why subjects exhibit large sway amplitudes during normal standing.

### 9.6.2. Intermittent- or bang-bang control

A possible source for confusion lies in the different uses of the term "intermittent". In this thesis, intermittent fluctuations of control errors are discussed. That is, the alternation between phases with small control errors and bursts of extreme fluctuations. Intermittent control, in contrast, denotes control strategies with discontinuous control actions. Such strategies have been discussed in the context of human motor control. The discussion becomes even more complicated since there are several different types of intermittent control strategies. Direct evidence for the intentional use of intermittent control has so far only been reported in very specific tasks.

In one such task, subjects stood on a balancing board and had to shift their weight in order to regulate a force in a virtual control task while simultaneously keeping their own balance [SFV+09]. The virtual task was to stabilise an inverted pendulum with friction. When the board was used with a low gain, subjects tended to apply forces that were bimodally distributed. Either a maximum force was applied in one of two possible directions, or the pendulum was allowed to drift freely. Velocities were still unimodally distributed.

In another experiment, a simulated unstable load with very high friction was investigated [LGLG11]. This task may be considered similar to balancing an inverted pendulum inside of a honey pot.

Subjects exerted a control force using a joystick. Subjects who were instructed to not use the joystick continuously, but to gently tap on the joystick with their fingers, preferred this method and performed slightly better than when using continuous control.

These strategies are similar to a bang-bang controller (on-off controller) which switches between discrete states like, for example, a thermostat. Such controllers can be optimal in certain certain scenarios. In the two experiments discussed here, however, their effectivity likely stems from the very specific task conditions.

Some studies argued that intermittent control has advantages over continuous control. However, such conclusions were usually drawn from the comparison with proportional-derivative control and therefore don't necessarily apply to all types of continuous control [KZJ11]. Others argued that strategies with a minimal change of acceleration were the least susceptible to multiplicative noise ([HW98, FS01], sec. 5.3). By this argument, some intermittent strategies like bang-bang control would have severe disadvantages in many human motor control tasks.

In principle, intermittent control and even bang-bang control is compatible with both predictive control and IAI, as shown by the model with continuous dynamics and pulsed control in figure A.14. The experimental observations in easy VSB tasks, however, are best described by smooth continuous control movements. This is consistent with vast parts of the motor control literature (sec. 5.3). That is not to say that subjects might not occasionally stop moving during a trial. However, only one subject in score condition (K) use a strategy involving frequent stopping (sec. 8.4). Also, subjects may sometimes perform ballistic movements when extreme errors have to be corrected quickly (sec. A.8). That is, when subjects realise that a large distance between C ans T has to be corrected to not lose control, they may use maximal velocities and accelerations over a very short period to catch up. Yet, subjects on average act more cautiously (sec. A.7, fig. 8.10 (b)).

Nevertheless, control with smooth movements may still be consistent with discontinuous force generation. For example, an intermittent activation of muscles may be an energy-saving alternative to impedance control (i.e. high stiffness) and therefore an epiphenomenon of effort minimisation [ATN13]. Muscle activations could be measured experimentally using, for example, electromyography. However, how muscle contractions are controlled in detail only marginally concerns the IAI which is initiated during movement planning.

### 9.6.3. Sequential effects & Bayesian estimation

In many behavioural tasks, subjects respond faster and more accurately to stimuli that are consistent with a recent pattern. Here we discuss one investigation of such sequential effects, and the connection of the adaptive models presented in this thesis to Bayesian estimation.

Sequential effects have been linked to the common tendency in humans and animals to mistake chance results for hidden patterns or causes. In [YC08] it is investigated whether they reflect the engagement of mechanisms that otherwise serve for adapting to a changing environment. To model sequential effects in a two-alternative forced choice task ([VNB+02]), two different Bayesian models are introduced. In both models, the probability $\gamma$ that a stimulus is repeated is a hidden parameter that has to be estimated. In a fixed belief model, $\gamma$ is assumed to be constant over all trials. In a dynamic belief model, $\gamma$ changes according to a Markov process. Only the latter model can explain the behavioural data. It is further shown that Bayes-optimal prediction in this case is well approximated by a linear filter that weights past observations exponentially.

As noted above, the presumably faster than optimal adaptation in naive subjects might be considered a sign that subjects initially seek robustness in potentially non-stationary tasks. This interpretation is consistent with the hypothesis proposed in [YC08] that humans apparently act sub-optimal in certain behavioural experiments because they are adapted to non-stationary environments. It is also consistent with the finding that subjects in VSB tasks can tolerate a frequently changing time constant $\tau$ ([PREP07], fig. 7.6). As a final note, applying recursive Bayesian estimation to the a discrete-time balancing problem with a hidden time constant leads to a Kalman-like filter. Adding a "locality bias", that is, making the estimator assume a rapidly changing $\tau$, yields the minimal control model described in section 7.3. This alternative derivation, however, is quite lengthy and does not add considerable new insights. It is therefore omitted.

### 9.6.4. Change Blindness

While we experience a detailed and stable visual world, evidence suggests that little information is preserved across views [SL97]. Unless a change in a visual scene evokes a localisable transient on the retina, people will generally not detect it. For example, in an experiment [SL98] one experimenter asked pedestrians for directions. This interaction was briefly interrupted by a door which was carried between the experimenter and the pedestrian. During this interruption, the occluded experimenter was replaced by a different experimenter. Only half of the pedestrians detected the change. This phenomenon is called change blindness.

In another experiment [TBHS03], subjects had to sort blocks of different sizes and colours in a virtual reality environment onto one of two conveyor belts. After picking up a block, subjects made a saccade (eye movement) towards the conveyor belts. Sometimes the blocks size was changed during this saccade. If size was a task-irrelevant feature, subjects generally failed to detect the change when they looked at the object again. Even though subjects were instructed to report possible bugs in the virtual reality program, 88% of the subjects never reported a change. If the block size was relevant for a decision that was made before the change, approximately one third of the changes were detected. If a decision based on the block size had to be made after the change, half of the changes were detected. The authors suggested that subjects extracted information from the fixation point only "just in time" when needed to solve the current goal.

Nevertheless, human memory is not generally bad. For example, when shown hundreds or thousands of photographs, humans can later accurately (sometimes 95% correct) remember which ones they have seen before. Despite this ability, subjects cannot detect whether they are shown a manipulated version of the image, as long as the gist of the scene remains intact (summary of several papers in [SL98]).

One might argue that these findings have a certain similarity to human behaviour in balancing tasks. This analogy becomes more obvious if the adaptive model is paraphrased as follows: Subjects internalise an abstract notion of the task of catching a target that constantly moves away from the cursor. The detailed information necessary to plan the corrective movements, however, is extracted from the observations just in time.

## 9.7. Conclusion and Outlook

We found that subjects in VSB tasks most likely employ highly dynamic forward models. That is, movement planning strategies which are based on information about predictable dynamics which is extracted from observations on the fly. Such a strategy presumably allows for behavioural flexibility with minimal demand on working memory. It was shown to be consistent with a minimisation of average errors. This type of controller naturally precludes its own ability to predict future dynamics of the control system. In particular, a tradeoff has to be made between the reduction of average errors by eliminating even local trends that arise by chance, and rare extreme missteps due to an increased susceptibility to noise. This tradeoff is not fixed. Instead, training under different task objectives can fundamentally alter a subject's behaviour in a balancing task.

These results highlight that action and perception are subsystems of a greater system. The interaction of these subsystems can give rise to emergent effects. For example, it can usually be expected that the environment stores information about itself. Internal models don't necessarily need to replicate more of this information than required for a decision at the very moment in time where it is made. This is particularly true if adaptive behaviour is favourable anyway due to incomplete information, non-stationarities, or random perturbations. However, as discussed above, while such strategies may be very flexible they might also lead to phenomena like sequential effects and change blindness. If interaction with the environment eliminates dynamics which are required to observe information on the systems structure, a controller runs into an IAI.

In contrast, many existing motor control models treat movement variability and perception as separate problems. For balancing in particular, the most prevalent model in the literature (sec. 6.3) assumes non-adaptive behaviour, and enormous amounts of multiplicative execution noise–a hypothesis that we can consider highly unlikely based to the results presented in the previous chapters.

Most other motor models are adaptive and feature some kind of optimisation (sec. 5.1). However, perception and control strategies are typically optimised separately. To estimate hidden system states, many models use Kalman Filters which, in order to calculate the optimal weight for new observations (Kalman gain), are constructed with detailed knowledge about the structure of the control system. In addition, a typical control strategy is optimised according to some criteria like, for example, minimising jerk. To account for experimentally observed movement variability, these models often include multiplicative execution noise. However, as discussed above, the common attribution of movement variability to execution noise alone is not sufficient to explain empirical findings in balancing tasks.

One possible future direction for motor control modelling is to consider interactions between the different subsystems that contribute to behaviour. This idea raises many questions. At which stages of the action-perception cycle is adaptivity necessary? Which cost functions lead to holistic strategies for perception and control that can reproduce experimental findings? Are ideal observers necessary or even possible without excessive training for any given task? Is it possible that the CNS generally uses generic forward models that are adapted to a given task as needed? Alternatively, might humans even perform some kind of model-free control using ad-hoc-predictions? If so, can predictions be made across different tasks?

Normative models usually have to be constructed for any given tasks by the experimenter. Instead, some of the above questions may potentially be studied in the context of Optimal Feedback Control (OFC) . This framework allows the calculation of control laws that optimise a mixed cost function which specifies accuracy and energetic costs [TJ02, FW11]. OFC predicts that feedback gains are task dependent, vary throughout the task, and that task-irrelevant fluctuations will be ignored. In contrast, gains in Kalman Feedback Control (KFC) models are adjusted to minimise perception errors independently of the task.

As more concrete follow-ups to the previous chapters, several studies could be conducted. First, the influence of the task difficulty determined by the system time constant $\tau$ may be studied quantitatively in more detail. In particular, how the scaling of control errors changes when $\tau$ is varied close to the reaction time $t_r$ could allow to estimate how much multiplicative noise arises during movement planning as opposed to movement execution. Similar results might be obtained by comparing the influence of different levels of artificial noise in a large number of trials. The differences in movement amplitudes between the two most extreme conditions should be as large as possible. However, according to the presently available results, the influence of execution noise appears to be very small.

Next, first- and second-order target dynamics, as well as one-, two-, and three dimensional balancing tasks could be compared. Of particular interests is a detailed comparison of VSB and real stick balancing. While the results presented in this thesis should in theory be very general, this hypothesis remains to be tested empirically. A further question is how far IAI is relevant for postural sway. Such a study may directly compare fluctuations during stick balancing and upright standing. It is very likely that different results would be obtained for upright standing under different conditions (e.g. on one leg or an unstable platform), and for different task objectives.

# A. Appendix

## A.1. Significance of power-law fits

Here we demonstrate that the CCDF for the continuous balancing model follows a true power law. When temporal correlations are properly taken into account, remaining deviations from an analytical power law can be explained by chance. This finding further justifies the use of the model as surrogate data for tests of significance in the experimental data. These and more results were published in the supplement to [PP11].

### A.1.1. Motivation

Whether differences between two CCDFs can be explained by chance alone can be assessed using significance tests like the KS-test. However, there are also systematic sources for deviations. Experimental data often is only approximated by a certain analytical distribution of independent random variables, but does not truly follow this idealisation. In very large data sets, even the smallest differences may become significant. A common source for such deviations are correlations, including higher order ones. Then, the measurements are not perfectly independent anymore and deviations are larger than analytically predicted for absolutely independent random variables. While probability distribution and autocorrelation are completely independent features, statistical tests usually assume Independent and Identically Distributed samples. Yet, every time-series with a sufficiently high sampling rate violates this assumption.

A common problem for testing power laws in finite data sets is that often only the tail of a distribution follows a power-law in the limit of very large events. When determining a cutoff above which to fit, a non-trivial trade-off has to be made. On the one hand, systematic deviations due to imperfect convergence to the power law are more prominent for smaller events. On the other hand, many fewer samples are available for very large event magnitudes. Last not least, it is possible that a distribution has no hard cutoff, but a slow gradual convergence towards a power-law.

### A.1.2. Reducing correlations

To assess how well an analytical power law fits the CCDF-tails of simulated time series, we proceed in two steps. First, we show how deviations from the power-law in the model time series depend on temporal correlations. Second, we introduce a method to compare the quality of fits between a given distribution and surrogate data. This is done by comparing how well model time series and power law distributed independent random variates are fitted by an analytical power law.

Figure A.1 (a) shows the CCDFs of balancing errors made by the control model (sec. 7.4), and a fit. For large events, the distribution is approximately linear in double logarithmic coordinates until it takes a sudden break at the largest event. The reason for this effect is that the largest few data points all belong to a single peak in the time series. To reduce the correlations between the analysed events, distributions of subsets of the same time series are shown where only one event every second (fig. A.1 (b)) and every 10 seconds (fig. A.1 (c)) were used.[1] Finally, figure A.1 (d) shows the distribution for independent random variates that follow a power-law above a threshold. A method to quantitatively evaluate the goodness of the respective fits is the KS-statistic. This nonparametric test calculates the maximum distance $D$ between two distributions. For independent random variates, the p-values can then be calculated independently of the actual shape

[1] Note how this choice of subsampling intervals relates to the correlation structure investigated in figure A.6

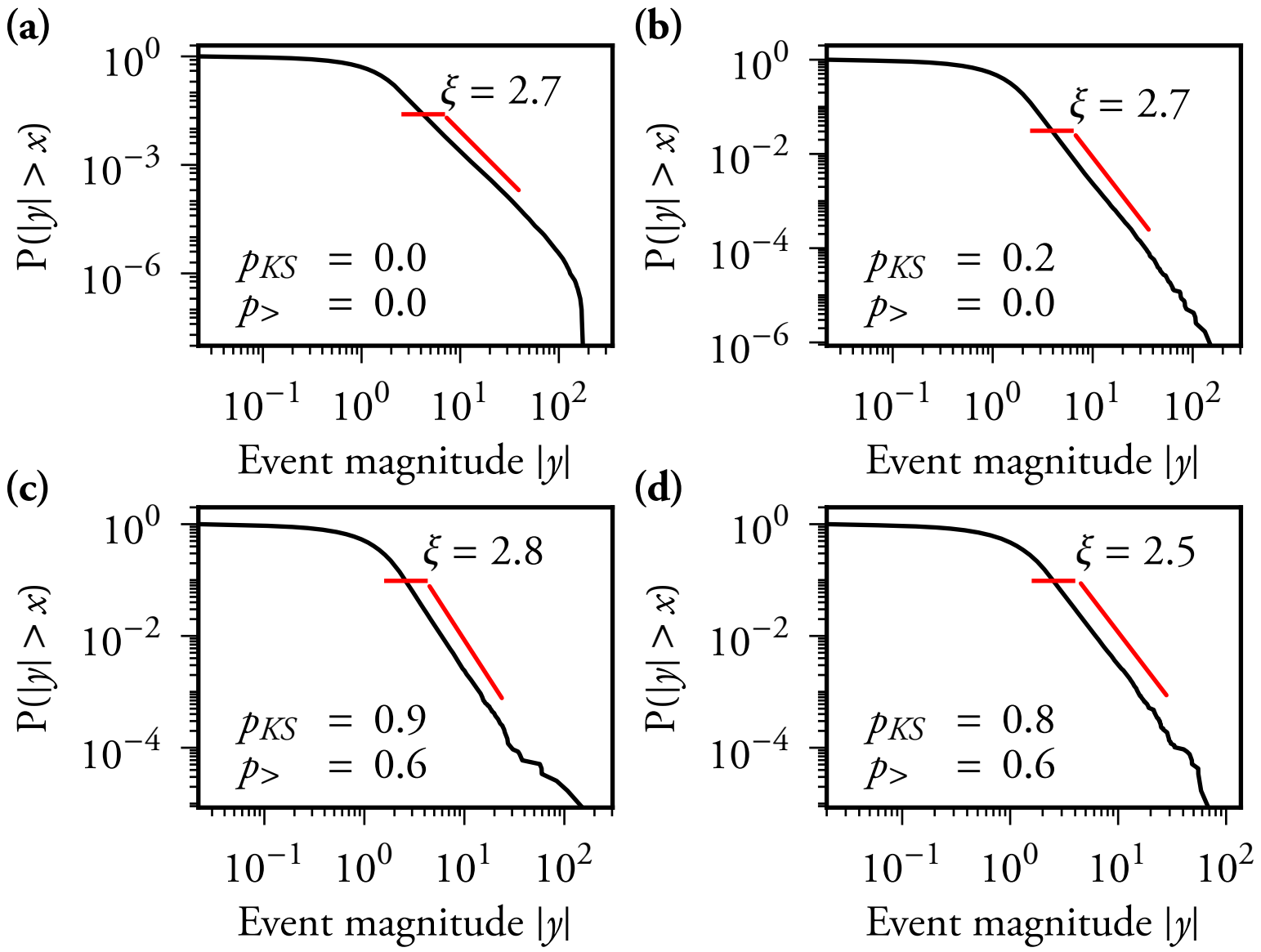

Figure A.1.: (**a**): Complementary cumulative distribution of balancing errors $|y|$ made by the control model described sec. 7.4 with reaction time $t_r = 170\,\mathrm{ms}$, gain $\gamma = 1.07$ and memory time constant $\tau_m = 140\,\mathrm{ms}$. The system time constant was $\tau = 250\,\mathrm{ms}$. Time discretisation was $h = 11.8\,\mathrm{ms}$. Red diagonal lines: power law fits. Short red horizontal lines: Hill-estimator cutoff optimising the KS-statistics. **(b), (c)**: Distributions of subsets of the same time-series as in (a), but with only one value of $|y|$ every $\Delta t = 1$ and $\Delta t = 10$ seconds used respectively. **(d)** Independent random variates that are distributed according to a Gaussian below a threshold $x_{th} = 2.5$ and according to to a power-law with $\xi = 2.5$ above. The exact analytical shape of the distribution was obtained by requiring the PDF and its first derivative to be continuous at $x_c$. The variates were then obtained from uniform ones using inverse transform sampling. $p_{KS}$-values refer to tabulated p-values for the Kolmogorov-Smirnov statistics. For the $p_>$-values, 1000 time series with random variates like in (d) were generated and a power-law was fitted to each one. The $p_>$-values refer to the fraction of times that the KS-statistics of a fit was worse than for the one distribution depicted in the respective subfigure. As the correlations in the model time series decrease, both goodness of fit tests converge towards those for the IID variates.

of the distributions that are tested against each other. Here, we call these common p-values $p_{KS}$ values. They are stated for each respective subplot in figure A.1. The probability that the deviations of model distributions from an analytical power-law are explained by chance alone increases as the correlations decrease. When only one event every 10s is taken into account, the $p_{KS}$-value for this particular time series even reaches 0.9. That is, even though the largest approximately 5 data points appear to deviate from the power law, even larger deviations are to be expected by chance alone according to the KS-statistics. On average, of course, one expects p-values values of 0.5 for two identical distributions. Commonly, p-values below 0.05 lead to the rejection of the hypotheses, that the observed deviations between two distributions can be assessed to chance alone.

### A.1.3. Testing for convergence

The previous method has several disadvantages. First, massive subsampling is required. Applied to real data sets containing at most $10^6$ events, only $10^3$ events are left over. Since only few of these events belong to the distribution tail, it spans less than one order of magnitude of control errors. Second, reducing the data set reduces the discriminatory power of the KS-Test. Third, to use as many data points as possible in smaller data sets and avoid fitting random kinks, it is useful to calculate fits using a range of cutoffs and then use only the one that minimises a loss function. We exclusively used the cutoff that minimises the KS-D-value. From our experience, various criteria generally tend to underestimate the cutoff if the tested range includes too small events. This additional complication may decrease the goodness of some fits.

To avoid these problems, we determine p-values by numerically calculating the probability that the KS-statistics of appropriate surrogate data shows bigger deviations from the respective fits than the tested distribution[2]. This p-value is called $p_>$ in the following, and in

[2]A similar method without respecting correlations is discussed in [CSN09].

figure A.1. Again, we expect this value to be 0.5 on average for data sets drawn from exactly the same distribution.

In principle, calculating $p_>$ requires no subsampling. Unfortunately, power-law-distributed random variates with appropriate correlations are not readily available. Therefore, we here compare very large subsampled model time series to independent random variates, allowing us to maintain a constant sample size while reducing correlations[3].

Figure A.2 (a) shows the average $p_>$ for model time series with different amounts of subsampling while the total number of events was held constant. That is, correlations are reduced without reducing the number of analysed events. Therefore, the observed convergence towards 0.5 for low correlations is not caused by a loss of power of the test. However, virtually identically results are obtained if the sample size is not kept constant [PP11]. Similar results are also found for different parameter sets including controller $\tau_m$ that minimise mean squared control errors (not shown).

We tested three different constant sample sizes. As the number of events is increased from the red curve in figure A.2 to the blue one, the test becomes more discriminative. Naively, one might have expected the opposite, that is, the finite correlation length to become insignificant for sufficiently large time series. However, the KS-Test assumes that the maximum distance $D$ between the empirical cumulative distribution function of $n$ IID samples of a random variable and the true underlying distribution scales as $D \propto 1/\sqrt{n}$. This scaling is violated for correlated samples, for which convergence is slower. Therefore, while $D$ still converges towards zero, calculated $p_{KS}$-values also converge towards zero instead of keeping an average of 0.5 as is the case for independent samples.

[3] Minor problems with the described method may still be expected because the exact correlations in the tested time series and the surrogate ones may differ. Also, distributions from model and experimental time series are convex for small $|y|$ and then usually become slightly concave when converging towards an approximately straight line in log-log coordinates. The distributions of the transformed random variates we use for comparison (fig. A.1 (d)) are convex up to the threshold and then exactly follow the power law by construction. However, these potential pitfalls turned out not to be very problematic.

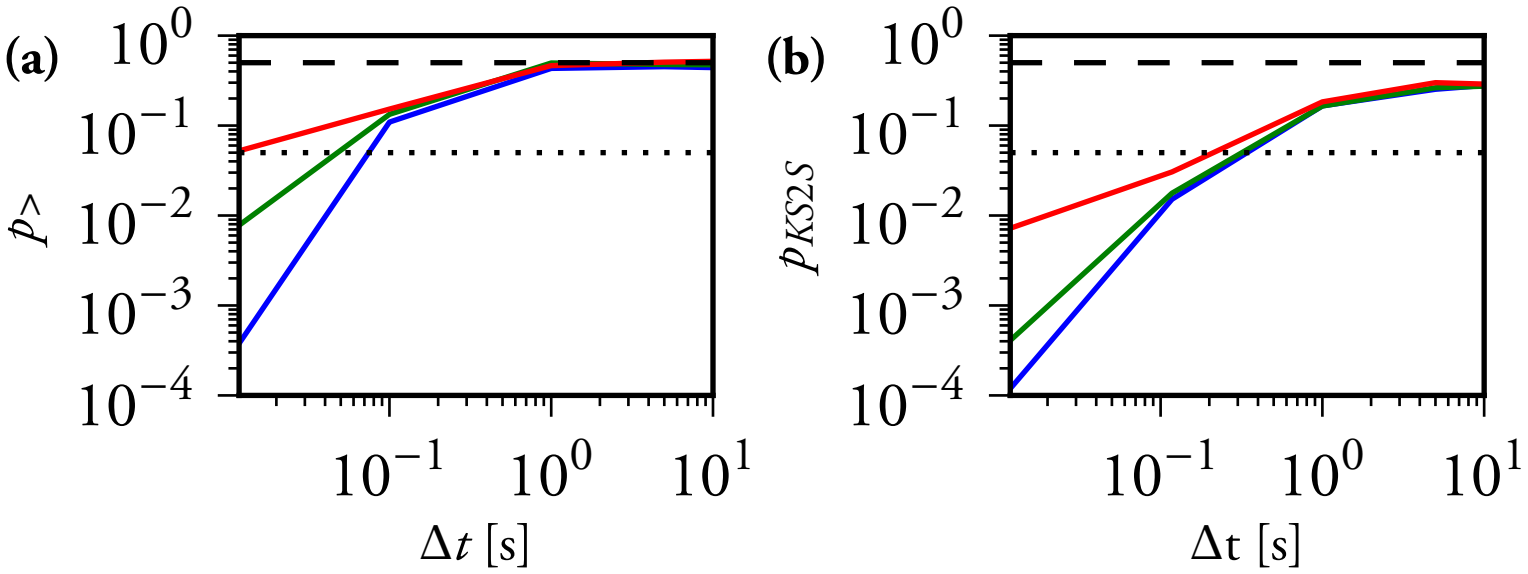

Figure A.2.: Significance tests with different amounts of decorrelation by subsampling. **(a)** Probability, that the KS-statistics for power law fits for independent random variates that follow a true power law above a threshold is worse than for control errors from the continuous control model. For each data point, $10^3$ model time series were compared to $10^3$ different sets of random variates each. **(b)**: Probability, according to the two-sided KS-statistics, that the tails of two simulations of the continuous control model follow the same distribution. For each data point, $10^2$ tuples of model time-series were compared to each other. In both **(a) and (b)**, parameters were chosen as in fig. A.1 except for trial lengths. Here the total number of tested samples from the model was held constant for each $\Delta t$ at $10^4$ (red), $10^5$ (green) and $10^6$ (blue) analysed control errors. For example, in order to get $10^6$ control errors that occurred 10s separated from each other, each simulated time series contained $8.5 \cdot 10^8$ steps before subsampling. To restrict the analysis to the distribution tails, cutoffs for the Hill estimator were set to use only the larger half of the log-range of the ccdf. For example, for sets consisting of $10^6$ events, only the largest $10^3$ ones were used. Dashed lines: A p-value of 0.5 is expected for a comparison of IID samples drawn from identically distributed populations. Dotted line: significance level $p_> = 0.05$. As correlations in the model time series decrease, KS-statistics converge towards those for the IID variates–albeit slower in (b) than in (a).

This effect can also be demonstrated numerically. However, since we don't have a method to generate power law distributed random variates with arbitrary correlations, figure A.2 (b) shows analogous results for a two-sided test. Here, the tails of two time series for the same model with identical parameters are compared. Just like in figure A.2 (a), correlations lead to an underestimation of the amount of deviations expected for a given sample size, and therefore low p-values. Convergence towards 0.5 for subsampled time-series is even slower than in the one-sided test. Most notably, for a fixed amount of subsampling p-values decrease when the number of samples is increased. This demonstrates that the KS-test indeed becomes more sensitive to correlations for larger sample sizes. We found similar results for correlated Gaussian random variates (not shown), but could not find an example where correlations become insignificant for large sample sets.

In conclusion, we cannot reject the hypothesis that model error distributions follow a power-law in the limit for large errors. This result justifies the use of the model as surrogate data for the experimental time series. Subsampling is not required in this case, since the model time-series exhibit temporal correlations that are very similar to those in the experimental time-series. P-values for VSB-experiments are discussed in sections 7.7, 8.5.3, and 8.8. Some more examples are shown in the supplement to [PP11].

## A.2. Model with increased gain and memory

As discussed in section 7.5, realistic time series for the balancing model are obtained for a combination of cautiously performed control with $\gamma$ slightly above one, and fast adaptation with $\tau_m \leq t_r$. Increasing either $\gamma$ or $\tau_m$ increases $\lambda_1$ which can even exceed $\lambda_2$ (not shown). In that case, the characteristic knee shape of the experimental spectra is not reproduced. Further, the power law scaling of the control errors $y$ vanishes if only $\tau_m$ is increased (fig. 7.5 (b)). Increasing also $\gamma$ amplifies control errors again, thereby reestablishing a power law tail. However, the characteristic shapes of the CCDF and PSD are not recovered, as shown in figure A.3.

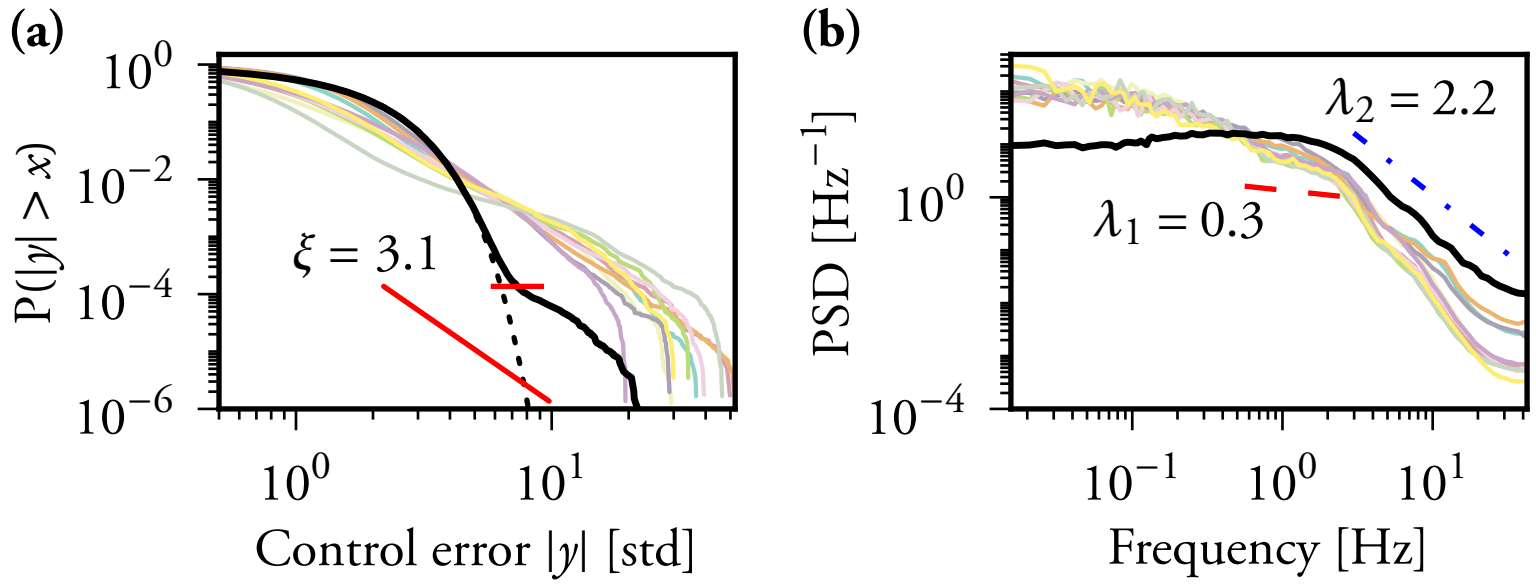

Figure A.3.: Comparison of experimental time series (thin coloured lines) with the balancing model in a slow adaptation, high gain regime (thick black lines). Parameters: $\tau_m = 1\,\mathrm{s}$, $\gamma = 5$, $t_r = 180\,\mathrm{ms}$, $\tau = 250\,\mathrm{ms}$. As opposed to the combination of smaller $\tau_m$ and $\gamma$, here **(a)** the onset of the power law in the probability distribution, and **(b)** the characteristic knee in the experimental power spectra is not reproduced.

## A.3. No adaptation and multiplicative execution noise

Figure A.4 shows a non-adaptive (yet predictive) controller which is driven into the critical regime by precisely tuned parametric execution noise. This model does not reproduce experimental spectra. It also features a softer onset of the CCDF scaling than observed in the experimental time-series and for the adaptive model. For this model, the shape of the error distribution depends on both the additive and multiplicative parts of the noise. Therefore, in contrast to the adaptive model, the overall scaling cannot be chosen independently from the tail exponent. Most significantly, the error distribution only resembles a power law for very specific parameter choices. No pa-

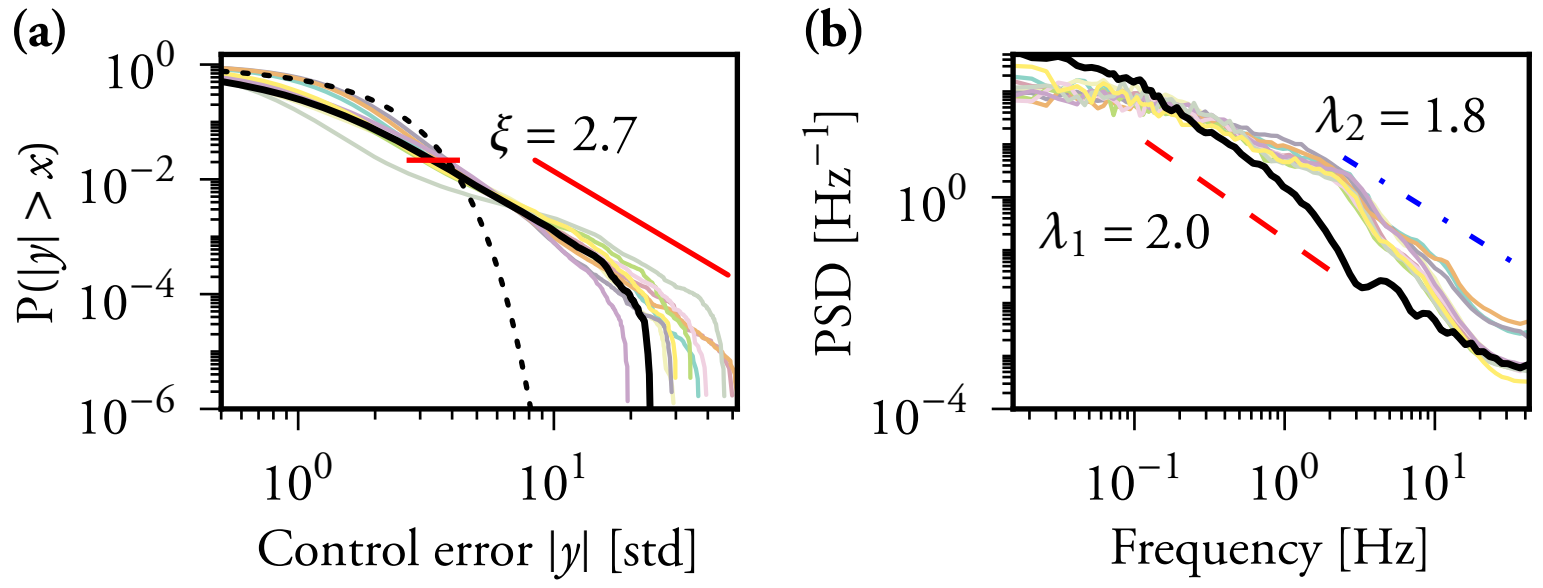

Figure A.4.: Comparison of experimental time series (thin coloured lines) and a modified balancing model where the estimator $1/\tilde{\vartheta}$ has been replaced by the true system time constant $\tau$ (thick black lines). To obtain power laws, parametric noise has been introduced by choosing the scaling for the driving noise as $\sigma(t) = 1 + 0.2 \cdot |\tilde{y}(t)|$. Other parameters: $\gamma = 1.1$, $t_r = 200\,\text{ms}$, $\tau = 250\,\text{ms}$. As opposed to the adaptive model (sec. 7.5), here both **(a)** the onset of the power law in the CCDF is slightly too shallow, and **(b)** the characteristic knee in the PSD is not reproduced at all.

rameter combination reproduces the experimental spectra. See also section A.8.

## A.4. Model high-frequency response.

Figure A.5 (a) shows the PSD for the balancing model (sec. 7.5) simulated with very short discretisation steps. The scaling converges towards values of 2 for high frequencies. Figure A.5 (b) shows the model driven by strongly low-pass filtered noise. The scaling exponent $\lambda_1$ is at most weakly affected while the high frequency scaling exponent $\lambda_2$ is doubled. This finding is consistent with the hypothesis, that $\lambda_1$ mainly characterises active control behaviour while $\lambda_2$ characterises intrinsic noise and passive damping properties of the system. The filtering also affects the small resonances close to the reaction time found in the experimental time series.

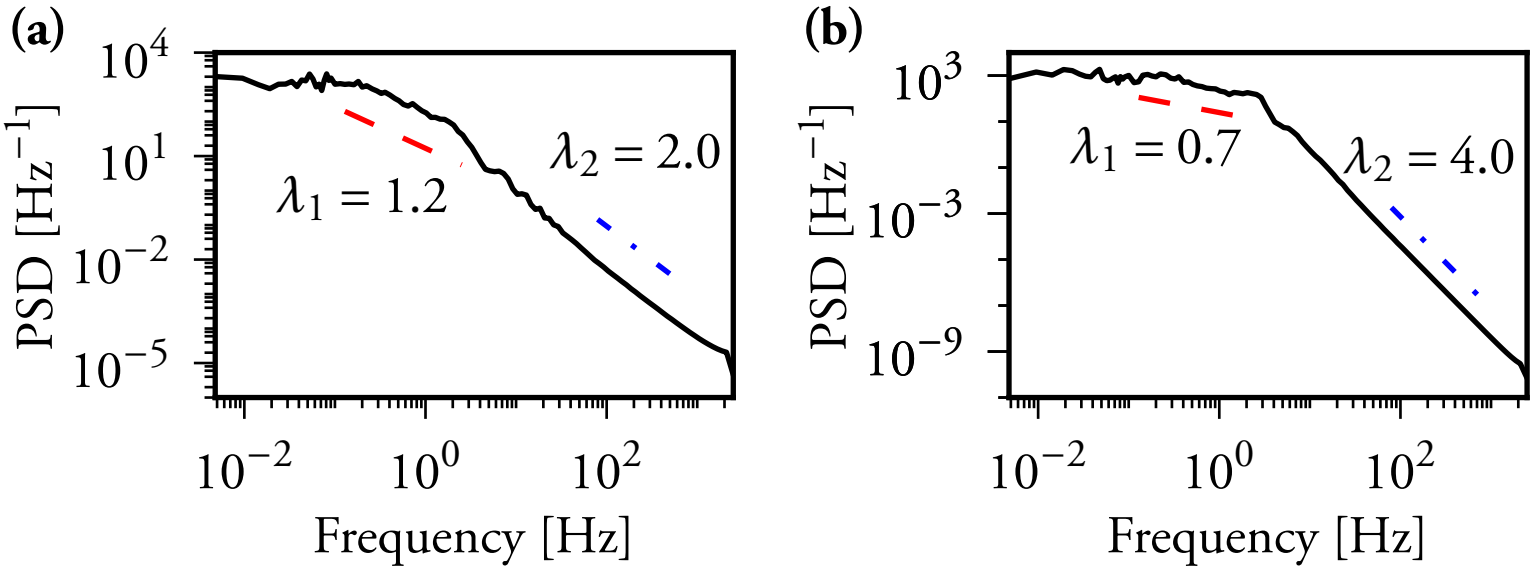

Figure A.5.: **(a)**: Power spectrum for the balancing model (sec. 7.4), but simulated with much smaller discretisation steps. Parameters: $t_r = 180\,\text{ms}$, $g = 1.1$, $\tau_m = 120\,\text{ms}$, $\tau = 250\,\text{ms}$. **(b)**: Power spectrum for a model with the same parameters, but driven by low-pass filtered noise with a time-constant $\tau_{\text{lp}} = 0.5\,\text{s}$.

## A.5. Variance- and displacement scaling

Multi-scaling in diffusive processes is sometimes investigated with different measures than those used throughout this thesis. Here, we apply two popular measures that allow to compare VSB to other processes. Figure A.6 (a) shows the dependence of the standard deviation of the series of increments $y(t + \Delta t) - y(t)$. For short increments, the scaling exponent of one characterises hyper-diffusive behaviour (i.e. positive correlations). Around increments of 10s the scaling approaches the expected value for an uncorrelated random walk of 0.5. Very similar behaviour has been reported for stock-market time series [MS00].

Figure A.6 (b) shows the dependence of mean square displacements between control errors $\mathrm{E}(y(t+\Delta t)^2 - y(t)^2)$ on the time $\Delta t$ by which they are separated. Again, hyperdiffusion is observed for small $\Delta t$. For the smallest timescales, experimental time-series are slightly more correlated then those from the model. Like the power spectra for very high frequencies, this deviation is most likely due to passive damping which is not included in the model. For $\Delta t$ in between the reaction

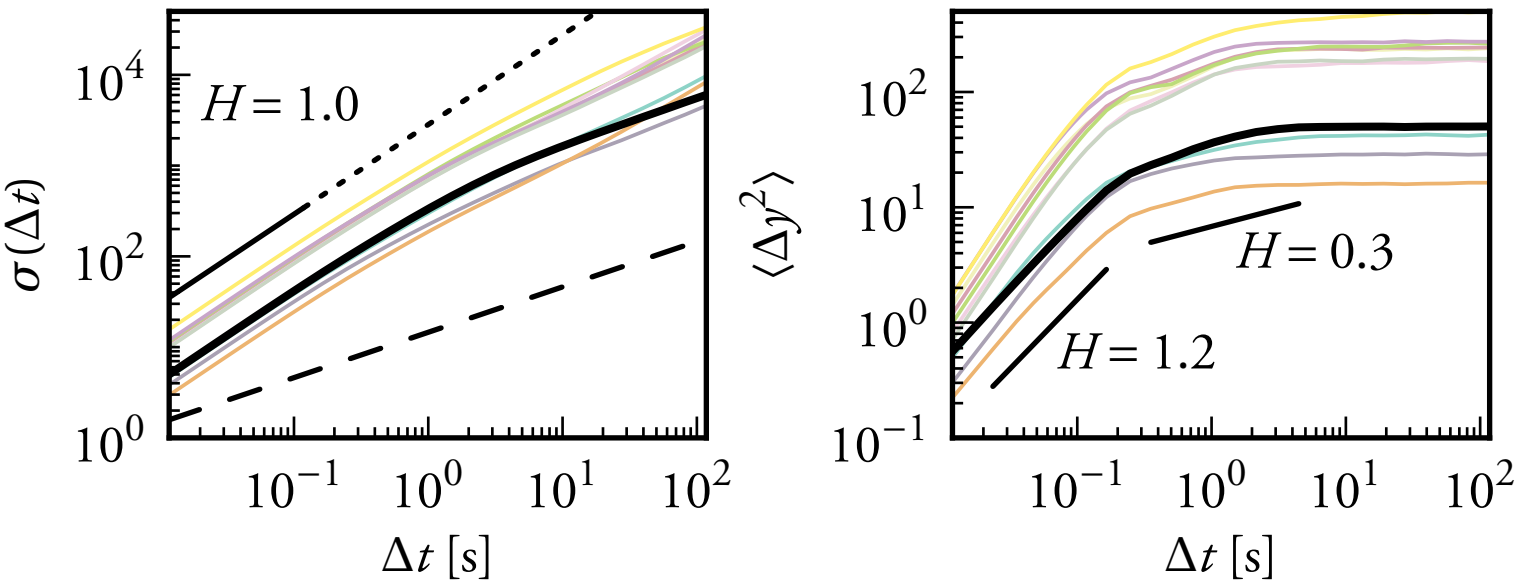

Figure A.6.: Scaling of **(a)**: the standard deviation of the cumulated magnitudes, and **(b)**: mean displacement for different time intervals $\Delta t$. The analysed time-series for model (thick black lines) and experiments (thin coloured line) are the same as in fig. 7.4. The dashed line in (a) indicates the scaling of a random walk (H = 0.5).

time and 10s, the scaling corresponds to subdiffusion. For even higher $\Delta t$, saturation is observed. This is expected for mean-reverting processes. Similar scaling regimes for displacements have been reported before for center of pressure trajectories of upright standing humans [CDL94, CC95]. They have been speculated to represent time scales dominated by open- and closed-loop control. While this idea is basically consistent with the structure of the adaptive control model, the latter one does not include a mechanism for open-loop control like damping of joints by reflexes or mechanical properties.

## A.6. Influence of the task difficulty

It was recently reported that easy VSB-tasks lead to more heavily tailed control error distributions than difficult ones ([MFC+11], Fig. 4). A more difficult task involves an unstable system with a shorter time constant. This corresponds to, for example, a shorter stick. As shown in figure A.7 this result is reproduced by the balancing model.

This finding may seem counter-intuitive and the opposite effect can be found for systems with very strong multiplicative execution noise (not shown). However, the model (fig. A.7) provides a simple explanation. In a very difficult task, subjects cannot keep the the controller as close to the target as in an easy task. Therefore, average fluctuations are bigger for more difficult tasks. However, the rare extreme missteps in easy tasks were caused by adaptation in situations where predictable dynamics are difficult to discern from random noise. This happens in particular if controller and target are very close to each other for prolonged periods of time (see secs. 7.2, and A.8). Therefore, in the model, the amplitudes of the most extreme control errors are less affected by an increase in task difficulty.

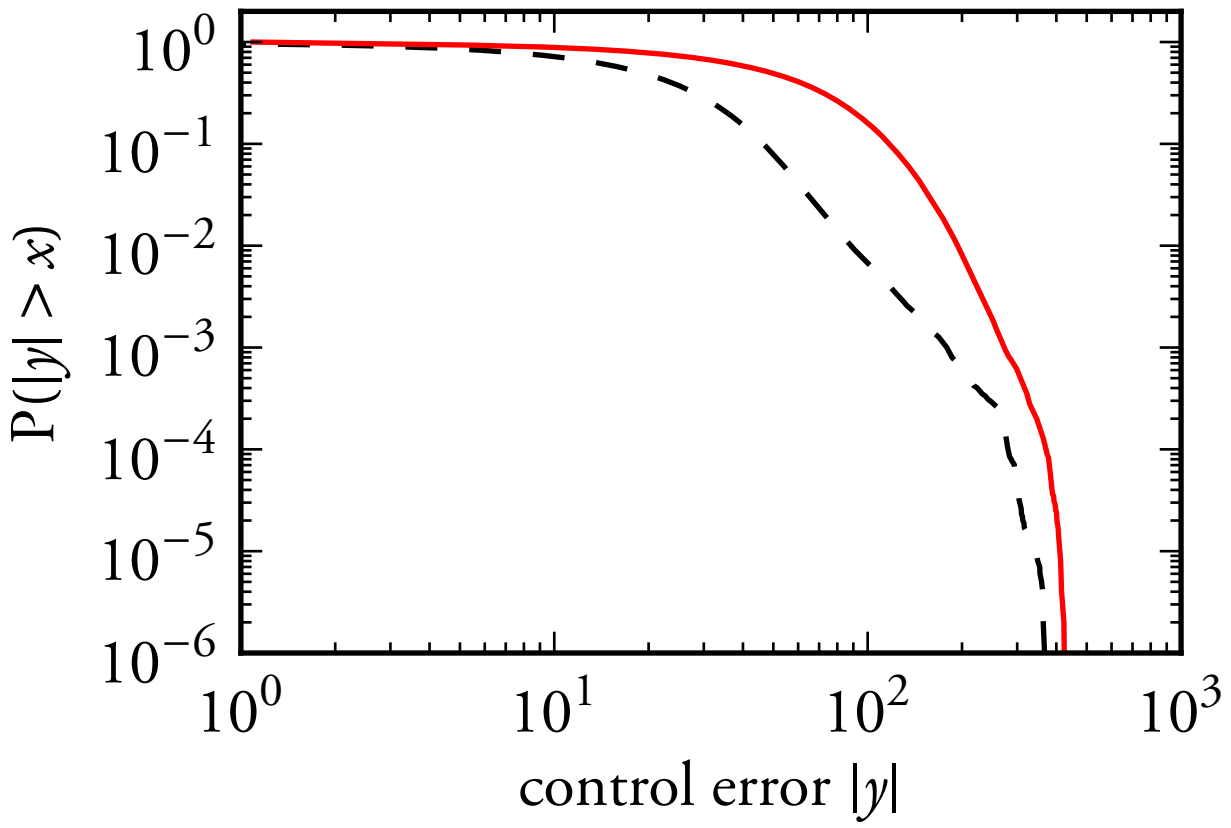

Figure A.7.: Comparison of CCDFs for the model (sec. 7.4) in a hard balancing task with $\tau = 100\,\mathrm{ms}$ (solid red line), and the same model in a more easy setting with $\tau = 250\,\mathrm{ms}$ (dashed black line). Other parameters: $t_r = 200\,\mathrm{ms}$, $g = 1.1$, $\tau_m = 140\,\mathrm{ms}$, $\sigma = 20$. For the purpose of this comparison, the two time series were not normalised.

### A.6.1. Rail trial completion

Figure A.8 shows trial completion rates for the experiments described in chapter 8. On each day and in each condition, 70% or more of the trials were completed. Completion rates for trials in (S) increase over

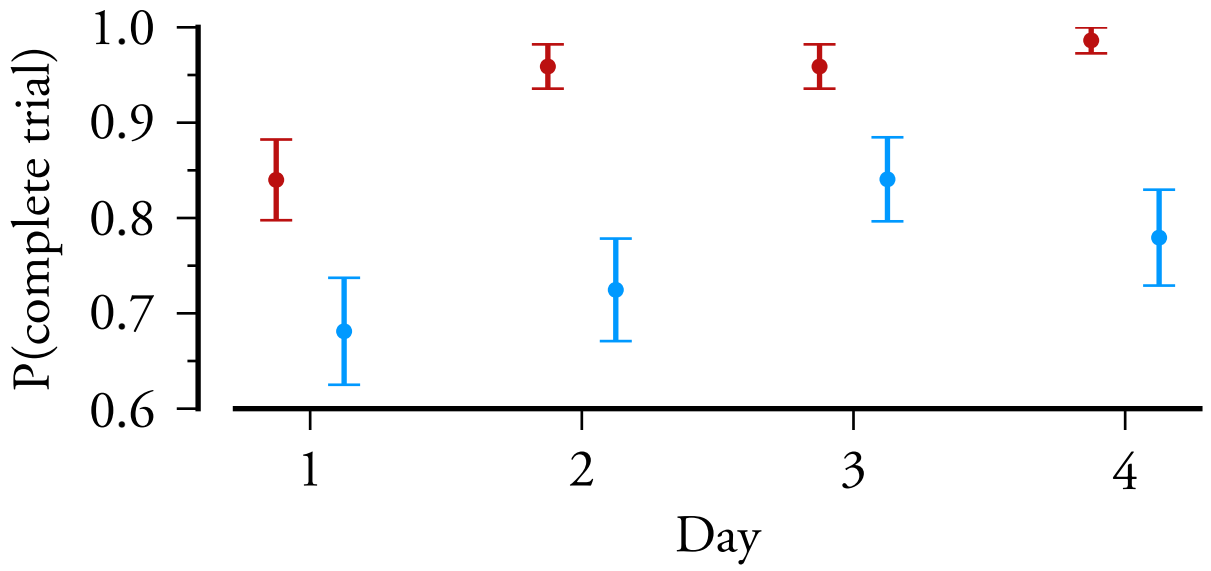

Figure A.8.: Trial completion rates in one dimensional VSB. Red: (S), light blue: (K). The y-axis starts at 60% for a better visibility of the means and standard errors. Since single trials can only be complete or incomplete, median statistics are inappropriate for this figure. Significance levels from Fisher's exact test are are consistent with the depicted standard errors (and therefore redundant and not shown).

time and are also significantly higher than those for in (K). In the latter condition, there are large differences between subjects, depending on their strategy.

## A.7. Measured and modelled linear responses

The noise trials in the one-dimensional VSB experiment described in chapter 8 allow for a calculation of the subjects' linear response functions. For each noise trial, the result shows a subjects average response to a $\delta$-pulse. An example is shown in figure A.9.

After the pulse, control errors initially grow. Only after the reaction time, the controller notices the unpredicted error and starts to correct it. Therefore, we obtain a direct measure of the reaction time. Result statistics are shown in figure 8.9. Some subjects' mean reaction times differ by more than two standard deviations (not shown). We can therefore conclude that individual differences exist. Automated model fits to the regular trials on a given day usually also reproduce the linear responses measured during noise trials well (fig. A.9).

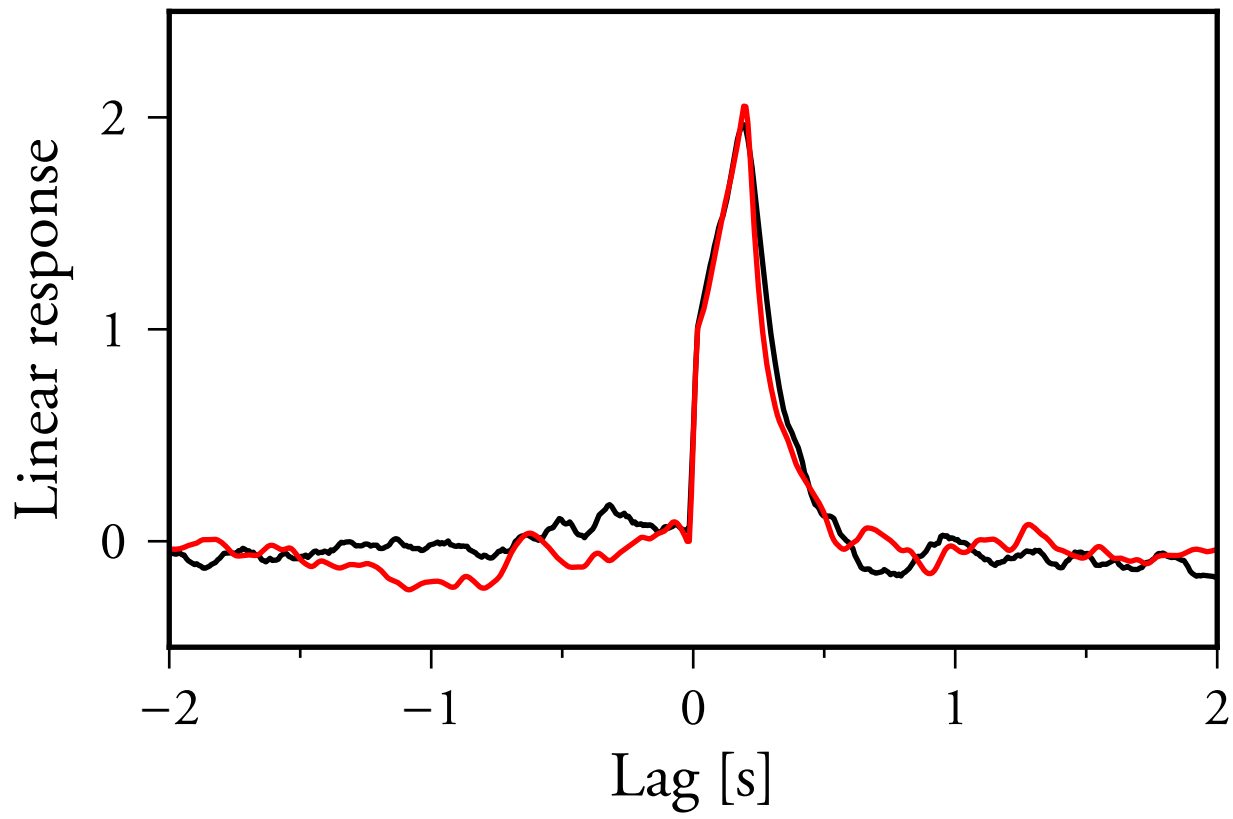

Figure A.9.: A linear response measured for subject 1 on day 3 (black), and for the model (red) using the same noise track and identical processing. Model parameters were set as the median of the automatically fitted regular trials for this subject and day (method in subsec. 8.7.1): $t_r = 204.3\,\mathrm{ms}$, $g = 1.0685$, $t_m = 152.5\,\mathrm{ms}$. Both linear responses were normalised to match a unit $\delta$-pulse.

As described in subsection 8.1.2, we used coloured test noise since white noise was too confusing in pretests of the experiment. Prior to calculating the linear responses, we filtered out noise with higher frequencies than present in the test noise (using a finite impulse response linear phase filter with a Kaiser window of order 10). Using the same method on different types of surrogate data revealed that the qualitative shape of the pulse responses was not affected (not shown). However, maxima of the linear responses were offset 10ms on average towards shorter times. This shift is caused by the applied filtering and noise present in finite time series. Under ideal conditions (white test noise and much longer time series), the peak coincides with model reaction times. Therefore, the reported reaction times (fig. A.9) correspond to the measured peaks of the linear response plus 10ms.

For the model, the decay of the pulse response for lags above $t_r$ depends on both $\gamma$, and $\tau_m$ in a complex way. Some additional modulation is caused by the use of coloured test noise. Therefore, unfortunately, the pulse responses do not provide a direct measure of further model parameters.

## A.8. Movement microstructure

The results in this part of the thesis focus on rapid online adaptation as a source for multiplicative noise. Therefore, the investigated balancing model includes an explicit high-level mechanism for movement planning. For movement generation, the simplest linear first order approximation is sufficient to explain many details of error statistics observed in VSB experiments.

The actual neuronal mechanisms for translating the planned movements into muscle commands and the physics of the arm were not considered. However, the data and models presented may provide a starting point for future work in this direction. While (average) pulse responses are well reproduced by the simple first order continuous model (sec. A.7), some differences are found for the shapes of the peaks in the actual time series. Therefore, we here discuss some representative examples from the experiments described in chapter 8.

Figure A.10 (a), and (b) both show parts of time series containing several error peaks of different amplitudes. Several of these "peak archetypes" may occur within the same trial. Smaller errors which are just above the baseline fluctuations are characterised by a sudden rise and a slower decay ((a): 12s, (b): 6s). Larger errors sometimes are counteracted very fast involving a small amount of oversteering ((a): 23s). If oscillations occur, they typically decay quickly. Other times, subjects understeer leading to a short series of peaks with the same sign ((a): 45s, (b): 75s after $y$ changes the sign once). Many peaks initially start with a smaller movement in the opposite direction, although this is not always the case (fig. A.10 (c), fig. 8.4 (a)). Plateaus with large errors for several seconds are less common (fig. A.10 (d)). For the model

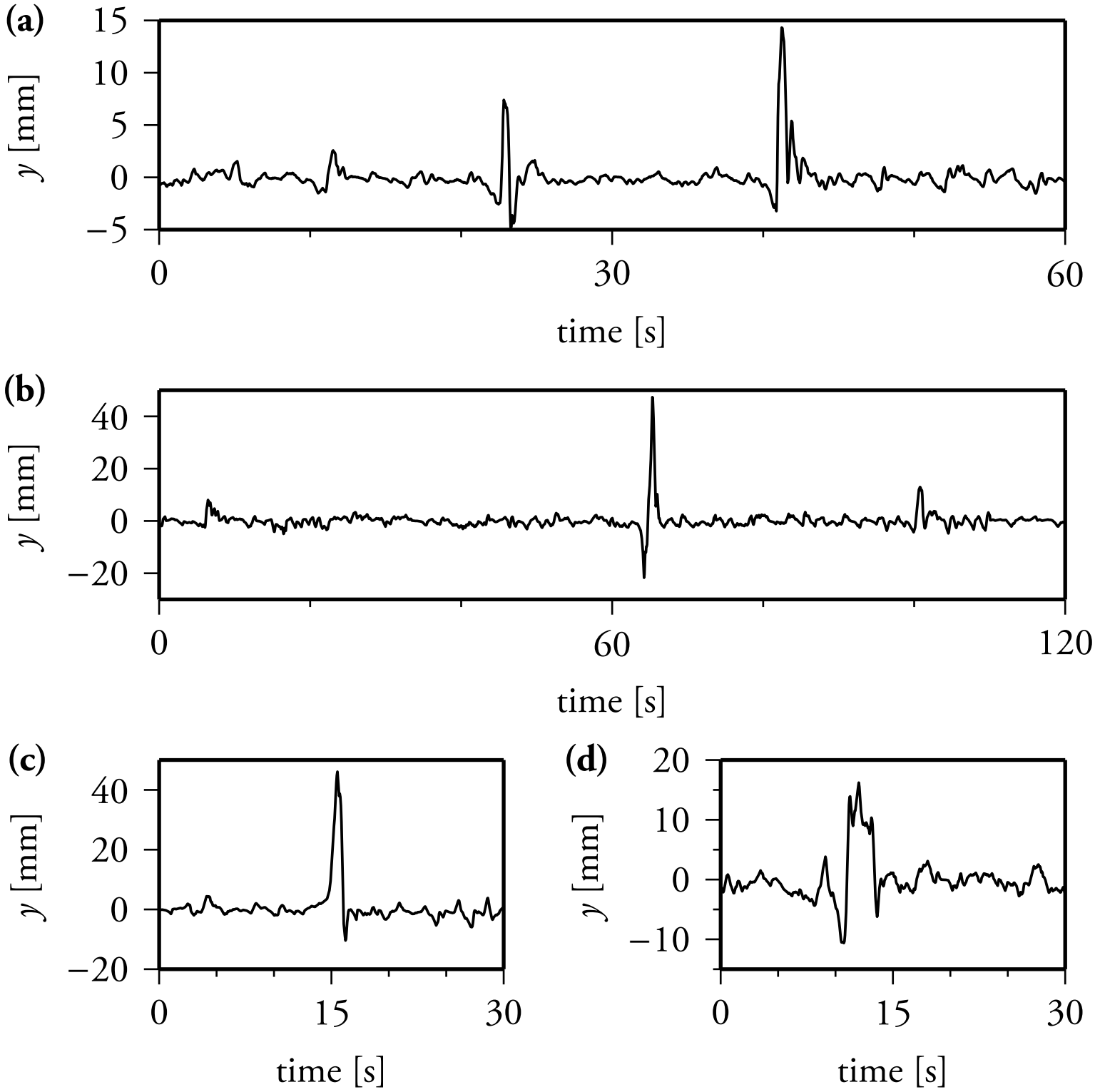

Figure A.10.: Parts of control error time series for one-dimensional VSB experiments (chap. 8). The examples represent different archetypes of error peak shapes in condition (S) and were taken from: **(a)**: subject 1, day 1, **(b)** subject 15 day 4 **(c)**: subject 6, day 1 **(d)** subject 5, day 1.

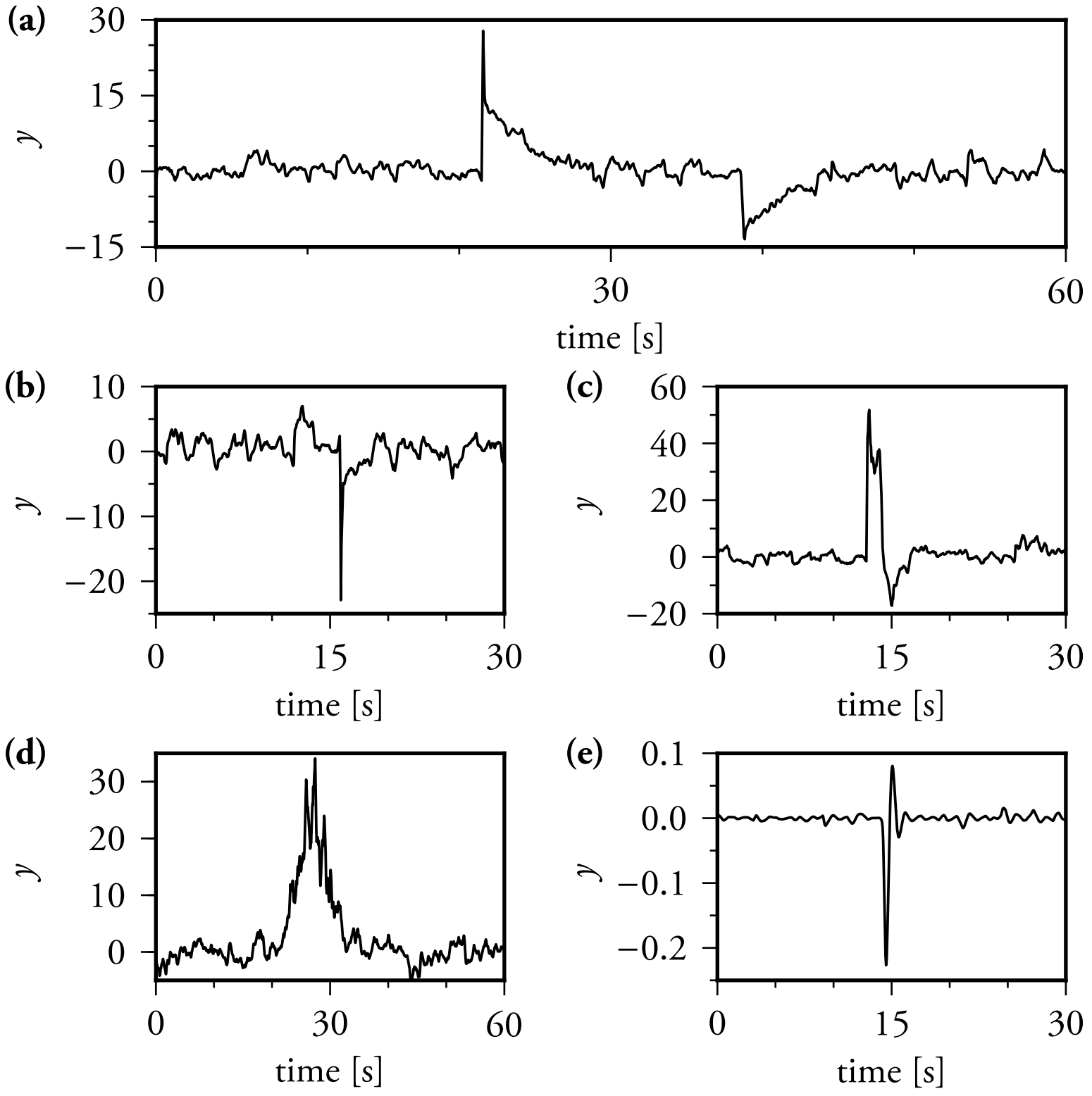

Figure A.11.: Details of simulated time series for different model variations. First, the basic continuous model with median fitted parameters (see fig. 8.10) for **(a)**: day one, and **(b)**: day three. Modifications include: **(c)**: a model with additional multiplicative noise with $t_r = 200\,\text{ms}$, $g = 1.1$, $\tau_m = 130\,\text{ms}$, $\sigma = 1\,\text{s}^{-0.5}$, $\sigma_y = 0.2\,\text{s}^{-0.5}$. **(d)**: a non-adaptive model with multiplicative execution noise with parameters like in figure A.4. **(e)**: A just underdamped second-order adaptive control model (sec. 7.8) with parameters $\vartheta = 10$, $t_r = 200\,\text{ms}$, $\gamma_a = 5$, $\gamma_v = 3$, $\tau_m = 160\,\text{ms}$.

(sec 7.4), peaks are typically corrected with strong understeering. This is shown in figure A.11 (a). The initial sharper correction can be understood intuitively: since control errors grew faster than expected, the estimator for a brief moment severely overestimates how fast control errors will grow in the future. After some more observations far away from the baseline noise become available, model predictions get very precise. Dynamics then are dominated by an exponential decay with a time scale set by the controller gain $\gamma$. Understeering, however, can be much more subtle than in figure A.11 (a). This is especially true for model parameters fitted to later days as shown in figure A.11 (b). Like in the experimental time-series, peaks in model time series often start with a small movement in the opposite direction.

Model variations also influence the shape of the error peaks. Including multiplicative execution noise in addition to adaptation leads to irregular clusters of errors and occasional plateaus (fig. A.11 (c)) with steep flanks. In contradistinction, a model with only multiplicative execution noise exhibits a slow buildup of large fluctuation as shown in figure A.11 (d). The same effect, which is in conflict with experimental observations, is also found for the model introduced by Cabrera and Milton (sec. 6.3). Finally, the second-order adaptive model exhibits wider peak tops than the first order one. It also allows for more oscillatory behaviour (fig. A.11 (e)).

In summary, control error time-series from both experiments and the adaptive balancing model exhibit isolated, extreme peaks. In contrast to processes with multiplicative execution noise only, these peaks do not build up slowly, but instead arise suddenly. However, while the adaptive balancing model reproduces statistical features very well, the shapes of single peaks are only approximated. Unfortunately, it is not quite certain how to interpret these findings. For example, the rare plateaus with large errors for several seconds could be caused by additional multiplicative execution noise (fig. A.11 (c)). Alternatively, they might occur intentionally (fig. 8.4 (d)) if a subjects wants to move the target, for example, towards the center of the screen. It is also unclear how much of the over- and understeering in the experimental time series is planned, or arises due to noise or the inertia of the arm. Nevertheless, the basic mechanism of IAI and the scaling statistics of the error time series appear to be largely invariant against such details of movement generation.

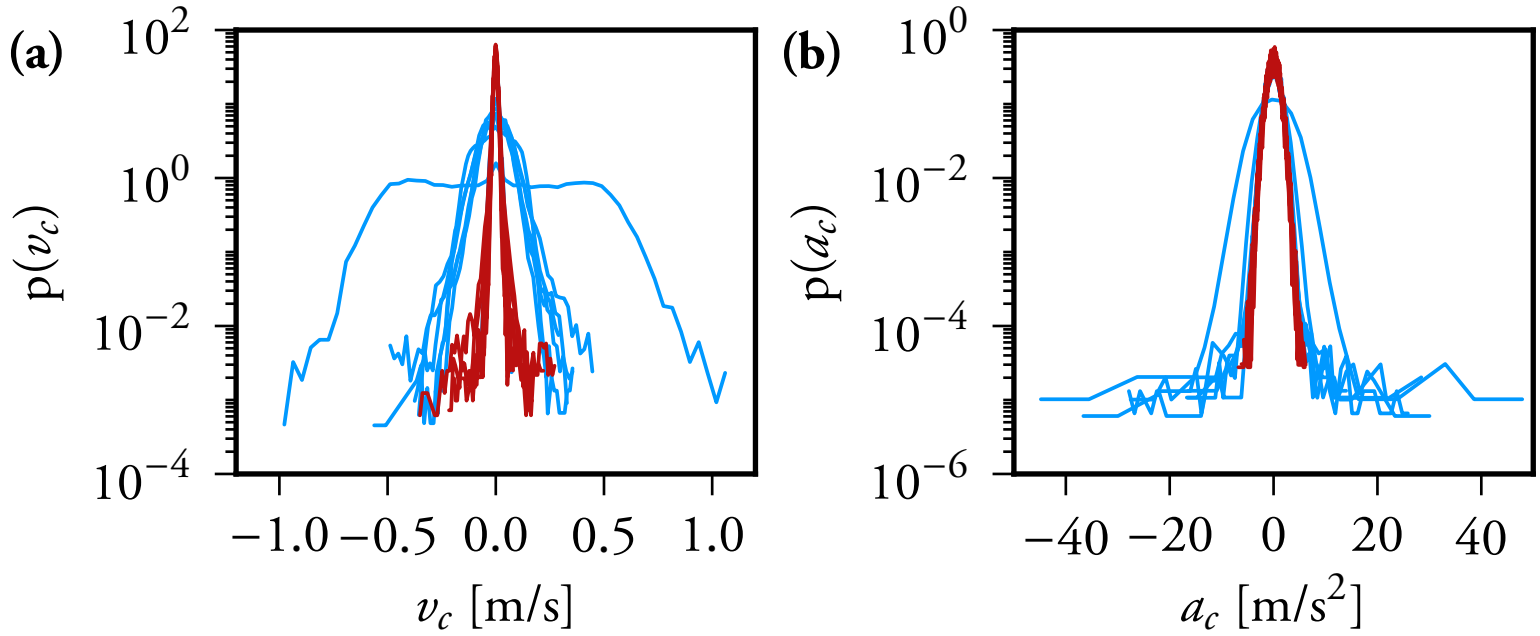

Figure A.12.: PDFs for **(a)** cursor velocities, and **(b)** cursor accelerations on day four in the experiment described in chap. 8. Score condition (S): red lines, (K): light blue lines. Time-series were downsampled to 100Hz to reduce high frequency noise.

## A.9. Hand velocities and accelerations

In chapter 8, subjects who minimise Kurt($y$) (K) keep larger distances $y$ between the target T and the cursor C than subjects who minimise Std($y$) (S). T moves faster for larger $|y|$. Therefore, subjects who don't keep $y$ as small as possible have to speed up their hand movements as shown in figure A.12 (a)[4]. In order to keep T on the screen, the direction of its movement has to change frequently. This also requires stronger hand accelerations in (K) than in (S) as shown in figure A.12 (b).

[4] One subject in (K) exhibits velocities that are comparable to (S). This particular distribution is difficult to see because it is mostly covered by the data from (S).

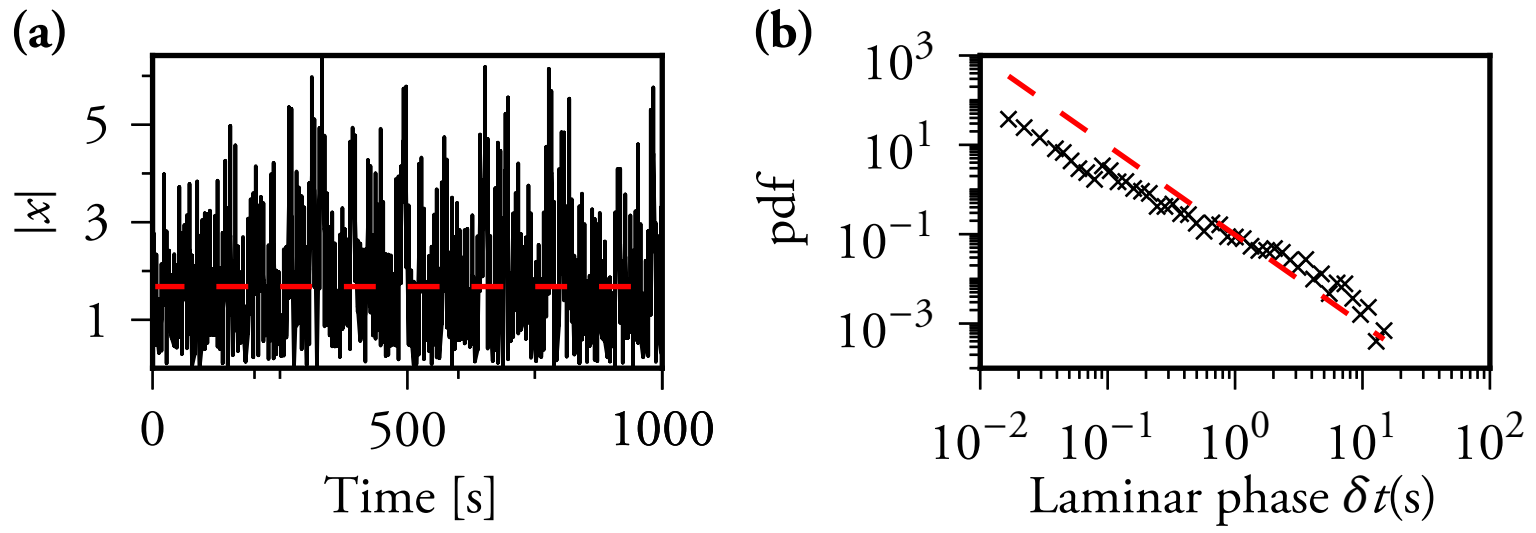

Figure A.13.: Simulation of a random walk with time-delayed drift back to the origin: $\dot{x}(t) = -0.1x(t-200\text{ms}) + \eta$, where $\eta$ is normal distributed white noise. **(a)**: Time-series and a threshold at 1.0005 E(|$x$|). **(b)**: PSD for the times between threshold passings from above $\delta t$ (crosses), and a power law with slope 3/2 (dashed line).

## A.10. Laminar phases

It has been reported that during stick balancing, the times between crossings of a small threshold from bigger control errors towards smaller ones scales like a power law with slope 3/2 [CM02]. This measure often yields results that are very similar to the so-called laminar phases, that is, the lengths of the phases below a threshold. This seems to have led some authors to use the terms interchangeably (which for simplicity is done here, too). The observed scaling was interpreted as evidence that the intermittent fluctuations during stick balancing are an instance of OOI (see also sec. 6.1).

In this thesis, laminar phases are not considered for two reasons. First, the author found this measure to be less useful in distinguishing between different control models than CCDFs and PSDs. Second, very simple stochastic processes that are detached from control can lead to laminar phases which appear to scale with exponent 3/2. A most primitive example is shown in figure A.13: a random walk with a time-delayed drift back to the origin. This process involves neither an instability nor multiplicative noise. Yet, the laminar phases don't look more different from a 3/2 power law than the examples in Figs. 1 and 3 in [CM02] do. Also, the laminar phases are often shorter than the delay, which was argued to be a beneficial feature of multiplicative noise in the same publication.

## A.11. Proof that the continuous estimator is unbiased

The expectation value of equation (7.6) is:

$$
\begin{aligned}
\mathrm{E}\left(\tilde{\vartheta}(\dot{y},y)\right) &= \mathrm{E}\left(\frac{\dot{y}}{y}\right) \\
&= \int_{-\infty}^{\infty}\int_{-\infty}^{\infty} \frac{\dot{y}}{y} p(\dot{y},y)\ d\dot{y}\,dy \\
&= \int_{-\infty}^{\infty}\int_{-\infty}^{\infty} \frac{\dot{y}}{y} p(\dot{y}|y)p(y)\ d\dot{y}\,dy \\
&= \frac{1}{\sigma_{\dot{y}}\sqrt{2\pi}}\int_{-\infty}^{\infty}\int_{-\infty}^{\infty} \frac{\dot{y}}{y} \exp\left(-\frac{(\dot{y}-\vartheta y)^2}{2\sigma_{\dot{y}}^2}\right) p(y)\,d\dot{y}\ dy \\
&= \int_{-\infty}^{\infty} \frac{\vartheta y}{y} p(y)\,dy \\
&= \vartheta\int_{-\infty}^{\infty} p(y)\,dy \\
&= \vartheta.
\end{aligned}
$$

Therefore, the estimator is unbiased.

## A.12. Time-discrete limit of the continuous model

In the following, we demonstrate how the minimal time-discrete IAI model (sec. 7.3) can be obtained as a limiting case of the more realistic continuous model (sec. 7.4). We first recapitulate the prerequisite continuous model equations for the readers convenience. This proof was published in the supplement to [PP11].

### A.12.1. The continuous model

The observable dynamics of the continuous control system are

$$\dot{y}(t) = \frac{1}{\tau} y(t) - \gamma\, \tilde{\vartheta}(t)\ \tilde{y}(t) + \beta(t) \tag{A.1}$$

with time constant $\tau$ and Gaussian white noise $\beta(t)$. The second term on the right-hand side is the controller's contribution, which is proportional to the expectation value (prediction)

$$\tilde{y}(t) = e^{\tilde{\vartheta}(t)\,t_r}\Big(-\gamma \int_{t-tr}^{t-0} e^{\tilde{\vartheta}(t)\,(t-t_r-t')}\,\tilde{\vartheta}(t')\tilde{y}(t')\ \mathrm{d}t' + y(t-t_r)\Big) \tag{A.2}$$

of $y(t)$ given observations up to time $t - t_r$ where $t_r$ is the controllers reaction time. Here,

$$\tilde{\vartheta}(t+t_r) = \frac{\int_{t-t_m}^{t} y(t')\Big(\dot{y}(t') + \gamma\tilde{\vartheta}(t')\,\tilde{y}(t')\Big)\mathrm{d}t'}{\int_{t-t_m}^{t} y(t')^2\ \mathrm{d}t'.} \tag{A.3}$$

is the continuous record maximum likelihood (ML) estimator for $1/\tau$. To make $\dot{y}$ negative on average, the control term contains a gain factor $\gamma > 1$. $t_m$ is the controller's memory length. In section 7.4, we modified equation (7.17) by using an exponentially decaying integration window to obtain a differential formulation. We will omit this step here because because we would have to revert it later on to obtain the desired limit anyway. However, the discrete limit can be performed using the form in the main paper, too.

### A.12.2. Performing the limit

Now consider a special case where the controller is not active all the time. Instead of a constant gain, let

$$\gamma\,\tilde{\vartheta} = \sum_{i=1}^{\infty} \delta(t - i\Delta t), \quad \Delta t = k_r\, t_r, \quad k_r \in \mathbb{R} \tag{A.4}$$

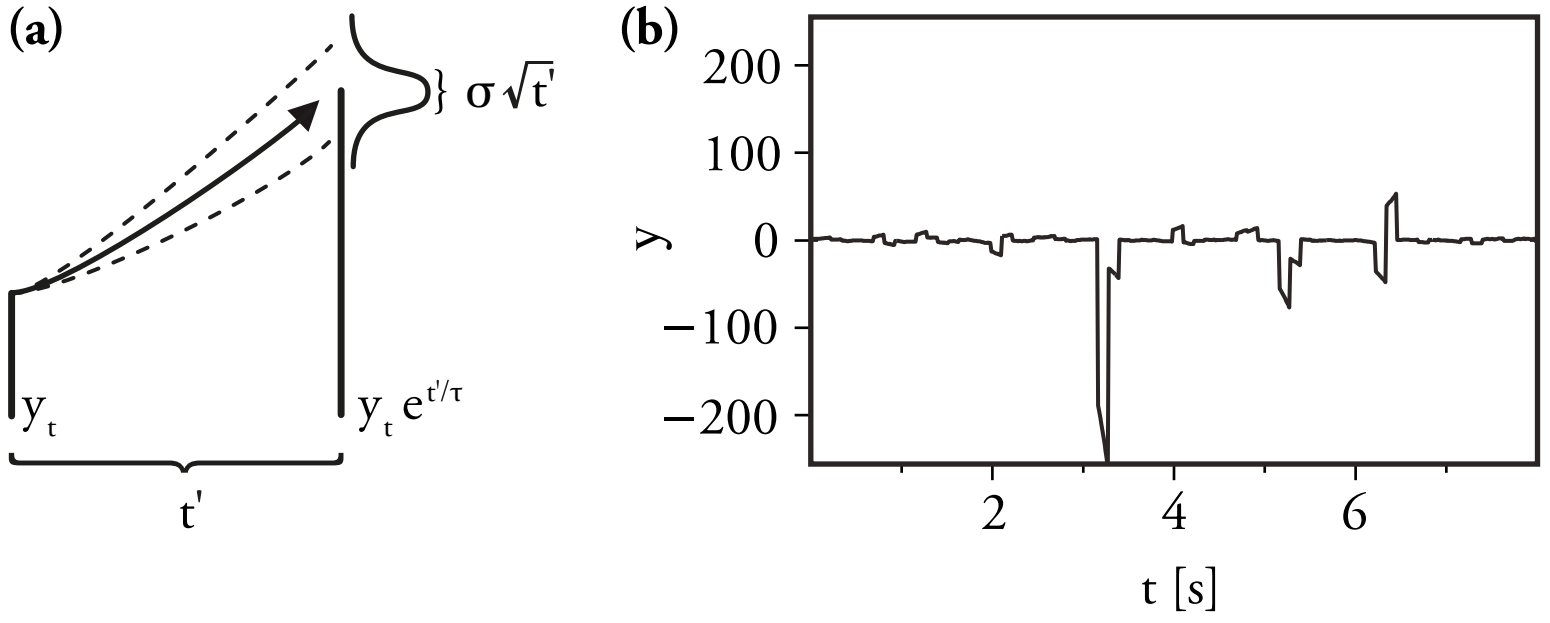

Figure A.14.: (a): Conditional probability density $p(y(t')\,|\,y(t))$, $t < t' < t + \Delta t$. (b): Short time-series of a hybrid continuous model with pulsed control defined by equations (A.1), (A.4), (A.7) and (A.8). Time constant $\tau = \frac{1}{3}$ s, memory length $m = 2$ steps, reaction delay $k_r = 1$ step, noise level $\sigma = 1$, pulse interval $t_p = 10/85$ s.

thereby removing the predicted $y$ completely during short control pulses. In this case, the stochastic differential equation (A.1) can be solved in between two control pulses (see figure A.14):

$$y(t + \Delta t - 0) = e^{(\Delta t - 0)/\tau}\; y(t) + \int_t^{t+\Delta t - 0} \beta(t')dt' \qquad \text{(A.5)}$$

where the zeroes indicate that the solution describes the system at time $t + \Delta t$, but before the control pulse at this time has been applied. $y(t)$ which is given includes the control pulse at time $t$. To obtain $y(t + \Delta t)$ after the control pulse has been applied, the latter is simply added. Hence, using the simplified notation

$$
\begin{aligned}
y_k &= y(k\Delta t) \\
k &= 1,2,\ldots \\
\alpha &= e^{\Delta t/\tau} \\
\tilde{\alpha} &= e^{\tilde{\vartheta}\Delta t} \\
\tilde{y}_k &= \tilde{y}(k\Delta t) \\
m &= \min\{n \in \mathbb{N} \mid n \geq 2, n \geq t_r/\Delta t\} \\
\beta_k &= \int_{k\Delta t-1}^{k\Delta t} \beta(t')dt'
\end{aligned}
$$

we obtain

$$y_{k+1} = \alpha\, y_k - \tilde{y}_{k+1} + \beta_k. \tag{A.6}$$

The prediction (A.2) can be expressed similarly after inserting (A.4):

$$\tilde{y}_{k+k_r} = \tilde{\alpha}_{k+k_r}^{k_r}\, y_k - \sum_{i=1}^{k_r-1} \tilde{\alpha}_{k+k_r}^{k_r-i}\, \tilde{y}_{k+i}. \tag{A.7}$$

Finally, we assume that the controller only observes the system at the times when the control pulses are applied. We then have to replace the continuous record ML estimator $\tilde{\theta}$ for $1/\tau$ (A.3) by the discrete ML estimator for $\alpha$:

$$\tilde{\alpha}_{k+k_r} = \frac{\sum_{i=0}^{m-2}(y_{k-i} + \tilde{y}_{k-i})\; y_{k-i-1}}{\sum_{i=0}^{m-2} y_{k-i-1}^2}. \tag{A.8}$$

Here, the integral over the observed time interval $[t-t_m, t]$ has been replaced by the sum over the past $m$ observations. Detailed information on this limit can be found e.g. in [PY09]. For an intuitive understanding, note the large bracket in the numerator in equation (A.3) containing the observed velocity corrected by the controllers action at the time of observation. This term is changed analogously to (A.6) since interaction between controller and system now takes place only during control pulses. In (A.8), the bracket in the numerator therefore contains the observed $y$ corrected by the control pulse at that time.

Equations (A.6) - (A.8) represent the discrete-time limit of the continuous model equations (A.1) - (A.3) under the condition (A.4) that the controller interacts with the system only during short control pulses. Choosing the reaction delay $k_r = 1$ and memory $m = 2$ (therefore dropping the sums), the discrete system reduces to the minimal model equations (7.10), (7.11) (renaming $k$ to $t$).

### A.12.3. Discussion

We performed the discrete time limit such that during each pulse the expectation value of $y$ is removed completely. Hence, the controller is optimal given the pulse times and memory length. However, this controller is not able to faithfully reproduce many of the experimental findings. Also, the time scale of the discrete steps cannot be related to any time scale in the continuously recorded experimental time-series. A more detailed discussion is found at the end of supplement two to [PP11].

## A.13. Predictability of extreme events

Here we demonstrate that the Information Annihilation Instability does not necessarily lead to completely unpredictable dynamics. Inserting equation (7.10) into (7.11), the minimal IAI model takes the form

$$y_{t+1} = -\frac{\beta_{t-1}\, y_t}{y_{t-1}} + \beta_t. \tag{A.9}$$

Assume that an observer with knowledge of $\mathcal{Y}_t = \{y_t, y_{t-1}, y_{t-2}, ...\}$ and equation (A.9) attempts to predict $y_{t+1}$. Assume further that the observer already has an estimation $\tilde{y}_t$ which includes all information contained in previous observations–a prediction that is so good that it perfectly predicts the first term in equation (A.9). Then, the probability density for $y_{t+1}$ is

$$p(y_{t+1}|\mathcal{Y}) = \mathcal{N}(\tilde{y}_{t+1}, \sigma_\beta^2) \quad \text{where} \tag{A.10}$$

$$\tilde{y}_{t+1} = -\frac{\beta_{t-1}\, y_t}{y_{t-1}}, \tag{A.11}$$

since the observer knows everything except for $\beta_t$, which did not influence the observations $\mathcal{Y}_t$.

In the next time-step, the actual realisation of $y_{t+1}$ is observed. This allows the observer to calculate the previously unknown noise term $\beta_t$:

$$\beta_t = y_{t+1} - \tilde{y}_{t+1} \tag{A.12}$$

even though it cannot be observed directly. Since now the observer knows $\mathcal{Y}_{t+1}$ and $\beta_t$, the task to predict $y_{t+2}$ at time $t+1$ requires the same operations that were necessary to estimate $y_{t+1}$ at time $t$. An optimal prediction doesn't even require all events $\mathcal{Y}_{t+1}$. The previous two observations of $y$ and the previous optimal prediction suffice:

$$\tilde{y}_{t+1} = -\frac{y_t}{y_{t-1}}(y_t - \tilde{y}_t). \tag{A.13}$$

Figure A.15 shows a realisation of the minimal model and the corresponding predictions. For each time step $t$, the prediction error is exactly the noise term $\beta_{t-1}$ which was unknown at time $t-1$ when the prediction was made. Although we assumed perfect knowledge about the initial state when deriving the predictor, it is robust against small errors in the initialisation.

An alternative to the above method is to consider (A.9) as an MA(1)-Prozess

$$x_t = b_t \beta_{t-1} + \beta_t. \tag{A.14}$$

with one time dependent parameter $b_t = y_t / y_{t-1}$. Then, a general solution to the prediction problem can be obtained even for completely unknown initial values. For large $t$, this solution converges against (A.13). An overview over the necessary methods, and the more general class of ARMA-processes can be found in [Hon90]. If the process to be predicted cannot be observed directly, Kalman-filters (which can also be considered recursive Bayesian estimators) can be used. Unfortunately, these considerations are not trivial to transfer to more complex balancing models. Attempts to predict the continuous model in a similar way were not successful as of yet.

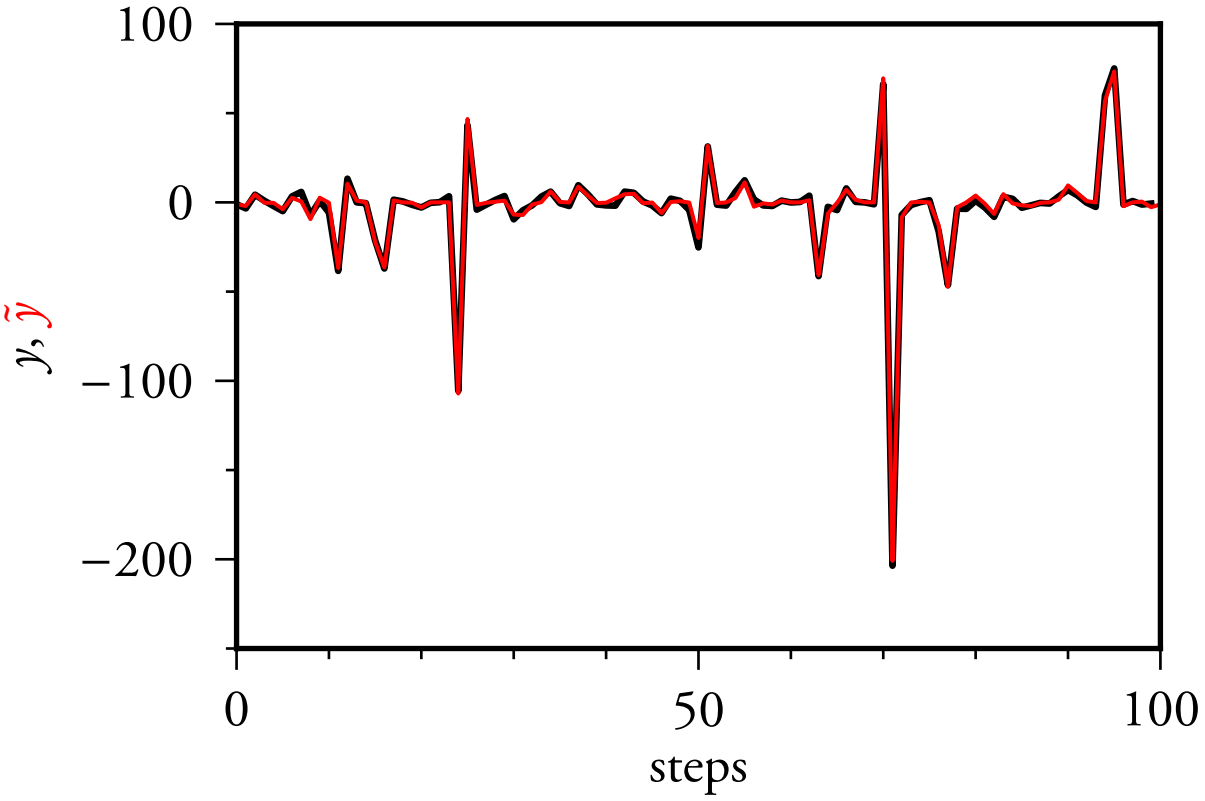

Figure A.15.: Minimal IAI model described by (7.10) (black line), and a prediction of each time step based on past observations according to (A.13) (red line).

## A.14. Documents for 1-D-VSB experiments

Universität Bremen | Zentrum für Kognitionswissenschaften | Institut für Theoretische Physik

# Versuchsteilnehmer Gesucht / Participants Wanted

Zur Erforschung von Grundlagen der menschlichen Bewegungsplanung und -kontrolle werden insgesamt 20 gesunde Versuchsteilnehmer **zwischen 18 und 35 Jahren** gesucht.
Im Experiment wird am Computer virtuell ein instabiles System balanciert. Jede Versuchsperson versucht an **vier aufeinander folgenden Tagen** für jeweils eine Stunde (inkl. Pausen) einen möglichst hohen Highscore zu erzielen. Die Teilnahme wird mit

**8 € / Stunde**

vergütet.

To investigate fundamental principles of human movement planning and -control, we are looking for 20 healthy testing subjects **aged 18 to 35**.
During the experiments, participants balance a virtual instable system displayed on a computer screen. Each participant tries to reach the best highscore possible on **four subsequent days** for one hour (incl. breaks) each day. Payment for participants is

**8 € / hour.**

Kontakt / Contact: **Felix Patzelt**

**Cognium**, AG Pawelzik (Theoretische Neurophysik), Zimmer / Room 2430
**Hochschulring 18** (nahe Fallturm / close to Drop Tower)

E-Mail: **balancierversuche@neuro.uni-bremen.de**
Tel.: +49 421 **218 62005**

**Felix Patzelt**
E-Mail: balancierversuche@neuro.uni-bremen.de
Tel.: 218 62005

Universität Bremen

Universität Bremen · **Fachbereich 01** · Postfach 33 04 40 · 28334 Bremen

**Institut für Theoretische Physik Abt. Neurophysik**

Fachbereich 01
Physik/Elektrotechnik

Zentrum für Kognitionswissenschaften

Arbeitsgruppe
**Prof. Dr. Klaus Pawelzik**

Hochschulring 18
Cognium, Raum 2440
28359 Bremen

Dipl. Phys.
**Felix Patzelt**

Cognium, Raum 2430

Telefon (0421) 218 - 62005
eMail felix@neuro.uni-bremen.de

## Aufklärungsbogen für Versuchspersonen

Vielen Dank, dass Sie sich bereit erklärt haben, an einer Studie zur menschlichen Motorkontrolle teilzunehmen. In dieser Studie wird untersucht, auf welchen Funktionsprinzipien und -mechanismen die Planung von Kontrollbewegungen beim Balancieren beruht. Dazu werden Sie an vier aufeinanderfolgenden Tagen für jeweils bis zu einer Stunde mittels eines Schiebereglers ein auf einem Computerbildschirm dargestelltes System balancieren (tägl. 10 Trials à 3 Min. + Pausen). Weiterhin informiert Sie ein Highscore über Kontrollfehler, die Ihnen dabei unterlaufen. Sie haben die Aufgabe, Ihren Highscore soweit es Ihnen möglich ist zu verbessern. Vor Beginn der Versuche wird ein Datenblatt ausgefüllt. Bitte erscheinen Sie möglichst ausgeruht zu den Versuchen. Eine Teilnahme unter Einfluss von Alkohol, Drogen oder Medikamenten, die die Fahrtüchtigkeit beeinflussen können, wird nicht gestattet.

Die im Rahmen der Studie durchgeführten Experimente sind nichtinvasiv, d.h. es werden zu keinem Zeitpunkt direkte physische Veränderungen am Körper vorgenommen. Medikamente werden nicht verabreicht. Daher sind von den Experimenten ausgehende Gesundheitsbeeinträchtigungen und Risiken kaum vorstellbar.

Die Vergütung beträgt 8,-- Euro pro Versuchsstunde.

**Die Teilnahme an der Studie ist freiwillig. Sie können jederzeit und ohne Angabe von Gründen aus der Studie aussteigen, ohne dass dabei für Sie persönliche Nachteile entstehen werden. Ein Wiedereinstieg in die laufende Studie ist nicht möglich. Die Ergebnisse der Studie und ihre Daten werden vertraulich behandelt und unterliegen den Vorgaben des Bremischen Datenschutzgesetzes.**

Sie haben jederzeit die Möglichkeit, Fragen zu Ablauf, Thematik und Organisation der Studie zu stellen. Für Hinweise, Kritik, Verbesserungsvorschläge, Anregungen und Beobachtungen sind wir sehr dankbar.

Informationen, die ihre Identität mit ihren Daten in Verbindung bringen, werden von diesen getrennt und Dritten unzugänglich gesichert aufbewahrt und unmittelbar nach Abschluss der Studie gelöscht, spätestens nach einem Jahr.

Mit Ihrer Unterschrift bestätigen Sie, dass Sie über die Studie und Ihre Rechte vom Versuchsleiter aufgeklärt wurden. Darüber hinaus erklären Sie sich damit einverstanden, dass Ihre anonymisierten Ergebnisse der Studie für wissenschaftliche Veröffentlichungen verwendet werden dürfen.

Datum: ________________ Unterschrift: __________________________

**Sekretariat**
Agnes Janßen

Cognium, Raum 2470

Telefon (0421) 218 - 62000
Fax (0421) 218 - 62014
eMai ajanssen@neuro.
uni-bremen.de

Universität Bremen

Universität Bremen · **Fachbereich 01** · Postfach 33 04 40 · 28334 Bremen

**Institut für**
**Theoretische Physik**
**Abt. Neurophysik**

Fachbereich 01
Physik/Elektrotechnik

Zentrum für Kognitionswissenschaften

Arbeitsgruppe
**Prof. Dr. Klaus Pawelzik**

Hochschulring 18
Cognium, Raum 2440
28359 Bremen

Dipl. Phys.
**Felix Patzelt**

Cognium, Raum 2430

Telefon (0421) 218 - 62005
eMail felix@neuro.uni-bremen.de

## Subject information sheet

Thank you for for participating in this scientific study on human motor control. In this study, fundamental principles and mechanisms underlying the planning of control movements during balancing are investigated. On four subsequent days, you will use a slider to control a system displayed on a computer screen for up to one hour each day (10 Trials per day, 3 min. per trial + breaks). After each trial, a highscore will inform you about your performance. Your task is to improve your score as much as possible. Before you begin your trials, a data sheet is filled out. Please make sure to be in good health and to sleep sufficiently while you take part in this study. Participation under the influence of Alcohol or (even medical) Drugs which could influence ones driving ability is not acceptable.

The experiments conducted in this study are non-invasive. That is, at no point will physical changes be made to your body. No drugs are administered. Therefore, no risks or negative effects to your health are imaginable.

Payment for participation in the experiments is 8,-- Euros per hour.

**Participation in the study is voluntary. You may drop out of the study at any point in time without giving any reasons and without any personal disadvantages. Re-entering the study is not possible. All results and personal data will be treated discretely and are subject to the privacy law of the city of Bremen.**

You may ask questions regarding experimental procedures as well as concerns and organisation of the study at any time. We appreciate any kind of remarks, criticism, suggestions and observations.

Information linking your identity to your data is kept separate from said data and inaccessible to third parties. It will be destroyed when the study is completed, after one year at the latest.

By signing below, you confirm that you have been informed about this study and about your rights by the experimenter. You further consent with the usage of your anonymised results for scientific publications.

Date: ________________ Signature: __________________________

**Sekretariat**
Agnes Janßen

Cognium, Raum 2470

Telefon (0421) 218 - 62000
Fax (0421) 218 - 62014
eMai ajanssen@neuro.
uni-bremen.de

## Datenblatt - Institut für theoretische Neurophysik
ZKW, Universität Bremen, Hochschulring 18, 28359 Bremen

| | |
|---|---|
| **Code**: BALHS11-___________________ | **Datum**: _______________ |
| **Geboren** (Monat/Jahr): ______________ | **Geschlecht**: O m O w |

**Trifft eine der folgenden Eigenschaften zu?**

- **Vorerkrankungen** (neurologisch / ophthalmologisch / Herz-, Kreislaufsystem)
  *Beispiele nennen: Schlaganfall, Parkinson, Grauer Star, Herzinfarkt, Epilepsie*
- **Konsum von Medikamenten, die die Fahrtauglichkeit beeinträchtigen**
  *Wäre Konsum dieser Medikamente an den Versuchstagen nötig?*
- **Regelmäßiger, sehr starker Koffeinkonsum**
  *Anzeichen: Zittern, Nervosität, Schlafstörungen, erhöhter Puls, Extrasystolen („Herzstolpern")*
- **Regelmäßiger Alkoholkonsum tagsüber für das Wohlbefinden nötig.**
  *Versuche müsse nüchtern durchgeführt werden. Dabei dürfen keine Entzugserscheinungen auftreten.*
- **Starker Nikotinkonsum** (deutlich mehr als 1 Schachtel / Tag)
  *Treten starke Beeinträchtigungen auf, wenn eine Stunde lang nicht geraucht werden kann?*
- **Regelmäßiger Konsum anderer Drogen.**

Ja führt zum Ausschluss (ohne Speicherung der Daten) O nein

| | |
|---|---|
| **Computernutzung** (Stunden/Tag), **-spiele**: ___________________ | **Sport**: ___________________ |

**Sehhilfe**: O keine O Brille O Kontaktlinsen

**Stärke** (dpt): O Links _________ O Rechts _________

*Fehlsichtigkeit muss auskorrigiert sein. Zu Versuchsbeginn noch einmal nachfragen, ob der Stimulus klar zu erkennen ist.*

**Händigkeit** (Tätigkeiten vormachen, X: Präferenz, XX: Andere Hand wird nie benutzt):
*Zeilen 1-5: Score Berechnen nach Edinburgh Handedness Inventory, Oldfield 1971. Zeile 6: Experimentspezifische Fragen.*

| | | | | | |
|---|---|---|---|---|---|
| Schreiben | OO L | OO R | Mit Messer Schneiden (ohne Gabel) | OO L | OO R |
| Zeichnen | OO L | OO R | Mit Löffel essen. | OO L | OO R |
| (Ball) Werfen | OO L | OO R | Mit Besen kehren (obere Hand) | OO L | OO R |
| Mit Schere schneiden | OO L | OO R | Streichholz anzünden | OO L | OO R |
| Zähne putzen | OO L | OO R | Kiste öffnen (Deckel) | OO L | OO R |
| Computer-Maus / Touchpad | OO L | OO R | Schieber im Versuch | OO L | OO R |

**Probandengruppe / Highscore-Kriterium**
O Minimaler quadratischer Fehler O Minimale Kurtosis

------------------------------------------------------------ Hier abtrennen ------------------------------------------------------------

## VP-Identifizierung - Institut für theoretische Neurophysik
ZKW, Universität Bremen, Hochschulring 18, 28359 Bremen

| | |
|---|---|
| Code: BALHS11-____________ | Name: ________________________________ |
| E-Mail oder Telefon: ______________________________________________ | |

Getrennt vom Datenblatt zu Lagern!

# Part III.

# Speculative markets

# 10. Characterisation and modelling of financial markets

In the following chapters, analytical, numerical, and experimental results that concern the modelling of financial markets as dynamic and adaptive information processing systems are presented. Particular attention is placed on collective dynamics towards information efficiency, equilibria, bubbles, and market instabilities. This approach is explained in the next section after a brief introduction to the field and more classical approaches in economic research. The existing literature is explained and discussed in more detail in the rest of this chapter.

## 10.1. Introduction

> "It is not from the benevolence of the butcher, the brewer, or the baker that we expect our dinner, but from their regard to their own interest."
>
> Adam Smith

Social systems self-organise. Consequently, collective dynamics can emerge that pursue a common goal not present in the behaviour of the individual agents. This idea had a particular impact on economic theories where individuals acting out of pure self-interest were thought to substantially, if unknowingly, promote the public interest as if they were led by an invisible hand [Smi76].

The concept of equilibrium, in which every agent acts in their own selfish interest, became dominant [FG09] in modern economics [Sam09]. Equilibrium models exhibit desirable traits. For example, every agent's wish to buy or sell a good at the equilibrium price is fulfilled.

### 10.1.1. The Efficient Market Hypothesis (EMH)

In the following, we focus on financial markets. Here, one fundamental hypothesis is that competitive traders exploit information that enables profitable trades (sec. 10.2). In this picture, the market relaxes rapidly towards equilibrium prices that "fully reflect available information" [Fam70], or at least come close to this limit [GS80]. Consequently, risk-free profits cannot be made by (re-)using said information. In ideal financial equilibrium, current prices are the best possible measure of "fundamental values", as well as the best predictors of future prices, and thus provide accurate signals for allocating resources to their most productive uses [FG09, FL12].

If true, one of the implications of the EMH is that prices should fluctuate randomly [Sam65]. That is, prices would immediately adjust to relevant unpredictable news, and there would be no other systematic trends [otRSAoS13, Lux09].

Due to a significant number of theoretical and practical advantages, as well as consistency with some key empirical findings, game theory and the resulting equilibrium models are almost the only approach in mainstream economic theories [FG09]. Despite reoccurring reports of empirical "anomalies", many economists still consider market efficiency to be a valid starting point [BGV00, FG09]. These anomalies include the finding that early models can only explain a fraction of the variation of real prices from the variation of fundamentals [CS89]. Numerous model extensions were developed to improve this situations [otRSAoS13].

### 10.1.2. Empirical (and other) problems

Some features of price movements that previously were considered minor anomalies are in fact so prevalent and dominant that many non-orthodox researchers now consider them to be "stylised facts" [Far99, Lux09, MS00]. Here we focus on two of the most prominent observations.

Magnitudes of price changes ("volatilities"), are found to be correlated over long periods of time. That is, large price changes are typically followed by large ones and small changes by small ones [Man63, GVA+99] (see also figs. 10.2 (b), 10.5 in sec. 10.3). Furthermore, logarithmic price changes (log returns) exhibit heavy tails that are well described by power-laws (figs. 10.2 (a), 10.4). Hence, extreme events that are many times bigger than the standard deviation occur at a much higher frequency than what would be expected if they were Gaussian distributed [Man63, GVA+99, Far99]. In the natural sciences, phenomena similar to these stylised facts have been observed in so-called "critical" states in complex systems at the boundary of order and disorder. These highly fragile states self-organise in certain out-of-equilibrium systems (see part I).

Another finding may be considered even more devastating for the EMH: Most extreme price jumps are not caused by identified new information, and most identified new information doesn't cause large price jumps [JLGB08, CPS89, Fai02]. Furthermore, analyses of high-frequency trading data indicate a self-referential and rather incremental information processing operating with long memory (sec. 10.6, [Bou10, BFL09]) and over various time scales [AMS98, PS10].

These findings were associated with substantial market inefficiencies including herding effects [DW96], "bubbles", or the interactions of heterogeneous traders with limited rationality [LM99b] in a market exactly at a critical point [CMZ05a]. Therefore, more attention to out-of-equilibrium theories that include microeconomic interactions of traders has been called for [FG09, Lux09, HB10].

So far, however, there is no consensus on how the EMH is to be refined; some even argued that it should even be disposed of completely [LM99b]. One of the problems as we move away from a perfectly rational market is that we face an increasingly diverse zoo of possible approaches. There are even significant differences among leading economists within the field of equilibrium economics. For example, even two economists who won the Sveriges Riksbank Prize in Economic Sciences in Memory of Alfred Nobel in the same year (2013) disagree on whether markets are rational [Kes13]:

> “[Eugene Farma] and I seem to have very different views. It’s like we’re different religions.”
>
> Robert Shiller

Our focus on financial markets notwithstanding, solutions to the aforementioned problems are also potentially relevant for macroeconomics, that is, models of the aggregate economy. Here, equilibrium models and the lack of attention to microscopic heterogeneous interactions have been criticised as well [HB10]. It has also been argued that macroeconomic models should include a financial sector [BS14]. Furthermore, the use of new financial instruments made possible by the EMH, as well as a common disbelief in bubbles and market instabilities [Fel08] and thereby in their impact on macroeconomics, are frequently named as factors that contributed to the economic crisis that peaked in 2008 [Kru09, CFH$^+$09, HB10]–although some remain convinced that all causes were external to markets [FL12].

### 10.1.3. A different approach

In the following chapters, we investigate whether the apparent antinomy of stabilising information efficient control and a dynamics resembling systems operating close to criticality can be resolved. Here the term “control” is used to describe the hypothesised tendency of markets to absorb the impact of predictable information. This description allows for an abstract comparison with part II where balancing a stick on a finger tip and similar elementary control problems were investigated. The results show that power-law-distributed fluctuations can be a signature of an adaptive controller that minimises predictable local trends. For human subjects and our model, Complementary Cumulative Distribution Functions (CCDFs) of the arising fluctuations follow power-laws in the same range as those of log-returns.

Motivated by these results, the main ambition in the following chapters is to investigate if market dynamics that strive towards information efficiency can also give rise to instabilities. This task is not trivial since the variables in the previously investigated balancing models don't represent economically meaningful quantities. We therefore search for minimal ingredients that allow multi-agent market models to reproduce the stylised statistical features of real price changes. The latter were found in very different markets, and hardly changed over the last century (sec. 10.3, [Lux09]) or even since the 18th century [Har98]. Accordingly, we assume them to be caused by fundamental mechanisms of trading, and to be at least qualitatively independent of most details of the organisation of a particular market. The same argument can be made for many details of trader behaviour, which changed dramatically over time as new financial instruments were invented and computers took over much of the trading activity.

In addition, we focus on linking individual and collective behaviour. A primary concern is to to disentangle the effects of different means of adaptation that are intermingled in many existing models discussed below. Therefore, we differentiate between agents as an abstraction of either pure trading strategies or actual traders, and investigate the differences between some of the most elementary types of information in markets. The resulting insights are prerequisites for self-organised information efficiency and thereby for answering whether the latter can be related to market instabilities. At the same time, these results contribute to a better understanding of multi-agent models in general.

### 10.1.4. The course of action

In order to find appropriate assumptions for parsimonious dynamical market models, relevant fundamental results from the existing literature are discussed in the following sections. Unfortunately, due to the sheer volume of existing work, a complete overview is out of the scope of this chapter.

First, we discuss why markets should evolve towards equilibrium at all, how market efficiency is commonly defined, as well as some of the implications, benefits and problems of this reasoning. Next, example time-series are shown that exhibit the "stylised facts" introduced above. In the subsequent sections, we briefly discuss economic rationality, bubbles, and price formation. Even though this introduction has a limited scope, some aspects of the economic literature are discussed that are not necessary to understand the models introduced in later chapters. These details are, however, relevant to show how this work is related to the existing fields of research, why certain simplifications were made, and how diverse and full of controversies the existing literature really is. An especially large gap exists between standard economics and the interdisciplinary complex systems approach. In the final sections of this chapter, multi-agent models are introduced. One particular type of multi-agent model is discussed in more detail: minority games.

Equipped with these basics, and starting in the next chapter, we pursue the questions posed above. A working hypothesis following from part II is that instabilities and non-Gaussian fluctuations should arise in markets adapting to endogenous information. This use of past prices as an input makes sense if the market is supposed to self-organise towards unpredictable price changes.

Our first approach is directly motivated by the main result of chapter 7: a predictive and adaptive controller that successfully eliminates all predictable dynamics self-tunes to a critical point. In chapter 11, an instantiation of this effect is demonstrated in a parsimonious trading model. It is demonstrated that the success-dependent change of the impacts of different trading strategies leads to collective learning analogous to a neuronal network. If traders try to profit from complex temporal patterns in the over- or underreaction of their peers, however, equilibria are perpetually destabilised. In this regime, the model reproduces the heavy-tail distributed and clustered movements of real prices.

A different approach is taken in chapter 12, where we introduce a modified minority game as a minimal experimental model for collective behaviour in highly speculative markets. Subjects were predominantly information efficient with respect to the most recent past. A close link is found between the unbiased random walk property of prices, and bubbles. Combining the surprisingly simple stochastic process that captures the subjects' behaviour with the pricing rule of the trading model from chapter 11 also reproduces the aforementioned "stylised facts". Next, a slightly more complex experiment is presented that incorporates more features of said trading model. It allows for extreme price jumps even for small numbers of subjects. This is demonstrated in closed group experiments and in a public browser game.

## 10.2. Equilibria and efficiency

Consider, for example, two markets, $A$ and $B$, where apples are traded at different prices $p_A < p_B$. It is then profitable to buy apples at $A$ and immediately sell them at $B$. Such trades, however, increase the demand at $A$ and the supply at $B$. Therefore, $p_A$ is expected to rise while $p_B$ is expected to fall as long the aforementioned trade remains profitable. Neglecting friction (e.g. transportation costs), the prices should quickly relax towards an equilibrium where $p_A = p_B$. If this was not he case, there would be a "money pump" which could be operated at any scale. Generating immediate riskless profit like this is called (pure) arbitrage.

The above reasoning can be extended to speculative trading involving different points in time. For instance, if one had reason to believe that the price of a certain stock was about to rise tomorrow, a profit could be made by buying said stock today and selling it tomorrow. This trade, however, would increase today's demand and tomorrow's supply. Hence, if information about future values arrive at a highly competitive market, the price should immediately change to reflect them. Information that was already available should not lead to price changes.

Similar arguments can be made in a variety of economic settings. Yet, the descriptions in the previous two paragraphs are extremely simplified. First, the arguments only hold under certain conditions: besides neglecting friction, we implicitly assumed informed traders and mutually interchangeable (fungible) goods. For now, we also avoided the question how exactly price changes are caused. In fact, many economic theories argue the other way around: they assume that traders are price takers who accept the price as given and adjust demand and supply accordingly. Processes for finding equilibrium prices are discussed in section 10.6.

Moreover, trades at different points in time, as in the second example, involve risk due to the uncertainty about future price movements.[1] In these types of arbitrage opportunities, there is only an expected positive payoff and losses may occur in practice. For example, a certain trading strategy may be profitable on average over time.

As stated above, equilibrium theories assume the absence of arbitrage opportunities. This implies that the current price of any traded asset $i$ can be written as the expectation value

$$p_i(t) = \mathrm{E}\big(m(t+1)x_i(t+1)\big) \tag{10.1}$$

[1] Many real trading strategies involve not just buying and selling stocks, but also bonds, contracts on future rights or obligations, borrowing assets that are currently not owned, and combinations thereof. Thereby traders can limit their risk (e.g. by hedging), or magnify possible gains and losses (i.e. increase leverage). However, as stated before, we only focus on very elemental properties of trading and forgo evitable details in the following.

of the next payoff $x_i(t+1)$ discounted with a factor $m(t+1)$ [otRSAoS13]. For stocks, the payoffs are defined as the next period price plus dividends:

$$x_i(t) = p_i(t) + d_i(t). \tag{10.2}$$

The expectation value in (10.1) is calculated with respect to the possible states of nature at time $t+1$, which $m(t+1)$ and $d_i(t+1)$ may depend on. The discount factor is the same for all assets and closely related to the return rate $r_f$ for risk-free assets (safe interest rate):

$$\mathrm{E}\big(m(t+1)\big) = \frac{1}{1+r_f(t)}. \tag{10.3}$$

In general, temporal discounting refers to the tendency of people to discount the value of future rewards (see below).

Over a short time horizon we can neglect the safe interest rate and dividends. Assuming that $m$ also doesn't vary much across different states of nature amounts to $m \approx 1$. Then [otRSAoS13],

$$\mathrm{E}\big(p(t+1)\big) = p(t). \tag{10.4}$$

In other words, the price follows a martingale. That is, future price changes are unpredictable from available information. This suggests that there are no traders who can always beat the market by taking advantage of less informed ones. Beating the market refers to generating returns in excess of the market average. Equation (10.4) is highly relevant for the following chapters since they are concerned with models of speculative trading over short time horizons.

A market where price changes are unpredictable from available information is called informationally efficient. The term "available information", however, leaves some room for interpretation. A commonly followed suggestion is to distinguish between three different versions of the EMH [Fam70]. In its weak form, it is impossible to systematically beat the market using historical prices. In the semi-strong-form, it is impossible to systematically beat the market using publicly available information. In strong-form informational efficiency, it is impossible to systematically beat the market using any information. The last concept was generally deemed unrealistic and more or less impossible to test [otRSAoS13]. Therefore, researchers focused on the weak and semi-strong EMH.

Empirical findings in favour of the EMH include the general absence of exploitable autocorrelations among price changes in financial markets [Fam98]. Some event studies further show that information (e.g. the announcement of a stock split) is usually incorporated into asset prices rapidly [otRSAoS13]. There also appears to be a general consensus that stock prices are quite unpredictable in the short term, even though predictability increases over longer time horizons [otRSAoS13]. Consistently, fund managers generally don't outperform markets [otRSAoS13], at least not repeatedly over longer periods of time [Fam98]. Some evidence even suggests that active funds systematically underperform the market ([BGV00]; see also [Ode99], and sec. 10.4).

Nevertheless, as mentioned before, not all empirical findings are unproblematic for the EMH. For example, prices are not completely random. Some predictable patterns in price movements of stocks and other assets disappeared after being discovered, possibly because traders began exploiting them. Other reliable patterns, however, did not disappear even years after being published (see, e.g., [LM99a, FG09]). Some investment strategies based on such patterns yield sustained profits, seemingly challenging the EMH.[2] Economists disagree on whether the persistence of patterns represents a violation of market efficiency [FL99].

[2] Instead of patterns, classical risk arbitrage strategies [Kue12] include e.g. merger arbitrage where one company takes over an undervalued one. This bid is expected to rise to its "true fundamental value". Only the fastest traders can profit from such opportunities. These strategies basically use equilibrium theory to determine temporary misalignments of prices which are expected to vanish. Similarly, in liquidation arbitrage breaking a company apart and selling its parts yields a profit. In pairs trading, traders track highly correlated assets. If the prices drift apart, they are expected to converge again in the future.

The EMH (and many equilibrium models) face the problem that empirical tests are necessarily joint tests that include auxiliary assumptions such as a particular asset-pricing model. A rejection of the joint hypothesis may therefore be caused by a “bad model” [Fam91]. For instance, apparent excess returns of a particular asset or strategy may be due to unobserved risk factors [FG09]. Furthermore, if some information is costly to obtain, there should be an equilibrium state of disequilbrium where small inefficiencies remain that don’t allow for profits in excess of the costs [GS80]. As another example, (10.1) can be iterated forward to yield the expected discounted value of future dividends. Yet, directly testing such a prediction requires an explicit model for the discount factor $m(t)$. One such model is the Consumption Capital Asset Pricing Model (CCAPM) which assumes a representative rational investor maximising utility (see also sec. 10.4). Unfortunately, under this model, and especially on shorter time scales, prices move far too much to be explained by changes in dividends [Shi81]. Researchers remain divided over how to account for this problem [otRSAoS13]. Some extended the model of rational agent expectations to include more factors and therefore more parameters that can be calibrated. Others considered results from psychological experiments that showed how humans generally deviate from rationality as it is defined by economic models (sec. 10.4).

The controversy around the EMH stems in part also from the inability of equilibrium models to describe deviations from equilibrium. Especially weak form efficiency in principle allows temporary deviations from equilibrium prices as long as they cannot be exploited systematically. The question how efficient a market or economy is can, however, hardly be investigated within the mainstream framework. Furthermore, economic experiments outside of small-scale lab experiments are nearly impossible to conduct. In consequence, and in contrast to established theories in the natural sciences, there appear to be few unambiguous empirical findings concerning economic theories [FG09]. Even most stylised facts could hypothetically be consistent with equilibrium if price fluctuations were reflecting only the statistics of events external to the market, and if the resulting fluctuations could not be exploited systematically.

Nevertheless, there is substantial direct empirical evidence against markets in stable equilibria with fluctuations that are purely driven by external factors. Most of the largest price movements cannot be attributed to discernable news, and most news does not lead to large price movements [JLGB08, CPS89, Fai02]. Large price changes also occur far too often to be explained by new information on fundamentals [JLGB08]. Price volatility when markets are closed is also much lower than when they are open [FR86]. "Markets appear to make their own news" [FG09].

We end this section with a few comments on the relevance of efficiency for economics in a broader sense. Arbitrage equilibrium is a precondition for general economic equilibrium. The latter can often be shown to be either allocatively efficient, meaning that the economy is as productive as possible, or Pareto efficient, meaning there is no change in choices that would make everyone better off. Prices in an ideal market provide accurate signals for resource allocation. The primary role of stock markets in particular is allocation of ownership. That is, it allows companies to raise money by selling shares of ownership. An informationally efficient asset market, however, must not necessarily be Pareto efficient or generate allocative efficiency in the economy. This additional complication appears to leave ample room for speculation on whether government intervention can possibly improve on a free market economy or not. See also [FG09, BGV00], and the references therein. Note that many economists caution against a normative interpretation of efficiency.[3]

Despite all of the aforementioned controversies, general equilibrium theory helped transforming economics from a qualitative to a quantitative and formally rigorous discipline. These models, however, typically require very strong assumptions.[4] Beyond that, arbitrage efficiency and informational efficiency are particularly useful because they allow to draw conclusions from fewer assumptions. For example, no arbitrage and prices following a random walk are the only two assumptions needed to calculate a unique risk neutral price for derivatives in the Black-Scholes Model. In principle, this allows to identify and eliminate mispricing, and to reduce risk [FG09].

[3] Such remarks are found, e.g., in [Sam65, FG09]. See also, e.g., [Sti02]

As noted above, some information is actually incorporated into price changes. For example, one study found orange prices to predict future orange prices and even future weather better than weather reports, although the variability of the prices was found to be inexplicably high [Rol84]. Many useful properties, as well as the elegance of the EMH and of equilibrium economics, make these theories highly appealing. This might explain some of the impact neoclassical economics had also on economic policy, for example in the form of market deregulation [Kru09].

## 10.3. Stylised facts

Here two most prominent stylised facts are explained in detail, namely the high frequency of extreme price changes and the clustering of volatile and quiet phases. We first demonstrate these statistical regularities on the basis of daily price changes for stock- and Foreign Exchange (FX) markets, and discuss similar findings for other quantities in the literature. Overviews on this topic are also found, for example, in [Far99, Lux09, Con01, GPL$^+$00]. (See also table 1 in [Lux06] for a compilation of explanations for power-laws in finance and their respective shortcomings)

Our first three example data sets, daily closing values of stock market indices, are shown in figure 10.1 (a). These indices are weighted averages of stock prices. Since the Dow Jones Industrial Average (DJIA) time-series is the longest one, we use it as a benchmark in the following chapters. That is, we treat it like an idealised stock price, the fluctuations of which give us an idea as to how realistic output of stylised stock market models should look like. The other indices are shown for comparison and to demonstrate that the features we are interested in hardly depend on how an index is calculated in detail.[5]

[4] The prototype for general equilibrium is the Arrow-Debrieu model which assumes agents with rational expectations who maximise their own utility (see sec. 10.4), perfect competition (price taking), and market clearing (see sec. 10.6) [FG09].

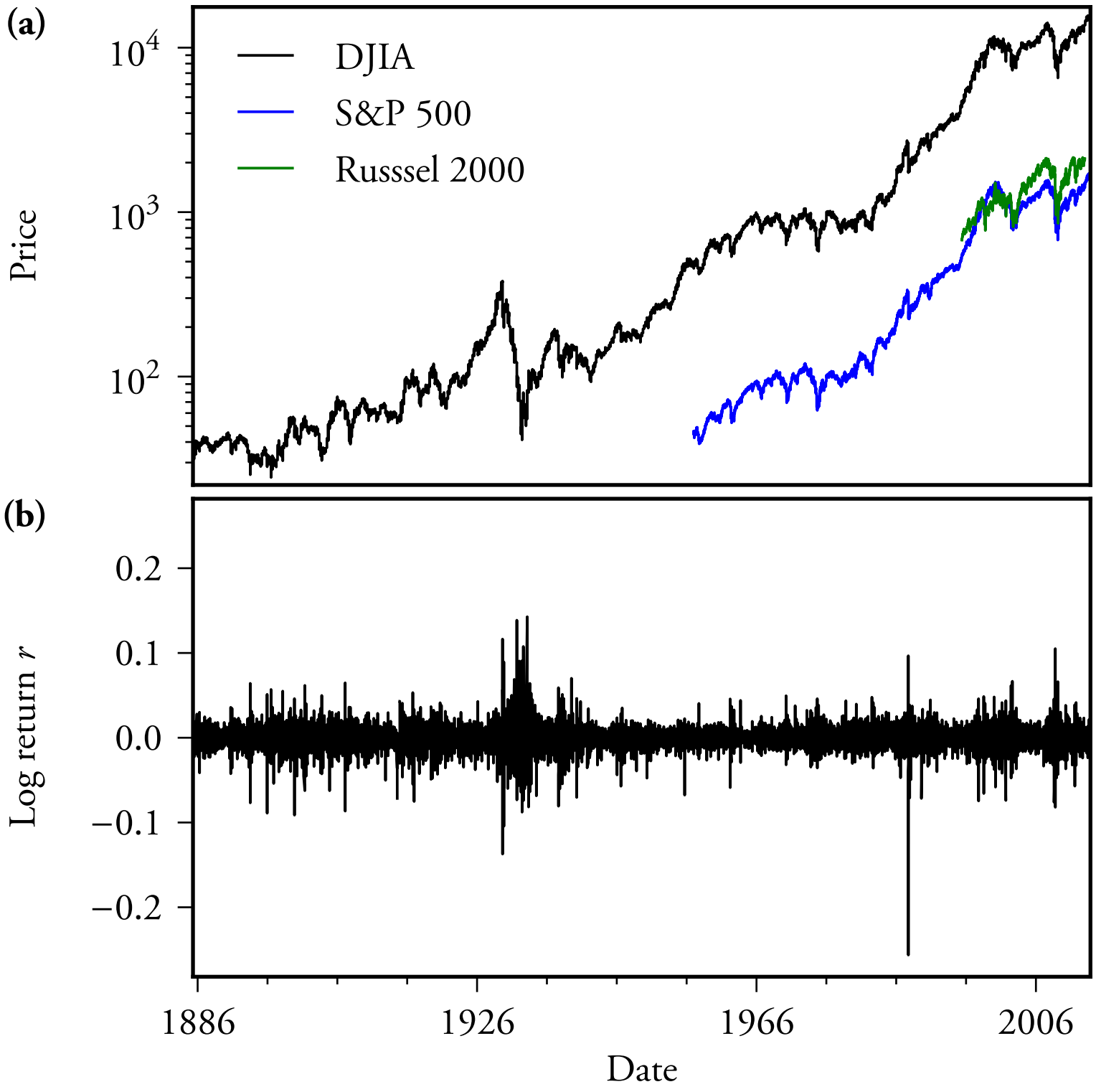

Figure 10.1.: **(a)** Historic daily closing values of three stock market indices. Dow Jones Industrial Average [Wil13] (black line), S&P 500 [Fed13a] (blue), and Russel 2000 without dividends [Rus13] (green). **(b)** Daily logarithmic price changes (log returns) for the DJIA .

The indices in figure 10.1 (a) exhibit a roughly exponential growth (note the logarithmic scale). We therefore need a scale-free measure to quantify price changes. The most natural choice is the log return

$$r(t+1) = \ln p(t+1) - \ln p(t), \tag{10.5}$$

where $p(t)$ is the price of an asset or the value of an index at a certain point in time $t$. Note that the return of investment $\exp(r(t+1))$ measures the relative price change, and hence the profit at time $t+1$ in relation to capital invested at time $t$.

A log return can be calculated over different time periods. Its value over a longer time period equals the sum over all logarithmic price changes during that period. For example, the sum over all all hourly log returns on one day is equal to the log return for the whole day. Hence, log returns are sums over random variables. Therefore, one might assume that the distribution of log returns should quickly converge towards a Gaussian, especially for heavily traded assets (sec. 2.2). Many economic models actually assume Gaussian log returns, including the Black-Scholes model variations of which are still being used.[6]

---

[5] The value of the DJIA is the weighted sum of the prices of one share for each of 30 large publicly owned companies based in the U.S. Each price is weighted by a divisor to compensate for stock splits or reinvested dividends. The S&P 500, on the other hand, is the weighted sum of the market capitalisation of 500 large U.S. companies. That is, the summand for each company is the total value of its publicly available shares, again weighted by a divisor. The Russel 2000 is also capitalisation weighted, but in contrast to the S&P 500 it comprises of 2000 companies with small market capitalisation.

[6] Note that even for log-normal returns the absolute price changes would exhibit quite extreme jumps.

Unfortunately, this assumption is false. Traders therefore use heuristic corrections (e.g. volatility smile) to compute approximately correct estimates of risk.[7]

Figure 10.2 (a) shows the CCDFs of the daily log returns for our example indices and a Gaussian distribution. As one might have guessed from the frequent occurrence of extreme price changes in figure 10.1, the distributions are heavy tailed.[8] These tails are well described by a power law $P(|r| > x) \propto x^{-\xi}$ with an exponent $\xi$ close to four. Log returns therefore have a finite variance, but the kurtosis diverges (albeit quite slowly for daily returns). This further implies that log returns over increasing periods of time should converge towards a Gaussian distribution if the log returns were Independent and Identically Distributed (IID).[9] In reality, however, this convergence is extremely slow: log return distributions are stable on time-scales from minutes to many days ([GPL+00, PGA+99]). This topic is discussed in more detail in section B.3.

Despite log return autocorrelations close to zero after 15 – 30min ([Far99], see fig. 10.2 (b) for daily returns), however, even longer-term log returns are not independent. They exhibit higher-order correlations that become visible in the so-called volatility clusters: Large returns are likely to be followed by large ones, and small returns are likely to be followed by small ones (fig. 10.1 (b)). This clustering is quantified by the autocorrelation of log return magnitudes as shown in figure 10.2 (b). In fact, there is a whole spectrum of clusters on very different time scales up to years. It has been suggested that the slow convergence of log returns is closely related to these long-range higher-order correlations.

[7] Analytical results for non-Gaussian fluctuations can be obtained as demonstrated in [Bor02].

[8] Consider, for example, the biggest one-day drop in the history of the DJIA: Oct. 19 1987 (Black Monday). On that day, the DJIA dropped almost 25%, eliminating hundreds of billions of dollars in stock value. No clear external event could be identified to have caused the crash, but several, mostly internal, factors were brought up that may have contributed [Shi87, Wal87].

[9] After the heavy tails of log return distributions were discovered [Man63], some researchers initially assumed the distributions might be Lévy stable and hence have a diverging variance. This would have been very problematic since most statistical properties then would be ill defined. The finding that in contrast the variance is finite might explain why many economists stuck with the mathematically convenient Gaussian distribution. For a detailed discussion, see e.g. [Far99].

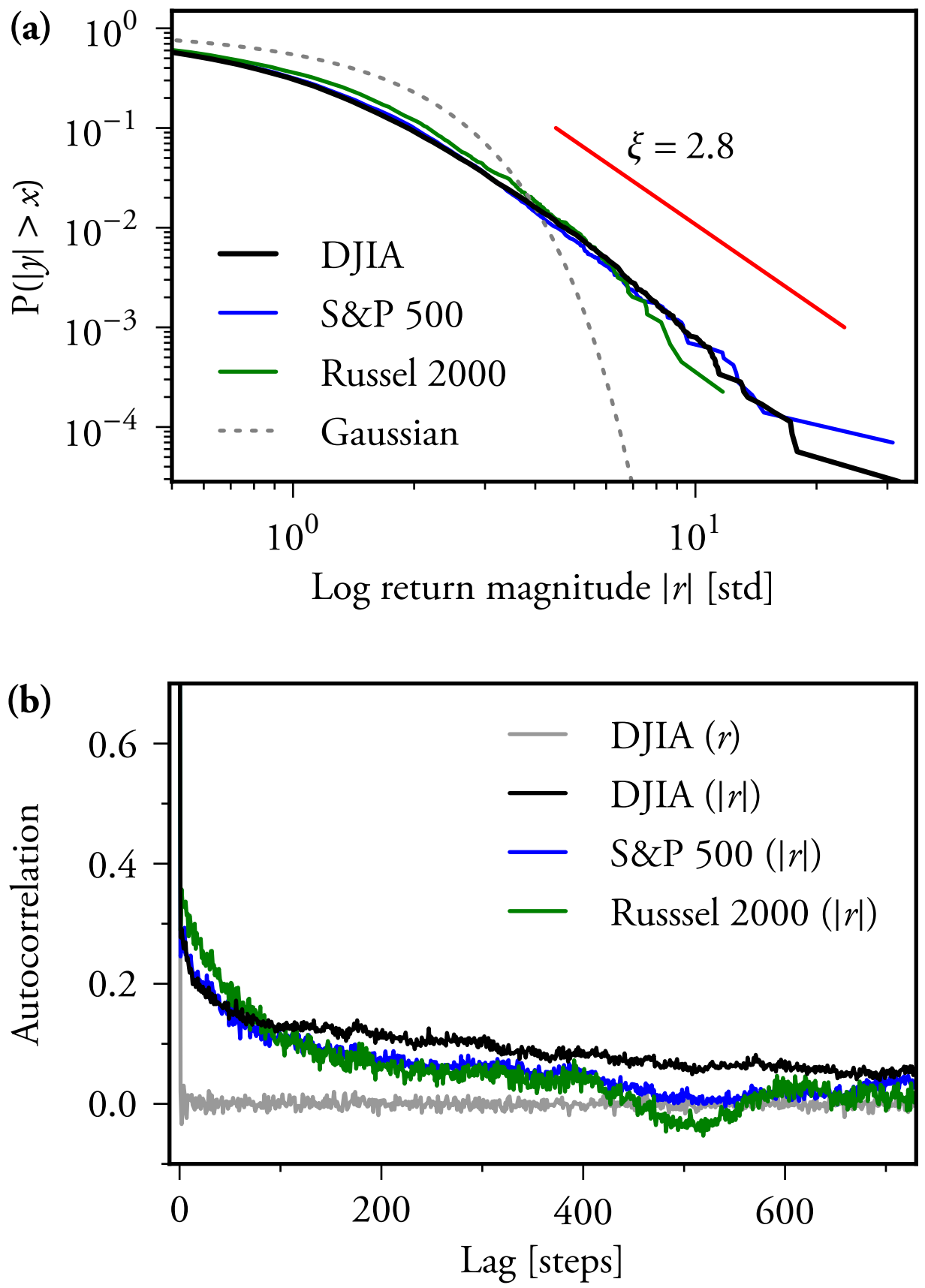

Figure 10.2.: Analysis of log returns calculated for the stock market price indices shown in figure 10.1. The different lengths of the three time series account for most of the differences. **(a)** CCDF for the normalised log return magnitudes. Red diagonal line: power-law fit for the DJIA (Hill estimator with KS-optimal cutoff, see section 2.5). Grey dotted line: Gaussian with unit variance. **(b)** Autocorrelations for the log return for the DJIA (solid grey line), and for the log return magnitudes for all three indices (line colours as in (a), methods as described in sec. 3)

Very similar results can be found in completely different data sets. Figure 10.3 shows the pound sterling and the yen, both measured in U.S. dollars, since the beginning formation of the modern FX market. This is a market where currencies with freely floating exchange rates can be traded under few conditions.[10]

In contrast to stock markets, the FX market is global and decentralised: brokers/dealers negotiate directly with one another without a central exchange or clearing house. Yet, after some initial transients, log return distributions closely resemble those of stocks. Very similar results are found for exchange rates measured in a different currency than the U.S. dollar (not shown).

CCDFs for several exchange rates, as well as a Gaussian and the DJIA since 1980 for comparison, are shown in figure 10.4. The tail exponents show some variation with $E(\xi) = 3.4$, and $Std(\xi) = 0.8$. However, all distributions are clearly heavy-tailed. The average exponent is close to the exponents for the indices discussed above.

The autocorrelations of the return magnitudes for the exchange rates vary across currencies too, as shown in figure 10.5. Yet, all of them exhibit positive correlations over hundreds of days. Time-series variability is also discussed in section 11.3.

Results consistent with those reported above were found for many different assets and for returns over different periods of time from minutes to days [Far99, Lux09, Con01, GPL⁺00]. For example, one study of the stocks of 1000 U.S. companies reported CCDF exponents $2.5 < \xi < 4$ for returns over time scales from 5 minutes up to 16 days [PGA+99]. On the time-scale of months, log return distributions become more Gaussian.

[10] The daily global trading volume in the FX market is by far higher than the GDP of national economies [KR10]. In 2000, the global daily turnover was 1.5 trillion U.S. dollar per day. 40% of the trades involved round trips within two days or less, and most of them were speculative [Ehr02]. By 2013, the daily turnover had reached even 5.3 trillion U.S. dollar [KR13]. Most trades contain the U.S. dollar on one side (ibid.).

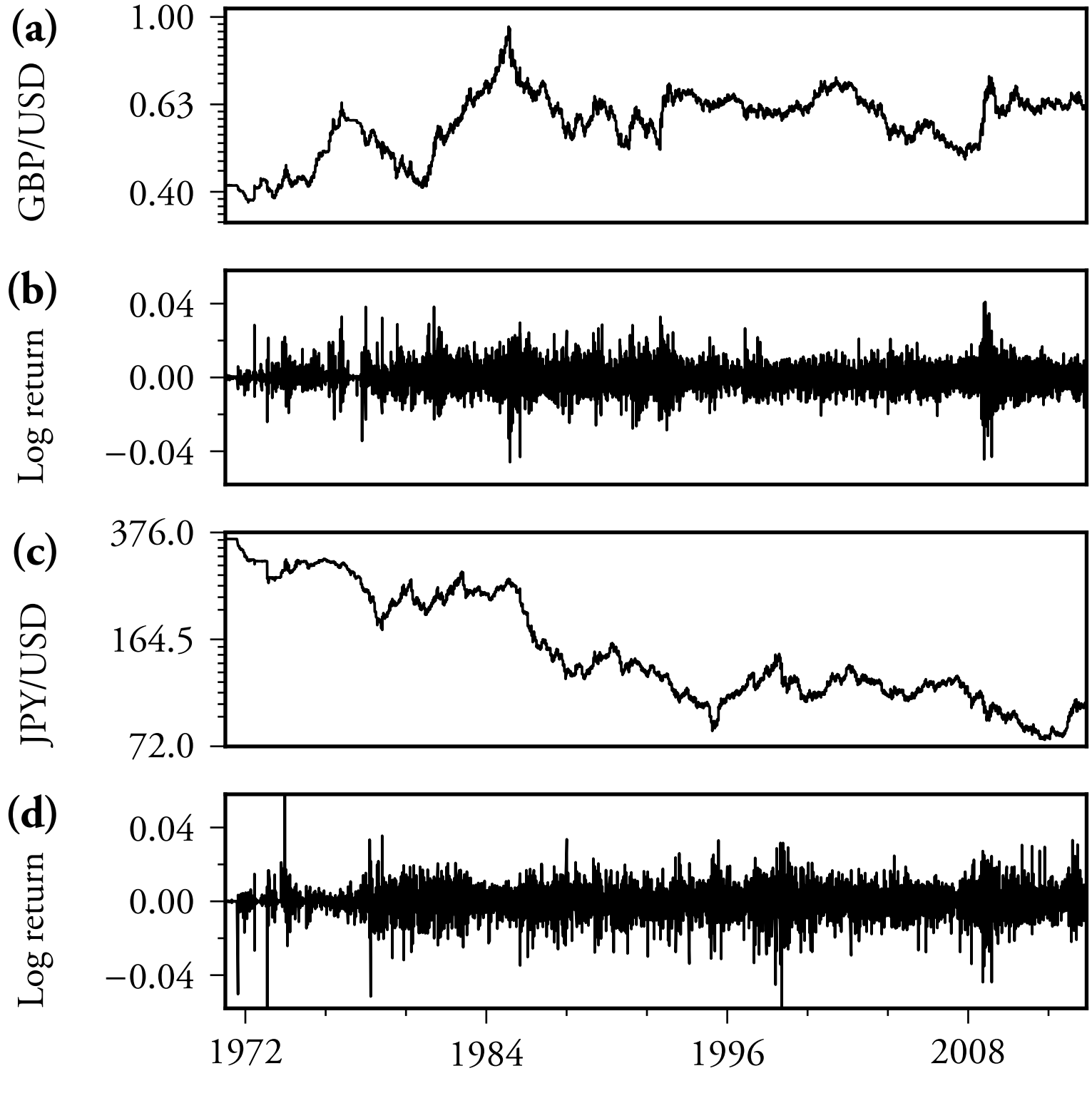

Figure 10.3.: Historic daily foreign exchange rates and corresponding returns. **(a)**, **(b)** British pound sterling per US dollar. **(c)**, **(d)** Japanese yen per U.S. dollar.

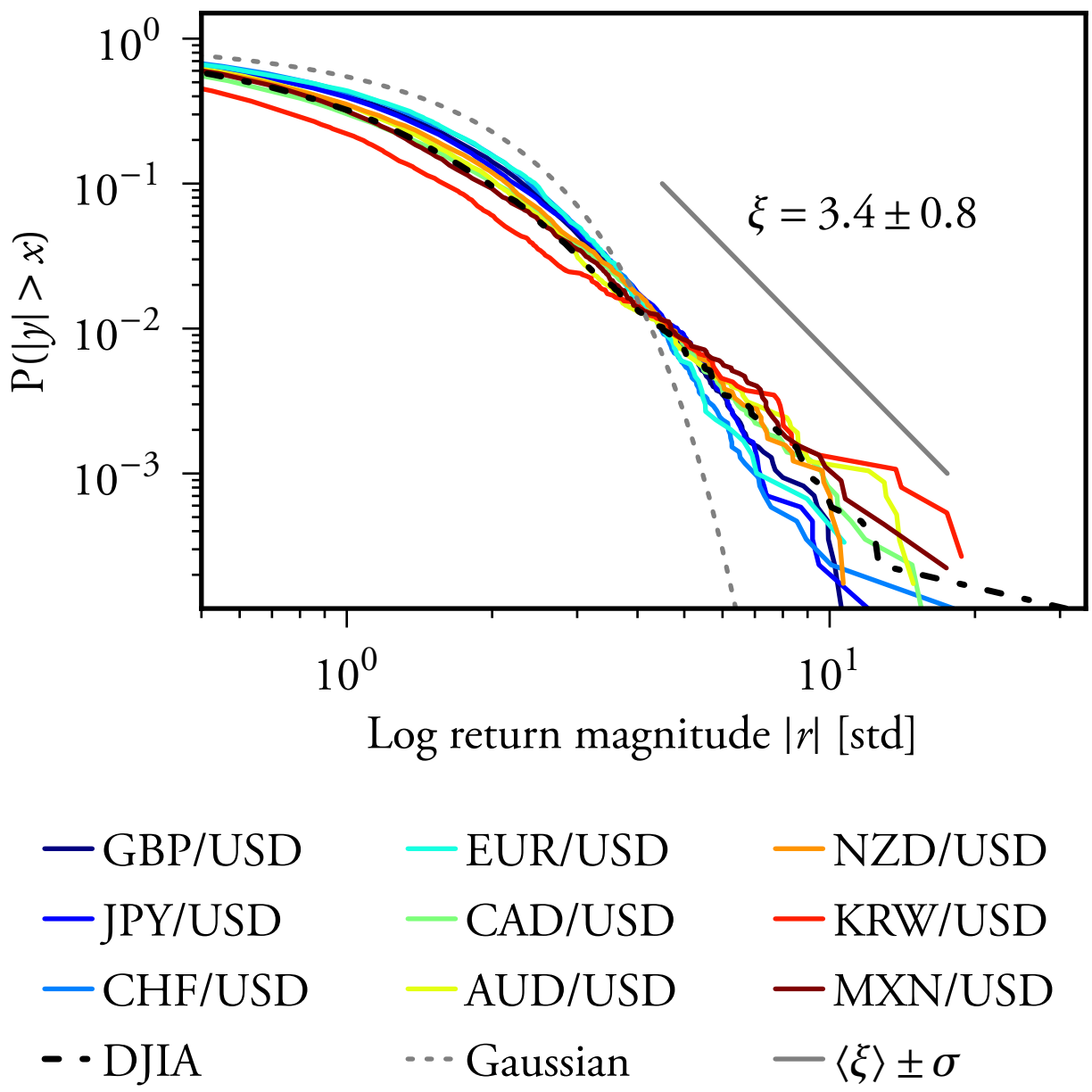

Figure 10.4.: CCDF for the normalised log return magnitudes for foreign exchange rates of several currencies measured in US dollar. For each time series, obvious transient effects after the rate became freely floating were discarded. That is, GBP, JPY, CHF, and CAD analysis starts in 1980 (see also figure 10.3). AUD and NZD start in 1991, MXN in 1996, KRW in 1999, EUR in 2002. The DJIA time series used here for comparison was also shortened and starts in 1980. Red diagonal line: average of the power-law exponents for the exchange rates (fitted as described in section 2.5). Grey dotted line: Gaussian with unit variance. **(b)** Autocorrelation of the log return magnitudes.

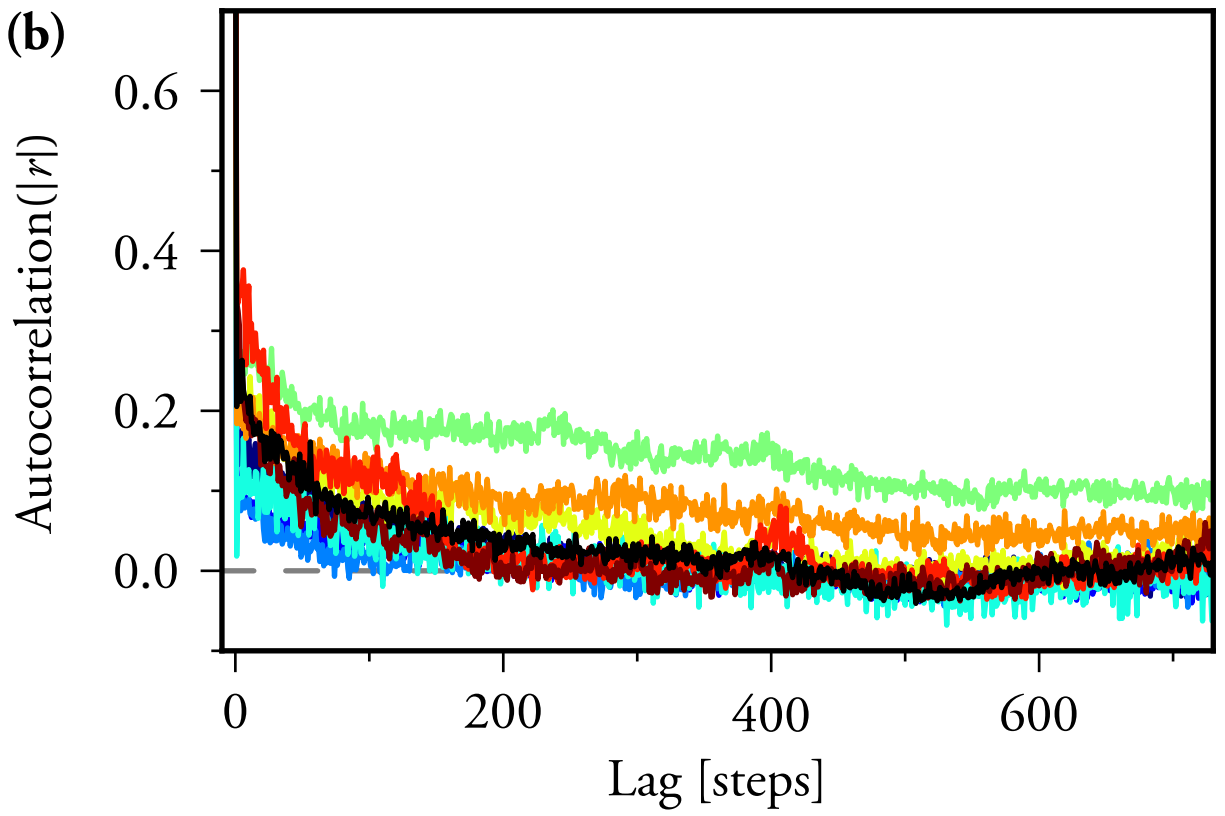

Figure 10.5.: Autocorrelations of the log return magnitudes for the same exchange rates also shown figure 10.4. Colors are identical.

Note that the stylised facts remained very stable over many decades (see figs. 10.1, 10.3, and [Lux09]). Even stock and FX data sets from the 18th century show the same properties [Har98]. Crashes are even documented for the very first stock market, in the 17th century [Pet11].

## 10.4. Rationality, heterogeneity, and behavioural economics

> “One of the things that microeconomics teaches you is that individuals are not alike. […] If we didn't have heterogeneity, there would be no trade.”
>
> Ken Arrow

A core assumption in the theories presented above is the selfishness of economic agents. For example, arbitrage opportunities are expected to vanish because somebody will exploit them. Many problems, however, require more specificity. A very common assumption in economic theories is that of a rational representative agent, sometimes called "homo economicus". Economists use the term "rational" to denote selfishly maximising utility or profit using all available information. Agents are often assumed to have rational expectations. That is, their predictions of the future are unbiased. In addition, many models (e.g. for general equilibrium) even require perfect information. These assumptions contribute to succinct and unique equilibrium models (see, e.g., foonote 4 in sec. 10.2), at the price of unrealistic requirements for the agents' abilities. Therefore, bounded rationality has been introduced to some models to account for limited cognitive abilities, and limited availability of information to individual agents. Unfortunately, there are so many possibilities in which agents could deviate from perfect rationality that additional constraints seem warranted. Such constraints may be found by studying how humans actually behave as opposed to how they should. [FG09, Sti02]

Behavioural economics studies bounds of rationality in economic decision making. Thereby, many common assumptions in prevalent economic models have been rejected based on behavioural experiments [FG09]. For example, humans deviate from simplistic utility functions, use possibly biased heuristics, and violate assumptions on temporal discounting. Humans also tend to deviate significantly from perfect selfishness. For example, they voluntarily share material payoffs with anonymous others or decline unfair splits. The extent of this behaviour is culturally dependent [HBB+01].

Not all behavioural research has to take place in lab experiments. Numerous studies investigated the real-world behaviour of economic actors, often with less than flattering results. For instance, financial analysts' predictions are often worse than a simple "no change" forecast; yet they agree with each other much more than with the actual result [GB05]. Investors also appear to trade much more than they should if the market were in equilibrium [FG09]. This is reflected by the global trading volume in financial markets which is two orders of magnitude larger than global production (ibid., see also foonote 10 in sec. 10.3). Investors even trade too much for their personal gain: trading less would, on average, increase their returns [Ode99].

These findings, however, are more problematic for common equilibrium theories than for the EMH on its own.[11]. As stated in section 10.2, the martingale property of price changes only requires that there is no systematic and predictable over- or underreaction. It was further argued, that active (short term) fund management is a zero-sum game before costs: "Good (or more likely just lucky) active managers can win only at the expense of bad (or unlucky) active managers" [FL12]. After costs, a passive index fund becomes extremely hard to beat in an efficient market.

If psychological biases that were found for individuals would systematically influence the market, this would create arbitrage opportunities that traders performing more careful analyses could exploit. Additional limitations are necessary under which such biases survive in the market [FG09]. Moreover, individual rationality or the lack thereof does not imply the same is true collectively in the market, and vice versa.[12]

While market psychology might actually explain some market failures, there is no known mechanism explaining how it could give rise to the stylised facts we are interested in here. As we shall see in the following, irrational behaviour per se is not sufficient to explain long-range

[11] This might, nonetheless, be due to the notorious difficulty to construct conclusive empirical tests of efficiency (sec. 10.2). For instance, the no arbitrage condition can be described without assumptions on utility. It thereby escapes the criticism on utility arising from psychological experiments. However, auxiliary assumptions necessary to test the EMH, like the CCAPM pricing model, often do depend on assumptions on utility.

[12] This argument was brought up against representative agent models before [Kir92]. However, it will become relevant in a very different sense in the following chapters

correlations and return power-laws. One might further argue that the impact of psychological biases should have changed over time after traders adopted more advanced theoretical methods, and especially after most trading became algorithmic. Option prices, for example, more closely match the predictions of the Black-Scholes model since it is used actively [FG09]. Many professional traders today also make conditional forecasts on possible future states of nature; they now actually behave more like the assumptions in popular economic models (ibid.). The very limited variability of the stylised facts over time and across markets implies that they are caused by some even more fundamental properties of markets.

Nevertheless, the importance of one aspect of human behaviour has to be reinforced at this point: individuals and their expectations are different from each other. Otherwise, there would be no trade (see e.g. [MS82], [Hom06, HB10] and the references therein). Already Adam Smith argued that conflicting interests of individuals, and specialisation due to cognitive limitations are essential for cooperation in the first place (as discussed in e.g. [FG09, Kir92]).

## 10.5. (Rational) bubbles

> “Markets can remain irrational far longer than you or I can remain solvent.”
>
> John Maynard Keynes

> “The word ‘bubble’ drives me nuts”
>
> Eugene Fama

A bubble is commonly used to refer to a "a period in which prices exceed fundamental valuation" [Sch13]. As discussed above, however, valuation is always a joint test that requires a pricing model. Valuation is also often ex-post wrong. Some efficient market proponents argue that the term "bubble" is misleading since it implies that one should have known that an asset was mispriced given the then-available information. This is again very hard to prove, since a "bubble warner" could be right by chance [Sch13, Kes13].

Nevertheless, the term "bubble" is used frequently by economists and laymen. Famous historical examples include the Dutch tulip mania (1634-7), the South Sea bubble (1719), or more recently the dot-com bubble (ca. 1997-2000) and the housing bubble (ca. 2000-2007) [Bru08, Sch13, Kru09]. There is considerable evidence that asset prices may be significantly misaligned for extended periods of time. For example, there appear to be no economic models which can explain more than a fraction of the movements of stock prices or foreign exchange rates (see, e.g., [BGV00, otRSAoS13], and the references therein).

One might assume that bubbles are caused by psychological biases or irrational herding (see [Kes13], and sec. 10.4). The opposite direction is pursued in the literature on rational bubbles. Here, situations are investigated where current owners of an asset have good reasons to believe they can resell the asset at a price above the fundamental value. There are four main strands of such models [Bru08]. If all investors have rational expectations and identical information, rational bubbles cannot emerge within an asset-pricing model. Overpricing that is present when the asset starts trading can persist under strict conditions. In an asymmetric information bubble, investors have different information and prices only partially reveal other traders' aggregate information. Then, under certain conditions like constrained short selling, finite bubbles can occur. Another model type investigates rational, well-informed traders who interact with biased ones. If arbitrage is limited (e.g. due to risk factors), the rational investors cannot prevent the bubble by going against it. On the contrary, there is incentive to "ride the bubble". That is, to continue to buy an overpriced asset, hoping that the already exuberant price will rise even further. Finally, bubbles can also emerge in a market with heterogeneous beliefs and short selling constraints. Then, even if traders are correct on average, optimists can influence the market more strongly than pessimists.

There appear to have been very few attempts to directly link mechanisms for bubbles with the stylised facts described above. The dynamics of rational expectation bubbles were investigated in [LS02]. Equation (10.1) allows for prices that grow with a constant rate $1/m > 1$. To turn this explosively growing bubble into a stationary process, a finite burst rate for an unexpected reset of the price is assumed. It is therefore possible to define multiplicative stochastic processes for prices that are consistent with rational expectations. These bubble processes are very similar to the Kesten process (sec. 4.4). They exhibit power-law tailed log return distributions, but the tail exponents $\xi < 1$ are far smaller than those for real log returns.

In one model with heterogeneous beliefs and short selling constraints, a bubble can form if the optimists have sufficient wealth compared to the asset supply for some period [Sch13]. Such bubbles are accompanied by large trading volume and volatility because the imbalanced price during the bubble is sensitive to the supply: if the latter is limited for some time, a bubble can form. Then, an unexpected increase in supply (e.g. due to sales from insiders) can cause the bubble to implode. A different but related model includes a non-linear market mechanic: the leverage cycle [Gea10, TFG11]. During good times, high leverage[13] allows those who value a certain asset more (for whatever reason), to drive up the price. In fact, leverage increases the profits of trend-following strategies during trending markets, creating evolutionary pressure towards even higher leverage. The problem arises when prices drop unexpectedly. The wealth of the leveraged op-timists drops, forcing them to sell to meet their margin requirements. Due to the crisis, margins are tightened, which further amplifies the downward spiral. Similar problems were identified during the recent economic crisis which followed leverage of up to 60 to one in 2006. Return distributions under this mechanism have been investigated in a multi-agent model implementation discussed in section 10.7. In the same section, other models are discussed where bubbles occur together with volatility clusters due to shifts in market ecology.

[13] That is, the use of credit to trade assets, thereby increasing both potential profits and risk.

Bubbles and crashes consistently arise in laboratory double auctions[14] even if participants had sufficient information to compute the fundamental value of the traded asset [SSW88]. Uncertainty about dividends seems to have no effect on the bubbles [NRR01], but uncertainty about the behaviour of others does: Repeated sessions of the same group of subjects under stationary conditions reduces the risk of bubble [PS03]. Other experimental paradigms focused more on the subjects' ability to predict future prices and used automated synchronous trading. It was found that some groups slowly converge towards fundamental prices while others fail to converge. This behaviour is consistent with subjects using heterogeneous adaptive learning strategies [HW09]. There seem to be no experiments in the literature where the scaling of log returns and long-range correlations were measured. Investigating these features in the existing paradigms would be quite challenging: Subjects in all of the experiments cited above had plenty of time to make their decisions. In consequence, only short time series with e.g. 15 periods in [SSW88] or 50 in [HW09] where recorded.

[14] That is, a process where potential buyers and sellers submit their bid- and ask prices to an auctioneer. In the continuous (asynchronous / intertemporal) case used in many experiments, trades are executed immediately if the highest bid price is above the lowest ask. Other orders are stored in a queue. Dividends in typical experiments are payed at the end of trading periods, which last several minutes.

## 10.6. Price formation

So far, we have covered a number of observations and theories on prices, but we have hardly discussed how prices actually form. A price is essentially a conversion rate at which two parties agree to exchange two different types of goods, one of which typically being fiat money. Yet, standard economics (including neoclassical finance) assumes "perfect competition" where all agents are price takers. That is, no agent on its own can influence the price. At the same time, equilibrium prices ensure market clearing. That is, demand and supply always match. Standard microeconomics also identifies how prices should move to eliminate excess demand: It is assumed that demand decreases with increasing prices while supply increases with increasing prices. Therefore, prices must increase to eliminate excess demand, and decrease to eliminate excess supply. There is, however, no trading at out of equilibrium prices since this would require at least one irrational agent who accepts a "bad deal".[15] Who adjusts the prices is not specified [Kir92].

One possibility to formalise a price finding process is the Walrasian auction. The auctioneer suggests different prices, and agents submit preliminary orders for each price. In a process called "tâtonnement",[16] the auctioneer keeps suggesting different prices and is thereby supposed to move closer to the price at which the market clears. When the equilibrium price is found, the actual trades take place. Unfortunately, there are few situations where tâtonnement converges towards an equilibrium even if it exists [FG09, Ly000].

An alternative is to assume omniscient agents who can determine the equilibrium themselves. The emergence of such an equilibrium is analogous to a Nash equilibrium in game theory [FG09]. In a game, each player's payoff depends on all the other players' actions. If each player knows the equilibrium strategies of all other players, there is at least one self-consistent equilibrium where no player can gain from switching their own strategy. Many models also use a representative agent to provide the unique and stable equilibrium which cannot be guaranteed by the underlying microeconomic structure [Kir92].

[15]As stated above, homogeneous and rational traders would actually never trade. The discovery of several no-trade theorems, however, seems to have done little to discourage the use of the assumptions from which they follow.

[16]The English term for "tâtonnement" is "groping" and seems to be less popular.

In practice, traders in financial markets are not omniscient, have finite cognitive and computational capabilities, and submit real orders. A stock exchange or a foreign exchange dealer matches buy- and sell orders and acts as a buffer for unexecuted orders. Most modern financial markets operate continuously and accept two basic order types.[17] Limit orders state the worst allowable price for the transaction. If the highest buy order limit price exceeds the lowest sell order limit price, the transaction is executed. Limit orders often fail to be executed immediately and are then stored in an order book. Traders can also submit market orders which are executed immediately at the best available price. Most financial markets depend on market makers who provide liquidity by accepting both buy and sell orders. On some exchanges, market making is institutionalised. An open order book allows any investor to effectively act as a market maker.[18] [19] For a more detailed discussion on market microstructure see, for example, [BFL09, Ly000].

[17] Some financial markets determine the opening (and closing) prices in special sessions (e.g. an auction). All orders are executed at the same price at the end of the session. That is, at the opening (or closing) price. Orders submitted in some pre-opening sessions can be retracted before the end of the session. See, e.g, [BHS99, Smi13]

[18] Market makers profit from offering to buy at a lower price then they offer to sell. If prices are not sufficiently mean reverting, however, market making strategies generate losses. A designated market makers may be compensated for this risk e.g. by charging fees.

[19] Some bilateral stock transactions are arranged outside of the open order book and reported publicly later. The FX market lacks a centralised exchange in the first place. These details, however, do not matter for the following more abstract view on markets.

In any case, each financial market has a clearly defined mechanism which determines the price for each transaction based on the orders submitted by the traders. The price at each point in time is therefore determined by demand and supply. The price changes step by step as orders are executed. In other words: prices emerge through trading, and trading takes place while the prices change gradually. Therefore, there has to be uncertainty and heterogeneity among the trading parties–otherwise, as discussed above, there would be no trading. Furthermore, for prices to carry information, orders have to carry information too. Since the number of orders at any given point in time is finite, so is the impact of a single order. In fact, the empirical market impact of an order is a strongly concave function of its volume.[20] Therefore, the relative impact of small orders diverges. [BFL09]

The mechanistic emergence of prices is clearly susceptible to any sufficiently large group of traders who submit orders based on whatever motivation, be it erroneous. To consolidate these results with the idea of prices reflecting fundamental values, one might presume that only accordingly informed orders have a permanent impact. A popular assumption is that markets include a small fraction of "noise traders" who submit random orders that can cause temporary deviations from fundamental values [BFL09, Ly000]. Well informed investors (e.g. market makers), however, should then quickly identify and eliminate the mispricing. If, however, new relevant information arrives at the market, these smart investors should adjust their expectations accordingly, which leads to a permanent price shift. This idea can explain excess volatility and overreaction to news, while prices still should reflect fundamentals on average [otRSAoS13]. A common argument for as to why the "fundamentalists" should ultimately dominate the prices is that they buy underpriced assets (thereby driving prices up) and sell overpriced ones (driving prices down). This strategy is ex-pected to be highly profitable. Noise traders or trend followers should be less successful and therefore driven out of the market [BFL09]. On the other hand, as discussed in section 10.5, there are possible scenarios where the situation is not as clear cut (see also footnote 18 in sec. 10.6).

[20]Numerous studies of continuous double auction markets suggested the best fit for the impact function (normalised by the liquidity) to be either a square root or logarithmic.

Empirically, however, there is little evidence that the "noise" making up the excess volatility is mean reverting. In contrast, there is no sign of mean reversion of prices on shorter time scales, and only weak evidence of mean reversion on the scales of years [BGPW04, BGV00]. This suggests that there may be a band of uncertainty around fundamentals as large as 100%; only larger deviations would lead to reverting dynamics [Bou10, Bla86]. It was indeed found that uninformed mechanical price pressure due to an imbalance of demand and supply appears to have much larger impact on price movements than information [Hop07].

These findings may be caused in part by the inherent difficulty to identify and interpret the relevant news. This problem even affects professional traders (see sec. 10.4). Furthermore, several surveys indicated that on short time horizons, investors tend to use "chartist" strategies that extrapolate short-term observations. Mean-reverting "fundamentalist" investment strategies appear to only dominate on longer time horizons [HW09].

Another problem was identified in the very trading mechanics of financial markets [BFL09, BGPW04, Bou10, FGL04]. Since buy orders lead to higher prices and sell orders lead to lower prices, market impact always influences prices to a traders disadvantage. A large buy market order, for example, would remove not only the lowest sell order, but also those at higher prices. Therefore, traders split up large transactions into so-called meta orders. These meta orders are then spread over a longer time interval (up to months) to avoid influencing the price adversely. Therefore, order flow is correlated over long times and the market should almost never be in equilibrium. Prices, however, appear to move unpredictably because traders try to conceal their true intents. Thereby, even highly liquid markets have a low revealed liquidity. Because the unpredictability of prices in this picture arises not from random orders but from a delicate balance of antagonistic market forces, a sudden correlation of market orders from different actors may lead to a liquidity crisis that cannot be counterbalanced by market makers. There is some evidence indicating that vanishing revealed liquidity may have a much larger effect on price fluctuations than volume [BFL09]. Furthermore, "large returns are not caused by large orders" [FGL04]. Note that these findings also constitute strong evidence against the idea that insiders reacting to private information cause large price jumps directly.

Summing up, information processing in markets may be slower than anticipated, markets may respond highly nonlinear to fluctuations, and market granularity may matter. Such microstructure implications can be long-lived and relevant on large scales (see also [Ly000]).

## 10.7. The rise of multi-agent models

As discussed above, equilibrium models fail to reproduce the magnitude, distribution shape, and temporal clustering of real price fluctuations. The desire of practitioners for useful risk-assessment tools led to the use of stochastic processes that can reproduce the statistics of real returns. The most prominent example for such a process is the Generalised Autoregressive Conditional Heteroskedasticity (GARCH) class of processes. They model log returns as a Gaussian white noise process with a dynamic variance which depends on previous returns as well as on its own previous states.[21] [MS00] After careful calibration these processes can, however, at best reproduce the stylised facts on the phenomenological level; they cannot explain how these features arise in marktes. For this purpose, models are required that include actual market mechanisms and link them to the observed phenomena.

[21] To correctly reproduce the long-range correlations of volatility clusters, GARCH processes require explicit dependencies on a large number of previous states.

From the late 1980s on, the idea gained momentum that market psychology, bounded rationality, and the interaction of different trading strategies may be responsible for excess volatility and overreaction. This idea, which goes against mainstream neoclassical economics, led to an explosion of multi-agent models which, in addition, was fueled by the increasing availability of computing power during the 1990s. In these models, market dynamics arise from the explicitly modelled actions of boundedly rational heterogeneous agents.[22] Trading activity usually falls in one of two broad categories: detailed modelling of high-frequency trading using continuous double auctions, or more simplified blockwise trading for models that investigate the interaction of strategies on longer time scales. Prices are usually adjusted quickly to ensure immediate market clearing, or more slowly in order to incrementally reduce excess demand. Due to the enormous diversity of multi-agent models, we can only discuss a small fraction of examples in the following. Some more discussion can be found, for example, in [Hom06, Lux09, Con05, FG09].

Multi-agent models are imperfect information processing systems with internal feedback loops and, as such, can overreact to changes in externally provided fundamentals. In many models, clustered volatility arises from switching between low and high activity regimes [Con05]. This is usually caused by shifts in the composition of agent strategies in the market. A prominent example of this type of model is presented in [LM99b]. It contains three types of trader strategies and one asset. Fundamentalists buy the asset if the price is below the fundamental value and sell otherwise. The volume of the fundamentalists' orders grows with increasing mispricing. The fundamental value of the asset itself is determined by Gaussian white noise. There are two types of chartists, that is, traders who bet on trends: optimists, who always buy, and pessimists, who always sell a fixed amount of the asset. Traders stochastically switch between strategies and the transition probabilities depend on momentary profits earned by the different strategies. The price increases or decreases with a probability that depends on the excess demand. When the price is close to the fundamental value, no strategy has an advantage and a surplus of optimists or pessimists may arise by chance. This destabilises the market and prices drift away from the fundamental value. With increased mispricing, fundamentalists increase their trading volume and counter the trend. The chartists are then no longer successful and the mispricing collapses. Volatility and trading volume are higher during these misalignment phases[23] than in a balanced market.

[22] Many phenomena in financial markets–especially in high frequency data–can be quantitatively explained from structural constraints alone, even if orders are placed at random. Examples include average volatility, the shape of the average order book, anomalous diffusion, and market impact. These zero-intelligence models may be considered the conceptual opposite of equilibrium models. The resulting prices in these models, however, are not efficient. A more detailed discussion of this topic can be found, for example, in [FG09, BFL09].

Log returns for this model exhibit volatility clustering and heavy-tailed distributions. These reasonably realistic fluctuations do not reflect external news without bias. Instead, they are substantially driven by the endogeneous interaction of traders. This result was argued to contradict the EMH.[24] In fact, many similar models reproduce the stylised facts to some extent, but they disappear in equilibrium. The latter is often reached for special parameter choices or simply after sufficient simulation time when all strategies become equally profitable [FG09]. An important shortcoming of this class of models, however, is that the stylised facts are usually finite size effects [Lux09, Lux06]. Furthermore, many models feature several coupled equations and many parameters that have to be adjusted (about a dozen in [LM99b]). This impedes a detailed understanding of the mechanisms generating the stylised facts and of the dependence on parameters. Simpler models which depend on fewer parameters still exhibit several different dynamical regimes (see also sec. 10.9). Realistic intermittent fluctuations may be very sensitive to parameters that balance stabilising and destabilising strategies [GB03].

[23] Misalignment can occur in the form of bubbles with exuberant prices or anti-bubbles where the asset is undervalued.

[24] The authors, however, did not explicitly discuss which forms of the EMH (see sec. 10.2) precisely this statement referred to.

One particular critique of models with strategy switching is that the latter would have to occur on a broad range of time scales, including very short ones. In [TFG11], it was argued that strategy switching on many different time scales may be caused by leverage instead of a priory different trading strategies.[25] As discussed in section 10.5, unexpected price fluctuations can force traders who use credit to increase the profitability of successful strategies to sell into a falling market. This mechanism creates a highly nonlinear feedback loop. Log return fluctuations of a corresponding multi-agent model are highly asymmetric, apparently more so than those shown above. Their magnitudes, however, can approximate real return distributions. It is nevertheless not clear how specific asymmetries in some market mechanisms might account for the prevalence of very similar stylised facts in very different markets (e.g. in FX markets, or historic stock markets). The authors in [TFG11] suggested that the leverage cycle is only one example of how local risk regulation can generate systemic risk–a point to which we will come back later.

At this point, it appears as though the emergence of collective optimisation has almost been lost. Equilibrium models attempted to find an overreaching mechanism that could cut through the complexity of microscopic interactions. These models, however, demand much more from individual agents than the ideas of Adam Smith. He argued that selfish individuals unwittingly contribute to the common good through cooperation. Equilibrium theories usually require almost god-like agents (see sec. 10.4 or, e.g., [FG09]). Even "as-if rationality" as an avenue to efficiency relies on at least some agents who effectively behave close to rational. Only these agents are assumed to thrive in a competitive market, causing other strategies to eventually die out [LeB11]. Among the models discussed so far, however, only few can reproduce the stylised facts of section 10.3, In these models, the rational behaviour of value investors (the fundamentalists) gets lost in the complexity of agent interactions and market mechanics. In the next section, we discuss a different class of models which bring us closer to the idea that the whole can be greater than the sum of its parts.

[25] See also [PS10] for an empirical study on trend switching on different time scales.

## 10.8. Minority games

The desire to better understand the complex dynamics of multi-agent models sparked the creation of even more stylised toy models that can be treated analytically. The most prominent such model is the Minority Game (MG).[26] Aiming to provide a stylised theory of speculative trading and information flow, it has received much attention, especially in the interdisciplinary physics community, since it can be studied using methods from statistical mechanics. Some of its features, which are introduced below, were carried over to other models, including those presented in the next chapters. We therefore use, whenever possible, a consistent notation throughout this work which slightly differs from [CMZ05b].

In MGs, agents have to repeatedly make a binary choice. In each round, the agents whose choice is in the minority win. A simple intuition for this rule is that the two options may represent buying or selling an asset. When the majority of agents want to buy, the price is likely to be high and sellers may get the better deal. When the majority wants to sell, buyers may get the asset at a low price. High or low with respect to what? We will come back to this question in sections 10.9, and 11.8.

The MG sacrifices realism for simplicity to allow for a thorough understanding of very elemental dynamics. First, in high frequency trading, not everyone can be a winner[27]: the MG is a model for competition, and (geometrical) frustration. Initially, the MG was meant to be played by an uneven number of agents such that there is no nash-equilibrium. This requirement was dropped when it became clear that it has no influence on the results for the prevailing implementations of the game. It goes to show, however, that the MG is intended to study a scenario where standard game theory fails; it is in this respect diametric to equilibrium economics.

[26] An extensive overview is found in [CMZ05b]. For a brief introduction see e.g. [Mor04]

Another fundamental feature of MGs is heterogeneity of agent strategies–otherwise, there would be no minority. Even though the minority rule can be applied to different scenarios, the MG became associated with heterogeneous boolean strategies which were used in almost all studies (see below).

Finally, the minority game is a model for coordination and the breakdown thereof. It is a prime example for a phase transition. That is, a transition beyond which agents cannot be understood as independent elements anymore, allowing for coherent collective behaviours. However, as we demonstrate in the following, the mechanisms for extreme events in common MGs are specific to the respective implementations. We therefore take a closer look at these models in the following sections.

### 10.8.1. The Original Minority Game (OMG)

In the original formalisation of the MG, the two possible actions for each agent $i$ at time step $t$ are represented by 1 and $-1$. Agents base their decisions on public information states. In each time step one of $D$ possible states, which is denoted by an index $\mu(t) \in \{1, \ldots, D\}$, is conveyed to the agents. A strategy maps each state onto one of the two possible actions. Hence, $2^D$ different strategies exist. Each agent, however, only "knows" about $S$ strategies. At each point in time, each agent $i$ uses their so far most successful strategy denoted by $s_i(t) \in \{1,\dots,S\}$. Most studies focus on the case $S=2$ because the qualitative results don't change for larger $S$.

---

[27] This statement may appear to contradict the notion that markets contribute to the common good, even though it is actually neutral in this respect. Trading per se does not create any wealth, and speculators are interested in short-term exploitation of mispricing, not long-term investments. Dividends and increased productivity due to efficient resource allocation have little influence on speculative trading with round trips within days or even less time. See also sec. 10.2 (eq. (10.4) in particular), sec. 10.4, and footnote 10 in sec. 10.3.

More precisely, the decisions for each agent $i$ are determined by a strategy vector whose elements $\sigma^{\mu}_{i,s_i}$ initially are drawn randomly out of $\{0,1\}$ and then kept constant. That is, each agent's decision (e.g. to buy or to sell) is predetermined at random for each $\mu$ and $s$.

The actions of each agent at each point in time are

$$a_i(t) = 2\sigma^{\mu(t)}_{i,s_i(t)} - 1. \qquad (10.6)$$

Distinguishing between decisions $\sigma$ and actions $a$ will be relevant for model extensions below.

The outcome of the game at each time step is

$$A(t) = \sum_{i=1}^{N} a_i(t) \qquad (10.7)$$

where $N$ is the number of agents. Each agent keeps track of the predictive power of their strategies using a utility score $U$ which is updated according to

$$U_{i,s}(t+1) = U_{i,s}(t) - a_i(t)A(t). \qquad (10.8)$$

The active strategy $s$ is determined by[28]

$$s_i(t) = \operatorname*{argmax}_s U_{i,s}(t). \qquad (10.9)$$

Note that agents update all of their strategy scores at each time step, not just the score of the active strategy.

[28] For analytical treatment, $s$ is usually determined probabilistically, which leaves the general features of the game intact.

We consider two different methods for the generation of the information states $\mu$. For endogenous information, agents possess a memory of the most recent $K$ winning decisions which are indicated by the step function $\Theta(-A)$. This information can take one out of $D = 2^K$ possible states:[29]

$$\mu(t+1) = \sum_{i=0}^{K-1} 2^i \, \Theta\big(-A(t-i)\big) + 1. \qquad (10.10)$$

For example, if agents had a memory of the past $K = 2$ returns, they could distinguish between $D = 4$ possible combinations of subsequent winning decisions: $(1,1)$, $(0,0)$, $(1,0)$, and $(0,1)$. The public information $\mu(t) \in \{1,2,3,4\}$ then encodes which one of these 4 states describes the immediate history of the system.

For exogenous information, the information states are determined exclusively by factors that are external to the market (i.e. the states of nature). The $\mu(t)$ here are drawn randomly and independently with equal probabilities $P_{\text{ext}}(\mu) = 1/D$.

For both types of information, moderate numbers of agents cooperate despite selfishly following their utility score. Figure 10.6 (a) shows the relative strength of outcome fluctuations $\text{Var}(A)/N$, which was termed "global efficiency" in the literature [CMZ05b], for different numbers of agents. If the latter were determining their actions in each round by flipping a coin, $\text{Var}(A)/N$ would be constant and equal to one.[30] However, starting with few agents and then incrementally increasing their number actually reduces the relative fluctuations up to a certain point. For even larger $N$, $\text{Var}(A)/N$ rises again and eventually surpasses one.

[29] This notation is used to clarify that the sequence of winning decisions $\Theta(-A(t-K+1)),\ldots,\Theta(-A(t))$ is the binary representation of the integer $\mu(t+1)-1$. In numerical simulations, it is often more convenient to use the equivalent form $\tilde{\mu}(t+1) = (2\tilde{\mu}(t) + \Theta(-A(t))) \mod D$, where $\tilde{\mu}(t) = \mu(t) - 1$.

[30] This result follows from the variance of the binomial distribution which has to be scaled to account for actions $a \in \{-1,1\}$.

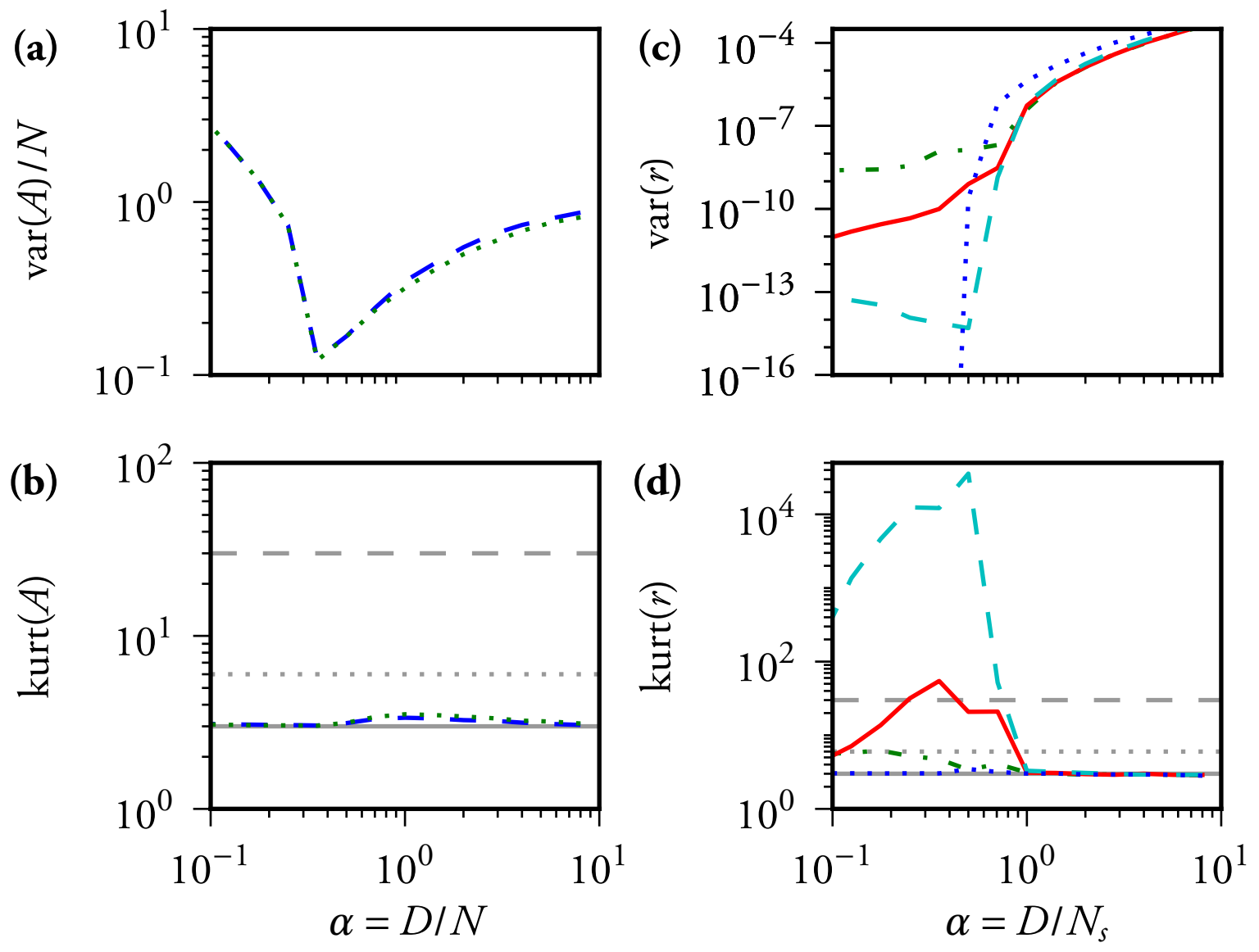

Figure 10.6.: Variance and kurtosis for different numbers of speculative agents in several minority games for $D = 2^9$. $10^7$ time steps were discarded to remove transients. **(a), (b)** Original minority game with $S = 2$. Blue dashed: extrinsic information. Green dotted: intrinsic information. **(c), (d)** Minority game with dynamic capitals, $N_p = 16$, and extrinsic information. Green dash-dotted: $S = 2$, $\gamma = 0.001$. Solid red: $S = 2$, $\gamma = .01$. Dashed cyan: $S = 2$, $\gamma = 0.1$. Dotted blue: $S = 1$, $\gamma = 0.01$. The latter model's variance levels around $10^{-30}$ for $\alpha < 0.5$. The horizontal grey lines in (b) and (d) indicate expected kurtoses for: a Gaussian (3, solid line), an exponential distribution (9, dotted line), and a very rough estimate for a real daily return time series of the same length (30, dashed line).

It was found that this behaviour is characterised by the ratio $\alpha = D/N$ of the number of information states $D$ and the number of agents $N$. When there are very few agents, they behave almost as if they were flipping coins to make their decisions. As more agents are added, the chance increases that agents can cooperate and counteract each others actions. There is a reason for this behaviour: if many agents use similar strategies, they will lose more often. Therefore, the minority rule causes agents to differentiate. Beyond a critical point $\alpha_c \approx 0.34$, however, agents start to become correlated due to the limited number of possible independent strategies.

We conclude that the MG is a minimal model for self-organised cooperation among selfish heterogeneous agents. It further shows how the ability to coordinate breaks down when the agents' strategies become too correlated. This phase transition can actually be treated analytically [CMZ05b].

The game never settles into an dynamic equilibrium: $\mathrm{Var}(A)$ is always finite. It can, however, be efficient according to a definition of information efficiency that is especially tailored to the MG:

$$H = \frac{1}{D} \sum_{\mu=1}^{D} \mathrm{E}(A|\mu)^2, \tag{10.11}$$

which measures how well the outcome $A(t)$ can be predicted from $\mu(t)$ on average. For fixed $D$, $H$ is a decreasing function of $N$. Therefore, even games with strong herding ($\alpha > \alpha_c$) are informationally efficiency in this sense (not shown).

What the MG still lacks at this point is a link to observables in financial markets. If agents were to submit only market orders, then $A$ would correspond to the excess demand. It was argued that if trading took part via an order book which contained a logarithmically uniform density of limit orders, execution of the excess market orders should displace the log price by an amount that is linearly proportional to $A$ [CMZ05a]. In other words,

$$r(t) \propto A(t). \tag{10.12}$$

For the OMG, however, these log returns do not reproduce the heavy tails and volatility clustering of real log returns. Figure 10.6 (b) shows the kurtosis of $A$ (and therefore $r$) which is very close to 3 for all $\alpha$. That is, return distributions are always very close to Gaussian distributions.

It was argued that the reason why the OMG does not exhibit the stylised facts is that the trading volume is always constant. Therefore, two different mechanisms for volume fluctuations were introduced, which we discuss in the next sections. Note that these extensions predate the empirical evidence we cited above, indicating that volume fluctuations alone are insufficient to explain the high frequency of extreme price jumps (e.g. [BFL09] and the references therein). Furthermore, note that many features in the OMG are independent of the mechanism which generates the information states. We will, however, show in chapter 11 that this statement doesn't hold for other models with similar agent strategies, including some extensions of the OMG.

### 10.8.2. Dynamic capital

One method to incite non-Gaussian log-return fluctuations in the MG is to modulate the weight of the agents' actions according to their previous success [CCMZ01]. In this model, agents are endowed with some capital $c_i$, a fraction $\gamma$ of which they "invest" at each $t$:

$$a_i(t) = \gamma c_i(t) \left(2\sigma^{\mu(t)}_{i,s_i(t)} - 1\right), \tag{10.13}$$

where the strategy vectors $\sigma^{\mu}_{i,s}$ are defined as in the OMG. Positive actions $a$ are interpreted as buy orders, and negative $a$ as sell orders.

The outcome $A$ is equal to the excess demand (eq. (10.7)). The log returns for this model are defined as[31]

$$r(t) = \frac{A(t)}{V(t)}, \tag{10.14}$$

where

$$V(t) = \sum_{i=1}^{N} |a_i(t)| \tag{10.15}$$

is the trading volume.

Finally, we distinguish between two different types of traders. The $N = N_s + N_p$ agents are divided into $N_s$ speculators and $N_p$ producers. The speculators' utility scores and capitals adapt according to their success:

$$\left.\begin{array}{lcl} U_{k,s}(t+1) & = & U_{k,s}(t) - a_k(t)r(t) \\ c_k(t+1) & = & c_k(t) - a_k(t)r(t), \end{array}\right\} \quad 0 < k <= N_s. \tag{10.16}$$

The producers have only one trading strategy and their capital is constant:

$$c_j(t+1) = c_j, \quad N_s < j <= N_p. \tag{10.17}$$

The distribution of the producer capitals has little influence on the game dynamics [CCMZ01]. We therefore only consider the case $c_j = 1$ in the following.

Due to the simplifications in this model, capital in the case with only speculators is not conserved, but vanishes over time. Therefore, producers are absolutely necessary to maintain a nonzero trading volume. Furthermore, $\gamma$ needs to be sufficiently small to ensure that the agents' capitals remain positive at all times. Consequently, previous studies focused on the case $\gamma \ll 1$. The dependence of the game dynamics on $\gamma$ was, however, so far apparently not investigated in detail. It was even claimed that $\gamma$ has little influence as long as it is small enough (without showing results to back this claim). This is somewhat surprising, since the heavy return tails vanish for $\gamma \to 0$, a case which has been investigated before [CCMZ01].

[31] In [CCMZ01], the log returns are defined as $r(t) = -A(t)/V(t)$. This is a quite peculiar choice since an a positive excess demand would then lower the price. We therefore here adopted the convention used in [GZ09] for the log returns, utility scores, and capitals (see below).

Figure 10.6 (c) shows the log return variance plotted against ratio $\alpha$ of the strategy complexity and the number of speculators for four different parameter combinations. Three curves were obtained for $S = 2$ and different $\gamma$. All three exhibit a phase transition which is shifted towards larger $\alpha$ compared to the OMG, and which lacks the steep increase of fluctuations for small $\alpha$. Strikingly, larger $\gamma$ lead to less variance for small $\alpha$, that is, for large numbers of speculators. For $S = 1$ the phase transition is less shifted, but $\alpha < 0.5$ becomes an absorbing phase even for small $\gamma$.

Figure 10.6 (d) shows the kurtoses for the same simulations also shown in (c). For $S = 2$, the kurtoses exhibit maxima for $\alpha < 1$. These maxima are both higher and located closer to $\alpha = 1$ for larger $\gamma$. For $S = 1$, there is no notable excess kurtosis.

Several of these findings have not been reported in the existing MG literature. On the contrary, several contradicting claims were made. First, we have clearly shown that $\gamma$ has a much stronger influence on the game dynamics than claimed in [CCMZ01].[32]

Furthermore, it was claimed in [GZ09] that the model exhibits heavy tailed log returns close to the phase transition also for $S = 1$. The authors show two examples, at least one of which they admit contains transients. This author was, however, not able to confirm that the parameters in question, or any other set of parameters for $S = 1$, give rise to any heavy log return tails that are not artefacts due to transients or averaging over repeated simulations. There are actually strong arguments why there should be no heavy tails for $S = 1$. This discussion, however, requires the results from chapter 11. It therefore has to be postponed. Either way, the Probability Density

[32] There, the log return variance was shown only on a linear scale, and the kurtosis was not investigated.

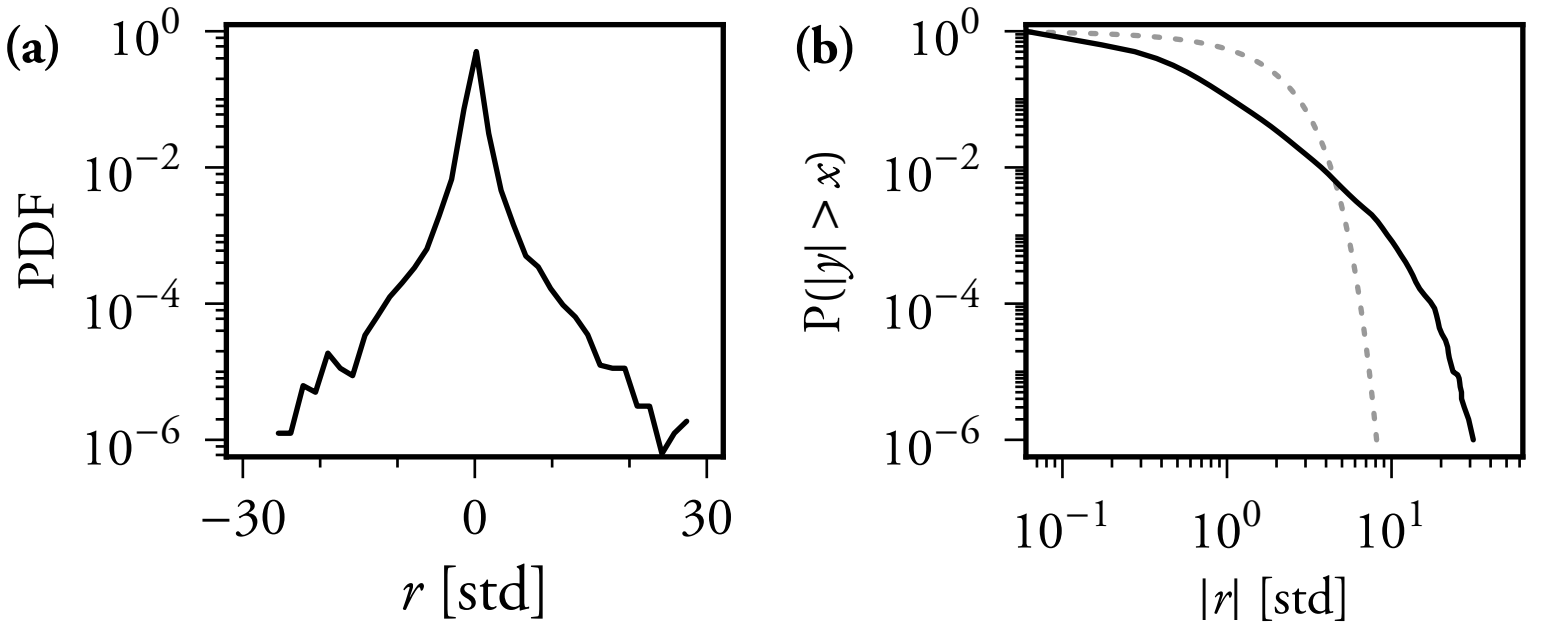

Figure 10.7.: Distribution of log returns for the minority game with dynamic capitals at the critical point. $D = 2^9$, $N_s = 1506$, $N_p = 16$, $S = 2$, $\gamma = 0.01$, extrinsic information. $10^7$ time steps were discarded to remove transients. **(a)**: PDF. **(b)**: CCDF.

Functions (PDFs) shown in [GZ09] appear to resemble exponential distributions much more closely than power laws. This impression leads to another shortcoming of the existing literature: The log returns in nearly all publications on MGs are investigated using PDFs in semi logarithmic coordinates. This makes it very difficult to judge whether the distributions resemble power-laws. Many of these PDFs don't exhibit strongly convex tails, which would be expected for power-laws.

Figure 10.7 shows both the PDF in semi logarithmic coordinates and the CCDF in double logarithmic coordinates for the MG with $S = 2$, $\gamma = 0.01$. The number of speculators maximises the kurtosis for these parameters (see fig. 10.6 (d)). The return distributions are heavy tailed and may exhibit a short scaling regime, but large returns are exponentially truncated. When increasing the sample size, only the truncated regime grows, not the power-law.[33] Other parameter sets did not yield more realistic results, except for very short time series where the truncation is not as apparent. A pronounced truncation was not found for real returns in section 10.3. Nevertheless, the return fluctuations for this model are quite promising when considered in relation to the model's simplicity. Note that Figure 10.7 (a) closely resembles figure 3 in [CCMZ01]. Log returns for the parameters used in figure 10.7 also exhibit correlated return magnitudes consistent with [CCMZ01] (not shown).

[33] In part II, we found the opposite result for true power-laws that only appear to be truncated due to temporal correlations. See also fig. B.5 for an example of a market model which generates power laws without severe truncation.

As a preliminary conclusion, we found that strategy switching and capital redistribution are distinct adaptation mechanisms with somehow different properties. The latter mechanism seems to be more capable at balancing the impacts of different strategies. Capital redistribution on its own can also overcome the herding due to correlated strategies in the OMG. However, heavy tailed log return distributions and volatility clustering in the MG presented in this section only emerge for a combination of both mechanisms, and close to the critical point. "What happens at the critical point awaits further investigations."[CCC+13] Note that we did not investigate intrinsic information in this section, and neither did the authors in the existing literature on MGs with evolving capital. We will discuss these models again in section 11.8.

### 10.8.3. The Grand Canonical Minority Game (GCMG)

Another mechanism for volume fluctuations is to allow unsuccessful agents to choose not to play. In distinction to the OMG, the agents' in the GCMG [CM03] have only one strategy. Their actions are determined by

$$a_i(t) = \theta(U_i(t)) \left(2\sigma_i^{\mu(t)} - 1\right). \tag{10.18}$$

The utility scores are

$$U_{i,s}(t+1) = U_{i,s}(t) - a_i(t)A(t) - \epsilon_i, \tag{10.19}$$

where $\epsilon_i$ is a threshold below which agents don't participate.[34] The agents are again divided into $N_s$ adaptive speculators, and $N_p$ non-adaptive producers:

$$\epsilon_j = \epsilon, \qquad 0 < j <= N_s \tag{10.20}$$

$$\epsilon_k = -\infty, \qquad N_s < k <= N_p. \tag{10.21}$$

The return is defined as

$$r(t) = A(t). \tag{10.22}$$

That is, it is identical to the outcome given by equation (10.7). Therefore, the log return in the GCMG is equivalent to the excess demand, and the volume is equivalent to the number of participating agents.

Figure 10.8 shows the model dynamics after transients for the same parameters used to showcase the models properties in figure 4.5 in [CMZ05b]. The GCMG exhibits characteristic volatility clusters that grow slowly and end abruptly. The return distributions for the GCMG in the this dynamical regime exhibit power law tails [CM03].

The dynamics of the GCMG can be understood intuitively. First, a very large number of agents is required.[35] In each round, due to the minority rule, most agents lose. This causes the speculators to leave the market. Because the nearly empty market is highly predictable, the inactive speculators begin to increase their scores again. This causes an influx of many agents with correlated strategies who give rise to strong fluctuations. The speculators start to lose again; the cycle repeats itself.

The GCMG is essentially a model for irrational fads. In fact, the herding described above vanishes as soon as the speculators consider that they themselves have an impact on the market [CMZ05b]. Does the model correctly capture empirical results? First, as discussed in section 10.6, several studies found that volume alone is not a good explanation for extreme price jumps. Second, in the GCMG, the volatility is high before new price extrema, and low afterwards (fig. 10.8). This appears to be the opposite of the volatility profiles shown in [PS10] (figs. 2 and 3 ibid.). Nevertheless, it might be possible that the GCMG describes dynamics which do occur in real markets, but don't cause the majority of the extreme price fluctuations in the same way as they do in the model.

[34] Similar to the OMG, analytical calculations are performed for probabilistic participation.

[35] Several authors, e.g. in [CM03], have stated that the stylised facts emerge near the critical point. Best results, however, require many more speculators. The reader may confirm by e.g. comparing figures 1 and 2 in [CM03] that heavy tails start to emerge approximately a factor 10 away from the phase transition.

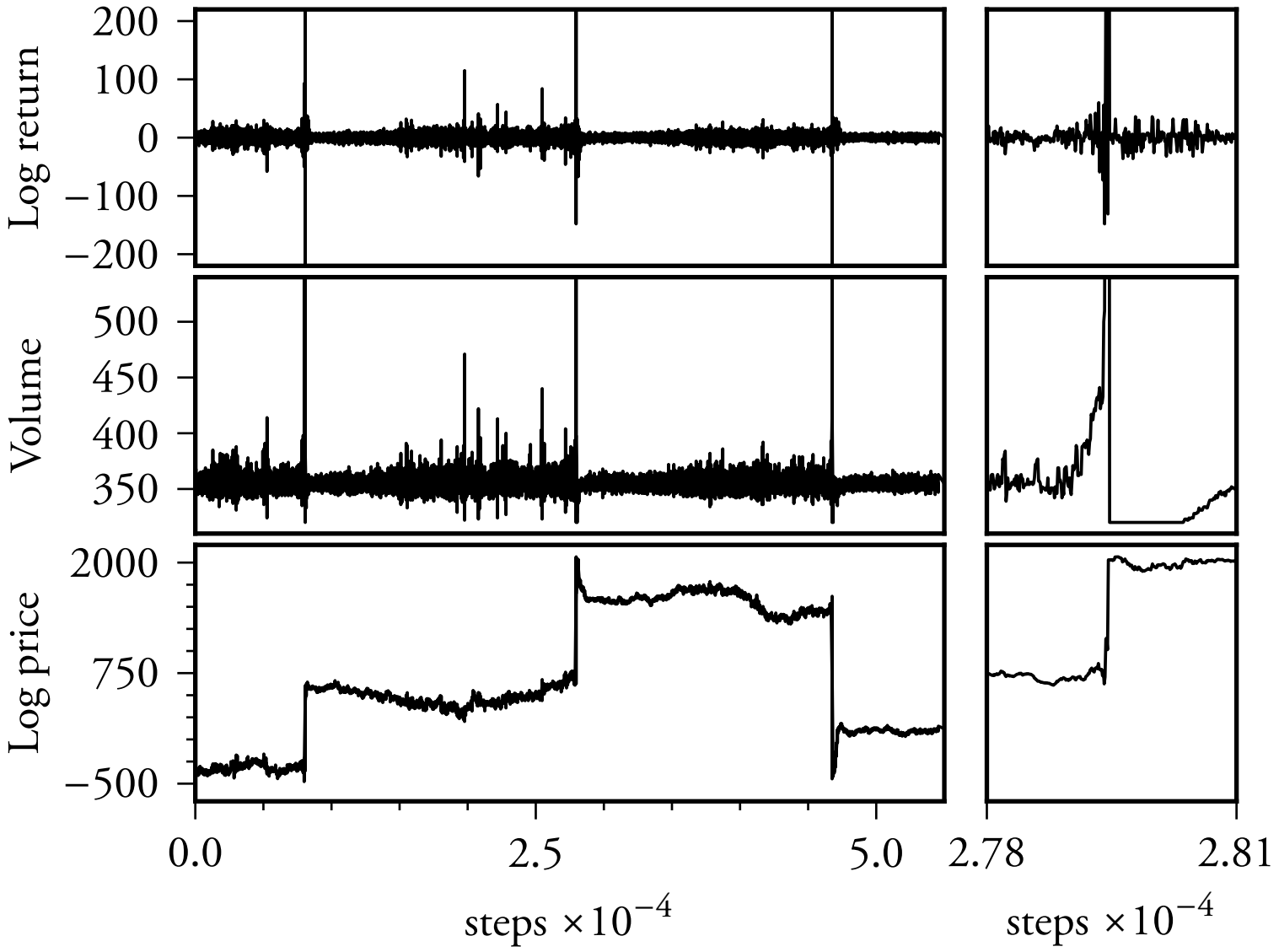

Figure 10.8.: Log return, trading volume, and log price for the grand canonical minority game with $D = 2^5$, $N_s = 9600$, $N_p = 320$, $\epsilon = 0.01$, and extrinsic information. $10^6$ time steps transient were discarded. The smaller right columns show 250 steps before and after an extreme price jump. Vertical axes are shared along each row. The highest peaks for the return and volume are cut off in order to keep smaller fluctuations visible.

The GCMG also suffers from several more formal problems. The herding behaviour described above is only a finite size effect. It vanishes for $D > 2^5$, or if noise is introduced to the model [CM03]. Even for small systems, the dynamics depend on the particular initialisation of the agent strategies. For about half of the simulations, the speculators will stay inactive after they first left the market. This latter problem, however, can be remedied by introducing a forgetting term in the utility function [CDMMC06]. Yet, this author found that even this latter model requires extensive parameter tuning to yield stable dynamics. Furthermore, the return distributions in this model tend to exhibit either early truncated tails, or bimodal price jumps. This can also be seen clearly in figures 3 and 4 in [CDMMC06].

The literature contains few MGs where agents can choose to not participate in addition to having dynamical capitals. These models seem to be able to generate heavy tailed log return distributions and volatility clusters. There appears, however, to be no detailed study of the interaction of the different effects in these models [CCMZ01, CMZ05a].

Despite certain problems with specific implementations, the MG greatly advanced the field of multi-agent modelling. The MG with dynamic capital seems to be a more promising starting point for a minimalistic market model than the GCMG.

## 10.9. Other statistical mechanics models

By demonstrating the applicability of statistical mechanics to socio-economic problems, the MG inspired many similar models. The first obvious variation is a majority game, which is highly unstable. Such a game corresponds to a market consisting entirely of trend followers. Therefore, several mixed minority-majority games were investigated which exhibit complex dynamics with stable and unstable phases [CMZ05a]. For some researchers, a major concern was that the MG payoff

$$g_i(t) = -a_i(t)A(t) \tag{10.23}$$

might be too simple to characterise trading success. Here, $a_i(t)$ is the action of agent $i$ at time $t$, and $A(t)$ is the outcome (see sec. 10.8). Payoffs in real markets follow from transactions at different points in time [AS03, CMZ05a] like, for example, buying an asset and selling it at a higher price. No consistent single-round payoff was found unless it is based on an agents expectation of future prices. Therefore, several alternatives to the minority rule were proposed. One such payoff was introduced in the $-game [AS03]:

$$g_i(t+1) = a_i(t)A(t+1). \tag{10.24}$$

This model includes speculators and a market maker. In contrast to the OMG, the $-game exhibits two phases: a majority phase where speculators profit, and a mean reverting phase where the market maker profits.

A much more complex model was introduced in [GB03]. It features an intertemporal payoff, different trader species who may chose to not participate or to temporarily act like a different species, dynamic capitals, and a set of rules for market clearing and asset conservation. The model dynamics exhibits three main regimes: oscillatory, stable, and intermittent. The authors suggested that the many different aspects of the model dynamics could be understood as thoroughly as the MG by investigating simplified versions of the full model.

Finally, the model investigated in [Bor01, KBF02, KB13, KBB12] features a two-dimensional grid of interacting agents, similar to the Ising model in solid state physics. These agents interact probabilistically via a "magnetic field". The local field consists of a majority term for the nearest neighbour interaction, and a minority term with respect to the global magnetisation. The latter is essentially a normalised excess demand. In addition, fundamentalist agents have knowledge of a "fundamental price" which is determined by a random walk. The fundamentalists have a stabilising effect on the price, and scale their volume proportionally to the mispricing. This model exhibits quiet phases with little magnetisation, and bubble phases with strong magnetisation and large fluctuations. These dynamics resemble those of the model presented in [LM99b] (see also sec. 10.7). The spin model, however, is formally much simpler. It is also one of the more robust models for the generation of realistic stylised facts (as they are defined above). However, since the spin model structurally is much closer related to physical models than to economic ones, it is somehow difficult to interpret. For example, agents in this model have no real incentive to act in a certain way since there is no payoff, utility, or capital. It is also not clear to which real-world interaction the grid corresponds. Nevertheless, there is a surprising connection between this model and the results presented in chapter 12, which we discuss in chapter 14.

In the next chapter, we implement some elements from the MG without an explicit minority rule. The resulting trading model improves the realism of several aspects of the MG with minimal formal complexity.

# 11. An inherent instability of efficient markets

> “Everything should be made as simple as possible,
> but not simpler.”
>
> Albert Einstein[1]

In this chapter, we show that speculative markets which absorb self-generated information can exhibit both evolution towards efficient equilibrium states and their subsequent destabilisation. In other words, we pursue the novel approach to modelling price fluctuations outlined in section 10.1.3.

First, however, we briefly explain the motivation for this approach based on the insights gained in the last chapter. We then introduce a minimal agent-based market model where the impacts of trading strategies naturally adapt according to their success. This model can quantitatively reproduce the stylised facts introduced in section 10.3. The model dynamics are investigated analytically and numerically in great detail. These results were published in [PP13]. Finally, we discuss similarities and differences compared to MGs.

## 11.1. Motivation

In chapter 10, we reviewed considerable evidence that financial markets are approximately information efficient in the sense that prices adjust to reflect certain information and cannot be easily predicted. Obvious mispricing, e.g. between different markets, is eliminated rapidly. More subtle statistical inefficiencies like predictable return patterns exist, but they tend to be eliminated over time (although new ones may appear).

[1] This may be a paraphrase [O'T14]

The EMH, though, is commonly understood to imply much more. Prices are also thought to reflect the "right information" in the "right way". That is, information on the future earning potential of an asset, (eq. (10.1)) which depends on fundamental factors relating to the real economy. Prices in an ideal market should rapidly relax towards their equilibrium values, thereby absorbing the fundamental information such that only relevant new information causes significant price changes.

Attempts to explain prices from "fundamental values", however, were largely unsuccessful: Prices move too much and exhibit considerable jumps far too often to represent the arrival of "fundamental" information. These price jumps are not caused by single large orders either. Traders–even those with private information–should avoid such orders anyway in order to minimise market impact to their own disadvantage.

Alternative explanations face several challenges. Collective dynamics may emerge that don't trivially follow from the behaviour of individual agents. On the one hand, if psychological biases and naive trend following dominated the market, this should create arbitrage opportunities. On the other hand, real traders are heterogeneous and not hyper-rational, as assumed by many economic theories. Otherwise they would not even trade. Furthermore, the "stylised facts" for empirical log returns have been observed in very different present and historical markets. They should therefore robustly emerge from very elemental properties of speculative trading. Yet, they cannot be easily explained by simple market failure.

Here we investigate whether the seemingly contradictory features of price changes arise in speculative markets which absorb self-generated information. It was demonstrated in chapter 7 that adaptive control of a dynamical system can generically lead to an instability where the susceptibility to noise dramatically increases close to the point of perfect balance. This principle would apply to markets if two requirements were fulfilled: First, markets need to absorb information about predictable price changes. As we discussed above, this is a rather common view in economics. Ideally, this property should hold independently of the rationality of the individual traders, which cannot be guaranteed. Second, a self-referential market would have to become susceptible to residual noise once all locally relevant information has been exploited. This property is actually intuitive, too. As traders try to detect trends or patterns in the price dynamics, they effectively predict how the market will react to available information. However, once the agents' actions have led to a balanced equilibrium, it becomes increasingly difficult to distinguish predictable price fluctuations from random noise. If traders then continue to act upon the random fluctuations as if they would hold meaningful new information, their actions will not be balanced anymore. That is, it may be impossible to predict whether a group of traders will overreact to the supposedly new information and to attenuate the resulting price jump by exploiting it. Therefore, atypically large price movements may become much more likely than expected for a Gaussian distribution.

## 11.2. A minimal trading model

As a concrete example of the fundamental dynamical instability arising from information absorption as it may be realised in financial markets, we introduce a minimal agent-based trading model. Each agent $i = 1, \ldots, N$ is representative of one trading strategy and possesses two types of assets which are called money $M_i(t)$ and stocks $S_i(t)$ in the following. For simplicity, we consider a coarse-grained price time series where one step $t$ could be considered as e.g. a day. In each trading step, each agent either offers an amount of money to buy stocks or an amount of stocks in exchange for money. The decision which action to take depends deterministically on a public information state, and on the agent's strategy. In each time step one of $D$ possible states, which is denoted by an index $\mu(t)$, is conveyed to the agents.

In a speculative market, traders react to information that can originate from both past prices, and any kind of external news. Both may potentially influence the agents' behaviour and therefore might ultimately lead to price changes which could be exploited. Therefore, we use the same consistent representation for both endogenous and exogenous information as in section 10.8.

Here, however, endogenous information encodes the most recent $K$ signs of the log returns which indicate whether the prices $p(t-K)$, $\dots, p(t-1)$ decreased or increased with respect to their predecessors:

$$\mu(t) = \sum_{i=0}^{K-1} 2^i \, \Theta\big(r(t-i-1) + \eta(t-i-1)\big) + 1 \tag{11.1}$$

where $\Theta$ is the step function, and $\eta$ is an arbitrarily small symmetric random variable with zero mean.[2] [3] This information is publicly available and can be considered to provide an important information about the state of the market.

For exogenous information we use the same notation but envision a binary encoding of external information. For simplicity, the $\mu(t)$ here are drawn randomly and independently with probability $P_{\text{ext}}(\mu)$. Unless stated otherwise, all $\mu$ have equal probabilities $P_{\text{ext}}(\mu) = 1/D$. It is possible to generalise the encoding to allow for mixed information. The results are similar to the endogenous case (sec. B.1.2).

[2] The noise term is included only to prevent the theoretical possibility of the game "getting stuck" in one information state. Simulation results do not depend on $\text{Var}(\eta)$ as long as it is small enough.

[3] For an intuitive example, consider again the case $K = 2$. Agents could then distinguish between $D = 2^K = 4$ possible combinations of subsequent price movements: (up, up), (down, down), (up, down), and (down, up). In other words, the public information $\mu(t) \in \{1,2,3,4\}$ encodes the direction of the price change from time $t-3$ to $t-2$, and of the price change from $t-2$ to $t-1$.

Each agent's decisions are determined by a strategy vector whose elements $\sigma_i^\mu$ initially are drawn randomly out of $\{0,1\}$ and then kept constant. These two possible decisions here correspond to trading an amount $m_i(t)$ of money or an amount $s_i(t)$ of stocks for the respective other asset in the next time step. Orders are placed with a constant use parameter $\gamma$:
Case $\sigma_i^{\mu(t)} = 1$ (agent $i$ buys stocks):

$$\begin{aligned} m_i(t) &= \gamma M_i(t) \\ s_i(t) &= 0 \end{aligned} \tag{11.2}$$

Case $\sigma_i^{\mu(t)} = 0$ (agent $i$ sells stocks):

$$\begin{aligned} m_i(t) &= 0 \\ s_i(t) &= \gamma S_i(t) \end{aligned} \tag{11.3}$$

Demand and supply are the sums of all buy and sell orders respectively

$$\delta(t) = \sum_{i=1}^{N} m_i(t) + \epsilon \tag{11.4}$$

$$\varsigma(t) = \sum_{i=1}^{N} s_i(t) + \epsilon \tag{11.5}$$

where $\epsilon \ll 1$ is a small positive number.[4] At each time $t$ the price $p(t)$ is determined from the ratio of demand and supply, respectively:

$$p(t) = \frac{\delta(t)}{\varsigma(t)} \tag{11.6}$$

with demand $\delta$ and supply of stocks $\varsigma$. This price naturally has the correct unit. Finally, all trades are performed at this price $p(t)$, similar

[4] This ensures that prices and returns are always well defined. The cases with zero demand or supply are, however, irrelevant for all practical purposes. A sufficiently small $\epsilon \ll 10^{-3}$ does not influence simulation results to a meaningful degree. All figures were generated using $\epsilon = 10^{-10}$.

to market orders except that here the exchange of money and stocks is executed synchronously for all orders of all agents at the same time $t$. This is a fair rule that could be used in a real market with only market orders executed at distinct points in time. In contrast to other minimalistic market models (secs. 10.8.2, and 10.9), this simple trading rule conserves the total amounts of the assets that are exchanged in a trade, as it should if trading fees can be neglected. [5] [6]

We focus on markets that are dominated by $N_s \leq N$ speculators who can only win or lose assets by betting on price changes within the market.

$$\left.\begin{array}{lcl} M_k(t+1) & = & M_k(t) - m_k(t) + s_k(t)\,p(t) \\ S_k(t+1) & = & S_k(t) - s_k(t) + m_k(t)/p(t) \end{array}\right\} \quad 0 < k <= N_s. \tag{11.7}$$

To investigate the effect of a small number of traders that convey new assets to the market or draw out their profits, we further allow for $N_p = N - N_s$ producers. In contrast to the set of agents representing the speculators, the producers' resources are defined to remain constant.

$$\left.\begin{array}{lcl} M_j(t) & = & M_j(0) \\ S_j(t) & = & S_j(0) \end{array}\right\} \quad N_s < j <= N_p. \tag{11.8}$$

[5] Another rule with asset conservation is presented in [GB03], but it requires a much more complex set of equations.

[6] This rule can be further justified by the following argument. For a market including stochastic limit orders gathered over some period of time, consider the hypothetical price $p^*(t)$ at which trades would take place if all agents scaled their orders by a common factor. Then the volume would change, but to preserve market clearing the price cannot be affected; that is $p(t)^* = p(t)$. Therefore, the price is a function of the ratio of demand and supply. After linearisation of this function for small small excess demands the price is proportional to the aforementioned ratio which justifies this choice of the pricing rule also as an approximation of the mean prices obtained from limit orders. Note that the price is also invariant against a fraction of unexecuted orders, if this fraction is the same for demand and supply.

Since the strategies of the producers are chosen in the same way as for the speculators, they also deterministically depend on the information state. Therefore, the producers contribute predictable amounts of liquidity and stocks to the market. These demands and supplies can then be exploited by the speculators, who are competitive and redistribute their possessions. Initially, all agents are provided with equal amounts of assets $M_i(0) = S_i(0) = 1$.

## 11.3. Stylised facts

A log return time-series for the model with exogenous information is shown in figure 11.1 (a). A strong reduction of initial fluctuations is observed, leaving only a narrow band of Gaussian distributed returns after the transient. Figure 11.1 (b) shows the endogenous case. Here, in contrast, initial return magnitudes are reduced only in the mean. The magnitudes of the few most extreme returns, however, are less reduced. The remaining fluctuations are analysed in figure 11.2 (a), where cumulative distributions of return magnitudes are shown for both cases and compared to the DJIA. The latter serves as an example for a typical price time series. For the endogenous case, the distribution tail is well described by a power-law and in good agreement with the DJIA. The latter exhibits a slightly steeper slope than the average simulation for the model (dark grey area in figure 11.2 (a)), but this may be a chance result.[7] We don't present more detailed significance tests here, since we can at best expect a rough correspondence between a real market index and a model at this level of abstraction anyway. The distribution tails for the model are not truncated and are as stable under accumulation as those for real log returns, as shown in the section B.3.

[7] In [PP13], we used the DJIA time series obtained from [Dep08]. This time series contains several presumably spurious extreme returns that are not present in any other data source for the DJIA ([Wil13, Fed13b], and others). Therefore, we here show the data obtained from [Wil13]. As a side effect, the very end of the CCDF slope for the DJIA here is slightly steeper than [PP13].

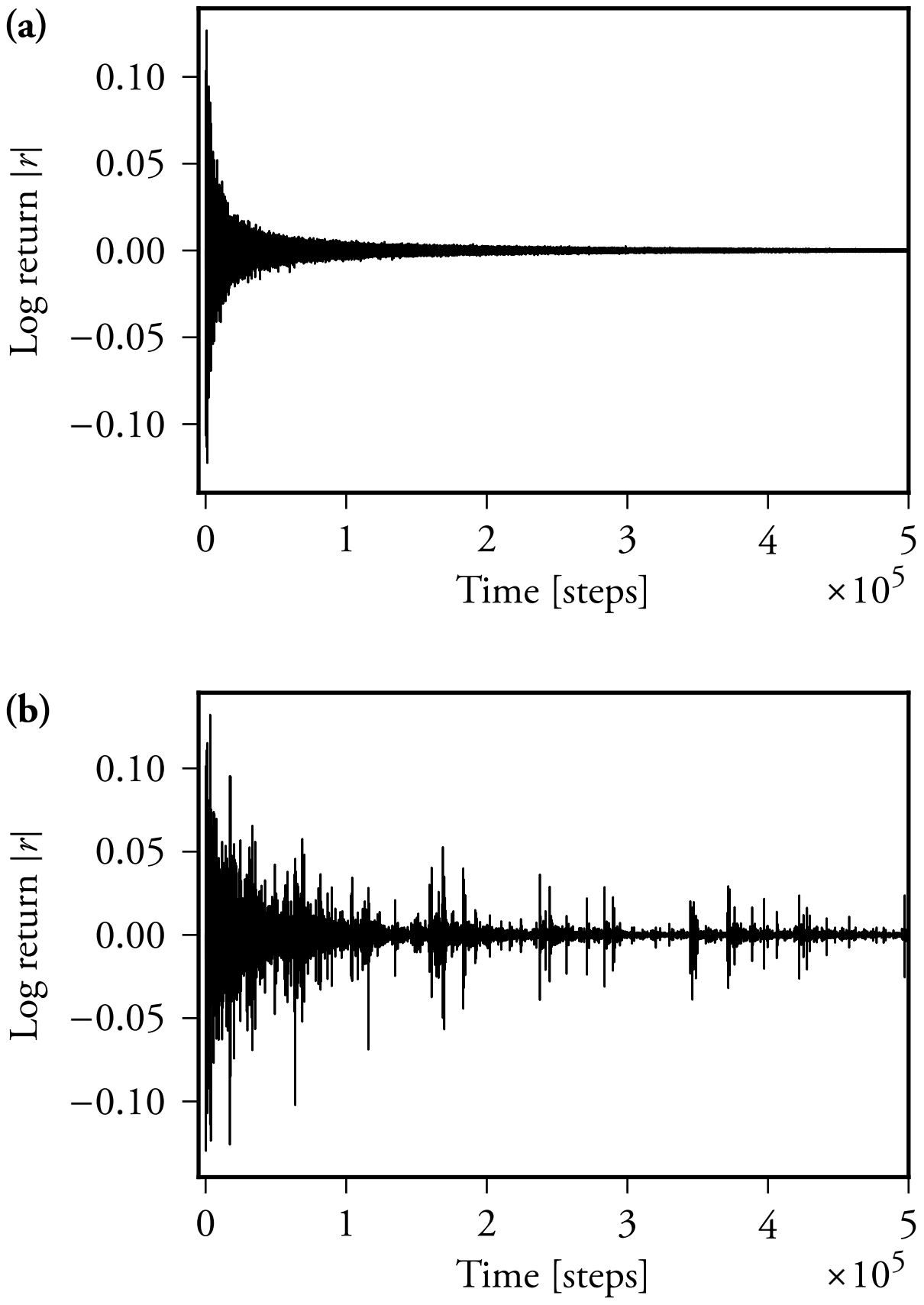

Figure 11.1.: Log returns for the trading model. **(a)**: Time series of the model with uniformly distributed exogenous information. Parameters: $N_s = 2^{10}$, $N_p = 0$, $D = 2^9$, $\gamma = 0.85$. **(b)**: Time series for the same model, but with endogenous information.

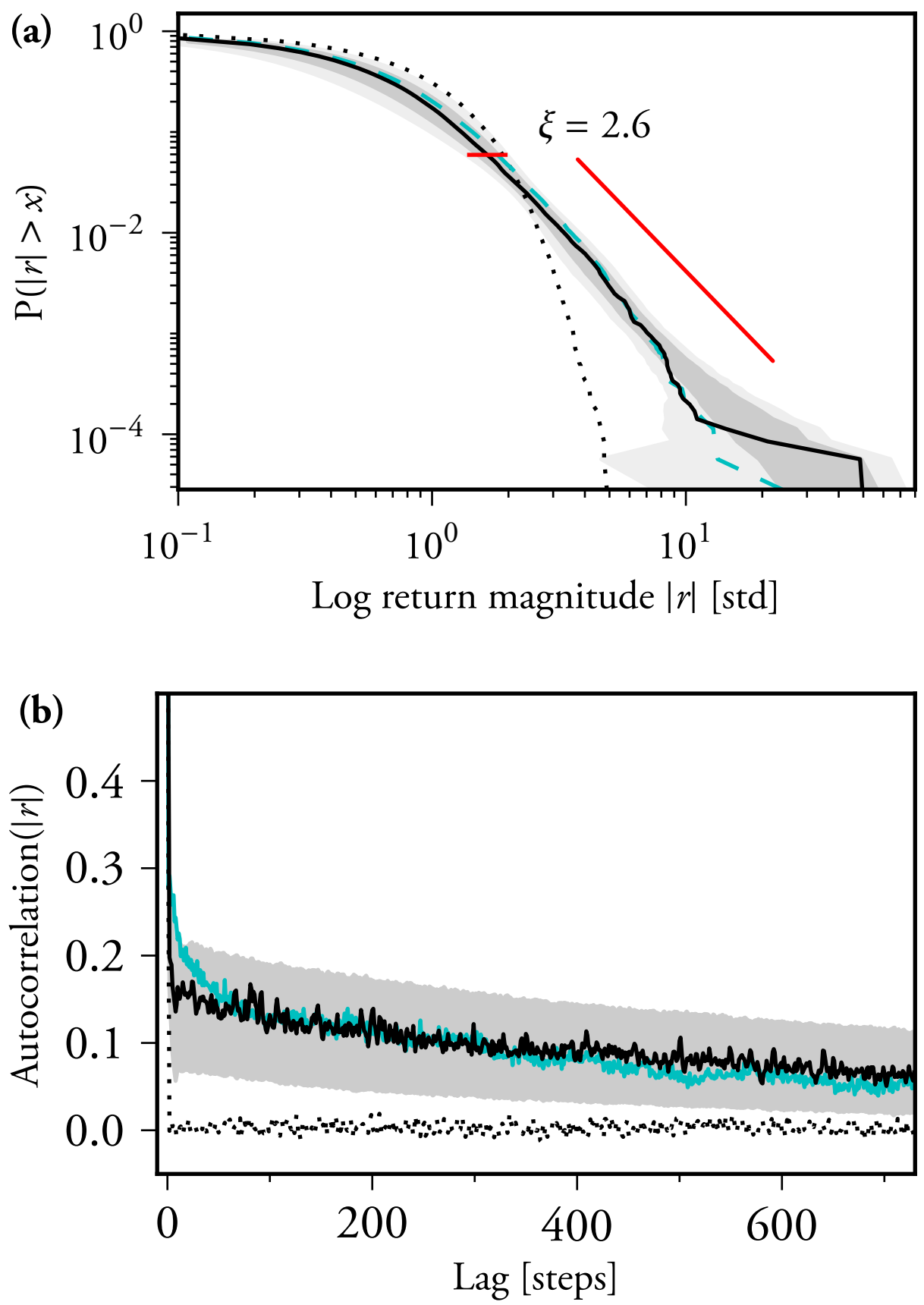

Figure 11.2.: Analysis of the log returns (methods as in sec. 10.3). **(a)** CCDF. Dotted black: The same simulation as in fig. 11.1(a). Solid black: the same simulation as in fig. 11.1(b). Dashed cyan: Daily returns for the DJIA. Short red line: power-law fitted to the simulation. For a fair comparison, only the last $3.5 \cdot 10^4$ of $10^7$ total time steps were analysed for the simulations to match the length of the DJIA time series. The dark and light grey areas cover one and two standard deviations, respectively, around the average of 100 simulations with the same parameters. **(b)**: Autocorrelations of the log return magnitudes $|r|$. Line styles are identical to (a). Here, both the example simulation and the DJIA stay within one standard deviation of the model average (dark grey area).

Return fluctuations in the endogenous case tend to form clusters in time. This effect is quantified by long-range temporal correlations of return magnitudes shown in figure 11.2 (b) and is also consistent with the DJIA. The only systematic difference is the initial decay of the correlations for lags of few days. This decay is slower for the DJIA than for the model, which might be due to the synchronous trading on each time step in the model.

## 11.4. Market efficiency as a form of collective learning

To understand the model dynamics, we first consider the exogenous case, which is fully analytically tractable. As we show in the following, the rules of asset redistribution by trading are equivalent to a learning rule related to gradient descent where $\gamma$ is a learning rate. Therefore, the market as a whole minimises predictable price changes. The reason for this stabilising control is that trading success increases the impact of agents whose actions contribute to a reduction of price fluctuations. Note that since agents represent trading strategies and not individual traders, $\gamma$ determines how fast relative impacts of different strategies are adjusted; it does not necessarily reflect how much of their resources actual traders would put at risk.

A phase transition with respect to the critical parameter $\alpha = D/N_s$ is identified at $\alpha = 1/2$, the point where random binary vectors (the agents) with positive weights (the assets) form a complete basis in the $D$-dimensional strategy space in the limit $N_s \to \infty$. Beyond this point, a speculative market without producers evolves the distribution of assets onto a manifold where the price is invariant to trading. That is, agents still trade and exchange assets, but the price remains constant. Markets that also include producers still exhibit finite returns for $\alpha < 1/2$. Otherwise, for $N_p \ll N_s$, return distributions only depend weakly on $N_p$ (see sec. B.1.1).

### 11.4.1. Invariant Manifold

Here we show that if one distribution of resources $(\overline{M},\overline{S}) = (\overline{M}_1, \ldots, \overline{M}_N, \overline{S}_1, \ldots, \overline{S}_N)$ exists for which the price $p(\overline{M},\overline{S},\mu) = \overline{p}$ is independent of the information $\mu$, this price is invariant with respect to any resource redistribution due to trading in a purely speculative market. That is, there is a manifold $Q = \{(\overline{M}',\overline{S}') \mid p(\overline{M}',\overline{S}',\mu) = \overline{p} \ \forall\ \mu\}$ of distributions of stocks for which the price is independent of $\mu$ and this manifold is closed with respect to trading according to equation (11.7). For the proof, assume that at some point in time the system is in a suitable state such that

$$\frac{\delta(\overline{M},\mu)}{\varsigma(\overline{S},\mu)} = \overline{p} \quad \forall\ \mu(t) \Leftrightarrow \tag{11.9}$$

$$\begin{aligned} \delta(\overline{M},\mu) - \overline{p}\,\varsigma(\overline{S},\mu) &= \\ \gamma \sum_{i=1}^{N} \left(\sigma_i^{\mu} \overline{M}_i - \overline{p}(1-\sigma_i^{\mu})\overline{S}_i\right) &= 0 \quad \forall\ \mu(t). \end{aligned} \tag{11.10}$$

Then, denoting the distributions of stocks and money after trading by $\overline{M}_i'$ and $\overline{S}_i'$ we obtain:

$$\begin{aligned} &\frac{1}{\gamma}\left(\delta(\overline{M}',\mu') - \overline{p}\,\varsigma(\overline{S}',\mu')\right) = \sum_{i=1}^{N} \sigma_i^{\mu'} \overline{M}_i' - \overline{p} \sum_{i=1}^{N} (1-\sigma_i^{\mu'})\overline{S}_i' \\ &= \sum_{i=1}^{N} \sigma_i^{\mu'} \left(\overline{M}_i - \gamma\sigma_i^{\mu}\overline{M}_i + \gamma\overline{p}(1-\sigma_i^{\mu})\overline{S}_i\right) \\ &\quad - \overline{p}\sum_{i=1}^{N}(1-\sigma_i^{\mu'})\left(\overline{S}_i - \gamma(1-\sigma_i^{\mu})\overline{S}_i + \frac{\gamma}{\overline{p}}\sigma_i^{\mu}\overline{M}_i\right) \\ &= \sum_{i=1}^{N}\left(\sigma_i^{\mu'}\overline{M}_i - \overline{p}(1-\sigma_i^{\mu'})\overline{S}_i\right) - \gamma\sum_{i=1}^{N}\left(\sigma_i^{\mu}\overline{M}_i - \overline{p}(1-\sigma_i^{\mu})\overline{S}_i\right) \\ &= 0 - 0 = 0 \qquad \square \end{aligned} \tag{11.11}$$

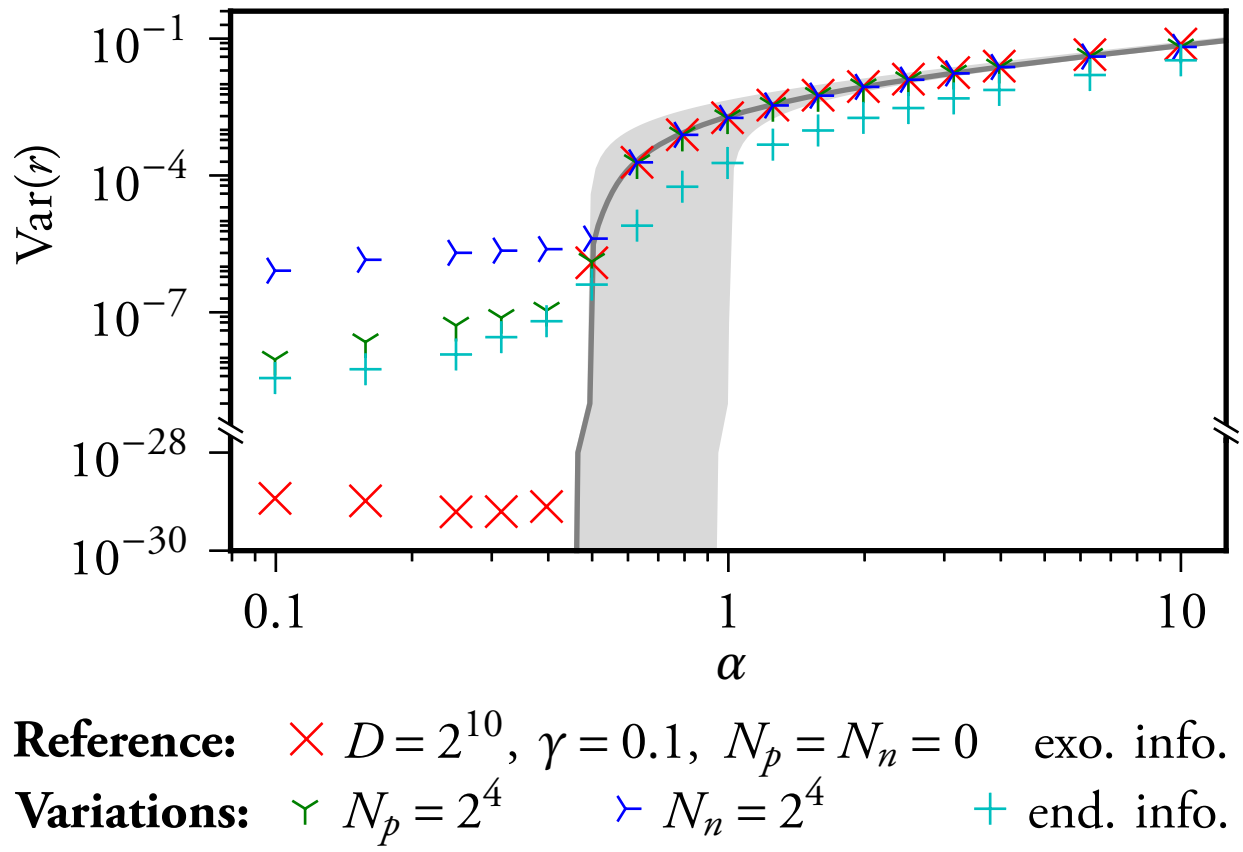

Figure 11.3.: Average log return variances for different values of $\alpha = D/N_s$. The model with exogenous information and only speculators serves as a reference. For comparison, simulations with a small numbers of either deterministic ($N_p = 2^4$) or random producers ($N_n = 2^4$, see main text) are shown as well as as the model with endogenous information. Grey area: Analytical upper and lower limit for exogenous information. Dark grey line: heuristic interpolation.

### 11.4.2. Completeness of the Strategies

As shown above, finding a resource distribution $(\overline{M}, \overline{S})$ for which the price is independent of the information is a sufficient condition for complete suppression of all price changes. That is,

$$p(\mu, \overline{M}, \overline{S}) \quad = \quad \overline{p} \;\forall\; \mu \tag{11.12}$$

which is equivalent to equation (11.10). To fulfil this criterion, we need enough agents to form a complete basis in the strategy space, which has $D$ dimensions. Then, the deviation from $\overline{p}$ caused by each agent can be cancelled by a superposition of the other agents for every $\mu$. This can be guaranteed if the number of speculators $N_s$ exceeds $2D$.

For an insufficient number of speculators, we can still calculate an upper and a lower bound for the variance of the log returns given $D$ and $N$ for a perfect superposition of speculators with exogenous information. Numerical and analytical results for this case are shown in figure 11.3. The mean variance is found to drop dramatically at $\alpha = D/N_s = 1/2$, with an increasingly sharp transition for large D (not shown). This phase transition can be understood by considering the probability that a random binary vector can be cancelled by an optimal superposition of $N-1$ random binary vectors with positive weights.

As an interim step, consider superpositions of random vectors with arbitrary weights. One such vector creates a one-dimensional subspace. Adding a second vector expands the dimensionality of the subspace to $d_2 = 2$ if it is linearly independent of the first one. Adding further vectors one by one, the probability that the $i$th vector does not lie in a $d_{i-1}$-dimensional submanifold is

$$P(d_i = d_{i-1} + 1) = 1 - 2^{d_{i-1} - D}. \tag{11.13}$$

We can therefore iteratively calculate the probability distribution $P(d_{N_s-1})$ of $d$ after adding $N_s - 1$ agents and the probability

$$P(d_{N_s} = d_{N_s-1} + 1) = \sum_{d=1}^{D} P(d = d_{N_s-1}) \; (1 - 2^{d-D}) \tag{11.14}$$

that one out of $N_s$ agents is linearly independent of the others. If a vector is linearly independent of the other agents in $d$ dimensions, it cannot be cancelled by a linear combination of the other agents for all $\mu$. However, it may still be possible to cancel this agent's impact for a subset of all possible $\mu$, i.e. for a smaller subspace. Therefore, the probability that an agent cannot be cancelled in any given time step is

$$P_{\text{c.c.}} = \sum_{d=1}^{D} P(d = d_{N_s-1}) \left(1 - 2^{d-D}\right) \left(1 - \frac{d}{D}\right). \tag{11.15}$$

The last term weights each summand with the fraction of dimensions in which the agent's impact is not cancelled. Finally, to relate the fraction of not-cancelled agents to log returns, we need to consider the fluctuations prior to any resource redistribution. Since all strategies and $\mu$ are chosen randomly, agents initially contribute to the demand or the supply at random. These fluctuations of demand and supply then follow a binomial distribution with $N_s$ trials and equal probability for buying or selling:

$$\delta \propto \mathscr{B}(N_s, 1/2) \tag{11.16}$$

$$\varsigma \propto \mathscr{B}(N_s, 1/2). \tag{11.17}$$

Since

$$\mathrm{E}(\delta) = \mathrm{E}(\varsigma) = N_s/2, \quad \text{and} \tag{11.18}$$

$$\mathrm{Var}(\delta) = \mathrm{Var}(\varsigma) = N_s/4, \tag{11.19}$$

we can approximate the price for small deviations:

$$p = \frac{N/2 + \Delta\delta}{N/2 - \Delta\varsigma} \approx 1 + 2\frac{\Delta\delta - \Delta\varsigma}{N_s}. \tag{11.20}$$

Therefore,

$$\mathrm{E}(p(0)) = 1, \tag{11.21}$$

$$\mathrm{Var}(p(0)) \approx 4/N_s, \tag{11.22}$$

and finally

$$\mathrm{E}(r(0)) = 0, \tag{11.23}$$

$$\mathrm{Var}(r(0)) \approx \frac{8}{N_s}. \tag{11.24}$$

Combining equations (11.15) and (11.24), we obtain the expected variance of the log return for an optimal superposition of agents without the positivity constraint on the resources

$$\mathrm{Var}(r) = \mathrm{Var}(r(0))\ P_{\text{c.c.}}. \tag{11.25}$$

Since resources cannot be negative, they form a positive cone. Each agent that is linearly independent of the others spans a half space. Therefore, $2N_s$ agents are necessary to completely span the strategy space. Yet for small numbers of agents, each agent still represents a full degree of freedom, since the probability that two agents lie on the same 1-dimensional submanifold is vanishingly small. However, as the number of agents is increased such that $\alpha \to 1$, an increasingly large number of new agents only converts a halfspace into a full one. Therefore, equation (11.25) represents a lower limit for the variance of the log returns, which is a good description for $N_s \ll D$. An upper limit is obtained by changing equation (11.15) such that each agent increases $d$ by $1/2$. This is a good approximation for $N_s \approx 2D$. The area in between these limits is shown in figure 11.3 (shaded grey). The lower limit has a phase transition at $\alpha = 1$ while the upper limit has a phase transition at $\alpha = 1/2$. A phase transition at $\alpha = 1$ is already present in equation (11.14). The gradual convergence for the true variance of the system from the lower to the upper limit is captured by a simple heuristic interpolation: for the dark grey line in figure 11.3, the probability for a new linearly independent agent to increase $d$ by one is $P_1 = \min(1, N_s/2^{m+1})$ while the probability to increase $d$ by $1/2$ is $P_{1/2} = 1 - p_1$. The presented theory describes the numerical results (figure 11.3) for the model with endogenous information very well for $\alpha \leq 1/2$. For full markets, the residual error for simulations with only speculators is determined by the numerical precision.

When producers are present, the residual error is noticeably higher. This is due to the fact that producers push the system off the invariant manifold. This error depends on the agents' use and vanishes for small $\gamma$. Still, predictable producers are cancelled much better than chance because speculators can successfully predict their choices. For comparison, in figure 11.3, results are shown also for $N_n$ modified "noisy" producers, who buy or sell randomly with equal probabilities.

For endogenous information ($D = 2^K$), the phase transition appears smoother and slightly shifted towards larger $\alpha$. A stronger reduction of average returns for $\alpha < 1/2$ occurs due to the more localised adaptation. That is, for prolonged periods of time, agents only have to adapt to subsets of all possible $\mu$. The phase transition is independent of $\gamma$ and the agents initial capitals, as shown in Figures 11.6 and B.4.

### 11.4.3. Gradient Descent

We now investigate how the system evolves towards the invariant manifold. We focus on large numbers of agents and small $\gamma$. The resource redistribution due to subsequently trading the two assets for one another is found to be a special case of a learning rule which minimises log return magnitudes. Even more generally, we consider the error function

$$e = r^2 \tag{11.26}$$

and show that its gradient

$$\frac{\partial e}{\partial X} = 2r\frac{\partial r}{\partial X}, \quad X \in \{M_1, \ldots, M_N, S_1, \ldots, S_N\} \tag{11.27}$$

with respect to the agents' resources is dominated by terms with the opposite sign as the change in the agents' resources. Therefore, any scaling of the agents' resources, which keeps the sign of the return for money and the opposite sign for stocks, corresponds to minimising log return magnitudes similar to a gradient descent.

To begin with, consider two subsequent time steps where the information takes the states denoted by $\mu$ and $\mu'$ respectively. We again examine a market consisting of speculators only. The derivative of the return with respect to the resources of an agent $k$ is

$$\frac{\partial r(M,S,\mu,\mu')}{\partial M_k} = \frac{\sigma_k^{\mu'}}{\delta'} - \frac{\sigma_k^{\mu}}{\delta} + O(\gamma) \tag{11.28}$$

$$\frac{\partial r(M,S,\mu,\mu')}{\partial S_k} = \frac{1-\sigma_k^{\mu}}{\varsigma} - \frac{1-\sigma_k^{\mu'}}{\varsigma'} + O(\gamma), \tag{11.29}$$

with

$$\delta = \delta(M,\mu), \qquad \delta' = \delta(M',\mu'), \tag{11.30}$$

$$\varsigma = \varsigma(S,\mu), \qquad \varsigma' = \varsigma(S',\mu'). \tag{11.31}$$

The change in resources after trading twice is

$$\Delta M_k = M_k'' - M_k \tag{11.32}$$

$$= \gamma\left(S_k\left((1-\sigma_k^{\mu})p + (1-\sigma_k^{\mu'})p'\right) - M_k(\sigma_k^{\mu'} + \sigma_k^{\mu})\right) + O(\gamma^2) \tag{11.33}$$

$$\Delta S_k = S_k'' - S_k \tag{11.34}$$

$$= \gamma\left(M_k\left(\frac{\sigma_k^{\mu'}}{p'} + \frac{\sigma_k^{\mu}}{p}\right) - S_k(2-\sigma_k^{\mu} - \sigma_k^{\mu'})\right) + O(\gamma^2). \tag{11.35}$$

We are interested in

$$\Delta r_k = \left(\Delta M_k \frac{\partial r}{\partial M_k} + \Delta S_k \frac{\partial r}{\partial S_k}\right) \tag{11.36}$$

and continue only with leading terms in $\gamma$.

For now, we also assume that agents can only perform round-trip trades (indicated by "RT" in equations). The general case will be discussed later. This means that agents buy in one step and sell in the next or vice versa:

Case $\sigma_k^{\mu}(1-\sigma_k^{\mu'}) = 1$:

$$\frac{\Delta r_k^{RT}}{\gamma} \overset{\gamma\ll 1}{\approx} \frac{M_k - p'S_k}{\delta} - \frac{M_k/p - S_k}{\varsigma'} \tag{11.37}$$

$$= \frac{M_k}{\delta}\left(1 - \frac{\varsigma}{\varsigma'}\right) + \frac{S_k}{\varsigma'}\left(1 - \frac{\delta'}{\delta}\right) \tag{11.38}$$

Case $\sigma_k^{\mu'}(1-\sigma_k^{\mu}) = 1$:

$$\frac{\Delta r_k^{RT}}{\gamma} \overset{\gamma\ll 1}{\approx} \frac{pS_k - M_k}{\delta'} + \frac{M_k/p' - S_k}{\varsigma} \tag{11.39}$$

$$= \frac{M_k}{\delta'}\left(\frac{\varsigma'}{\varsigma} - 1\right) + \frac{S_k}{\varsigma}\left(\frac{\delta}{\delta'} - 1\right). \tag{11.40}$$

Above, we used $p = \delta / \varsigma$ and $p' = \delta' / \varsigma'$. Then,

$$\frac{1}{\gamma} \sum_{k=1}^{N} \Delta r_k^{RT} = \frac{\varsigma'}{\varsigma} - \frac{\varsigma}{\varsigma'} + \frac{\delta}{\delta'} - \frac{\delta'}{\delta} \tag{11.41}$$

$$= \frac{\varsigma'}{\varsigma}(1 - \frac{p'}{p}) + \frac{\varsigma}{\varsigma'}(\frac{p}{p'} - 1) \begin{cases} < 0, & r > 0 \\ > 0, & r < 0\,. \end{cases} \tag{11.42}$$

Therefore, the change in the total error function

$$\sum_{k=1}^{N} \left( \Delta M_k \frac{\partial r^2}{\partial M_k} + \Delta S_k \frac{\partial r^2}{\partial S_k} \right) \overset{\text{RT}}{\leq} 0 \tag{11.43}$$

can never be positive if agents only perform round-trip trades.

On average, this result holds even for the general case. The reason we have to consider averages is that those agents who buy or sell two times in a row always decrease the amount of money or stocks they own after two time steps. These agents' resources are therefore expected to change in the opposite direction of the gradient half of the time. That is, for every given pair of informations $(\mu, \mu')$, a quarter of all agents is expected to have their resources evolve such that they contribute to a future increase in $r(\mu, \mu')^2$. For large systems, however, the actual influence of these agents is negligible. This is shown in the next paragraph.

Demand and supply are well described by binomial processes for sufficiently large systems, as shown above. Here, we express demand and supply as:

$$\delta = \frac{N}{2} + \Delta\delta, \qquad \mathrm{E}(\Delta\delta) = 0, \qquad \mathrm{E}(\Delta\delta^2) \leq \frac{N}{4} \tag{11.44}$$

$$\varsigma = \frac{N}{2} + \Delta\varsigma, \qquad \mathrm{E}(\Delta\varsigma) = 0, \qquad \mathrm{E}(\Delta\varsigma^2) \leq \frac{N}{4}. \tag{11.45}$$

The relative fluctuations around the mean demand $N/2$ are only $\sqrt{N}/2$ and hence small for large N. We can therefore expand equations (11.28) and (11.29) for small fluctuations:

$$\frac{\partial r(M,S,\mu,\mu')}{\partial M_k} \approx \sigma_k^{\mu'}\left(\frac{2}{N}-\frac{4\Delta\delta'}{N^2}\right)-\sigma_k^{\mu}\left(\frac{2}{N}-\frac{4\Delta\delta}{N^2}\right) \tag{11.46}$$

$$\frac{\partial r(M,S,\mu,\mu')}{\partial S_k} \approx (1-\sigma_k^{\mu})\left(\frac{2}{N}-\frac{4\Delta\varsigma}{N^2}\right)-(1-\sigma_k^{\mu'})\left(\frac{2}{N}-\frac{4\Delta\varsigma'}{N^2}\right). \tag{11.47}$$

As the above equations show, agents who perform round-trip trades contribute a term of order $N^{-1}$ to the gradient with respect to each asset. When agents buy or sell twice, they only contribute a term of order $N^{-1.5}$ for one asset. Therefore, the influence of these agents vanishes for sufficiently large $N$. By a similar argument, approximately a quarter of all agents performs either one of the actions (buy, sell), (sell, buy), (buy, buy), and (sell, sell) and fluctuations can be neglected for large $N$. As we have just shown, the impacts for the last two actions vanish for $N \to \infty$. Even in finite systems, as discussed above, these actions contribute towards increasing future returns only with probability $1/2$ and towards decreasing returns otherwise.

In conclusion, the expected change in $r(\mu,\mu')^2$ over repeated trades with the same information

$$\mathrm{E}\left(\sum_{k=1}^{N}\left(\Delta M_k \frac{\partial r^2}{\partial M_k}+\Delta S_k \frac{\partial r^2}{\partial S_k}\right)\right)_{(\mu,\mu')} \leq 0. \tag{11.48}$$

is always negative given a sufficiently large number of agents.

## 11.5. Information annihilation instability (part two)

The above results establish that the market as a whole tends to attenuate its responses (i.e. the magnitudes of the returns) to information states that are presented in random order, Independent and Identically Distributed. Because this is straightforwardly related to learning via the adaptation of the agents' possessions, we term this effect "information annihilation": a repeatedly presented information state no longer carries information about exploitable price changes. The relation to Shannon information is discussed in section 11.7.

When the $\mu(t)$ are endogenously generated, the same mechanism of information absorption present in the exogenous case ensures that the system relaxes towards local price equilibria and returns vanish, but only transiently. To illustrate this basic principle, figure 11.4 (a) shows the price time series of a simulation with a very small use $\gamma$. At any point in time, the system moves towards a certain price which characterises a local equilibrium. As the system approaches this equilibrium, price fluctuations are reduced. These fluctuations generally consist of complex oscillations like the one shown in figure 11.4 (b). The equilibria become unstable once all predictable information is exploited by the speculators. Then, even little overshooting of the adaptation dynamics or noise can lead to price changes corresponding to information states that were not predicted by patterns in the immediate past. Because the market is not well adapted to these new states, the possibility of large price changes increases dramatically. Compared to [PP11], we here discovered an instability due to information annihilation in a mathematically different way, which demonstrates that this concept is even more general.

For larger $\gamma$, this behaviour is not as obvious. This is shown in figures 11.5 (a) and (b). Time series appear random and distinct oscillations are rarely visually recognisable. Still, phase diagrams from extensive simulations demonstrate that return distributions are largely unaffected by these effects over wide ranges of $\gamma$. This is shown in the next section.[8] [9]

[8] In modified models which include a noise source, even price time series with small $\gamma$ appear much more similar to fig. 11.5 than fig. 11.4. That is, for example, models with some agents who choose their action randomly, or models with mixed information as it is defined in sec. B.1.2 (not shown).

[9] Volatility clustering, however, is influenced by $\gamma$, as shown in fig. 11.9.

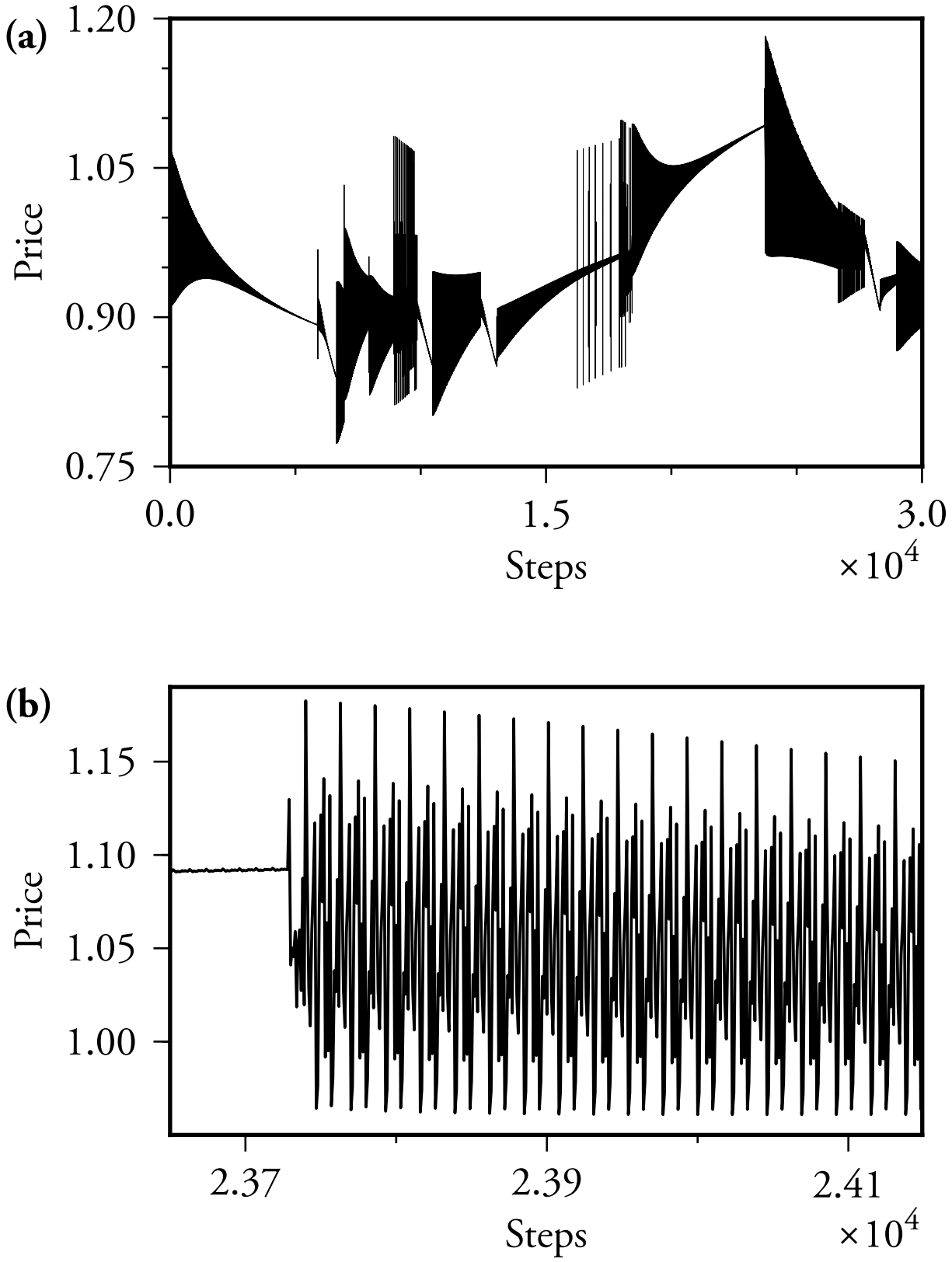

Figure 11.4.: **(a)**: Price time series with a very slow rate of resource redistribution (use) $\gamma = 0.01$. Other parameters: $N = 2^{10}$, $N_p = 2^4$, $D = 2^9$.
**(b)**: A zoom in on the time series shown in (a).

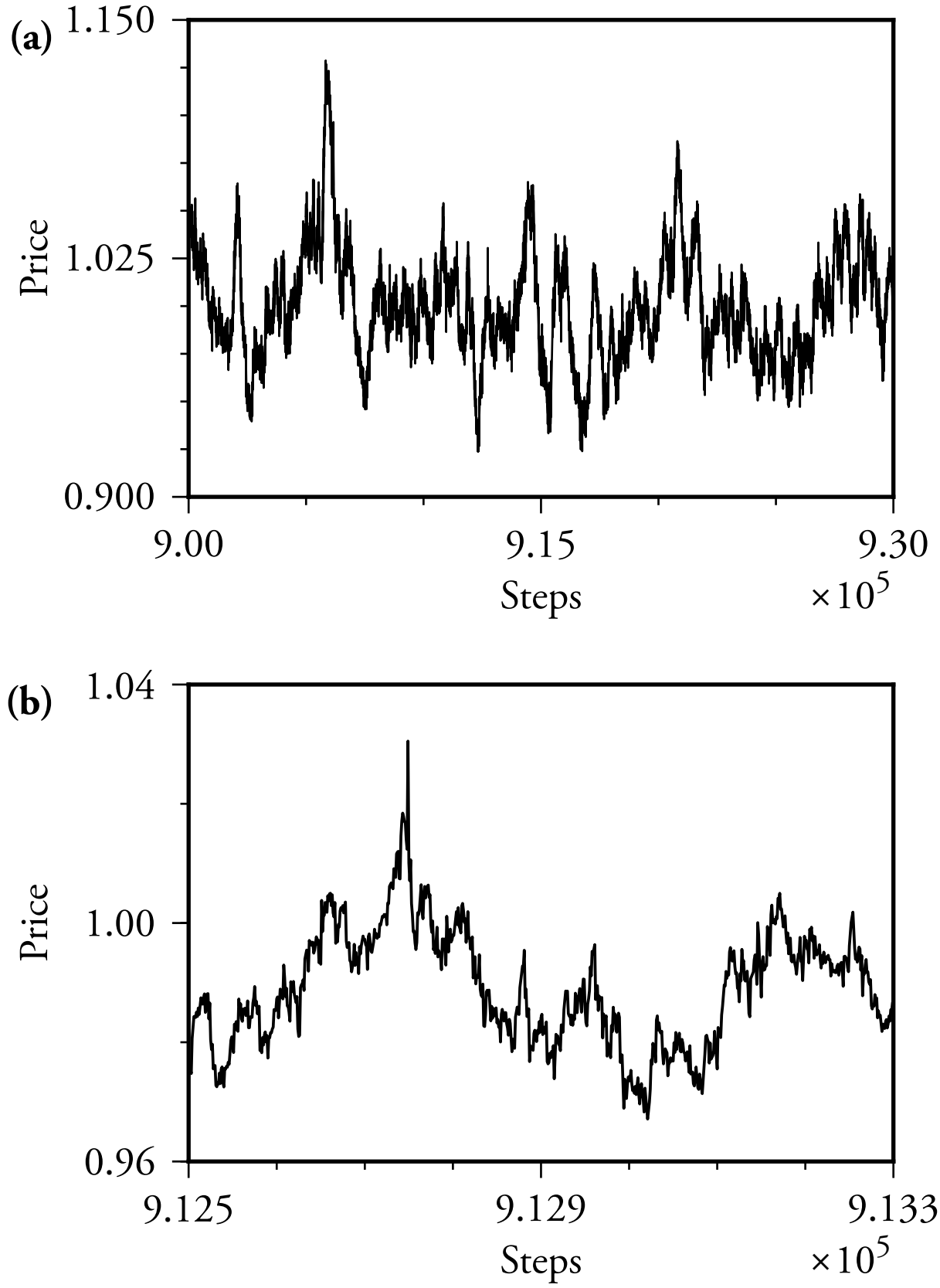

Figure 11.5.: **(a)**: Price time series with $\gamma = 0.8$, other parameters equivalent to fig. 11.4 (a). **(b)**: A zoom in on the time series shown in (a).

## 11.6. Phase space

For exogenous information, figure 11.6 (a) shows the ratio of initial and final mean log return magnitudes in simulations for different $\alpha$ and $\gamma$. This measure of the attenuation of average fluctuations during transients is invariant to the absolute amount of fluctuations.[10] Results reflect the phase transition identified above.[11]

Figure 11.6 (b) shows the impacts that infrequent extreme returns have on the remaining variances, which are measured by kurtoses after transients. Return distributions for exogenous information are close to Gaussians[12], except for $\gamma$ slightly below one (this is explained later, see footnote 14).

For endogenous information, the return attenuation is shown in figure 11.6 (c). The results are similar to the exogenous case (see also fig. 11.3). The kurtoses, however, show a very different behaviour. For endogenous information, the stronger the reduction of return magnitudes (fig. 11.6 (c)), the heavier tailed the return distributions are (fig. 11.6 (d)). This establishes a clear link between local information annihilation and extreme returns in our model for the whole parameter space. The phase transition, however, is distinguished by a maximum in the speculator income (sec. B.2). Perhaps not coincidentally, the model best reproduces real data in the same parameter range.[13]

[10] The latter depends on the absolute number of agents, as discussed above.

[11] Values below one correspond to an increase in fluctuations during transients. This is observed only for markets with few speculators and a large use (upper right hand corner in figs. 11.6 (a) and (b)).

[12] The kurtosis for a Gaussian distribution is three, see sec. 2.1

[13] As one might expect from fig. 11.6(d), the shape of the log return distributions depends on $\alpha$. PDFs are close to exponential for empty markets. Around the phase transition, power laws are observed. For overcomplete markets, results depend on the presence of producers. Without them, prices in the over-complete phase are constant (as proven above). Producers create a dense band of baseline activity above which the heavy tailed fluctuations due to the speculators arises. For very small alpha, return fluctuations visually resemble more of a sausage with toothpicks in it. The activity of the producers also reduces long-range

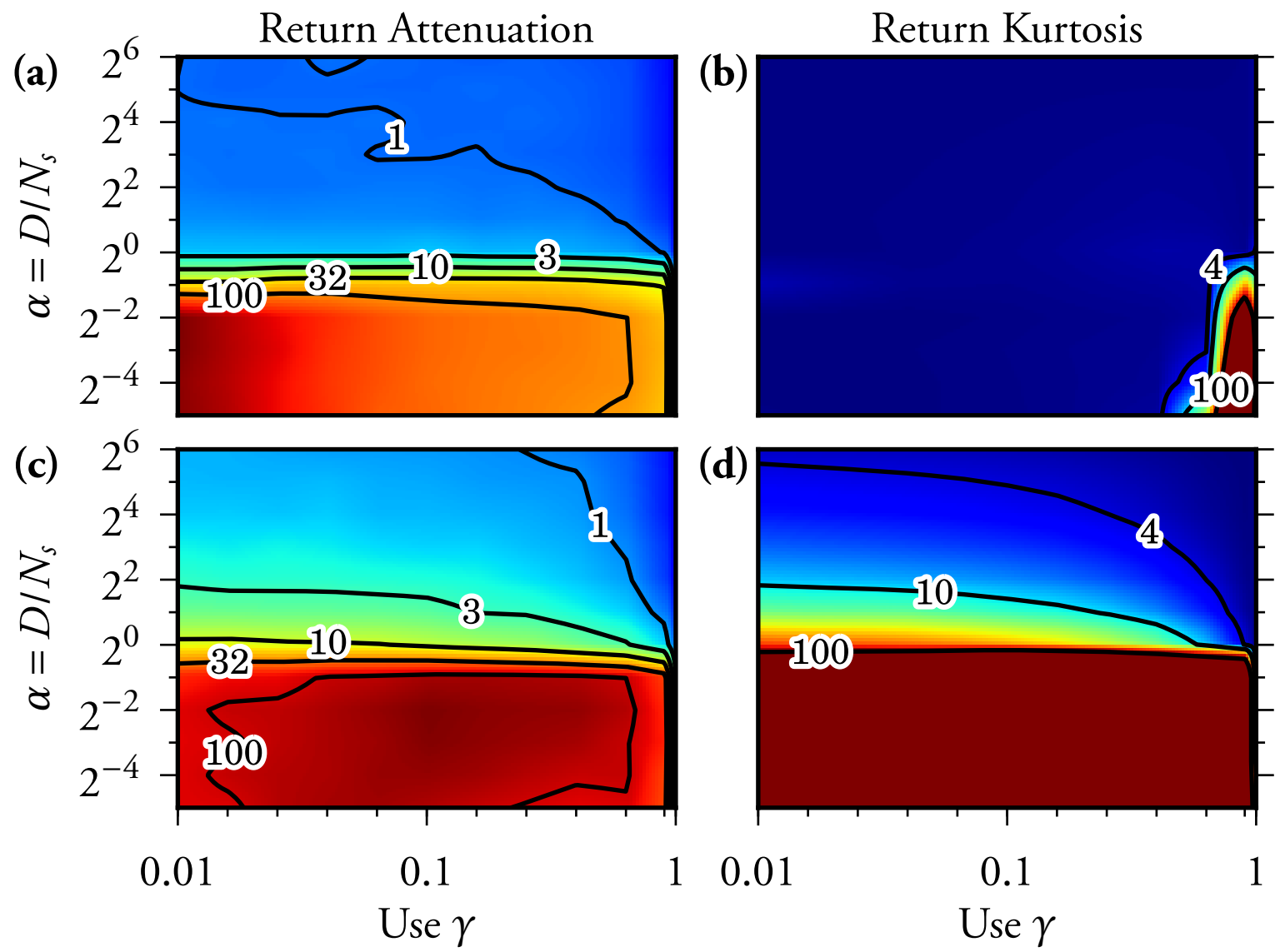

Figure 11.6.: left column: attenuation of average log return magnitudes during transients for different $\alpha$ vs $\gamma$. Right column: kurtoses for the same simulations. **(a)**, **(b)**: exogenous information with $P_{\mathrm{ext}}(\mu) = 1/D$. **(c)**, **(d)**: endogenous information according to eq. (11.1). The system size for all simulations was constant at $N_s = 2^{10}$ speculators and $N_p = 2^4$ producers. For each time series, $2 \cdot 10^7$ time steps were simulated. Reductions are measured as the ratio between the mean log return magnitudes during the first 10 and the last $10^7$ time steps. The kurtoses were calculated for the last $10^7$ time steps. Simulations were performed on a grid. All axis ticks correspond to node positions. For each node on the grid, 50 time series were analysed and the results were averaged. Linear interpolation and colour mapping were performed after logarithmising the values at each node; contour line labels are the actual values for attenuations and kurtoses. For $\alpha \leq 1/2$, the kurtoses reach extreme values that can not be reliably estimated from finite time series. Therefore, the colour scale in (b) was set to not extend to values above 100.

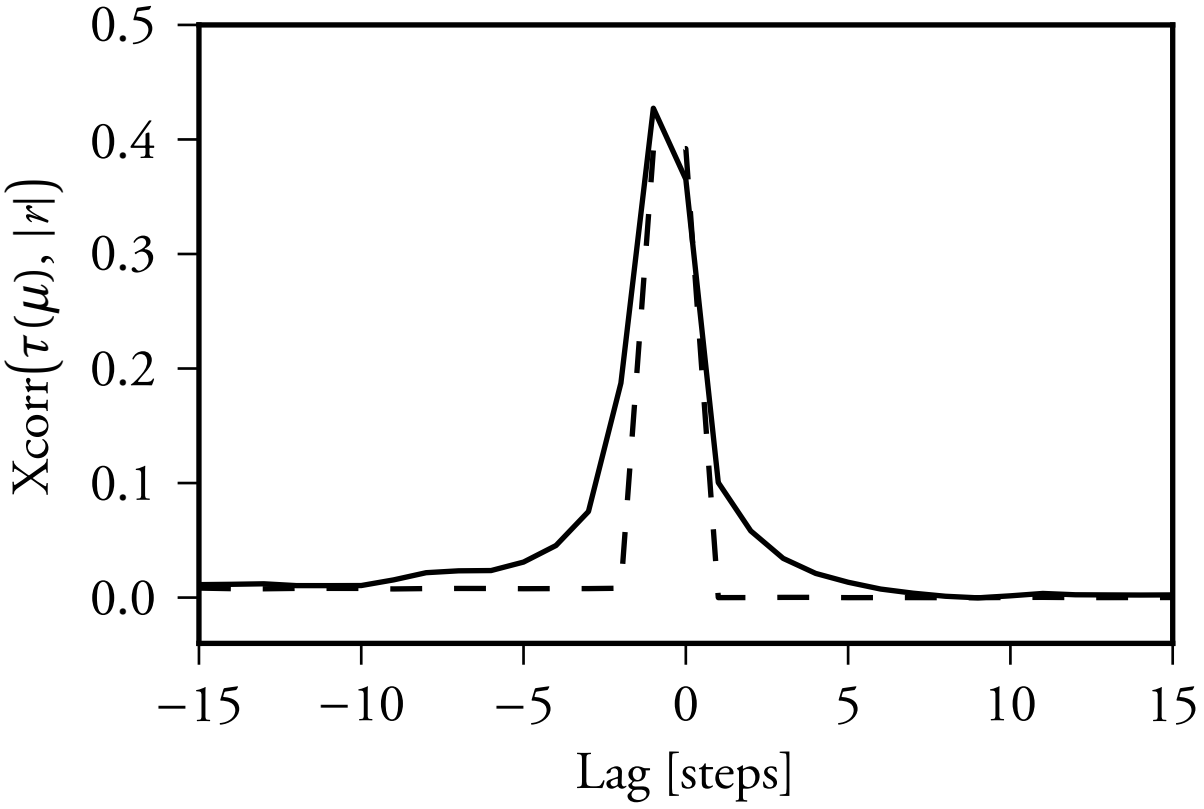

Figure 11.7.: Correlation of return magnitudes $|r|$ with the time $\tau$ since the corresponding information states occurred last. Model parameters: $D = 2^{10}$, $N_s = 2^{11}$, $N_p = 0$, $\gamma = 0.5$. Simulation length: $T = 2 \cdot 10^7$; the first $T/2$ time steps were discarded for the analysis. Black line: Endogenous information. Dashed line: Exogenous information with $P_{\text{exo}}(\mu) \propto \exp(-0.02\,\mu)$, leading to $P(\tau) \propto \tau^{-2}$. Both lines are averages over 10 simulations.

## 11.7. Returns encode surprise

The log return magnitudes can be quantitatively related to the novelty of the corresponding information states. Intuitively, states which did not occur in a long time are more surprising and therefore carry more information than the ones visited more recently – a concept that is closely related to Shannon information. We indeed find that large returns are caused by information states that have not occurred for a long time: the more surprising an information state is, the higher the corresponding log return. The correlation between log return magnitudes and the times $\tau$ that have passed since the respective information states occurred last is shown in figure 11.7 for endogenous information (solid line).Here, the absorption of local information in combination correlations. Over time, however, speculators become richer and therefore slowly regain influence. See also sec. B.2.

with rare jumps leads to a strongly inhomogeneous distribution of visiting frequencies over the information set: the probability distribution $P(\tau)$ is power-law tailed with an exponent of approximately 2.5 (see sec. B.1.3).

This suggests that the self-reflexive dynamics for endogenous information generates a characteristic distribution of information states that ultimately underlies extreme price fluctuations. We tested this hypothesis by using inhomogenously distributed exogenous information states that lead to similarly distributed $\tau$. Then, as in the endogenous case, return magnitudes are strongly correlated with $\tau$ (figure 11.7, dashed line).[14]

## 11.8. Relation to minority games

When an agent trades one asset for another, the total value of their endowment has not changed. Only round-trip trades, where the assets are traded again at a later point in time, permanently change the wealth of an agent. At the very least, when evaluating the apparent desirability of a trade, the prices at two points in time at which the round trip could have been completed have to be considered. Buying shares of a stock, for example, might be considered a good decision if the price rises later on.[15]

As we have shown in section 11.4.3, round-trip trades over two subsequent time-steps dominate the dynamics in the trading model. Therefore, a minority rule with respect to the return (of investment) $R = \exp(r)$ can induce dynamics that minimise return fluctuations.

---

[14]For very large $\gamma$, even uniformly distributed $\mu$ are occasionally not repeated for a sufficiently long time to be "surprise" the market (see fig. 11.6 (b))

[15]Note, however, that the profit has not been realised until the shares have actually been sold at a higher price. An agent who keeps on buying assets in a market with an upwards trend, for example, appears to be successful according to the minority rule if it is applied to a short time interval. If the price suddenly falls, however, the very same agent can loose everything. Therefore, the minority rule holds only on the time scale of a round-trip trade.

### 11.8.1. The MG approximation to trading

Consider a model where the information $\mu$, strategy vectors $\sigma_i$, and prices $p$ are defined as in section 11.2, but only the total capital $C_i(t)$ of an agent is kept track of. Each time-step, each agent invests a fraction $\gamma$ of their capital:

$$c_i(t) = \gamma\ C_i(t)\ \left(2\sigma_i^{\mu(t)} - 1\right) \tag{11.49}$$

where we interpret positive $c_i$ as buying and negative ones as selling an asset. Thus, demand and supply follow as:

$$\begin{aligned} \delta(t) &= \sum_{i=1}^{N} c_i(t)\,\sigma_i^{\mu(t)}, & (11.50)\\ \varsigma(t) &= \sum_{i=1}^{N} c_i(t)\,(1-\sigma_i^{\mu(t)}). & (11.51) \end{aligned}$$

Success, just like in the trading model, depends on the ratio

$$R(t+1) = \frac{p(t+1)}{p(t)}, \tag{11.52}$$

except for the simplification that only prices in direct succession are considered. The trading model further conserves the total amount of money and stocks individually, but not the combined value of all assets measured in units of one asset. Therefore, when approximating trading as an MG, the change in capital for speculators $k = 1, \ldots, N_s$ follows as

$$C_k(t+1) = C_k(t) + |c_k(t)|\left(\alpha(t)\,R(t+1)^{1-2\sigma_i^{\mu(t)}} - 1\right). \tag{11.53}$$

were the normalisation

$$\alpha(t) = \frac{\delta(t) + \varsigma(t)}{\delta(t)/R(t+1) + \varsigma(t)\,R(t+1)} \tag{11.54}$$

ensures that each agent's profit is in each time step is proportional to the return, like in the trading model, and that trading also conserves the total capital in the market.[16]

The MG described above exhibits similarities and differences when compared to the trading model. Figure 11.8 shows the phase transition in $\alpha$ for different $\gamma$. In contrast to the trading model, the over-complete phase for the MG is absorbing even when producers are present. This finding can be understood intuitively. As shown in section 11.4.1, trading in the absorbing phase without any producers evolves the system onto an invariant manifold. On this manifold, both assets are redistributed in each round due to trading, yet the price remains constant. When producers are present, they push the system away from this manifold. With just one asset, constant prices only require one distribution of capitals for which the price is independent of the information state. The manifold in the trading-like MG is reduced to a point in asset space. The impact of noisy producers, however, still cannot be cancelled (not shown).

Another difference from the trading model is that the MG rule requires lower "learning rates" $\gamma$ to work effectively. This is reflected in the difference in return attenuation between figure 11.6 (c) and figure 11.8 (a) for large $\gamma$. Furthermore, the MG only shows significant excess kurtoses close to the phase transition. The log return distributions on the critical point also follow power laws only for use parameters up to approximately $\gamma = 0.2$.

Yet another difference is that the MG shows almost no long-range volatility clustering. The autocorrelations of log return magnitudes for both models are shown in figure 11.9.

### 11.8.2. Comparison of different MGs

In the MG approximation to trading, an agent $i$ wins if $\sigma_i^{\mu(t)} = \Theta(-r(t+1))$, and loses otherwise. This is clearly a minority rule based on the change in price. Since we defined the latter as the ratio of demand and supply, this minority rule is also based on the change in demand and supply, and therefore on the change in agents' choices. In contrast, the common minority rule (sec. 10.8.2) is based only on the outcome at a single point in time. The same is true for the returns in these models. Therefore, prices changes in common MGs happen even if no agents change their actions. Hence, to minimise price changes in classical MGs, the sum over all actions (which all have the same unit) has to be zero. There is no notion of a price at which the market clears like in the trading model, or as in economic theories. This characteristic appears quite peculiar considering the argument that the minority rule is supposed reward buying low and selling high (see sec. 10.8). Nevertheless, this type of equilibrium can be learned by adjusting agent capitals according to the minority rule as well.

[16]Without normalisation, the total capital in the game diverges quite rapidly.

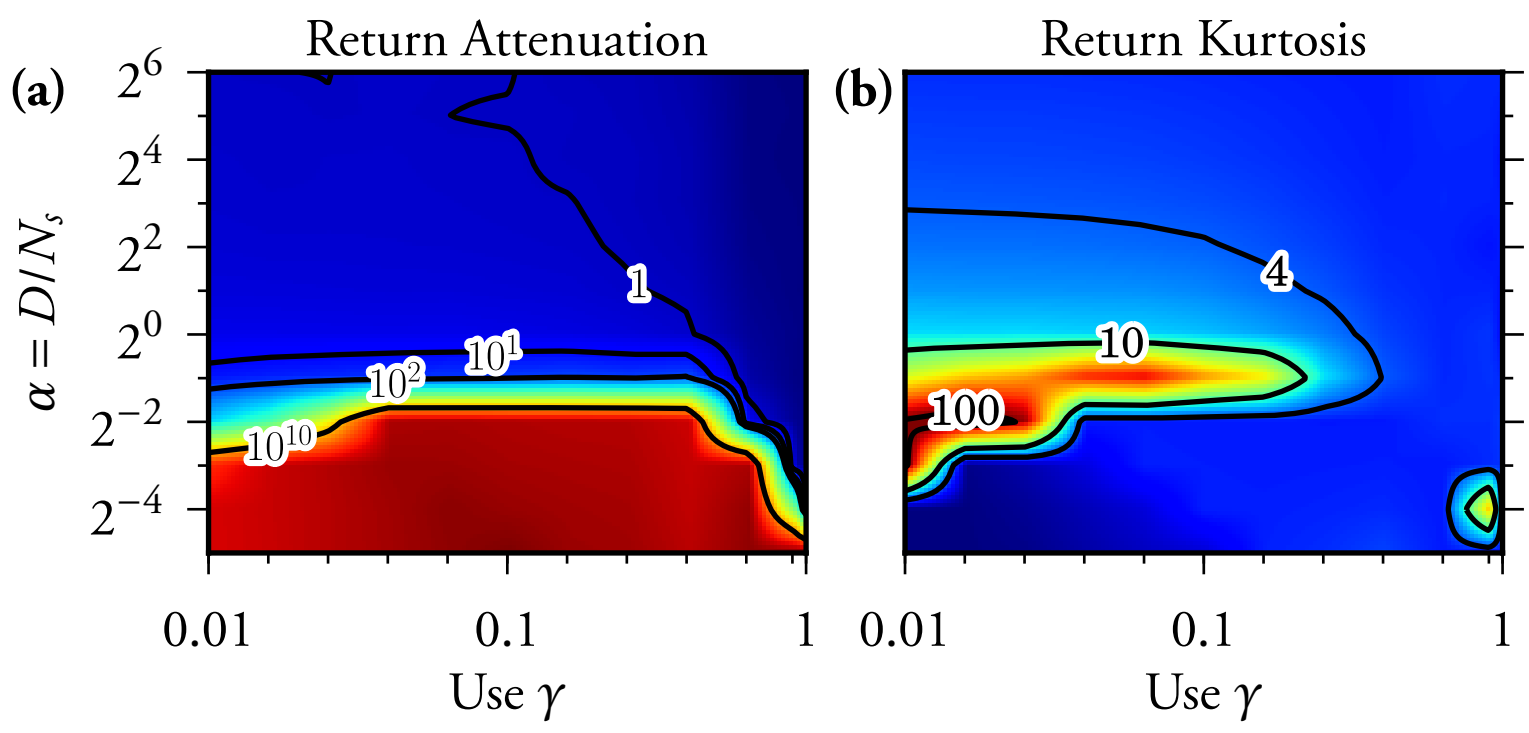

Figure 11.8.: Analysis of log return fluctuations for the MG approximation to the trading model. Parameters: $N_s = 2^{10}$, $N_p = 2^4$, intrinsic information. Figure generation like in figure 11.6. **(a)**: log return attenuation during transients. The over-complete phase is absorbing within the numerical limitations: log return variances after transients are below $10^{-28}$. **(b)**: log return kurtoses after transients.

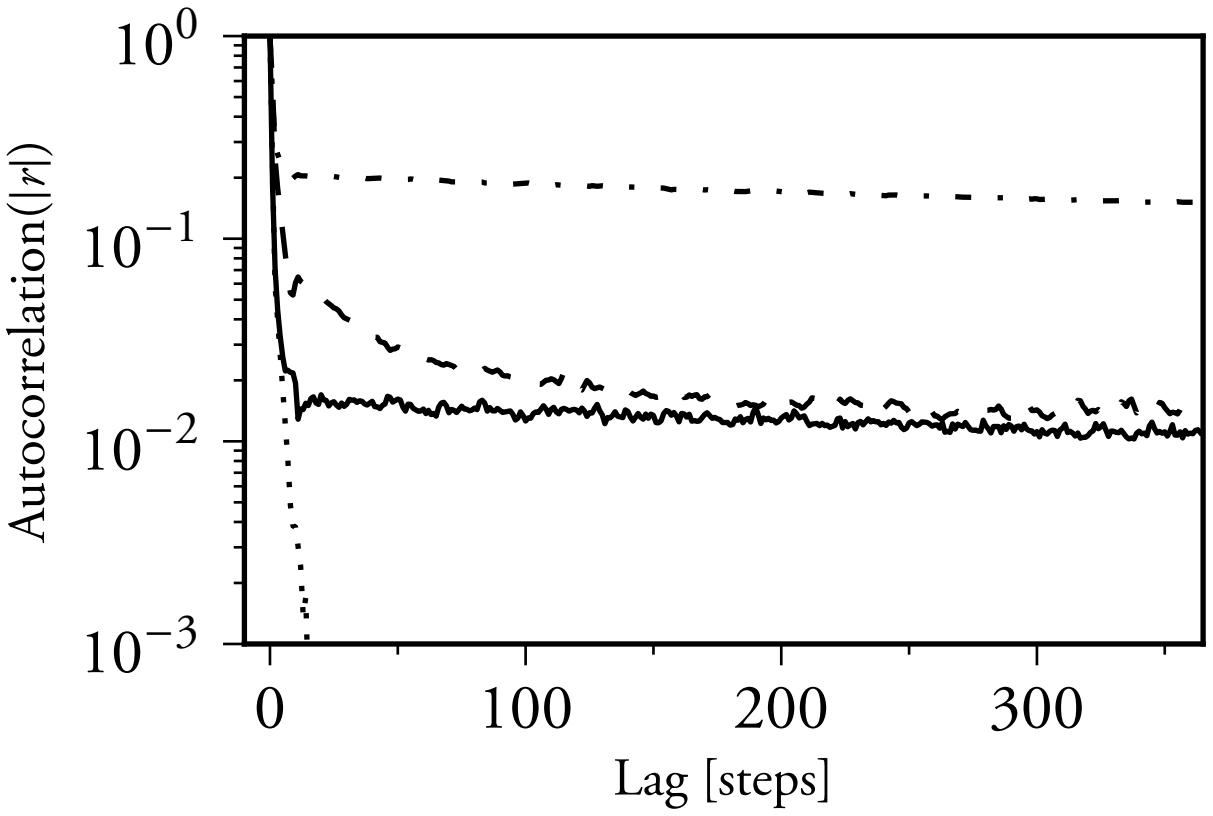

Figure 11.9.: Average autocorrelations of log return magnitudes for different models at the critical point ($D = 2^9$, $N_s = 2^{10}$, $N_p = 0$). The trading model for $\gamma = 0.2$ (dashed) exhibits long-range correlations which are nearly absent for the MG approximation (solid line). For $\gamma = 0.85$, the trading model exhibits even more long-range correlations (dash-dotted). The MG in this case, however, exhibits no long-range correlations at all (dotted). Averages were calculated over 50 repeated siulations.

Figure 11.10 shows the dynamics for a price calculated from the log returns for the minority game with dynamic capital discussed in sec. 10.8.2, but with only one strategy. The model has a strong tendency to get stuck in local attractors (this hardly happens in the trading model even without noise). This problem can be resolved with a few noisy producers. Then, however, the total capital in the market vanishes over time due to the lack of asset conservation in this model. Therefore, a small noise term in the information update rule is a much better solution. Nevertheless, the accordingly modified model exhibits information annihilation similar to the trading model (figure 11.4). Adaptation is, however, slower than in the trading model. High $\gamma$ are not possible in this MG since they lead to negative capital. Furthermore, prices over time can drift to arbitrary values since they have no meaning in this model. In particular, many simulations (except in the absorbing phase) show perpetual trends due to small imbalances in the initialisation of the agent strategies.

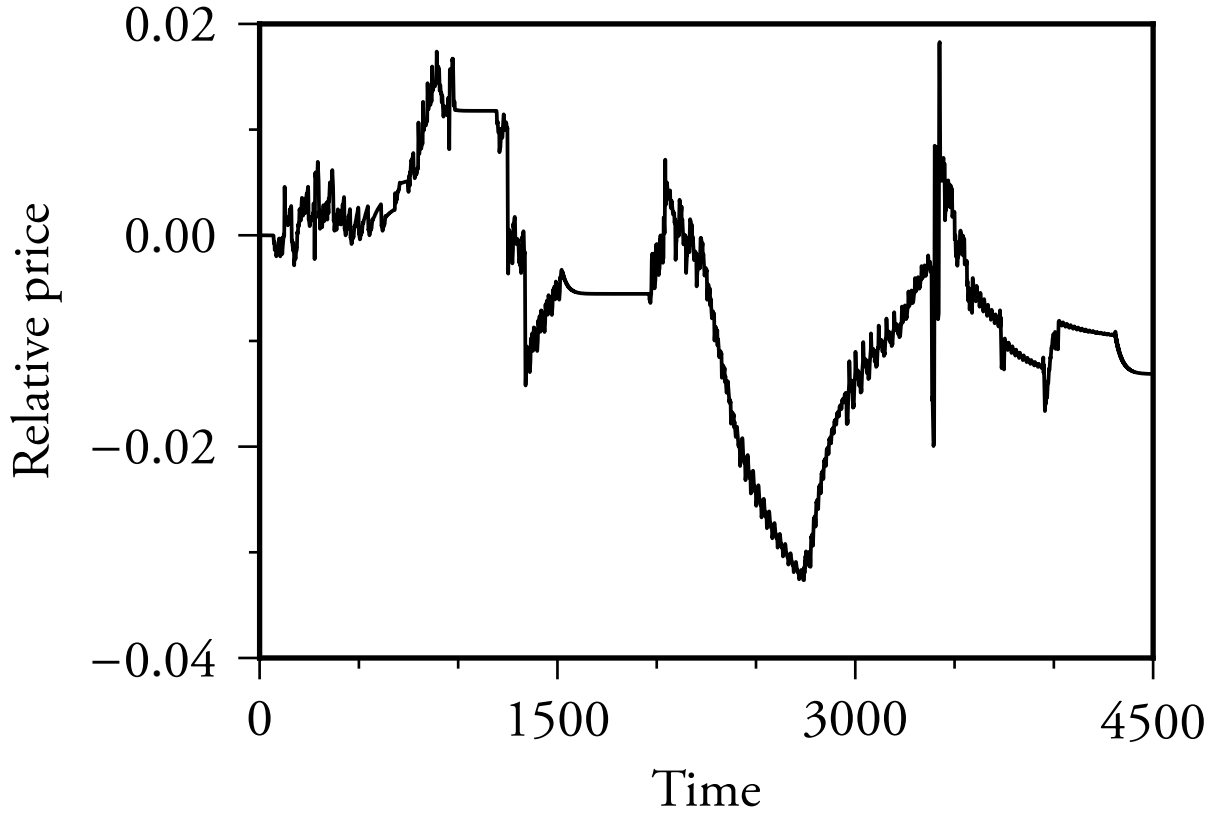

Figure 11.10.: MG with dynamic capital like in sec. 10.8.2, but with only one strategy, and information generation according to eq. (11.1). Other parameters: $D = 2^9$, $N_s = 2^{10}$, $N_p = 2^4$, $\gamma = 0.1$. This model does not contain an explicit price. However, a relative logarithmic price is obtained from the cumulative sum of the log returns. The result is shown above for a short extract from the model time series after transients.

The insights gained in this chapter further allow for a more intuitive understanding of models with dynamic capital and strategy switching like the one in sec. 10.8.2. In this case, unsuccessful agents will change their strategy. The strategies, however, are the very basis functions whose weights are learned by the market such that information is absorbed. Changing the basis functions forces the whole system to re-learn. Therefore, fluctuations in such systems, even in the over-complete phase, never vanish completely. Average fluctuations are smaller for larger $\gamma$ because the system re-learns faster (see fig. 10.6 (c)).

In the next chapter, we investigate the behavior of actual human subjects in the a MG paradigm where success depends on the change in collective behavior.

# 12. Bubbles and jumps from properly anticipated prices

> “And now for something completely different”
>
> Monty Python

A major obstacle to testing economic theories is that markets–let alone economies–are intractable to strict scientific experiments. The rules of a commercial exchange cannot be easily manipulated, and the ongoing activities depend on abundant uncontrolled variables. Laboratory experiments are a viable alternative. This method might be seen as analogous to the Petri dish in biology. Such an experimental model allows for studying fundamental elements of the system of interest, yet greatly reduces uncontrolled interactions. There is a risk that the elements in this model environment behave differently than outside of the laboratory. If, however, the necessary interactions that give rise to effects observed in the full system can also be included in the model, these interactions become much easier to study. Thereby, experiments can also help determine which elemental properties should be included in computational models.

Today, behavioural experiments are an established method in (non-orthodox) economics (see secs. 10.4 and 10.5). Such experiments, however, are limited to many fewer participants than real markets. This restriction may appear to preclude the study of large-scale collective phenomena in laboratory experiments.

In the following, we study the behaviour of real human subjects in experimental toy models of speculative markets. In particular, we investigate whether subjects exhibit some form of efficiency, and which information determines their behaviour. An intuitive visualisation of MG-like interactions is presented, which greatly facilitates the subjects' understanding of the game dynamics after little training. A modified minority rule based on the change in collective behaviour (see sec. 11.8) is found to lead to both coordination and bubbles. Then a multi-agent model is introduced where, in contrast to the models discussed so far, the agents mimic the behaviour of real subjects. This analytically tractable model exposes a systematic link between price efficiency, bubbles, and crashes. It persist in large systems, where log returns obtained from the pricing rule introduced in section 11.2 exhibit the notorious stylised facts (sec. 10.3). Finally, we briefly study how the effects discussed in this chapter relate to the trading model of chapter 11.

## 12.1. The seesaw game(s)

Economic group experiments commonly involve continuous double auctions, which robustly exhibit bubbles and crashes (sec. 10.5). These experiments, however, typically require substantial amounts of time for each trading period. Therefore, it appears, little emphasis was put on studying the statistics of the resulting price fluctuations. Simpler models may help to facilitate the collection of data on more trading periods. The MG might seem to suggest itself as such a minimal experimental paradigm, but the literature on this topic appears to be limited to few studies [CMZ05b]. In these experiments, subjects were found to coordinate better than chance, leading to reduced fluctuations. Their ability to exploit complex patterns as defined in equation (10.10), however, appears to be limited to $K \approx 3$. There seem to be no reports of bubbles, extreme price changes, or phase transitions in MG experiments.

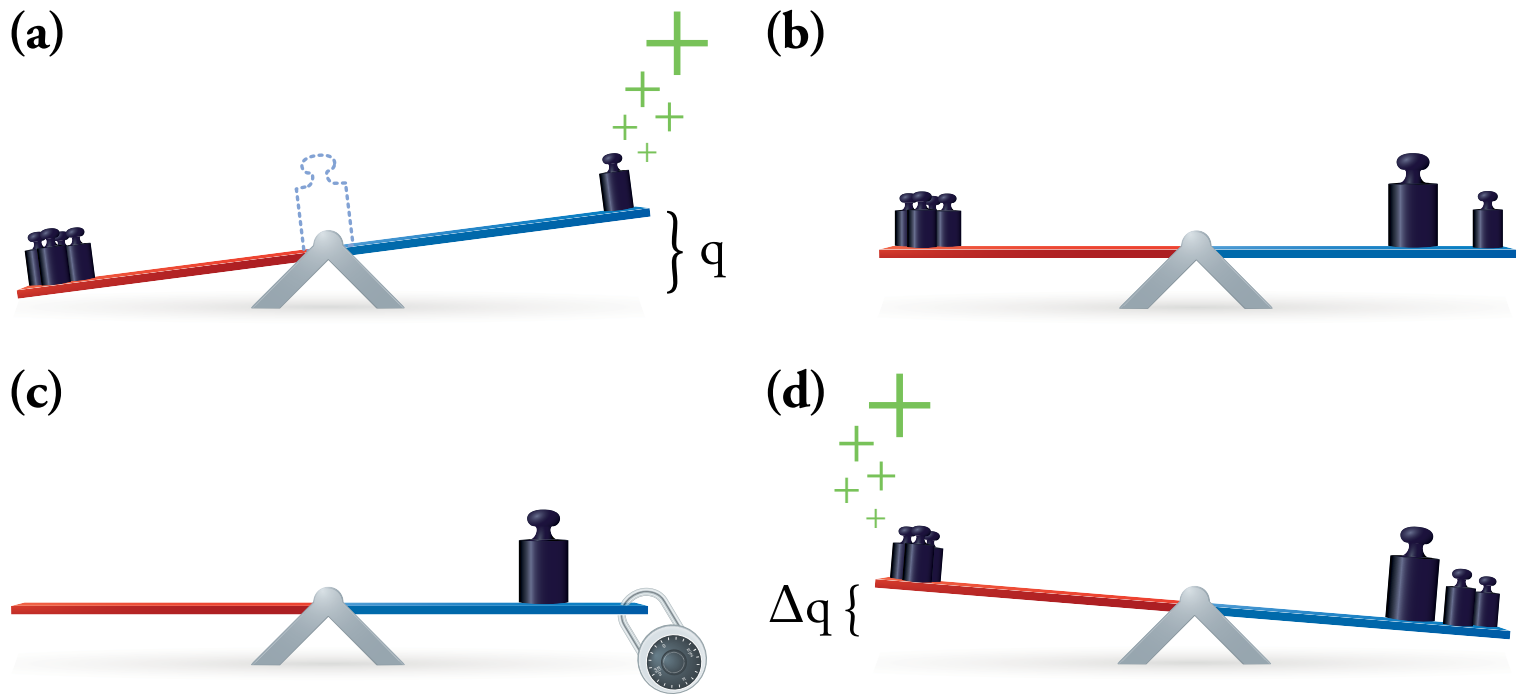

Figure 12.1.: The seesaw game (linear version). **(a)**: A Minority Game is visualised by a seesaw upon which agents place their bets. The winning side goes up. With only the agents' bets on the seesaw, the angle would indicate the excess demand $q$. **(b)**: As a modification, however, we balance the seesaw after each game turn (i.e. trading period). **(d)**: The seesaw is locked until the bets for the next turn have been placed. The balancing weight remains in place. **(c)**: The minority rule, therefore, is applied to changes in excess demand. After evaluating the outcome, the seesaw is balanced and locked again for the next turn (not shown).

Here we investigate the dynamics in a modified minority game. It was argued in section 11.8 that success in short-term speculation should be based on changes in the collective behaviour and not on the momentary state at just one point in time. Selling shares of a stock, for example, is only profitable if the stock was bought at a lower price. Figure 12.1 shows an intuitive visualisation of this rule. Note that the game generates outcomes at discrete time steps, but several stages lead up to each outcome. We therefore call the set of stages necessary for a new outcome a (game) turn. At the end of each turn the results for a new time step are appended to the respective time series for the outcome, actions, etc.

Agents in each turn place weights on either side of a seesaw.[1] Those agents win who put their weights on the side which goes up. In the first time step, this is the side with fewer total weights. Before each subsequent time step, however, the seesaw is balanced by an additional weight. This weight contributes to the next outcome. Therefore, the seesaw game implements a minority game with respect to changes in collective behaviour. If we interpret the agents' weights on one side of the seesaw as demand and the others as supply, the outcome $\Delta q$ indicated by the seesaw angle corresponds to the change in excess demand $q$.[2]

One interesting property of the seesaw game is that every outcome is a Nash equilibrium. A single agent who changes sides loses. Therefore, one might expect the seesaw to remain stationary after exactly one time step.[3] This, however, never happened during many experiments with different variations of the seesaw game. Nevertheless, one pathological case is possible. If all agents bet on one side of the seesaw and the balancing weight mirrors this maximal excess demand in the next turn, any agents changing their decisions will lose definitely. The only way to resolve this deadlock without an extra rule is for some agents to not play in the next step. Alternatively, to enforce a resolution, the turn may be repeated or the balancing weight may be moved to a random position.

[1] To be precise, subjects submit choices during a turn that determine which action they will perform at the end of that turn. No agents knows about the choices of the other agents before the turn is completed.

[2] One might argue that this implementation is technically very similar to a set of scales. A seesaw, however, is much more fun. We furthermore assume a small reverting force (as well as friction) such that the equilibrium angle of the seesaw is proportional to the imbalance of the weights on the seesaw. This is the case when the seesaw is, e.g., suspended below its pivot point or connected to a spring. The latter is the case on many playgrounds.

[3] This might be one of the reasons why no one tried this paradigm before. The MG was initially supposed to be a game without an equilibrium (sec 10.8).

## 12.2. A SIMPLE LINEAR SEESAW GAME

In the most simple instantiation of the seesaw game, agents $i$ at each time step $t$ chose an action $a_i(t) = 1$ or $a_i(t) = -1$ before a countdown runs out. No decision is registered as $a_i(t) = 0$. The weight that balances the excess demand $q(t) = \sum_i a_i(t)$ after each time step is moved according to

$$W(t) \begin{cases} = -q(t-1) & \text{for} \quad |q(t-1)| < N \\ \sim \mathcal{U}_\mathrm{d}(-n, n) & \text{for} \quad |q(t-1)| = N, \end{cases} \tag{12.1}$$

where deadlocks $|q| = N$ are resolved by moving $W$ to a random position drawn from the discrete uniform distribution. This case, however, did not occur in the trials analysed below. The outcome is

$$A(t) = \sum_i a_i(t) + W(t). \tag{12.2}$$

The agent payoff at the end of each time step is

$$g_i(t) = \Theta(-a_i(t)A(t)). \tag{12.3}$$

That is, the agents on the minority side win 1 point.[4]

We performed a series of experiments with voluntary naive subjects recruited in the Center for Cognitive Sciences Bremen.[5] Each subject played the game on a separate laptop, where the current outcome, all subjects' scores, and all actions were displayed. The history of the game was displayed for the past 20 time steps. For more details, see chap. 13. All subjects were seated in a large room and sufficiently far apart from each other to impede cheating. The game, including the subjects' scores, was also projected onto a large screen in front of the subjects.

[4] If the payoff rule were symmetric, such that the majority would lose one point, the average payoff would be negative. Most agents would then, after some time, hold negative scores. During early tests, this situation turned out to be highly discouraging for the subjects, who would stop playing to avoid further losses.

[5] Subjects received no payment but snacks were provided, as well as small prizes for the top-scoring subjects.

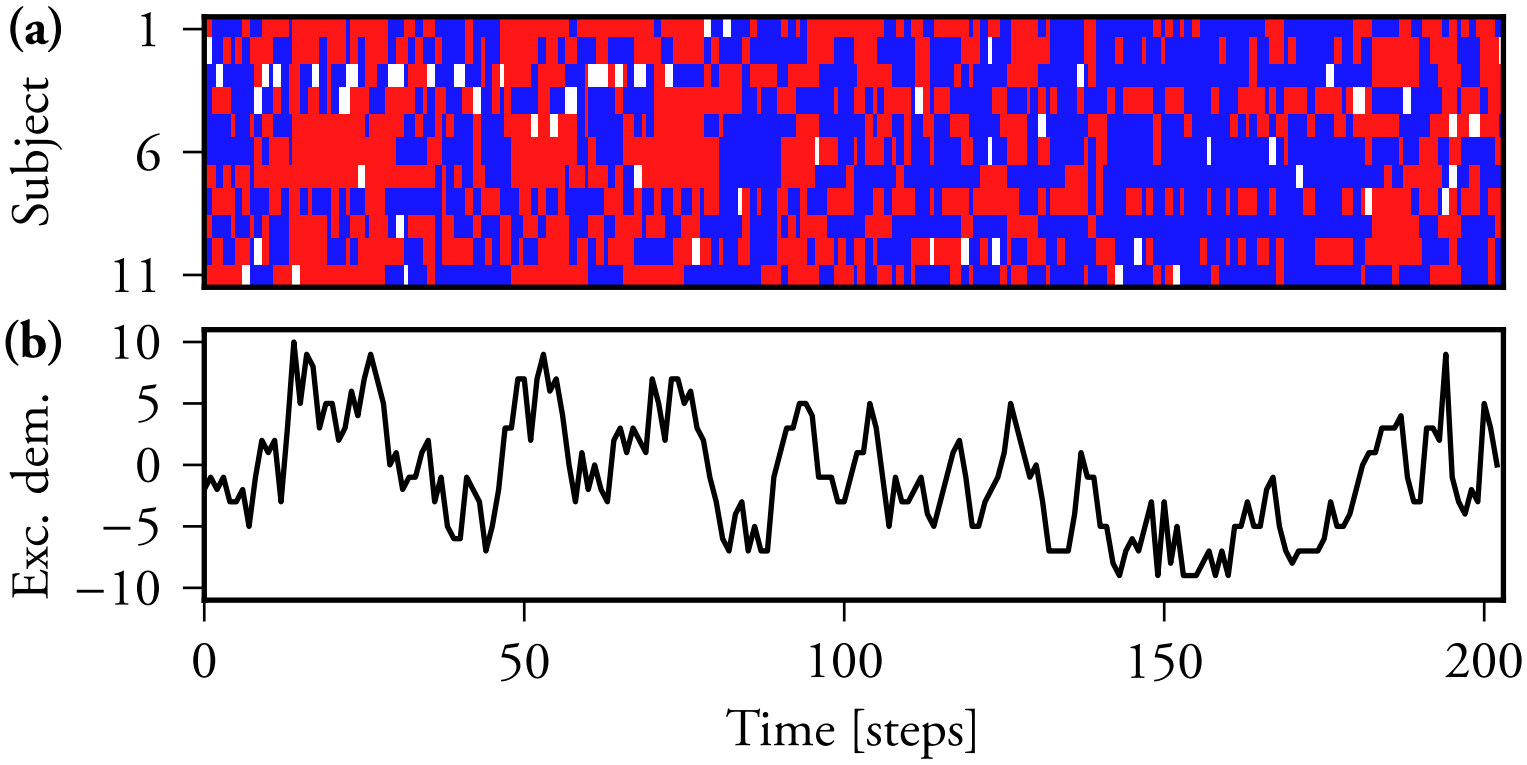

Figure 12.2.: Dynamics of the simple seesaw game with $N = 11$ subjects. **(a)**: Actions. Red and blue correspond to $a_i = \pm 1$, white to $a_i = 0$. **(b)**: Excess demand $q = \sum_i a_i(t)$.

The results for the largest group are shown in figure 12.2. Most of the time, the subjects seem to favour one side of the seesaw over the other. Several sudden changes in preference are observed. No deadlock occurred in this group. If the subjects had picked either side with equal probabilities in each turn, large excess demands would be less likely. Similar results were found for smaller groups (not shown). Result statistics are discussed in more detail in section 12.3. First, however, a model is introduced to explain the observed bubble-like herding behaviour.

## 12.3. Modelling the subjects' behaviour

If all agents flipped coins to determine their actions, the excess demand would be unpredictable. Its change, in contrast, would be easy to predict. If the expected excess demand was always zero, large demands would typically be followed by smaller ones, and vice versa. Profit seeking agents should bet against this mean reversion. In an efficient game, in analogy to equation (10.4), the expected new excess demand for each time step should be equal to its predecessor:

$$\mathrm{E}\big(q(t+1)\,|\,q(t), q(t-1), \ldots\big) \stackrel{!}{=} q(t). \tag{12.4}$$

The simplest way to fulfil this condition is to adjust the probability for the agents' actions in each time step accordingly. For such stochastic agents, demands are binomially distributed. Hence, equation (12.4) is fulfilled for each time $t+1$ if each agent chooses $a_i(t+1) = 1$ with probability $1/2 + q(t)/2N$, and $a_i(t+1) = 0$ otherwise.

For comparison with the experiments, a few extensions are of interest. First, subjects don't participate in each time step. This is captured by the probability to participate $\lambda = P(a_i \neq 0)$ (experiment: $\lambda = 0.955 \pm 0.007$). Furthermore, we introduce a parameter $o \in [0,1]$ which interpolates between coin-flipping agents ($o = 0$), and perfectly efficient ones ($o = 1$). We thereby obtain a measure for the efficiency of real subjects. In conclusion, the probabilities for all possible actions are

$$\begin{aligned} P\big(a_i(t+1) = 0\,|\,W(t+1)\big) &= 1 - \lambda \\ P\big(a_i(t+1) = 1\,|\,W(t+1)\big) &= \inf\left\{\lambda, \sup\left\{0, \frac{\lambda}{2} - o\frac{W(t+1)}{2N}\right\}\right\} \\ P\big(a_i(t+1) = -1\,|\,W(t+1)\big) &= \lambda - P\big(a_i(t) = 1|W(t+1)\big). \end{aligned} \tag{12.5}$$

All other rules are identical to the experiment (see above).

Actions for real subjects, agents who flip coins if they participate, and efficient agents are shown in figure 12.3. The experiment and the efficient model show clusters where one choice is preferred. These bubble phases are absent for coin flipping agents. The fluctuations for the subjects, the two model extremes, as well as a marginally inefficient case with $o = 0.8$ are analysed quantitatively in figure 12.4. Efficient betting decreases in the outcome variance (fig. 12.4 (a)) and removes mean reversion (fig. 12.4 (b)), but increases the probability of a bubble phase (fig. 12.4 (c)). The latter two statistics for the subjects are very close to the marginally inefficient model. The variance for the subjects is a bit closer to the perfectly efficient model. The coin-flipping model can, however, be rejected with high significance. Note that the real subjects are heterogeneous (figure 12.3 (a)) and may use other information which is not captured by the simple model. Nevertheless, subjects are very well described by stochastic agents that are almost efficient with $o$ close to one.

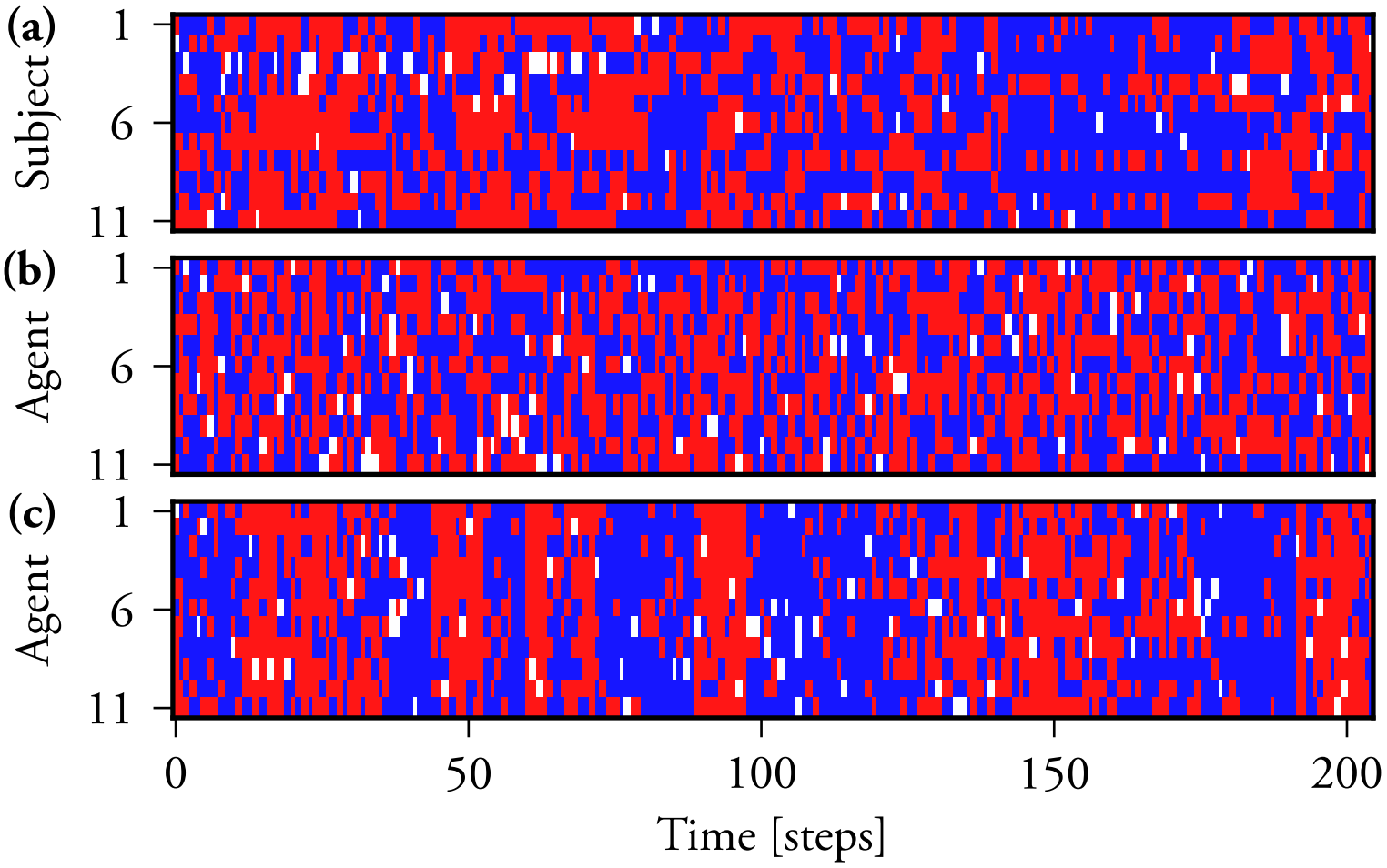

Figure 12.3.: Dynamics of a linear seesaw game for subjects and the two extreme cases of the model. Red and blue correspond to the actions $a_i = \pm 1$, white to not participating a time step. **(a)**: An experiment with 11 Subjects. **(c)**: model with equal probabilities for $a_i \pm 1$ (i.e. $o = 0$). **(c)**: demand-efficient model ($o = 1$). Agent participation in the models was equal to the experiments.

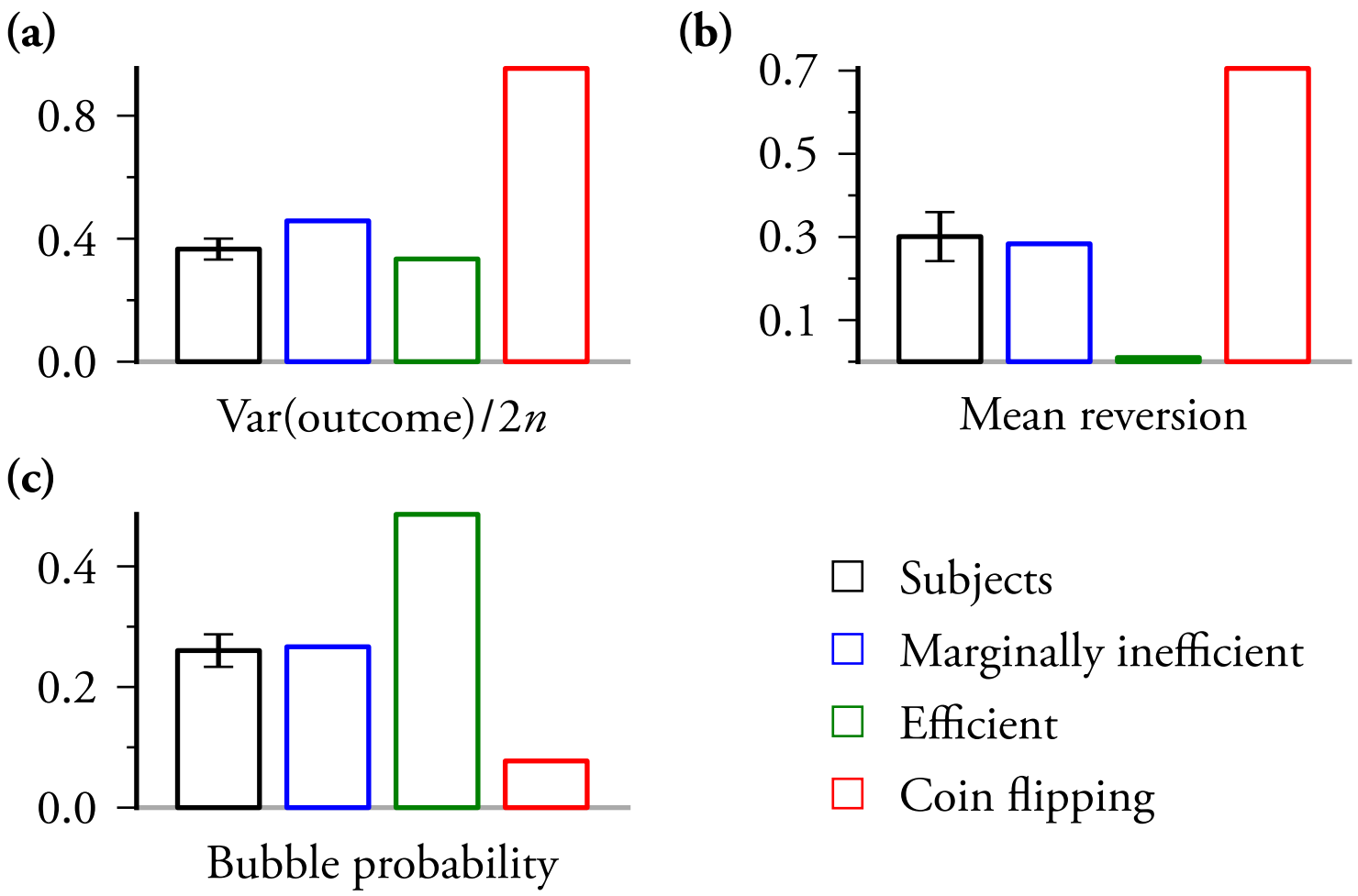

Figure 12.4.: Analysis and comparison of the choices in a linear seesaw game experiment with 11 subjects, and the model (eq. (12.5)) for three different efficiency parameters. Marginally inefficient: $o = 0.8$. Efficient: $o = 1$. Coin flipping: $o = 0$. The number of agents and participation are equal to subjects and simulations. For the subjects, bars represent the different sample statistics for three combined trials. Error bars represent the respective sample errors. For the simulations, the bars correspond to averages over 1000 simulations. No error bars are shown, since standard errors here are very close to the line width. **(a)**: Normalised variance of the outcomes $\mathrm{Var}(A)/2N$. The factor $1/2$ reflects that outcomes range from $-N$ to $N$. As in the OMG, the normalised variance for coin-flipping agents is one. **(b)**: the Pearson correlation coefficient $\rho\left(-A(t), \sum_i a_i(t-1)\right)$ of the outcomes and their preceding excess demands measures mean reversion. The sign of the outcome is inverted such that a positive mean reversion corresponds to a predictable drift towards the origin. **(c)**: the probability that one choice was made by at least twice as many subjects (agents) as the other choice.

If the system size is increased, the bubbles in the efficient model become even more extreme (not shown). The time scale of the dynamics increases with the system size, too. In the next section, we will reformulate the model to allow for a comparison with real markets. A pricing rule will be introduced to enable the study of log returns. Thereafter, we will compare price- and demand efficiency, and study the latter case analytically. Note that equation (12.5) does not depend on the agents' payoff at all. Therefore, a normative efficient model can be formulated independently of the minority rule (or any other payoff). Further simplifications are possible without changing the fundamental dynamics of the model. First, the possibility for agents to not participate can be dropped. Finally, including a very small number inefficient agents, who generate mean reversion only for the most extreme outcomes, removes the need for a special deadlock rule.

## 12.4. A minimal model for bubbles in efficient markets

Consider a market with $N$ agents: $N_s$ speculators and $N_n$ noise traders. At discrete times $t$ each agent places a market order to either buy or sell one unit of an asset (e.g. a stock). Thereby, agents contribute to either the demand $\delta(t)$ or to the supply $\varsigma(t) = N - \delta(t)$. We further require: 1. Increasing $\delta(t)$ increases the price while increasing $\varsigma(t)$ decreases the price. 2. Scaling all orders by the same factor yields the same market clearing price. Therefore, like in chapter 11, the price follows from the ratio of demand and supply

$$p(t) = \frac{\delta(t)}{\varsigma(t)} = \frac{\delta(t)}{N - \delta(t)}, \tag{12.6}$$

which naturally possesses the correct unit.

Agents make their decisions stochastically like in section 12.3, but we here postulate price efficiency: the probability for a speculator to buy at each time $t$ is chosen such that the expectation value of the new price given all previous observations

$$\mathrm{E}(p_t \,|\, p_{t-1}, p_{t-2}, \ldots) \stackrel{!}{=} p_{t-1} \tag{12.7}$$

is the same as the previous price. The correct buying probabilities can be obtained by numerical optimisation or in analytically closed form for large systems (see next section). This condition may be violated only if $\delta(t-1) > N_s + N_n/2$ or if $\delta(t-1) < N_n/2$. In these cases, it is impossible to be price efficient due to the discretisation of the traded assets. For $N_s \gg N_n$ and $N_s \gg 1$, however, we consider this boundary effect acceptable.

A model time series is shown in figure 12.5 (a). Large returns coincide with bubbles (fig. 12.5 (b)), or anti-bubbles (not shown). The distribution of log returns is power-law tailed (fig. 12.5 (c)). The exponent in the cumulative distribution approaches $\xi = 2$ for large systems, which is proved below. Finite size effects or large $N_n/N_s$ increase $\xi$ and may also truncate the log-return tails (not shown). Log returns are nearly uncorrelated[6] while their magnitudes are correlated for long periods of time (fig. 12.5 (d)), reflecting volatility clustering. A striking feature of efficiency bubbles is that the absolute scale of the fluctuations stays extremely high even for large systems, because a finite system without mean reversion will eventually reach its limits with probability one.

[6] There is a minuscule amount of anti-correlation, due to the mean reversion on the system boundaries, which is not visible here

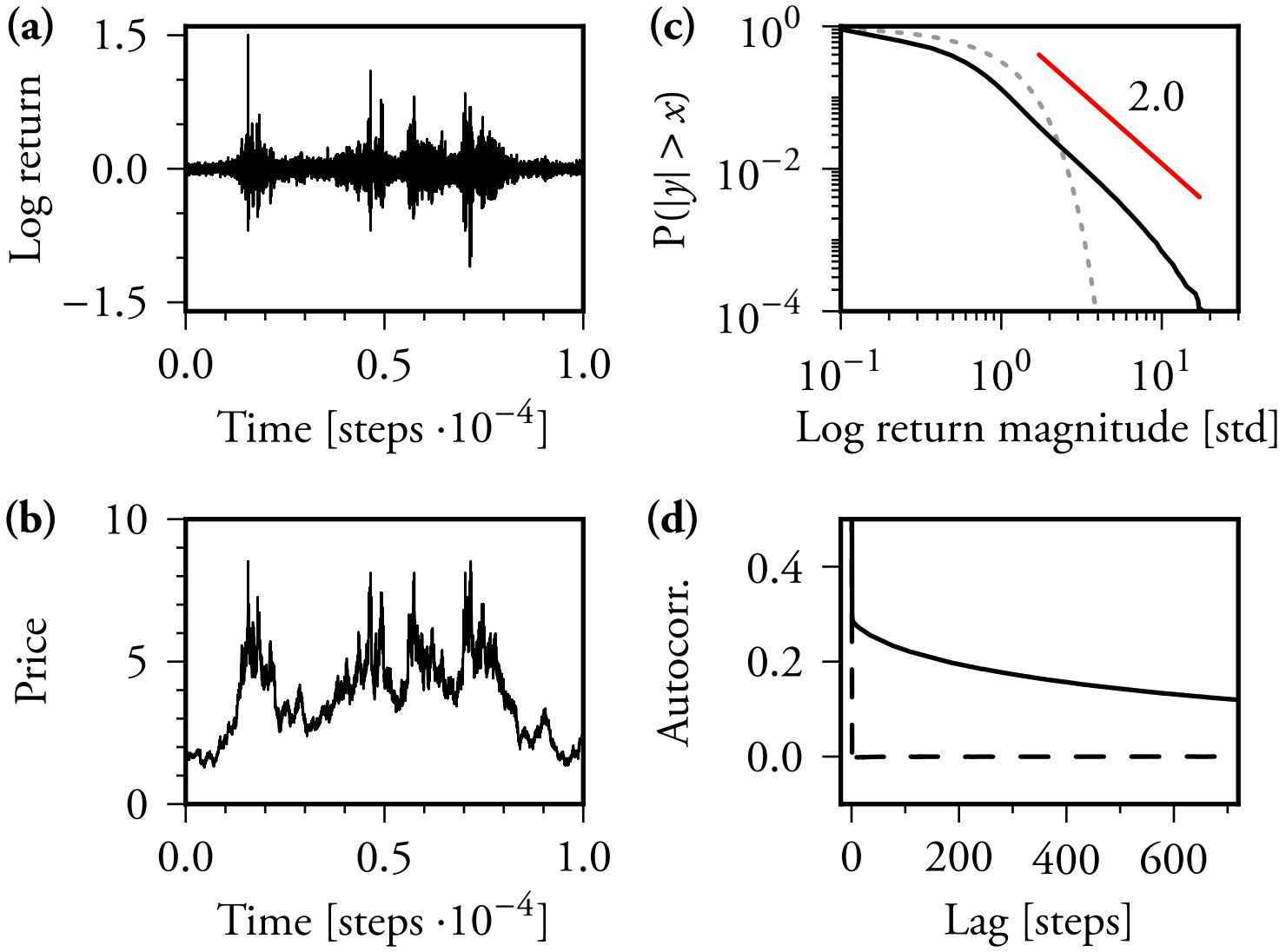

Figure 12.5.: price-efficient model with $N_s = 10^4$ speculators and $N_n = 10$ noise (coin flipping) traders. **(a)**: Log return time series. **(b)**: Price time series. **(c)**: CCDF of normalised log return magnitudes (solid black line) and a normal distibtion (dashed gray line). Straight line: analytical result. **(d)**: Average autocorrelation of the log returns (dashed line) and of their magnitudes (solid line). The lengths of the averaged segments matched those in fig. 11.2.

### 12.4.1. Analytical treatment

To obtain an explicit solution to equation (12.7), a good approximation for large $N$ is to require efficient demands instead of efficient prices:

$$\mathrm{E}\big(\delta(t)\,|\,\delta(t-1),\delta(t-2),\dots\big) \overset{!}{=} \delta(t-1). \qquad (12.8)$$

Since agents choose stochastically, the demands generated by the speculators and random traders each are binomially distributed. Equation (12.8) is fulfilled, where possible, if the probability for each speculator to buy at time $t$ is

$$P\big(\text{buy}\,|\,\delta(t-1)\big) = \frac{1}{2} + \frac{\delta(t-1) - N/2}{N_s} \qquad (12.9)$$

$$\text{for} \quad \delta(t-1) \in [N_n/2, N_s + N_n/2] \quad \text{(see above).}$$

Figure 12.6 (a) shows a comparison of equation (12.9) with a numerical optimisation with respect to equation (12.7). For efficient prices, there is a slight drift away from the system boundaries that is not present for efficient demands. This difference, however, decreases with an increased system size $N$.

### 12.4.2. Stationary solution

For large $N_s$, we can neglect the random agents and the difference between price- and demand efficiency. We thus consider $N$ agents who buy at each time $t$ with probability $\delta(t-1)/N$. The stationary demand distribution $\pi$ then satisfies

$$\pi_j = \sum_{i=0}^{N} \pi_i \pi_{ij}, \qquad \pi_{i,j} = \binom{N}{j} \left(\frac{i}{N}\right)^j \left(1 - \frac{i}{N}\right)^{N-j} \qquad (12.10)$$

where the probability to move from state $i = \delta(t-1)$ to state $j = \delta(t)$ is given by the transition matrix $\pi_{ij}$. For large $N$, equation (12.10) is satisfied by the uniform distribution. To show this, we first divide by $\pi_i = \pi_j = \text{const}$, and obtain

$$1 = \sum_{x=0}^{1} \binom{N}{j} x^j \, (1-x)^{N-j}, \quad \text{with} \quad x = \frac{i}{N}. \tag{12.11}$$

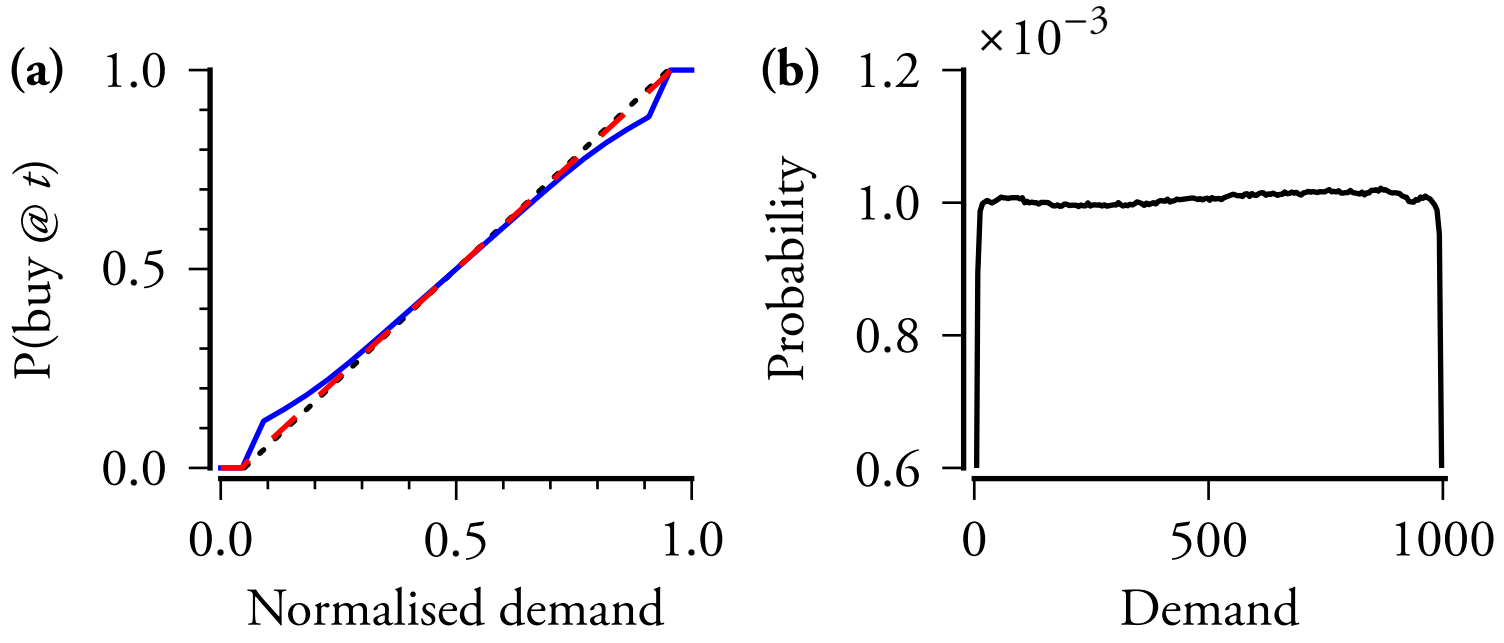

Figure 12.6.: **(a)**: Probability that an agent buys at time $t$ for different previous demands $\delta(t-1)$ normalised by the system size $N = N_s + N_n$. Fraction of noise traders: $N_n/N_s = 0.1$. The demand-efficient solution is given by eq. (12.9) (dotted black). The price-efficient solutions for $N = 22$ (solid blue) and $N = 110$ (dashed red) were obtained by numerical optimisation. **(b)**: Distribution of demands in a simulation of $N_s = 1000$ speculators and $N_n = 1$ noise trader for $10^8$ time steps.

For large $N$, we can replace the sum over $x \ll 1$ with an integral. The right hand side of equation (12.11) then reads

$$\binom{N}{j} N \int_0^1 x^j (1-x)^{N-j} dx = \binom{N}{j} N \frac{\Gamma(j+1)\Gamma(N-j+1)}{\Gamma(N+2)} \tag{12.12}$$

$$= \frac{N}{N+1} \xrightarrow{N\gg 1} 1 \qquad \square \tag{12.13}$$

Figure 12.6 (b) shows the demand distribution for a simulation of the price-efficient model. It is uniform except for the very edges where it drops sharply. For higher ratios $N_n/N_s$, the edges can also exhibit peaks.

### 12.4.3. Tail Exponent

The log return for two subsequent demands $\delta$ and $\delta'$ can be expressed as

$$r = \ln\left(\frac{\delta'}{N-\delta'}\frac{N-\delta}{\delta}\right) \approx \Delta\left(\frac{1}{\delta}+\frac{1}{N-\delta}\right), \quad \text{where} \quad \Delta = \delta' - \delta. \tag{12.14}$$

The approximation is obtained by expanding for small $\Delta$ up to the first order. This is possible, because the standard deviation for the binomial distribution is proportional to $\sqrt{N}$ and further vanishes for demands close to zero or close to $N$. Hence, the distribution of $\Delta$ will be very localised for large $N$. Fluctuations in $r$ are then dominated by fluctuations in $\delta$, especially for $\delta \ll N$, and $N-\delta \ll N$. By the same argument, the exact shape of $p(\Delta)$ is negligible for the scaling of the tail of the return magnitudes. Due to the symmetry with respect to $N/2$, we now analyse only the case $\delta \ll N$ where $r \approx \frac{\Delta}{\delta}$. The expected fluctuations in $r$ can be expressed by

$$\mathrm{E}\left(r^2 \mid \delta\right) \approx \left(\frac{\mathrm{E}(\Delta)}{\delta}\right)^2 \approx \frac{1}{\delta} := \tilde{r}^2. \tag{12.15}$$

Using the probability integral transform, and $p(\delta) = \text{const.}$ yields

$$p(\tilde{r}) \propto |\tilde{r}|^{-3}, \quad \text{and therefore} \quad p(r) \underset{\sim}{\propto} |r|^{-3} \tag{12.16}$$

for sufficiently large $N$ and $r$, and in agreement with simulations (fig. 12.5 (c)).

## 12.5. Learning price efficiency

The above results raise the question of why similar bubbles didn't arise in the trading model of chapter 11. The answer is that these trading agents did not receive any information on absolute prices. Therefore, no strategy could exploit predictable drifts that arise for high or low prices. We here briefly demonstrate that the trading model will learn to exploit and thereby countervail also this type of predictability. For simplicity, however, we will not pursue the question which strategies enable perfect price efficiency. Instead, a small but significant change in the definition of endogenous information with respect to (11.1) is studied.

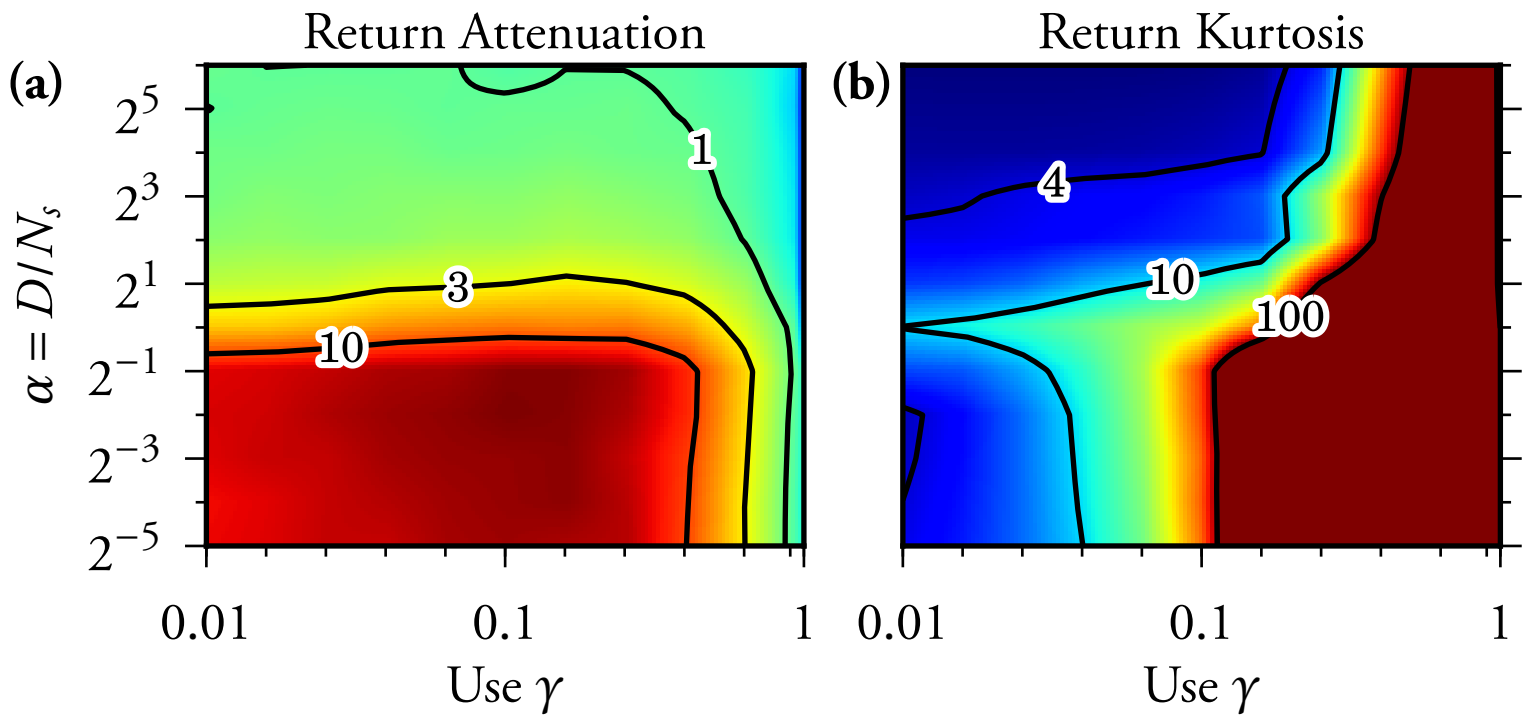

Figure 12.7.: Analysis of log return fluctuations for a trading model with absolute-price information with $N_s = 2^{10}$ speculators and $N_n = 2^4$ noisy producers. Figure generation like in figure 11.6. **(a)**: log return attenuation during transients. **(b)**: log return kurtoses after transients.

Allowing agents to distinguish between prices above and below one yields

$$\mu(t) = \sum_{i=0}^{K-1} 2^i \,\Theta\big(p(t-i-1) - 1 + \eta(t-i-1)\big) + 1. \qquad (12.17)$$

Figure 12.7 shows the phase space for the trading model with the above method of information generation. As in the minimal bubbles model, a small number of coin-flipping agents is required to allow the system to recover from extremely high or low prices. In contrast to return-based endogenous information, we here observe an interaction of the phase transition with respect to the completeness of strategies, and a dependence on the use parameter $\gamma$. For use parameters $\gamma \approx 0.1$, the phase transition along the $\alpha$-axis is similar to return-based endogenous information. For smaller $\gamma$, the kurtoses are maximal close to the phase transition because overcomplete markets are dominated by the noisy producers. For $\gamma \to 1$, the log returns exhibit both large average fluctuations and heavy tailed distributions for all $\gamma$. This result can be understood intuitively. For small $\gamma$, there is a competition between the predictability due to complex patterns, and short-term mean reversion. The latter type of predictability mainly depends on the latest bit of information, that is, on $p(t-1)$. For large $\gamma$, the time scale of resource redistribution becomes shorter than the length of the bit sequence encoded by the information index $\mu$. The system is then dominated by short-term destabilisation while the older bits mainly contribute noise. The completeness of strategies with respect to these bits is therefore less important. This effect is even more pronounced for mixed endogenous information with several return-based bits and one bit of absolute-price information (not shown).

In contrast to return-based information, equation (12.17) gives rise to the most realistic results for smaller use parameters. This the case particularly for $\gamma \approx 0.1$, where the dynamics in (over-) complete markets do not appear to depend on $\alpha$ at all. Figure 12.8 shows an example where agents only act on the most recent two prices. Clusters with large returns coincide with phases where average prices are either particularly high or low. For return-based information, only the novelty of information states and not the absolute price determines the local volatility (for comparison, see e.g. fig. 11.5).

In contrast to the minimal bubbles model of section 12.4, prices here don't wander arbitrarily far away from unity since agents only learn whether prices are above or below one (fig. 12.8(a)). There appears to be no reason why trading with a more nuanced encoding of prices should not yield results that are more similar to figure 12.5. A very simple example is presented as a part of the next chapter. There, however, we focus on an extended and more realistic seesaw game experiment, as well as ways to collect the choices of more subjects.

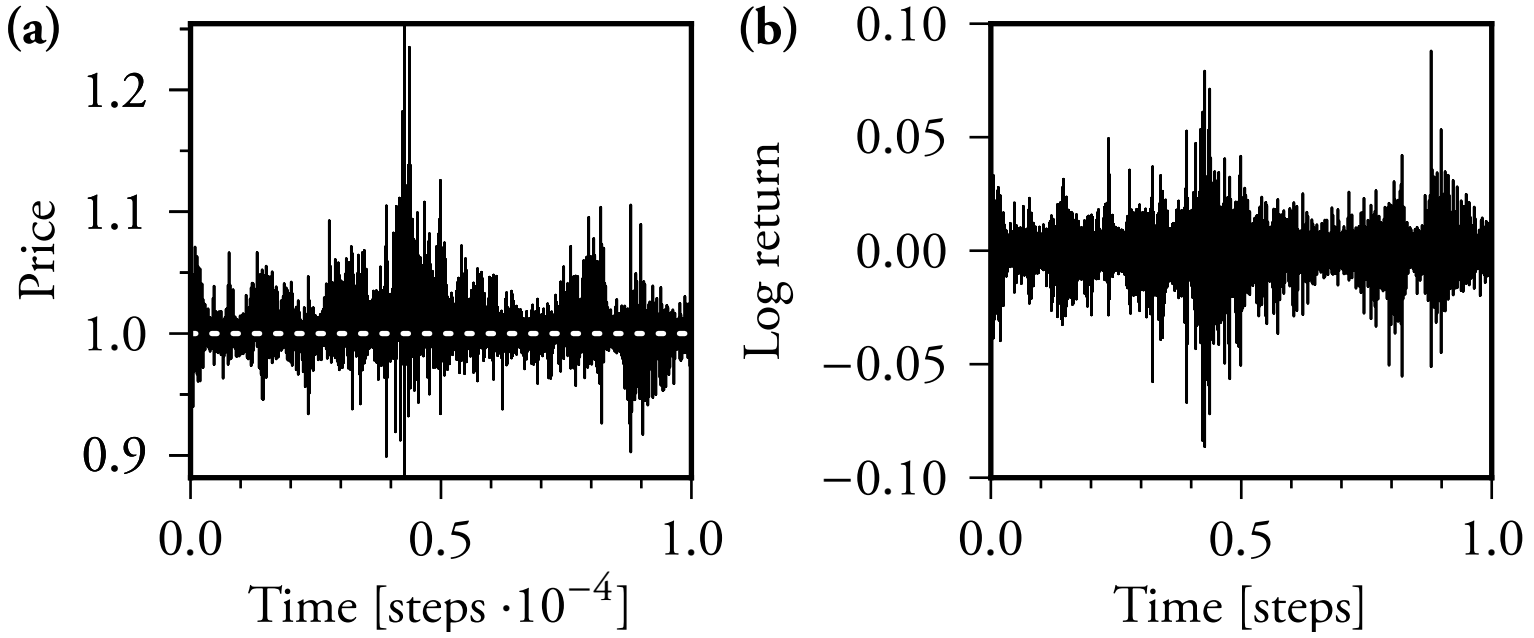

Figure 12.8.: Dynamics of the trading model analysed in fig. 12.7 for the parameters: $N_s = 2^{10}$, $N_p = 0$, $N_n = 2^4$, $D = 2^2$, $\gamma = 0.1$. **(a)**: Price. **(b)**: Log return

# 13. Extreme events in small-scale minority games

In the previous chapter, group experiments revealed dynamical principles that persist in large-scale theoretical models where they give rise to the notorious "stylised facts" introduced in section 10.3. In this chapter, we investigate whether extreme price fluctuations can also emerge in moderately sized groups of real subjects. For this purpose, the seesaw game introduced in the last chapter is extended. The new version encompasses many of the previous chapters' results. An intuitive web-browser-based implementation enables laboratory- and classroom experiments. Nearly arbitrary numbers of subjects can play the game online at seesaw.neuro.uni-bremen.de. Bubbles, local price patterns, power-law distributed price changes, and also, to a smaller extent, volatility clusters emerge for real subjects, for virtual agents that follow simple strategies, and in scenarios where both interact. Since the experiments continue as of this writing, this chapter only represents an overview of the project and a snapshot of the results so far.

## 13.1. The logarithmic seesaw

The seesaw game was introduced in chapter 12. A simple linear game was studied where all agents have the same impact and where the outcome is the change in excess demand between two subsequent time steps (game turns). Realistic log return distributions were obtained in a theoretical model of the subjects' behaviour in combination with the pricing rule introduced in section 11.2. This exact pricing rule can be calculated by the seesaw after a minor modification.

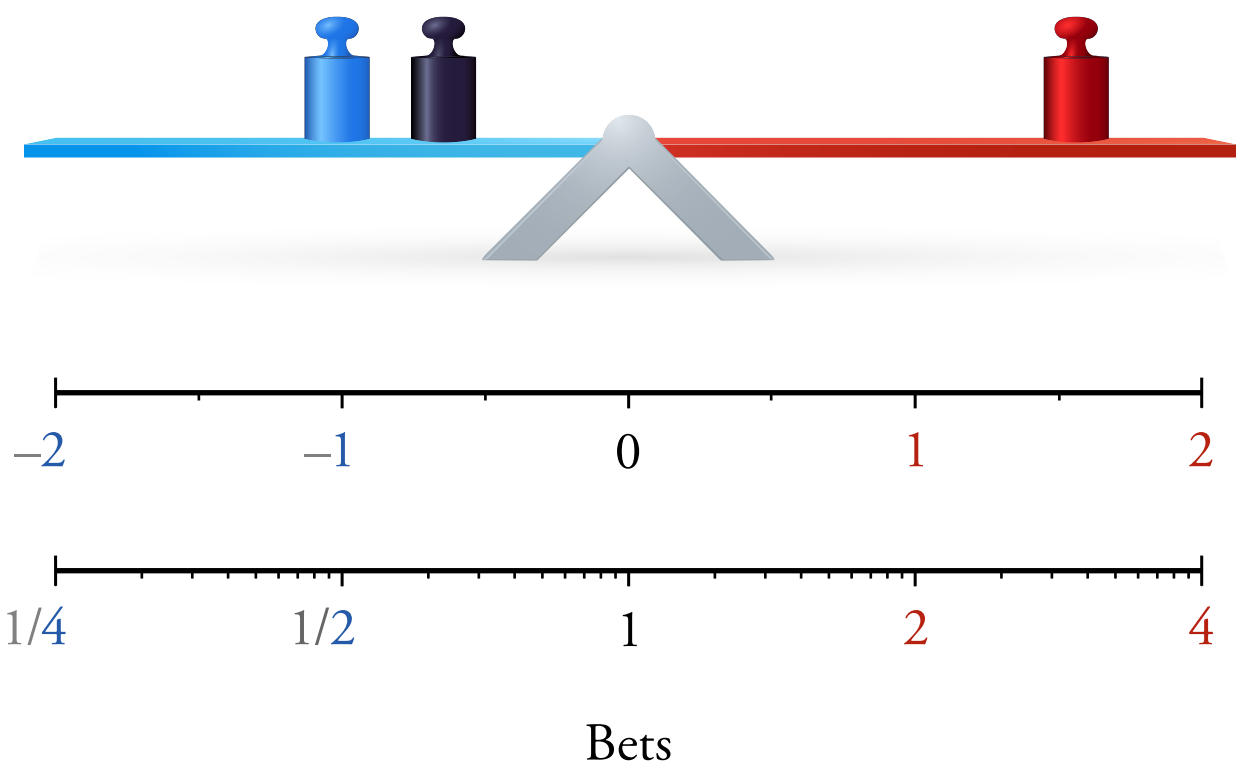

Figure 13.1.: Instead of each agent placing their own weight on the seesaw, coloured weights may be used to indicate the total demand and total supply. This representation allows for continuous bets, and also for different scales. For the linear scale (upper axis), two weights represent demand and supply (i.e. negative demand). A third weight balances the excess demand after each turn. Given a small reverting force, the angle of the linear seesaw after each round is proportional to the change in excess demand. For a logarithmic scale (lower axis), the physical positions of the weights on the seesaw are proportional to to ln(demand), and -ln(supply). Hence, the balancing weight here represents the logarithm of the previous price. The angle of the logarithmic seesaw after each turn is proportional to the log return.

Figure 13.1 shows a realisation of the seesaw where the totality of the weights bet on either side are embodied by a large coloured weight, respectively. These two weights representing demand and supply are moved along the seesaw in order to apply the correct forces, as is the balancing weight whose dynamics remain unchanged. This allows for the flexibility to change the scale which links demand and supply to the physical positions of the respective weights. For a linear scale, the effective torques are equivalent to figure 12.1. For a logarithmic scale, the angle of the seesaw after each turn is proportional to the log return.

Another advantage of the mechanism with three weights is that any number of agents can place arbitrarily heavy bets. This enlarges the space of possible outcomes for a given number of agents. Therefore, fewer agents are required in order to allow for a wide range of different possible returns. Furthermore, the situation on the seesaw becomes more transparent for large numbers of agents.

In the following, we focus on a logarithmic seesaw game where agents in each step place a fraction $\gamma$ of their capital on the seesaw. The payoff consists of two components. Wins and losses from speculation follow the MG approximation to trading described in section 11.8. It turned out, however, that a controlled growth of the total capital in the market is necessary to keep subjects motivated. Since the majority loses in each turn, most of the the players would eventually hold a negative capital and stop playing. A growing capital in the market successfully motivates subjects to increase their trading frequency even though their relative performance is unchanged.[1] In the current implementation of the game, a small risk-free interest is paid on the capital that subjects place on the seesaw. This is equivalent to multiplying the normalisation factor in equation (11.53) with a constant.[2]

---

[1] As discussed in sec. 10.4, speculative trading in real markets is, on its own, a zero- or even negative-sum game as well.

[2] We tested several other ways to implement capital growth. For a moderately sized, closed group of subjects, one easily predictable producer (or market maker) with a fixed income is a viable solution. When subjects can freely enter and leave the market, however, additional complications arise. Successful subjects will start to lose more often when their impact becomes too high. They, therefore, consistently withdraw capital (liquidity) from the game. During times where the website was highly frequented, the game would often be dominated by the producer after too much capital was withdrawn from the game. This scenario, where subjects for prolonged periods of time have to bet just against the producer until liquidity is restored, does not reflect the intention behind the game, to study the interaction of subjects. Another idea was to include virtual speculators (bots, see sec. 13.4) who leave after they lost capital and return with new capital. This solution required a rather extreme flow of bots, since the subjects were highly successful at exploiting them.

## 13.2. The seesaw browser game

All seesaw game experiments (including those in chap. 12) were performed using a custom browser-based implementation. Subjects can therefore participate on any computing device with a relatively modern web browser connected to the internet.[3]

The website currently comprises several sections. Content was added incrementally over time. To play the game, subjects have to create an anonymous account with a user name and a password. No personal information is stored. The game rules are explained concisely on the index page and again on a landing page that is displayed to new subjects who just signed up. A special section features more detailed explanations, as well as links to an interview with the radio station "Deutschlandfunk" and to scientific publications.

Subjects in the open online game are therefore not necessarily naive about the project. This is intentional since informed subjects may be considered the equivalent of informed traders in real markets. If required, however, the game can always be modified without revealing the true intent behind the changes. Furthermore, closed lab experiments were performed with naive subjects who were educated on the scientific background only after they played the game.

To increase long-term motivation, the website features a cumulative global high score. When subjects feel that their impact in the game is too high (or they just want to secure their profits), they can update their high score. Thereby, their in-game capital is set to ten points and the difference in capital is added to their cumulative high score. The cumulative score for a subject can become negative.

The seesaw game itself is presented on a single page shown in figure 13.2. The seesaw is displayed in the upper right corner. Below the seesaw, buttons are displayed that allow subjects to submit choices at any time during the countdown for each game turn. Alternatively, the arrow keys on a keyboard may be used. Subjects can adjust their use parameter $\gamma \in [0.1, 0.25, 0.5]$ which determines the fraction of their capital (points) that they place on either side of the seesaw. The capital of the subject is displayed right next to the buttons. At the end of each turn, the seesaw is animated to indicate the outcome as is shown in figure 13.3. The choices of each subject are revealed to the other subjects only at the end of the turn.

[3] Modern, as of this writing, refers to Firefox 3.5+, Chrome 4+, Safari 3+, Opera 11+, or IE9+. Therefore, only computers are excluded that, for security reasons, should have long been disconnected from the internet anyway. No plugins are required. The game is also playable on many touch-based mobile devices, but further optimisations should be made in the future.

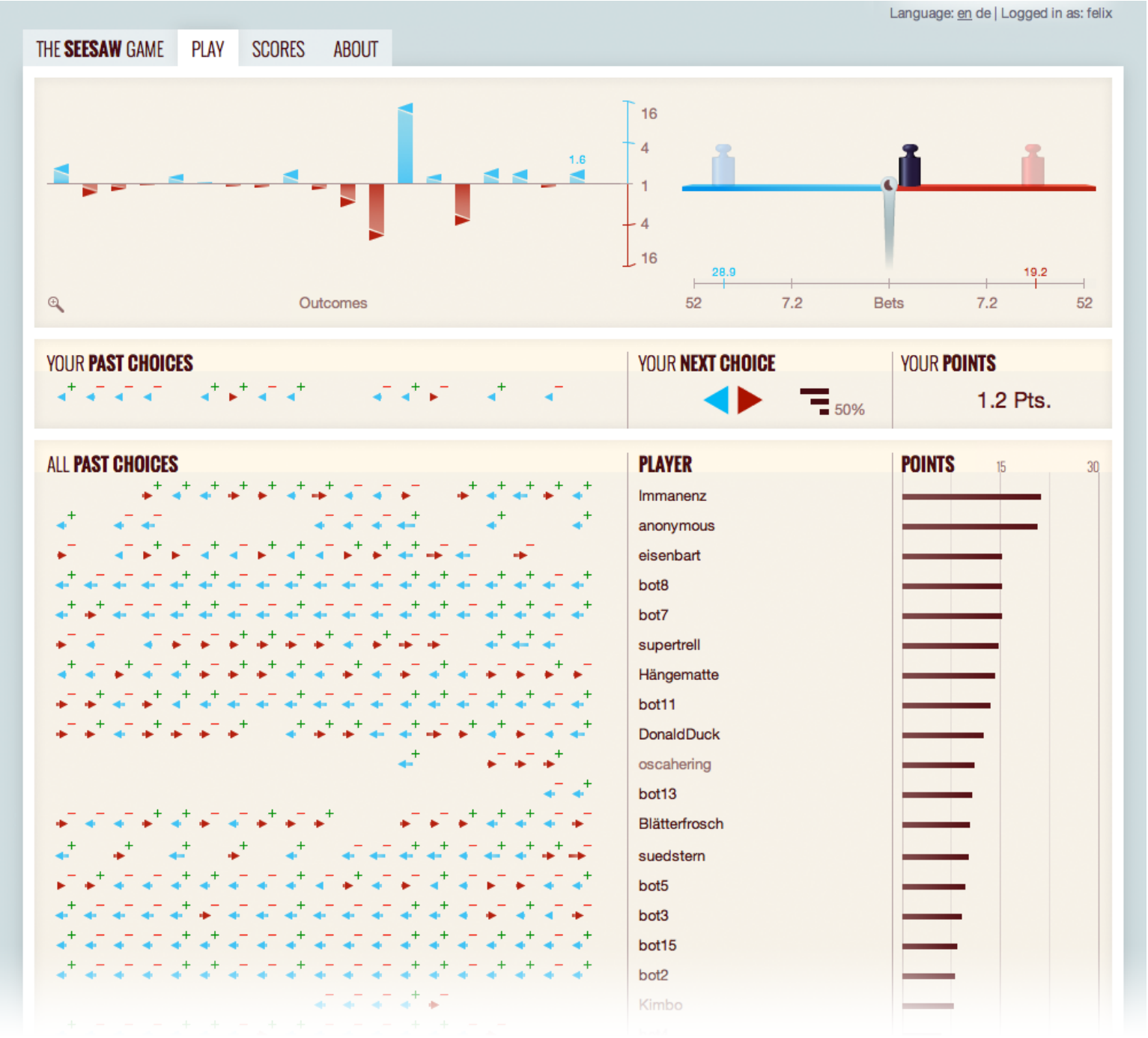

Figure 13.2.: The online seesaw game at seesaw.neuro.uni-bremen.de. The interface is divided into three main containers separated by whitespace: seesaw status (i.e. collective status, top), private status (middle), high score status for all players (bottom). Past results (outcomes, choices) are displayed on the left hand side of each container. The view scrolls to the right where new results are added. The oldest results disappear on the left hand side. The current state of the game is displayed on the right hand side of each container. The remaining time until the end of each turn is indicated by a pie timer, which is displayed in the center of the seesaw (top container). The weights which still indicate the preceding demand and supply are faded. Each subject can enter their choice by clicking on the buttons below the seesaw (middle container), or by using the arrow keys on a keyboard.

The majority of the screen area is dedicated to providing full transparency about the course of the game. This includes the most recent history of the outcomes and of all subjects' choices, as well as the current capital of each player. After each turn, the displayed list of active agents is sorted by capital in descending order to increase competition. Since the placement of a subject in this in-game high score might be below the content displayed in the browser window (fig. 13.2), the personal performance is also shown directly below the seesaw. Only subjects who were active during the visible game history are included in the list. The names of disconnected subjects are displayed in a lighter colour. Inactive yet connected spectators are listed at the very bottom of the page (fig. 13.3). Connection and disconnection events trigger asynchronous display updates. Special events are explained in-game as shown in figure 13.4.

The implementation as a browser game is much less controlled than typical psychophysical experiments, but offers the opportunity to experiment with many more subjects. The uncontrolled nature of this experiment may also be seen as an opportunity because it complements the perfectly regulated dynamics of artificial multi-agent systems. Two main aspects of the game need to be regulated tightly: the information to which subjects have access, and the choices which subjects can submit. Both requirements are ensured by architecture of the system.[4]

[4] It is very important in any web application that the server never trusts the client. Subjects can (and did) try to manipulate their web browsers.

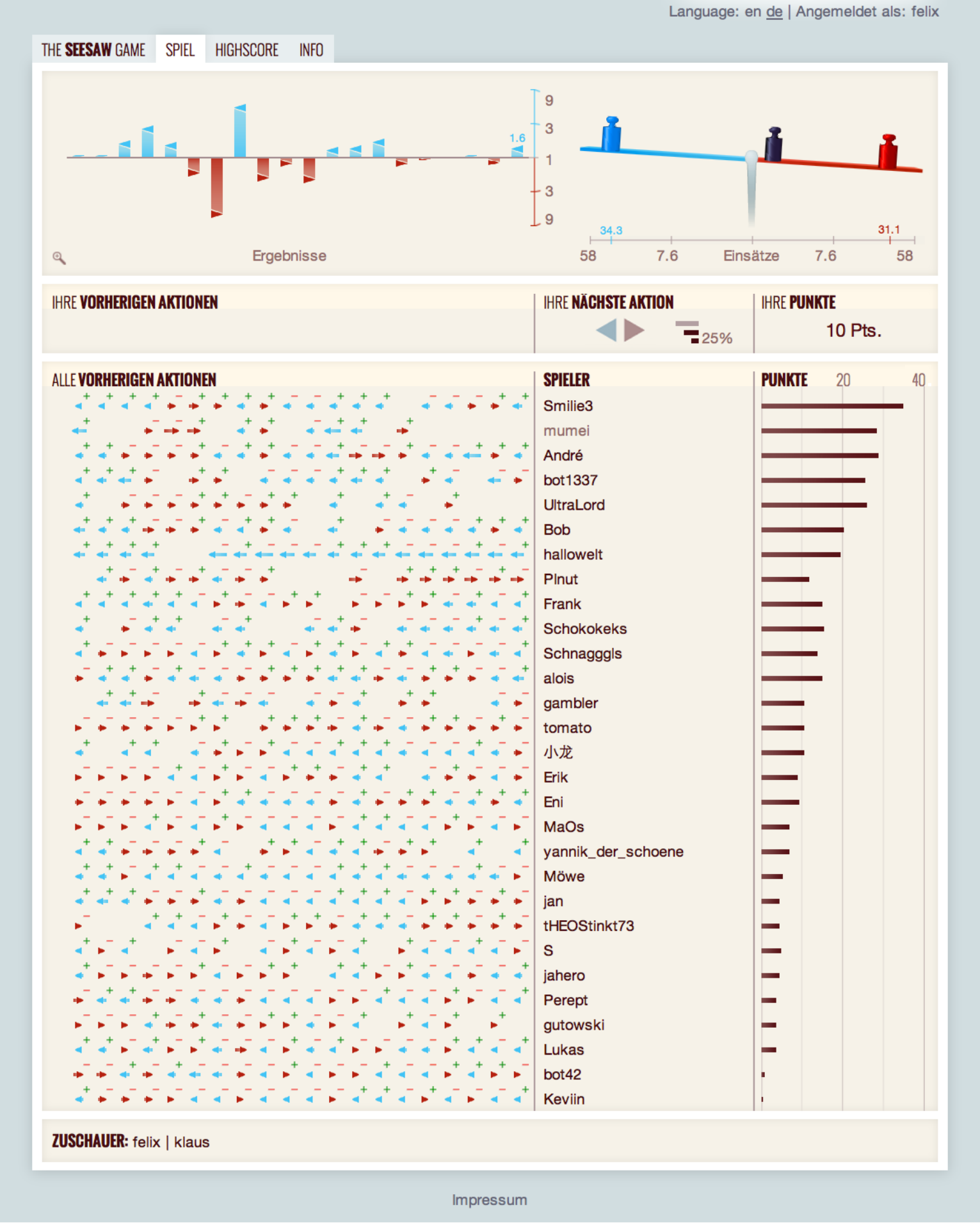

Figure 13.3.: A classroom experiment using the seesaw browser game. At the end of each turn, all subjects' choices are revealed and the weights on the seesaw are moved accordingly.

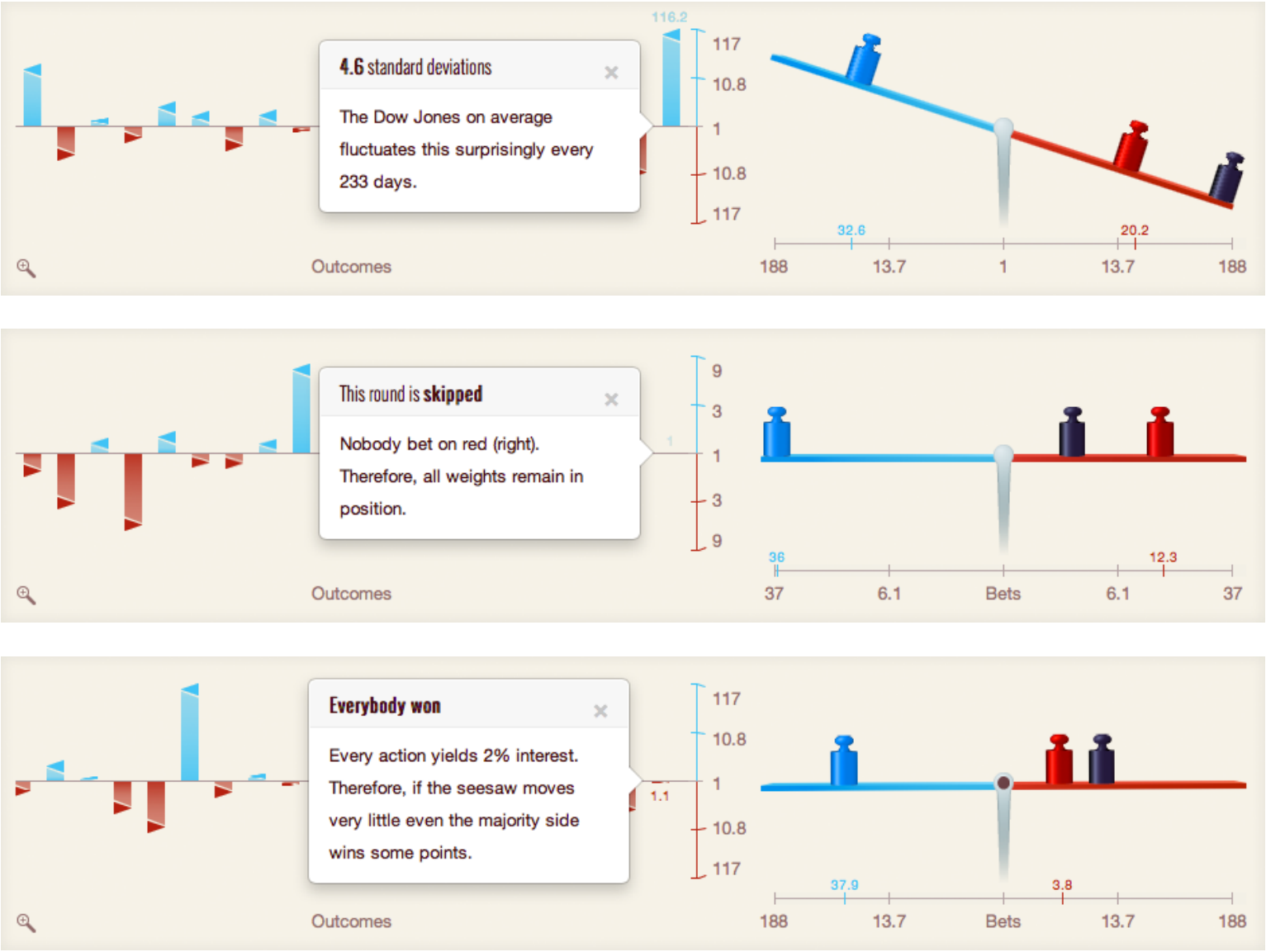

Figure 13.4.: Certain rare events are explained by overlays, which disappear automatically before the end of the next turn.

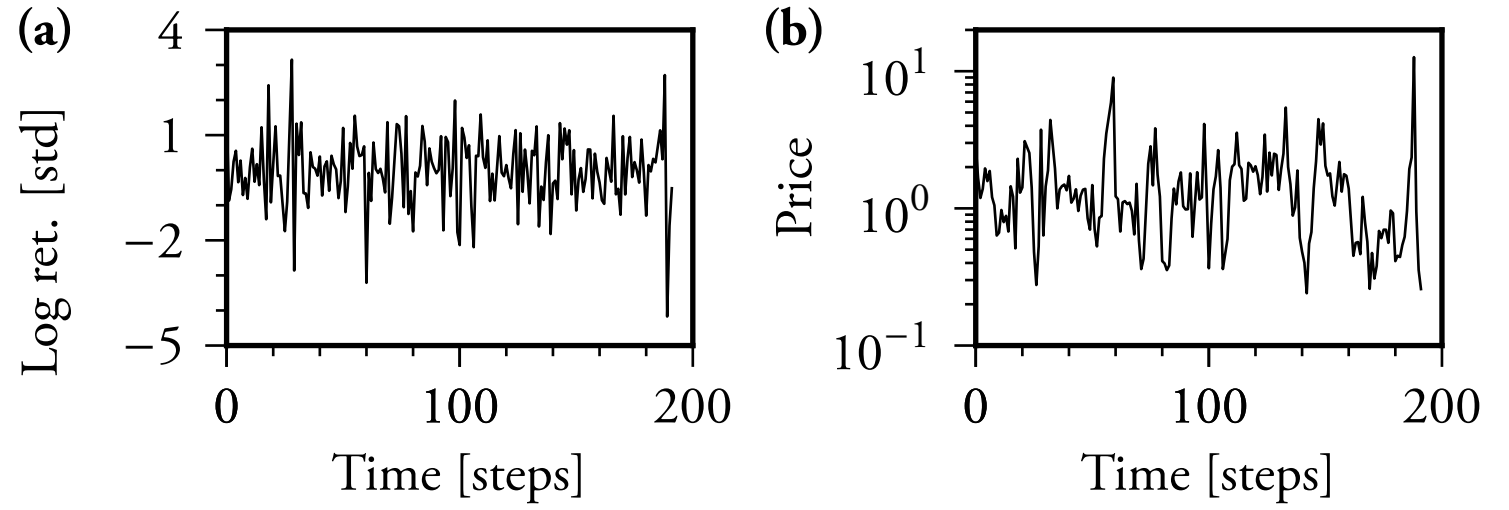

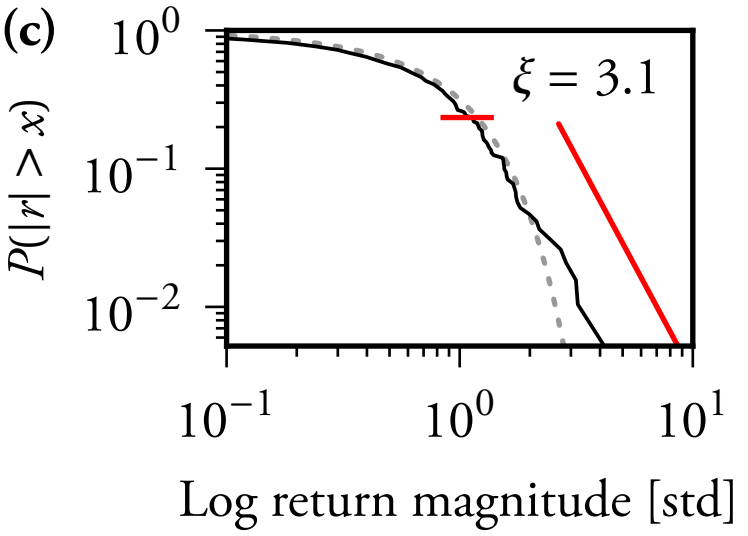

Figure 13.5.: Logarithmic seesaw game with 30 subjects. Time series: **(a)**: Log return and **(b)**: Price. **(c)**: CCDF of normalised log return magnitudes (solid black line) and a normal distribution (dashed gray line). Straight line: fit. Methods: sec. 10.3.

The server-side implementation currently runs on two servers in parallel to ensure redundancy and scalability. David Rotermund greatly assisted with the hardware setup, the installation of the Open BSD operating system, and the local networking setup. The real-time communication and game logic is implemented in separate processes using node.js [Joy]. The application state is stored redundantly in several instances of the redis data structure server. This setup allows for a lightweight, fast, reliable, and scalable distributed network application. The results of each game turn are also immediately written to disk in machine- and human readable form. The (near) real-time communication with the clients is performed using socket.io [Rau].

## 13.3. Results for a closed group

Results for a classroom experiment with naive subjects (students) are shown in figure 13.5. The payoff followed from equation (11.53)

except for an additional 2% risk-free interest on the capital placed on the seesaw as discussed above. The log returns unambiguously deviate from Gaussian white noise ($p \ll 0.01$ for both the Shapiro-Wilk test and for the one-sided KS-test). Despite the small number of samples, the distribution tail is clearly consistent with a power law. Log return fluctuations exceeding three standard deviations were also robustly observed in experiments with fewer subjects after less than half an hour of playtime (not shown). Significant results for the autocorrelation would require more samples (not shown).

## 13.4. Results for the open online experiment

To allow subjects to play the game on the public website at any time, virtual players (bots) were included. These bots are slightly predictable even collectively, and not perfectly price efficient for most of the time. This choice was made to motivate subjects to participate. Bots always bet 25% of their points. They can choose between three different strategies: Betting on the side opposite to the balancing weight, betting on the same side where the balancing weight is located, or flipping a coin. Bots stochastically switch to a random strategy, or, with a slightly higher probability, to the historically most successful one. In practice, one of the bots switches to a different strategy every few time steps. This setup leads to quite realistic return fluctuations for simulations with bots only, and for experiments where the bots interact with real subjects. The results are not very parameter dependent if the same bots stay in the game for a sufficiently long time. The game is stopped when no real subjects are logged in.

The game exhibits complex dynamics with volatility clusters and also sudden price jumps. Sometimes, local oscillations or patterns emerge. Examples are shown in figure 13.6.

Initially, the game was played only by a few subjects and the number of bots was kept constant. During this phase, no risk-free interest was necessary to stabilise the liquidity of the game. Instead, the capital of the bots was renormalised when no subjects were online. Results are shown in figure 13.7. The log returns are power-law distributed and some volatility clustering is observed. The significant correlation length is limited which is to be expected for small systems and given the number of analysed time steps. The prices exhibit several bubbles (and anti-bubbles since the game is symmetrical).

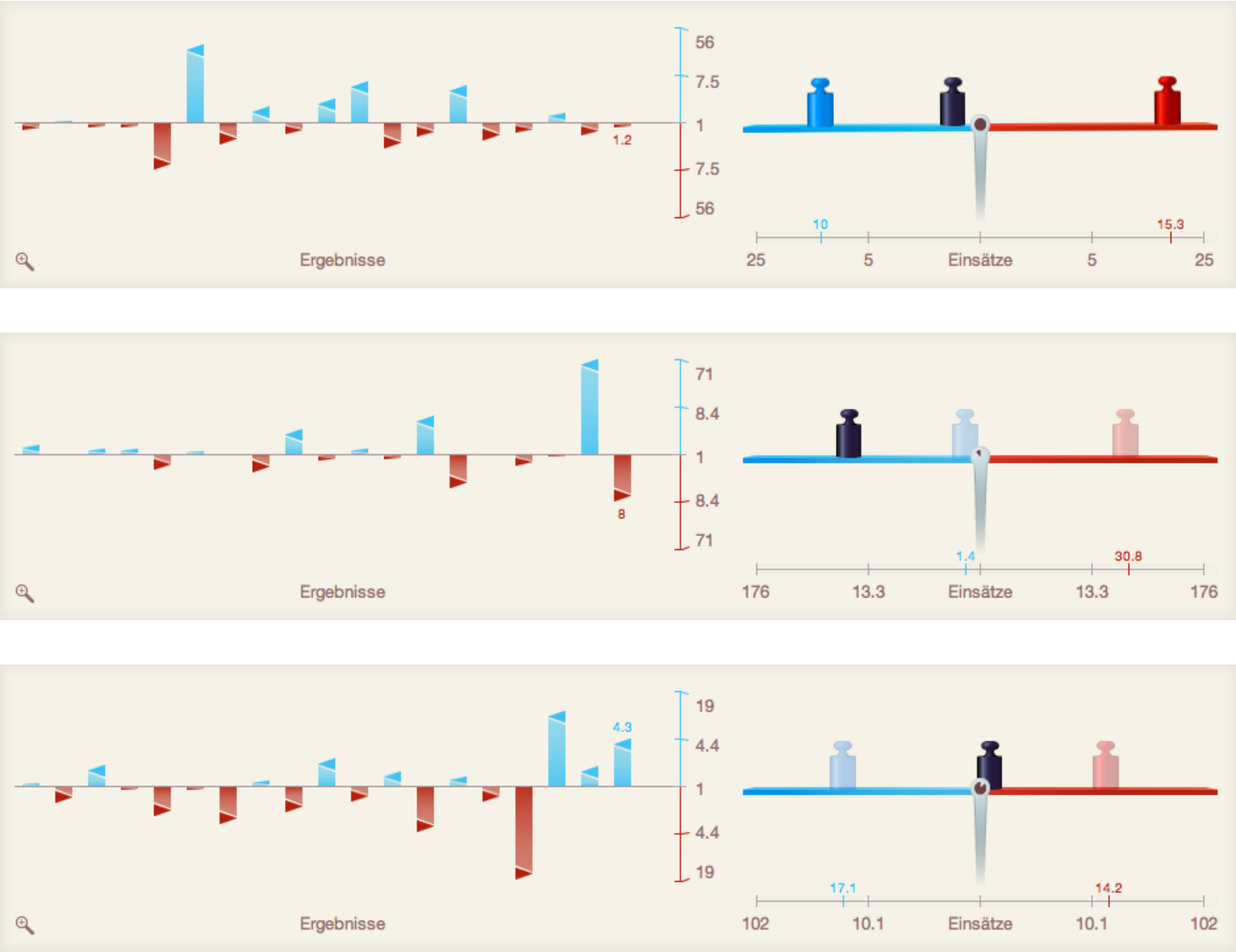

Figure 13.6.: Outcomes sometimes appear to follow patterns which are damped away or become instable. Note that there are several ambiguous possibilities to define sequences of repeating patterns in the topmost example. The three above screenshots, like figure 13.2 which exhibits less obvious oscillations, were taken when subjects and virtual players were active at the same time.

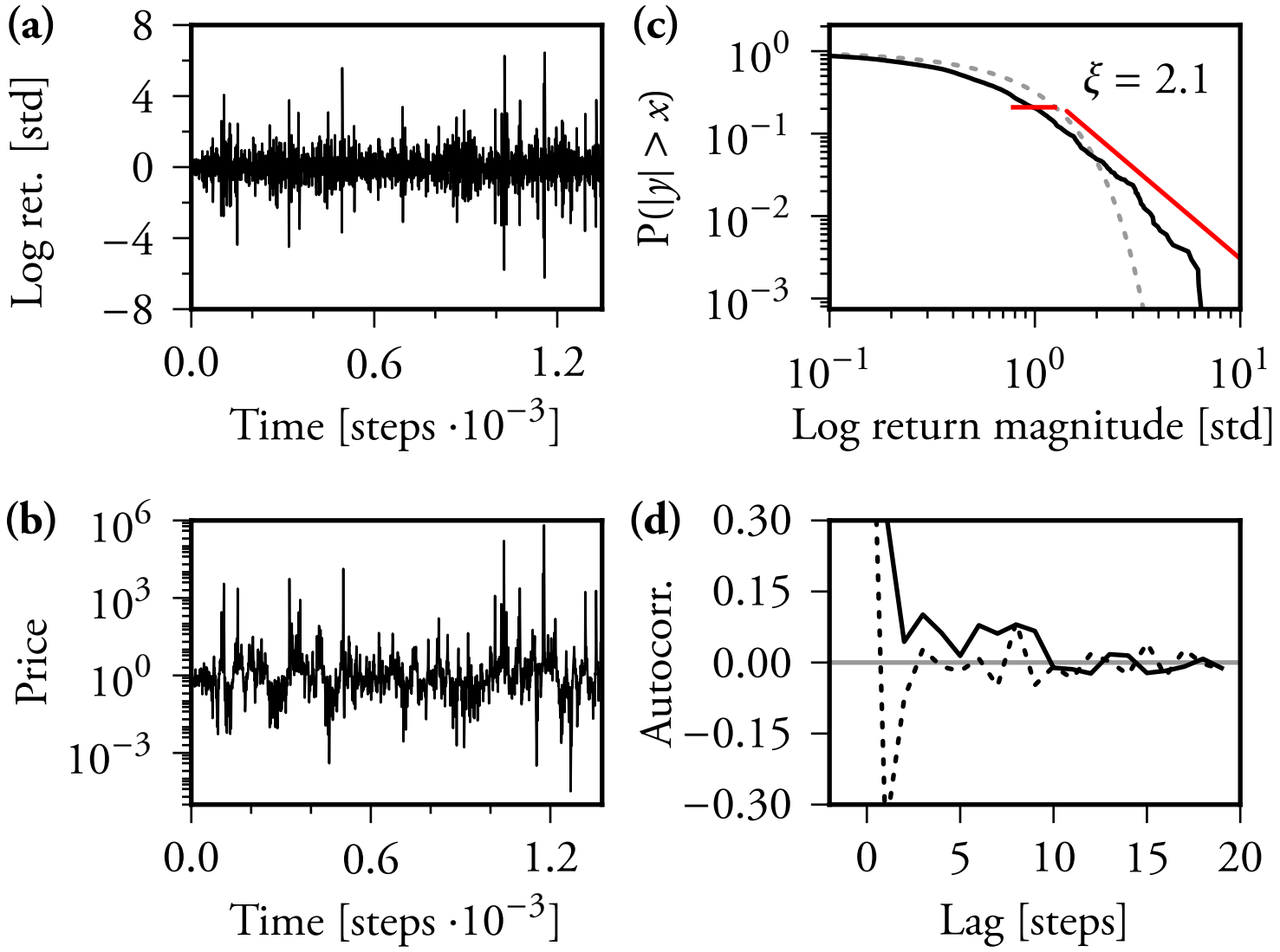

Figure 13.7.: An early online experiment with the logarithmic seesaw where 1-3 real subjects were logged in at each point in time. The number of virtual players (bots) was held constant at 15. **(a)**: A part of the log return time series. **(b)**: Price time series. **(c)**: CCDF of log return magnitudes (solid black line) and a Gaussian with the same variance (dashed gray line). Straight line: analytical result.

After some media coverage on the website, the number of subjects increased dramatically. Renormalising the capital of the bots during breaks is no longer an option since the game is often actively played for hours with only short interruptions during the night. Therefore, the small risk-free interest on the capital that is placed on the seesaw, which was discussed above, was implemented. The number of bots was made adaptive such that the number of active agents stays approximately constant unless more than 15 subjects are playing. At this point, bots that leave the game are not replaced and eventually only real subjects are active. This change decreases the influence of the bots if a sufficient number of subjects is active. Bots also stochastically leave the game (and are replaced shortly thereafter) when the total number of agents is not too high. This contributes to stabilising the total capital in the game.

Unfortunately, the collective dynamics of the bots is more parameter dependent when the bots are replaced too frequently. To turn this problem into an advantage, however, we can test whether the power law distribution of log returns is imposed by the bots. For several months, the game operated in a regime where the bots on their own would generate log return distributions whose tails decay slightly faster than in figure 13.7 (c). The log returns and prices for a part of the time series with moderate subject activity are shown in figure 13.8 (a) and (b), respectively. The log return distributions for high and low subject activity are compared in figure 13.8 (c)

The average number of total actions of all agents was $14 \pm 1$ per time step. Subject activity is characterised by the average number of subject actions per time step, estimated from a gliding window with a width of 20 time steps. We distinguish two cases in the following: Low participation (LP) where the activity is below 7, and high participation (HP) where the activity is at least 7. Results are robust with respect to changes of the window width and the threshold.

Since there are less samples for (HP) than for (LP), the latter were split into subsets that match the sample size for (HP). The CCDF average along the $|r|$-axis and the corresponding one- and two standard deviation intervals for (LP) are shown in figure 13.8 (c) as shaded areas. In comparison to the average for (LP), the CCDF for (HP) is significantly less concave: It is more than two standard deviations lower around $|r|/\mathrm{Std}(r) \approx 1.4$, and more than two standard deviations

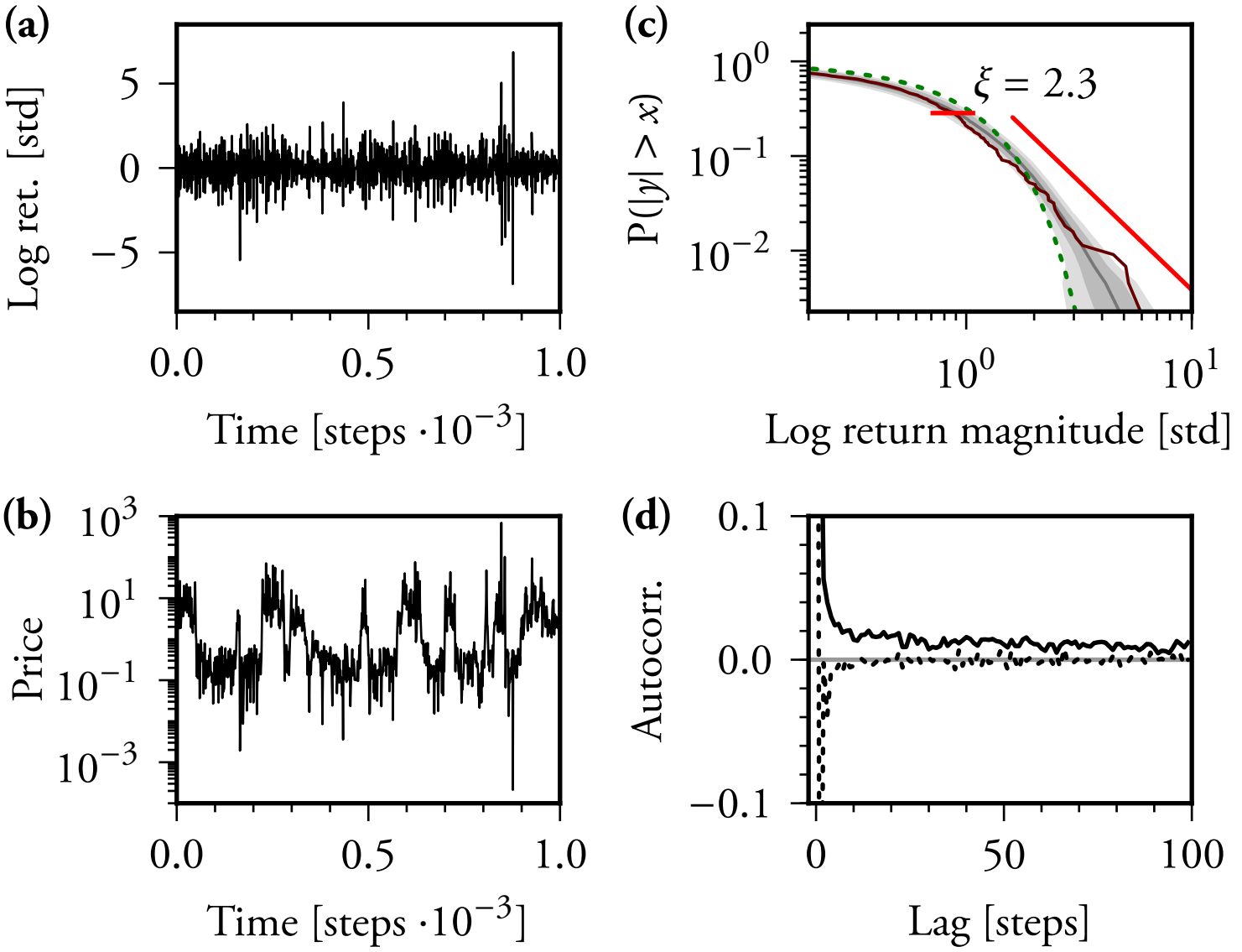

Figure 13.8.: Online experiment over several months where the number of subjects $N_s$ varied substantially. The number of bots $N_b$ was increased if $N_b + N_s < 15$ and decreased if $N_b + N_s > 15$. The bots' parameters here were adjusted such that the log return distribution tails decay slightly faster than in fig. 13.7 (c) if no subjects participate. **(a)**: Log return time series (excerpt). **(b)**: Price time series. **(c)**: CCDFs of log return magnitudes for two cases. Solid grey line: average for low subject participation (see main text). The dark and light grey areas cover one and two standard deviations, respectively, around the average. Solid dark red line: high subject participation. Straight red line: the high participation case is significantly better described by a power law than the low participation case (see main text). Green dashed line: normal distribution. **(d)**: autocorrelation of the log returns (dashed line) and of their magnitudes (solid line) for the whole time series. The minimum for the former (−0.38) was cropped to improve the visibility of the positive correlations for the latter. Methods: sec. 10.3

higher for $|r|/\mathrm{Std}(r) \approx 5$. Consistently, the CCDF for (HP) is also significantly better described by a power law than for (LP) ($p_{>} = 0.003$, that is, the probability that the KS statistic for the fit for (HP) exceeds that for a fit to a subset of samples for (LP)). The log return variance for (HP) is $0.62 \pm 0.08$. This is lower than the average log return variance for (LP), which is $1.4 \pm 0.2$. Since large kurtoses are very difficult to estimate from small samples, we only find a low presumption ($p = 0.07$) against the null hypothesis that (HP) does not exhibit a higher log return kurtosis than (LP).

Since splitting up the returns strongly affects autocorrelations, only the whole continuous time series was analysed. The results are shown in figure 13.8 (d). Subsequent log returns are short-range anti-correlated. The log return magnitudes exhibit a small amount of positive long-range correlations. Data collection continuous as of this writing. Possible future developments for the seesaw game are discussed in the next chapter, which concludes this thesis part.

# 14. Summary and Discussion

In the past chapters, seemingly contradictory features of financial markets were reconciled. On the one hand, considerable evidence suggests that financial markets satisfy some of the predictions of the EMH. Asset prices adjust, often rapidly, to reflect certain fundamental information. Remaining price fluctuations–at least on short time horizons–cannot be predicted easily. On the other hand, the vast majority of price fluctuations cannot be plausibly explained post-hoc by the arrival of new information about fundamentals. Logarithmic price returns further deviate from expectations according to the central limit theorem. Their statistics reveal ubiquitous scaling laws across very different markets and epochs that resemble critical phenomena in the natural sciences. This suggests that some elementary properties of speculative trading evolve markets into states of extreme susceptibility.

This apparent antinomy is resolved by considering the dynamical consequences of information efficiency with respect to self-generated (endogenous) information, that is, information on predictable dynamics within the market itself. At first, it might seem surprising that this approach leads to novel results. Even weak-form market efficiency implies that future price changes are nearly unpredictable from past prices. The (neo-) classical argument why this should be the case, however, is that significant price movements should always reflect the arrival of new fundamental (exogenous) information. Accordingly, any remaining fluctuations should be insignificant amounts of structureless noise. This point of view, while influential, is inconsistent with a growing number of empirical findings.

In this work, theoretical and experimental market models were presented which comply with empirical findings that contradict many existing economic theories. First, trades have a finite market impact. Second, the majority of trades do not arise as a reaction to new exogenous information, but predictable trends or patterns cannot endure nevertheless. The minimisation of arbitrage, therefore, reflects a dynamical balance of market forces. As a main result, this state was shown to robustly emerge from self-organisation based on simple elementary market mechanisms. Even more importantly, the successful elimination of predictable trends and patterns was found to leave highly information-efficient markets particularly susceptible to price bubbles and dynamical instabilities. Self-organised efficiency might have been expected according to established economic theories, but the latter result identifies a new and comprehensive explanation for the characteristics of financial markets described above. Existing explanations, in contrast, focus on specific inefficiencies as mechanisms for market failures. While we identified two different mechanisms that are plausibly relevant in real markets, results suggest the possibility of a convergence of the different models. They, furthermore, contribute to deeper understanding of several preexisting models as well. The main findings and their implications are discussed in more detail in the following.

## 14.1. Trading is collective learning and control

The parsimonious models and experiments presented in the previous chapters explain different aspects of information-efficient markets. In chapter 11, it was demonstrated that the simple yet plausible mechanism of success-dependent order size adaptation suffices to coordinate the impacts of diverse strategies such that information becomes absorbed in the price. The proof that the trading rules in our model correspond to an efficient gradient-based learning rule that minimises predictable return magnitudes provides a rigorous link of a fundamental market mechanism to adaptive control. It may therefore be considered highly plausible to prevail in real markets, despite the com-plexities of real pricing mechanisms and order size adaptations, as long as the latter correlate with trading success.

"Rationality", in the sense of an efficient adjustment of prices to new information, emerges in this model from the collective behaviour of many heterogenous traders. This differs from typical notions of rational markets in the literature. "As-if rationality" is commonly used to describe the idea that markets appear efficient because only "good strategies" are successful while the others eventually die out [LeB11]. This implies that trading success rewards agents who are "rational" in absolute terms and on their own, rather than trading success emerging only in relation to the behaviour of others. This neoclassical perspective has become increasingly criticised, especially from proponents of behavioural economics (sec. 10.4). Wealth evolution on its own also does not select for individual utility maximisation, which economists typically equate with rationality [LeB11]. We will come back to this point later.

A minority rule with respect to the returns is a dominating factor in the trading models' dynamics. That is, those traders profit most whose actions counteract the change of the actions of the majority. It was claimed in [CCC+13] that the trading model is equivalent to an existing MG with capital redistribution, which is incorrect. Even though the trading model and several MGs share a number of characteristics, there are many subtle and also some substantial differences between these models. Most importantly, the mechanisms for extreme events discussed for previous MGs are based on market failure and not on efficiency. Therefore, the results presented here are at least complementary to the published findings based on common MGs in several respects. We first discuss the differences between several payoff rules and means of order adaptation (see also sections 10.8 and 11.8.2).

### 14.1.1. Differences between payoff rules

A minority rule based on the sign of the log return is a good approximation of trading success in a roundtrip trade on the time-scale over which the return was calculated (sec. 11.8). Adjusting the impacts of traders accordingly minimises the log return magnitudes (sec. 11.4.3). This connection of trading and learning has not been demonstrated before.

The payoff in previously established MGs, however, is based on the excess demand, which in these models is assumed to be proportional to the log return. Learning based on either payoff optimises for different pricing rules. Furthermore, equation (11.6) adjusts the price such that demand and supply match in every time step. Prices only change if the agents change their orders. If the MG outcome defined by equation (10.7) is interpreted as a log return, there is no price in the model. It is implied, however, that prices change ad infinitum if all agents resubmit the same orders, except for the case where a set of orders is submitted that matches perfectly at any price. This is possible because all orders in the MG have the same unit.

Furthermore, the comparison of trading with two assets and the MG-approximation with just one asset revealed several differences. The trading model evolves onto a manifold of asset distributions where the price becomes invariant to the information states that are frequently conveyed to the market and therefore expected (sec. 11.4.1). The two assets, however, are still redistributed. For one asset, a single stationary asset distribution suffices for a state-invariant price. The latter model exhibits almost no volatility clustering and a significantly altered phase space in markets that include producers (sec. 11.8). Capital redistribution with just one asset furthermore violates the conservation of assets in a trade and therefore necessitate measures for compensation (see secs. 10.8.2 and 11.8.2, as well as eq. (11.54)). Moreover, it has been thoroughly discussed in the literature that a single-step payoff cannot be defined consistently (sec. 10.9). In conclusion, attempts to further simplify the trading model cause side effects that should be considered carefully when choosing a particular model for future research.

### 14.1.2. Differences between scaling and switching strategies

In common MGs, agents adapt either exclusively or at least predominantly by switching between different strategies. They therefore interact similarly to (frustrated) spins in solid-state physics. This differs substantially from agents with dynamical capital, who form basis functions for collective learning similar to neurons in a neuronal network. Many of the consequences are revealed in figure 10.6. Foremost, the phase transition for common minority games [CMZ05b] differs from the one for capital reallocation (sec. 11.4.2).

Further differences are most apparent in the over-complete phase. In the OMG, herding emerges due to correlated strategies with a fixed impact. If the impact of strategies is scaled instead, no herding emerges. The excess volatility in the OMG is therefore a result of the artificial lack of an appropriate mechanism of order adaptation that would allow for a coordination of correlated agents. The minority rule furthermore creates frustration. Agents who lose frequently will change their strategies. This eventually causes previously successful agents to switch their strategies too, since it is impossible for all agents to belong to the minority most of the time.

Frustration also explains why switching strategies increases fluctuations in models with dynamic capital: the very basis functions required for collective learning are replaced, necessitating re-learning. Faster capital redistribution reduces average errors because the re-learning is more effective. At the same time, the kurtosis is increased because the relative increase in fluctuations when a strategy is replaced is more severe. Whether this mechanism, which was not identified in the MG literature, is relevant for real markets possibly depends on the time-scale of interest. Do complex trading strategies suddenly appear and vanish completely, or does the market reallocate resources between strategies more smoothly? Note that even intra-day price jumps of several standard deviations occur frequently.

## 14.2. Dynamical instabilities

The trading model can reproduce important "stylised facts" of financial markets quantitatively to a surprising extent. Possibly more important, however, is how the extreme events in this model emerge. The fact that efficient information annihilation does not result in a unique and stable equilibrium, but instead can lead to local states that perpetually become unstable, refutes the prominent notion that these properties are mutually exclusive [LM99b].

While Information Annihilation Instability (IAI) was demonstrated in section 11.8 to also emerge in appropriately constructed MGs, the mechanisms for extreme events in common MGs are quite different. First, the OMG does not generate non-Gaussian fluctuations at all. It was argued in the established literature that a modulation of the trading volume is the crucial ingredient for reproducing the stylised facts [CMZ05a]. More recent empirical findings in real markets, however, challenge this view (sec. 10.6). As discussed in the previous section and section 10.8, extreme events in common MGs are caused by a breakdown of the efficient coordination of agents in over-complete markets where they become too correlated. This leads to an increase of average fluctuations and does not depend on how the information conveyed to the agents is generated. The detailed mechanisms, some of which are finite size effects, differ substantially among the different MGs (sec. 10.8).

In contradistinction, instability due to locally adaptive control is a unifying macroscopic principle that is not tied to specific microscopic interactions. In fact, it was first realised in a one-dimensional random map ([PREP07], sec. 7.3). The heavy-tailed distributions here are a direct consequence of the elimination of local trends or patterns which yields a net decrease in average fluctuation magnitudes. These fluctuations are therefore a sign of high efficiency and do not signal its breakdown. Instead they reflect surprising information to which the system is the more succeptible, the better it is locally adapted. Then, a closed loop involving endogenous information creates a dynamic instability leading to extreme fluctuations that are not caused by external events.[1] [2] This distinct role of endogenous information is not found in common minority games. On the contrary, many publications explicitly emphasised that the method of information generation is largely negligible (e.g. [CMZ05b]). The reason for this discrepancy is that IAI only becomes apparent in systems capable of highly efficient control in the first place.

## 14.3. Bubbles in efficient markets

In chapter 12, subjects in a toy market experiment were found to exhibit a tendency for herding caused by collective efficiency with respect to past outcomes. An analytically tractable model that mimics this behaviour relates price efficiency to bubbles, power-law log returns, and volatility clusters. This result is based on two factors. First, the lack of mean reversion in a simple bidding process leads to a uniform demand distribution. This bubble process incorporates no trading or payoffs: it simply implements price efficiency according to (10.4) (or alternatively demand efficiency) for a finite number of agents. Second, a non-linear pricing rule causes the system to be more susceptible in bubble phases. This is analogous to, for example, many buyers betting up the price of a scarce resource. Then, in absolute terms small fluctuations in the supply may lead to large relative price changes. The same effect, however, is also present for moderate imbalances of buyers and sellers in arbitrarily large systems.

[1] As shown in secs. 11.7 and B.1.2, surprising exogenous information causes large returns, too. Extreme market reactions to extreme external shocks, however, are expected from basic economic arguments. The main problem we solved here is the empirical finding that most extreme price jumps are not caused by external news (chap. 10).

[2] The "encoding" of surprise in the return is another parallel to a hypothesised neuronal code in the brain. See sec. B.4 for more details.

While the model is rather abstract and currently tied to a specific pricing rule, many qualitative and quantitative features of real returns are captured both numerically and analytically. For example, real prices are approximately diffusive over many time scales (sec. 10.6). Weak mean reversion on very long time scales (i.e. years in real markets) should not fundamentally change the results. The bubbles model actually exhibits mean reversion on time scales that are long compared to the time required to traverse the state space, and therefore much longer than the time scale on which speculation takes place.

Empirical studies (e.g. [BFL09]) indicated that diffusive prices indeed lead to states where market granularity matters due to micro-crises in revealed liquidity (sec. 10.6). There also appear to be examples for major bubbles with high volatility like the "dot-com bubble" in the late 1990s [PV04, Sch13], even though mixed opinions exist on the question which of the alleged bubbles exhibited increased volatility [AS04].

Furthermore, most existing models other than MGs that can generate clustered volatility and non-Gaussian log returns rely on a hidden variable that shifts between low and high activity regimes (sec. 10.7). In [LM99b], for example, the trading volume is coupled with the volatility and scales with price imbalance. The spin model from [Bor01] implements a similar principle in a much simpler way (sec. 10.9), and also involves a nonlinear pricing rule. Another nonlinearity with respect to market imbalance is a the leverage cycle (sec. 10.7). Price-efficient bubbles could potentially provide a unifying framework for many similar models. The bubbles model of chapter 12 may not be the most realistic market model, but it might be the "hydrogen atom" of the aforementioned model class. Another advantage of the efficient bubbles model over many others is, that the "stylised facts" here are no finite size effects[3].

[3] Prominent examples for finite size effects are found in secs. 10.7 and 10.8.3

While optimism, trend following, or "riding a bubble" were frequently blamed for market destabilisation (secs 10.5 ff.), no model seems to have systematically and consistently linked the non-vanishing impact of these strategies to the martingale property of prices before. Since price efficiency, given the necessary input, is readily learned from trading- or minority payoffs (sec. 12.5 and chap. 13), it may eliminate the need to manually tune the balance of different market forces in multi-agent models.

## 14.4. Experiments

In chapters 12 and 13, collective phenomena that also persist in large-scale models were demonstrated in group experiments with moderate numbers of interacting subjects. In particular, the seesaw game was introduced, which makes (modified) minority games highly accessible. The game combines information efficiency as in minority games with bubbles as in majority games in a simpler way than the $-game [AS03] [GB03], and without the necessity to fine tune to a phase transition. The version presented in chapter 13 exhibits many features of the theoretical models in this work, including self-organised efficiency, bubbles, local return patterns, power-law distributed log returns, and some volatility clustering even in small systems.[4]

The seesaw game combines many approaches towards understanding price fluctuations in a mathematically precise, particularly simple, and very illustrative way. Since it is entertaining, it not only serves as a scientific tool to investigate group dynamics, but also proved to be attractive for laymen to whom it playfully conveys basics of multi-agent modelling of extreme events. We also used the game successfully as an educational tool for students in a lecture on non-equilibrium systems in statistical physics. The game can be played online at seesaw.neuro.uni-bremen.de. Experience so far indicates the potential to collect even more data by including achievements to increase long-time motivation, further optimising the game for mobile devices, and especially by encouraging more international coverage.

[4] In many multi-agent models, including the trading- and the bubbles model presented above, the time scale of the collective dynamics increases with the system size.

### 14.5. Two answers to the same question?

The IAI of market equilibria and bubbles from price efficiency describe two distinct aspects of information efficient market models. Arguments were made above why both mechanisms are plausible in real markets. Curiously, a true martingale precludes return patterns. These, nevertheless, exist in real markets. An intuitive explanation is that the martingale property in a market must emerge from the balance of heterogeneous strategies. The result may be marginal inefficiency with subtle dynamical patterns that may be more difficult to detect or to exploit (e.g. because of costs or risks). Examples for the seesaw game are shown in figure 13.6. Even a combination of simple forces like fast diffusion and slow reversion creates complex patterns (see e.g. the references in [HB10]).

Both mechanisms scale volatility with an "imbalance" of market forces. In the trading model, this is a sudden reduction of decorrelation of the strategy impacts following surprising information states.[5] In the bubbles model, the imbalance applies to demand and supply, and extreme fluctuations require an additional nonlinearity. Whether such a nonlinearity could also stem from the a lack of decorrelation of agents in bubble states might be a worthwhile question for future research. As of yet, in particular the emergence of similar "stylised facts" from different mechanisms is formally a coincidence. This may indicate at least one of the following. On the one hand, power-law CCDFs and long-range correlations might under-determine the problem. Additional testable parameters could then be used to narrow down the true mechanism. On the other hand, there might be univer-sal features of information-efficient control that explain why similar results should emerge in formally different processes. We will come back to this idea in the final discussion (part IV).

[5] See also [Sor03] for a slightly related hypothesis on correlations, or "pockets of predictability".

## 14.6. Outlook

Future research should investigate information efficient control in more different model classes. A related question with respect to collective learning is: what can be learned under which conditions? Trading strategies without fixed thresholds might improve learning price efficiency, allow for bubbles superimposed with local patterns, and yield an invariance with respect to noise similar to the balancing models of part II.[6] Another important task is a clear characterisation of the payoff in continuous double auctions and its approximation by simpler rules like the ones considered in this work. Furthermore, since the minority rule was found to be most consistent on the time scale of roundtrip trades, maybe it should be considered in a paradigm where agent decisions correspond to going long and short.

Continuous-time martingale processes with limits may lead to a generalisation of the results of chapter 12. A particularly intriguing question is whether efficiency with respect to different nonlinearities in the price in the end leads to the same or to different return distributions. Obvious candidates are the nonlinearities of other models with high- and low activity regimes (see above). Another possible direction, related to the suggestions in the previous paragraph, is the question of how price efficiency may emerge in continuous double auctions (e.g. in an extension of zero intelligence models).

[6] For example, strategies could be constructed from linear combinations of lagged log prices or as linear predictors of returns like in [RP03]. Alternatively, a binary encoding may also feature several thresholds. Scale-free binary indicators could be constructed from the sign of the difference of two moving averages with differently decaying weights or simply from returns over different time periods. It may be worthwhile to research what kinds of strategies are actually used or have been used historically in practise.

A reoccurring motive throughout this work is that pure speculative trading and the minority payoff in particular are problematic when agents are not considered to be pure strategies, but utility optimisers. In the context of the trading model, agents that correspond to actual traders might be introduced as an additional model layer. These agents, or possibly even just one super agent, would then distribute their capital among the different trading strategies, similar to a portfolio. Success-dependent asset reallocation might then be consistent with both utility maximisation and collective learning. A relevant question is furthermore how the number of strategies can self-organise towards the phase transition. A possible objective function for such a process is average income per strategy (sec. B.2).

In the seesaw-game experiments with real subjects, the importance of a small risk-free growth of capital became apparent. Otherwise, most agents in a purely speculative market will lose and therefore, if possible, leave the market.[7] A promising perspective is the inclusion of dividends for this purpose, creating a controlled linear growth of capital which could potentially also be offset by consumption. First, different dividends or risks on multiple assets could be used to model asymmetries that are found in real prices. Furthermore, time dependent dividends that correlate with some information conveyed to the agents (e.g. a bit that indicates rising or falling dividends) allows to investigate how far multi-agent models collectively anticipate profits. This would also create a meaningful reference for bubbles in systems with more than one asset, and allow to discuss model dynamics in more traditional economic terms. It would, for example, allow to asses how far local price efficiency dominates and at which point going against a bubble becomes profitable. If the minority effect is offset by capital growth, a self-organisation of strategies, trading volume, or leverage based on trading success or expected profit may become viable.

[7] The trading model is robust with respect to a moderate adaptations of individual agent use parameters. Residual errors, nevertheless, increase.

An advantage of models with a pricing rule is that realistic costs like a transaction tax can be investigated. This was attempted before in GCMGs [BGMP09], where no price is defined. Therefore, the threshold for participation was raised to simulate a tobin tax (the basic model is explained in sec. 10.8.3). If the threshold is raised significantly, speculators stop trading. This reduces the volatility because the return in this model is equal to the sum of all actions. In particular, changing the threshold can move the system out of the GCMG's critical region where the finite size effect of volume herding emerges. A realistic transaction tax was tested briefly in the trading model of chapter 11 (not shown). The general effect is a gradual increase of average fluctuations without a significant change in the size of the most extreme ones. This is consistent with standard economics [GS80] and empirical data [Hau06]. The reason is that the model stops "learning" earlier which affects the minimal achievable volatility in local attractors. This result, however, may depend on specific model assumptions. In particular, the effect on bubbles was not tested as of yet. Further research is necessary to make statements that might potentially be valuable for economic policy. Note that transaction taxes might also serve other purposes such as financing bailouts.

Whether the mechanisms described in this work are indeed among the main causes of the notorious large jumps in real price time series can in principle be tested empirically. This would require identifying the information states that cause large price changes in a given market. Instead of the classic question, "What fraction of price movements can be explained by information on fundamentals?", the question would be, "Which information can explain most of the price movements and their statistics?" (and on which time scales). Further expanding the use of metaphors from neuroscience, this project may be considered as finding the "receptive field" of a market.

Difficulties may arise because any random variable could drive price fluctuations if agents expect it to be relevant. This problem of self-fulfilling expectations is known as "sunspot equilibria" [MS92].

One possible solution would be to use machine learning techniques to extract information that causes large returns from real price time series and various news sources. This would require access to high quality data sources, some of which may be expensive. A way to test appropriate methods in more controlled settings would be to use data from behavioural experiments, for example based on the seesaw game.

In the final part, we compare the results of part II and III, and discuss them in an even more general, interdisciplinary context.

# B. Appendix

## B.1. Trading model: additional figures

### B.1.1. Speculators and Producers

Figure B.1 shows the phase diagram for $\alpha$ versus the amount of producers in the market. As it turns out, a second phase transition with respect to the number of producers is found. This transition is independent of the one for the speculators. Small $N_p < 0.5 \cdot D$ only weakly influence return distributions.

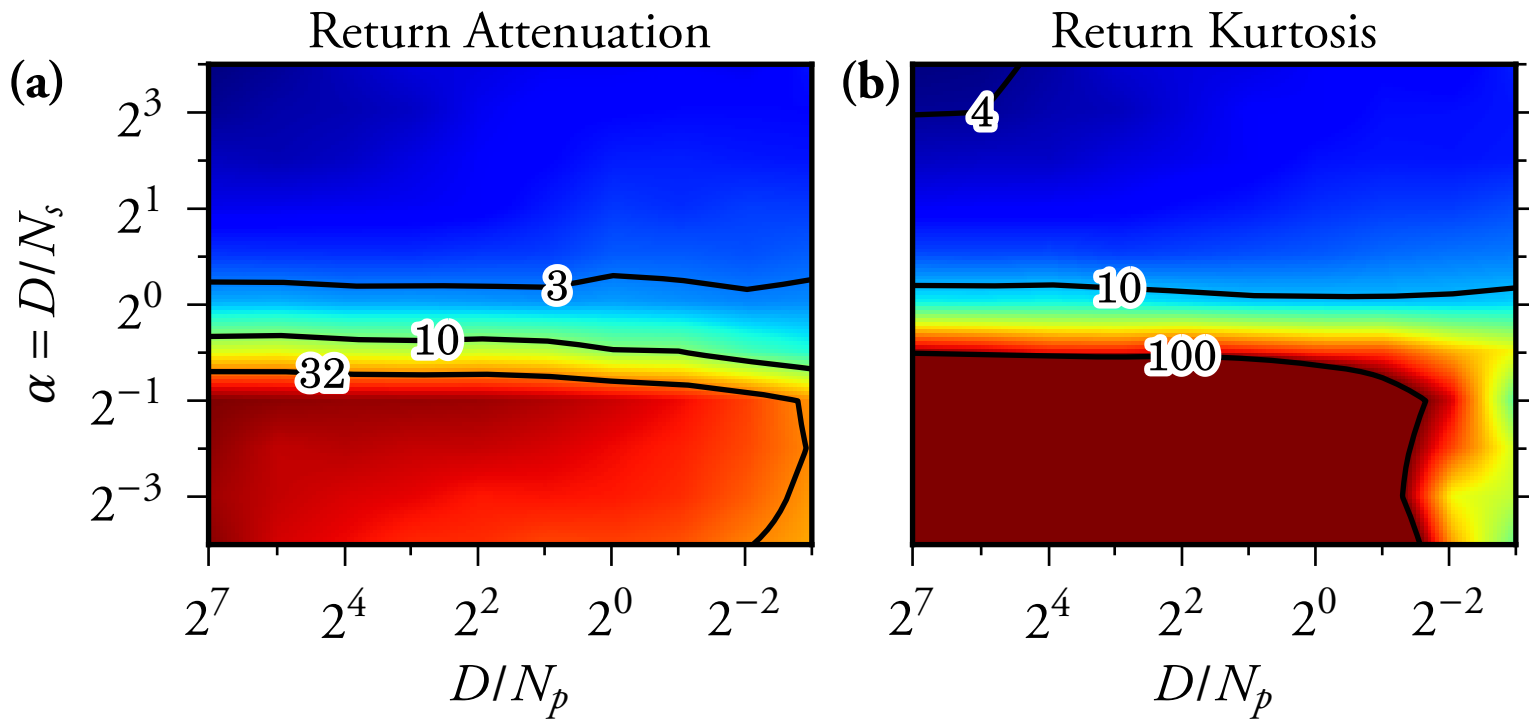

Figure B.1.: **(a)**: Reduction of average return magnitudes during transients, and **(b)**: kurtosis of log-returns of the model with endogenous information for different numbers of speculators $N_s$ and producers $N_p$ for constant memory $K =$ 10 (i.e. $D = 2^{10}$) and use $\gamma = 0.5$. Otherwise, the figure were generated exactly as fig. 11.6.

## B.1.2. Mixed Information

For a combination of endo- and exogenous information, results are similar to pure endogenous information as long as the endogenous part dominates. Generally, more exogenous information leads to a stronger reduction of fluctuations, less pronounced volatility clustering, and random time series without visible patterns even for small $\gamma$. The scaling of the remaining extreme returns remains unchanged. An example is shown in figure B.2.

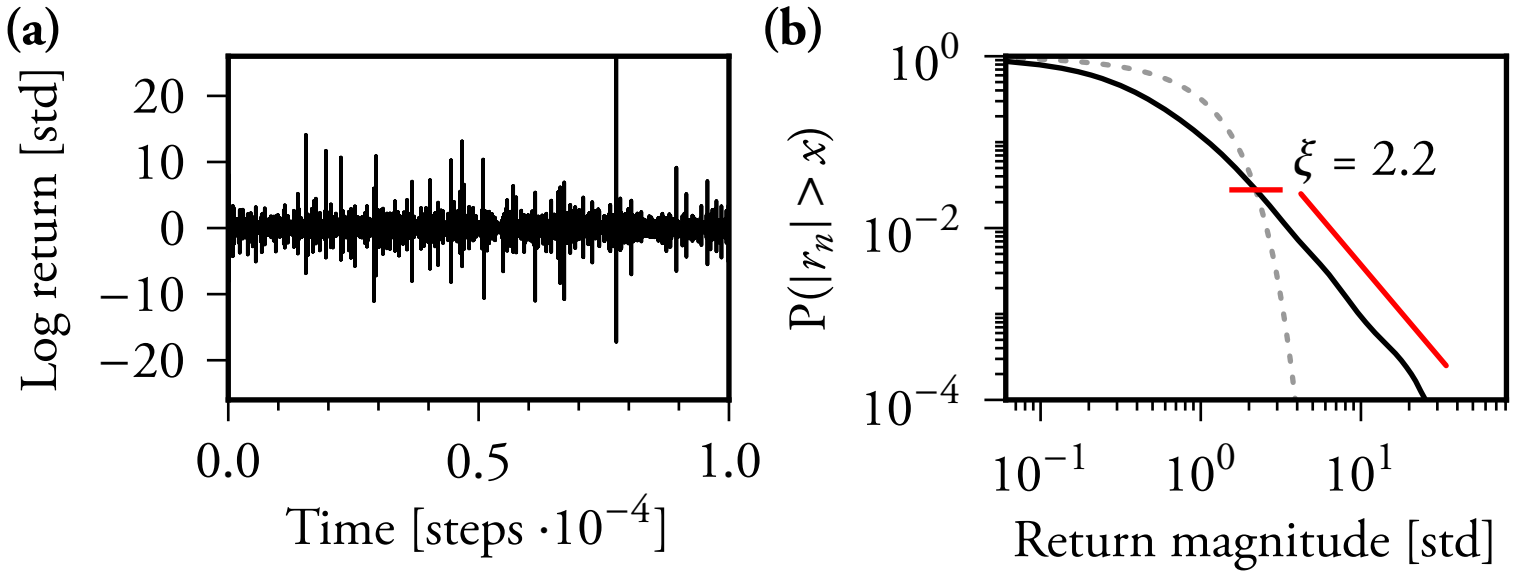

Figure B.2.: Simulation for a trading model with 3 bits of uniform exogenous information and 6 bits of endogenous information. $N_s = 2^{10}$, $N_p = 0$, $\gamma = 0.1$. The first $2^7$ time steps were discarded. Log returns $r_n$ are normalised by their standard deviation. **(a)**: Time series **(b)**: Solid black line: complementary cumulative distribution function. Short Red line: Hill estimator for the scaling exponent. Dashed grey line: normal distribution.

### B.1.3. Distribution of Information Ages (Surprise)

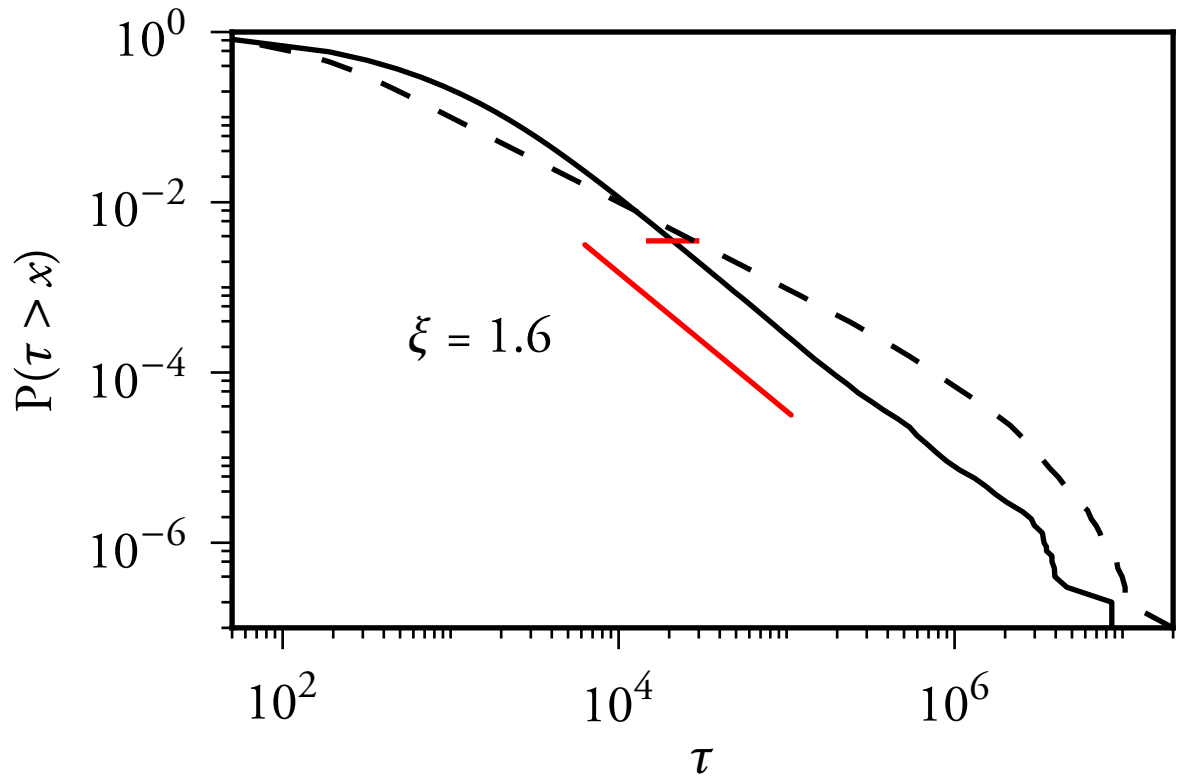

Figure B.3.: CCDF of the times $\tau(t)$ since the informations $\mu(t)$ occurred last. Solid black line: Model with intrinsic information and $D = 2^{10}$, $N_s = 2^{11}$, $N_p = 0$, $\gamma = 0.5$. Short red line: Hill estimator. Dashed line: Exogenous information with $P_{\text{exo}}(\mu) \propto \exp(-0.02\,\mu)$, leading to $P(\tau) \propto \tau^{-2}$.

## B.2. Trading model: income and the critical point

Figure B.4 shows the phase transition with respect to $\alpha$ in more detail. As in figure 11.3, we take one parameter set as a reference to which we compare simulations after initial transients for different parameters. For orientation, the log-return variances (figure B.4 (a)) and kurtoses (figure B.4 (b)), which have been discussed earlier, are shown again.

Mean speculator capitals

$$C_s(t) = \frac{1}{2N_s} \sum_{k=1}^{N_s} M_k(t) + S_k(t). \tag{B.1}$$

are not constant over time in markets that include producers. For empty markets, the ratio of average speculator and producer capitals quickly evolves towards an equilibrium. The more agents are added, the longer it takes for $C_s$ to saturate. In critical or crowded markets, a positive speculator income persists over long times. Then, average speculator capitals after transients are well described by:

$$C_s(t)^2 = C_s(t_0)^2 + a\,t, \quad t_0 < t. \tag{B.2}$$

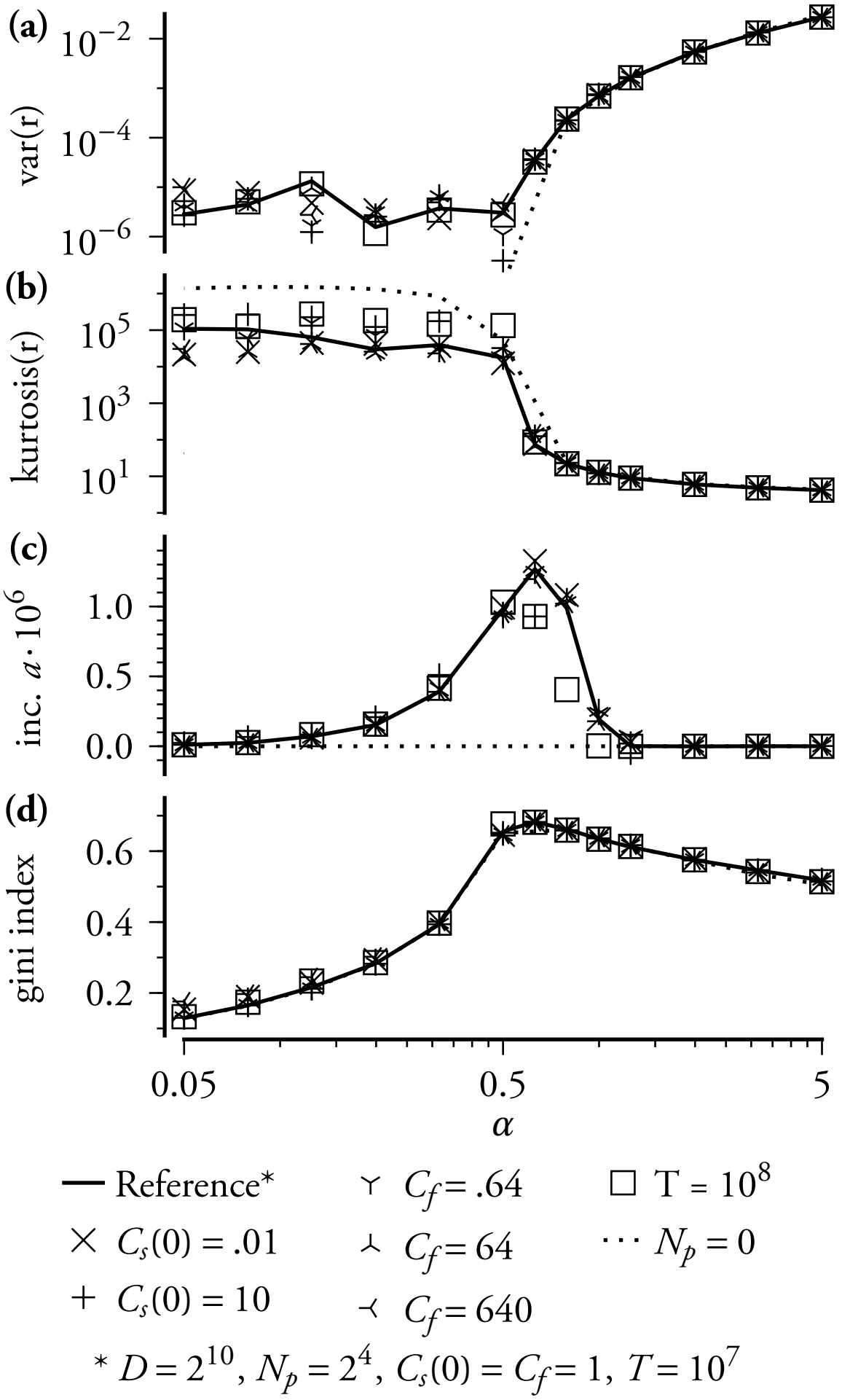

Figure B.4.: Properties of log-return and resource distributions for different $\alpha = D/N_s$, and endogenous information. Solid line: The model with unity (initial) speculator and producer capitals $C_s$ and $C_p$, and a small number of producers serves as a reference. For each other line or symbol, only one model parameters has been changed, respectively. Shown are averages from 50 simulations for each of which only the last $T = 10^7$ out of $2\,T$ time steps have been analysed. **(a)**: variances and **(b)**: kurtoses were calculated from log-returns. **(c)**: income factors $a$ according to equations (B.1) and (B.2). **(d)** Gini indices for speculator capitals after $2\,T$ time steps.

This result reflects that the impact of the speculators increasingly dominates the impact of the producers over time because the speculators get richer. It therefore becomes increasingly difficult to exploit the producers.

The income factor $a$ is shown in figure B.4 (c) and quantifies how well the speculators can exploit the producers. $a$ is found to be independent of the initial ratio between speculator and producer capitals. $a$ is maximal close to the critical point which can be intuitively understood: For empty markets, there is a finite chance for a producer strategy to lie outside of the space spanned by the speculators. Therefore, increasing the number of speculators increases their average income. For crowded markets, producers are already optimally exploited. Then, adding more speculators just distributes the maximal total income over more of them. An analogous maximum can be found in MGs [CMZ05b].

Figure B.4 (d) shows the Gini index, a common measure of wealth inequality.[1] Increased incomes coincide with increased capital inequality among speculators: the gini-index shows a maximum at the critical point. Thus, only a few speculators are most successful in exploiting the producers.

## B.3. Trading model: volatility and distribution tails

From a purely descriptive point of view, the distribution and autocorrelation of a random process are mathematically distinct features. For example, two processes can have the same probability distribution, but different autocorrelations. However, there are various ways of generating a random process where these two features are closely interdependent. Log returns in particular are sums over many logarithmic price changes. Therefore, if those price changes were IID, it would follow from the generalised central limit theorem that returns over longer time intervals should follow either a Gaussian or a Lévy stable distribution (sec. 2.2). Yet, returns have been found to be outside of the Lévy regime and still non-Gaussian, as well as remarkably stable as time intervals are increased [GPL⁺00, PGA⁺99, Lux09]. Since subsequent return magnitudes are not independent, it seems natural to assume that a dynamic volatility is to blame for the slow convergence of long-term returns towards the Gaussian distribution. A possibly related finding is that return distributions become less heavy tailed after normalisation by an estimate of the volatility at each point in time. The extent of this effect depends on the assumptions made about a hidden stochastic volatility process [ABDL00, FGV09].

[1] A Gini index of zero corresponds to all agents having equal wealth. A Gini index of one corresponds to one agent owning everything.

In the model described in sec. 11.2, jumps to a "surprising" part of the information (attractor) space creates large returns. Such events increase the probability for more large jumps while the market adapts to the new environment. Therefore, volatility in our model is a stochastic variable, and connected to non-Gaussian returns, but this connection is more complex than simply Gaussian noise with a time dependent amplitude. However, a detailed analytical characterisation of this volatility process is outside of the scope of this work. Here, we present two numerical analyses of the model returns.

First, distributions for the model are very stable when returns over more than one time step are calculated. Figure B.5 shows a comparison of single-step returns and cumulated returns over 1000 time steps for the model, and for similarly distributed independent random variates. The return distributions for the model are found to be more stable than the independent ones. Therefore, one time step in the model could also be interpreted as a shorter time interval than one day. This holds especially when increasing $N_s$ which leads to a slower decay of correlations of magnitudes over time. For a comparable analysis of high frequency returns, see e.g. [GPL+00].

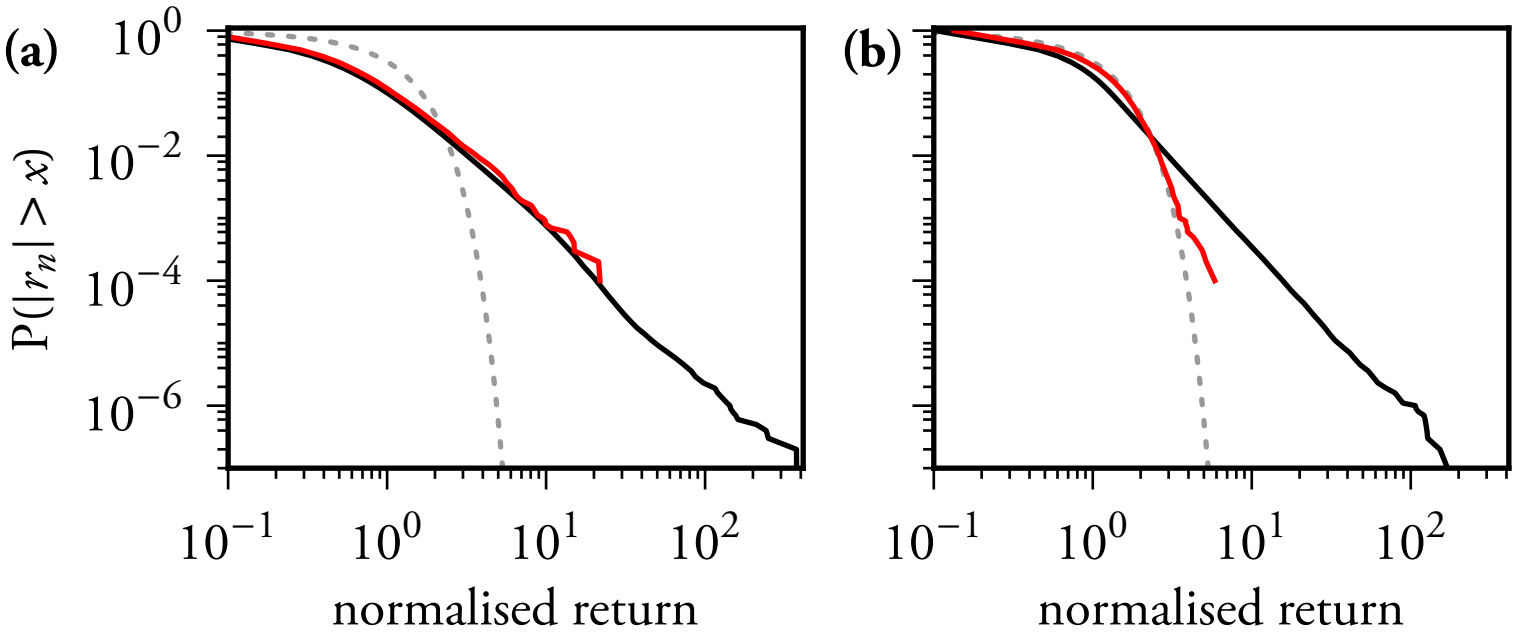

Figure B.5.: Comparison of log returns (black line) and cumulated log returns over 1000 time steps calculated from the same time series (red line). In both cases, the distributions have been normalised to unit variance. A normal distribution (dashed grey line) is shown for comparison. **(a)**: Log returns for the model (sec. 11.2) with $D = 2^{11}, N_s = 2^{12}$, $N_p = 0$, $\gamma = 0.8$. Transients were discarded. **(b)**: As surrogate log returns (black line), independent random variates were generated using inverse transform sampling. The distribution was chosen such that it follows a Gaussian below a threshold and a power law with a slope of 2.5 above the threshold. At the threshold, the PDF and its derivative are continuous.

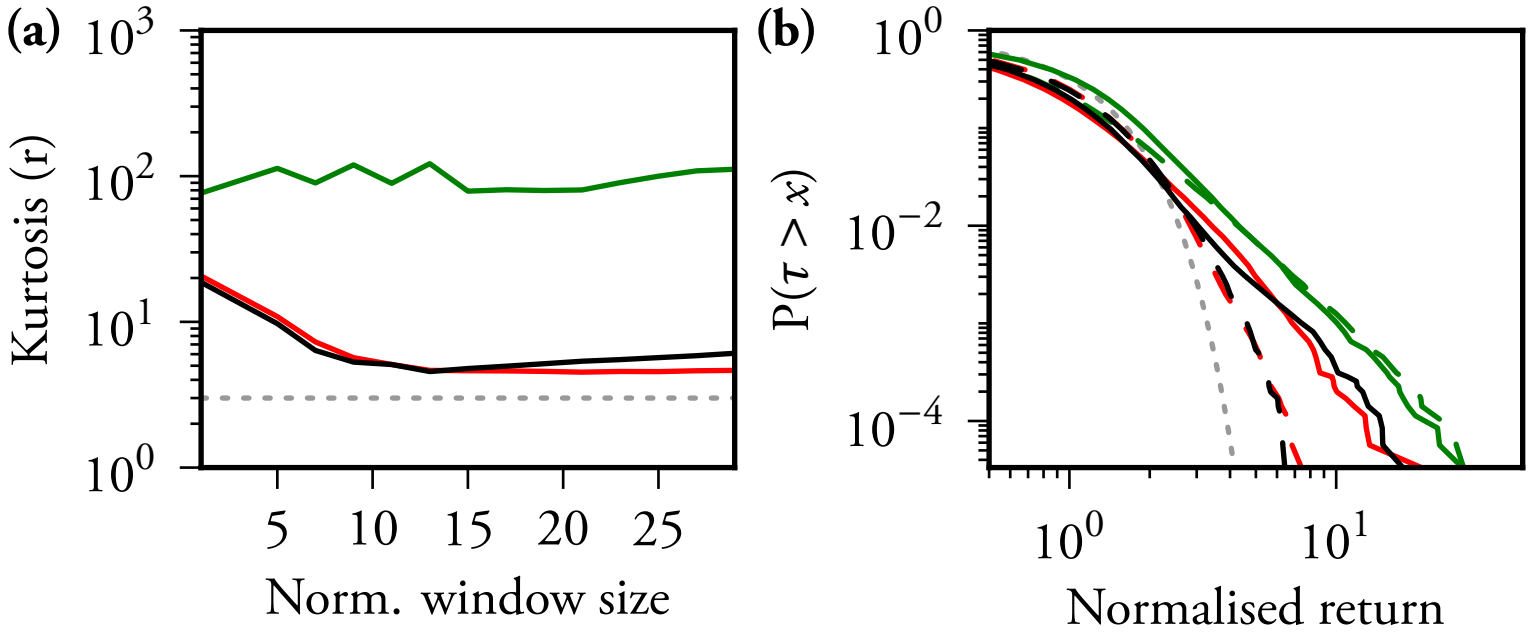

Figure B.6.: Normalisation of log returns with a sliding window reduces the heavy distribution tails for the model (red lines, $D = 2^9, N_s = 2^{10}$, $N_p = 0$, $\gamma = 0.8$), and for the DJIA (black lines). The reduction is not sufficient to make the distribution perfectly Gaussian (dotted grey lines). For independent surrogate returns generated as in figure B.5, the normalisation has no effect (green lines). **(a)**: Kurtosis of the log returns for different sizes of the normalisation window. **(b)**: CCDFs for the original (solid lines), and normalised (dashed lines) returns for the window sizes which minimise the respective kurtoses. All distributions are normalised to unit variance independently of the sliding normalisation to unit volatility.

Second, we consider a simple analysis for as to how the CCDF and local volatility interact. Returns are normalised by using sliding windows of different sizes (similar to [ABDL00]). The model behaves very similar to the DJIA in this analysis (figure B.6): Normalisation to constant estimated volatility reduces the kurtosis, but not enough to yield a Gaussian process. This effect is not found for independent surrogate returns.

In conclusion, simple model-free tests show that the model captures observed features of volatility clustering in real stock market returns very well, and that simulated time steps do not necessarily have to be interpreted as trading days.

## B.4. Market surprise vs. predictive coding in the brain

As shown in chapter 11, the success dependent redistribution of assets in a financial market is equivalent to a collective learning algorithm. Price returns were found to encode how surprising an information state (or state of nature) is to the market. This can be considered a predictive code.

Here we briefly substantiate the analogy to the nervous system. We argue, in particular, that the brain may utilise IAI to its advantage. Because neuronal coding is technically outside of the scope of this thesis, the introduction to the topic, the examples, and the discussion are kept to a minimum.

Neurons generate stereotypical electrical pulses, called "spikes" or "action potentials", which travel down axons (nerve fibers) to propagate signals. Information is encoded in sequences of action potentials, so called spike trains. A neuron's axon is connected to other neurons via junctions called "synapses". The arrival of an action potential on the pre-synaptic side causes the release of neurotransmitter molecules that trigger a post-synaptic potential in the post-synaptic neuron.

Most of the time, there is an excess of negatively charged ions inside a neuron. The neuron's cell membrane is essentially impermeable to most ions, except at specific ion channels, some of which are actively regulated. The membrane therefore acts like a combination of a capacitor and a resistor. Input spikes that increase the membrane potential are called "excitatory" and those that decrease the membrane potential are called "inhibitory". If a neuron's combined inputs during a small time window lead to a strong increase of its membrane potential, the neuron "fires", that is, it generates an action potential. Neurons, therefore, are often described as threshold elements.

For most parts of the brain, it is unknown how exactly information is encoded. Spike trains in the cortex are typically described as highly irregular and statistically very similar to a Poisson process [DA01]. A proposed cause for these firing statistics is a detailed balance of excitatory and inhibitory inputs close to the threshold such that firing depends on small fluctuations in the input [SN94]. Excitation and inhibition were experimentally found to be closely balanced not only on average but also during ongoing activity [OL08]. This raises the question of the functional role of the detailed balance. If only the firing rate were relevant, there would appear to be no functional reason to operate at the point with the highest output variability.

A related hypothesis is that spikes, in some conditions, signal unexpected stimuli or unexpected stimulus variations [Den08]. This is called predictive coding. Spikes then appear, by definition, unpredictable. Advantages of such a sparse use of action potentials include energy savings and data compression. A related finding is that balanced excitation and inhibition emerge for model neurons where spikes report unpredictable deviations from the mean input (ibid.).

We here show that the balance of excitation and inhibition may lead to a predictive code with high sensitivity to unexpected stimuli. We proceed in two steps. First, it is illustrated how spike trains can encode the degree to which excitation and inhibition are balanced. We then demonstrate that that a neuron which learns to balance excitation and inhibition for a specific predictable stimulus becomes a highly sensitive detector for unpredictable (i.e. unknown) stimuli.

First, consider an ordinary leaky integrate-and-fire neuron [DA01]. The sub-threshold membrane potential is described by

$$\tau_m \dot{V}(t) = V_r - V + R(I_e + I_i), \quad V(t) < V_{th}, \tag{B.3}$$

where $\tau_m$ is the membrane time constant and $R$ is the membrane resistance. When $V$ reaches the threshold $V_{th} = 1$, an action potential is fired and $V$ is reset to the reset potential $V_r = 0$. The neuron receives an excitatory input current $I_e > 0$ and an inhibitory input current $I_i < 0$. We use normalised units for simplicity. Both inputs are modelled as Gaussian white noise. The effective input is the sum of these two signals. Therefore, the average effective input is determined by the sum of the mean signals. The effective variance of the input depends on the cross correlation of the inputs. It vanishes for perfectly anti-correlated (i.e. balanced) excitation and inhibition. Figure B.7 shows the model results for an effective input with a constant mean and a time dependent variance. The mean input alone is just below the firing threshold. At first, $I_e$ and $I_e$ are highly anti-correlated. Then the correlation is reduced, leading to a sharp increase in the effective input variance that drives the neuron across the firing threshold. Therefore, a neurons firing rate in the balanced state is highly sensitive to fluctuations in the degree to which excitation and inhibition are balanced.

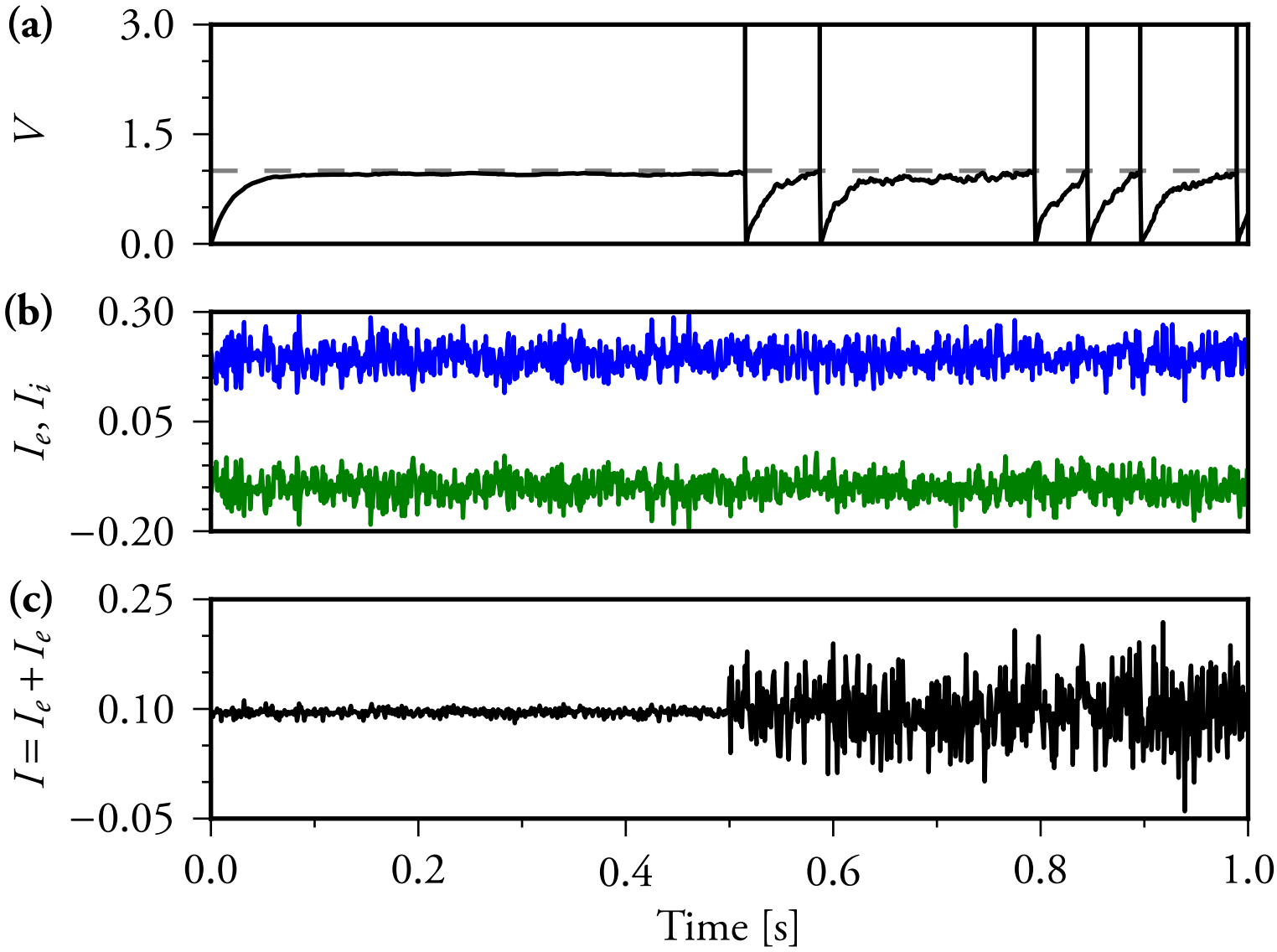

Figure B.7.: Leaky integrate-and-fire neuron with time-dependent balance of excitation and inhibition. Parameters (in normalised units): membrane resistance $R = 10$, membrane time constant $\tau = 20\,\text{ms}$, firing threshold $V_{\text{th}} = 1$. Excitatory input current $I_e \sim \mathcal{N}(0.195, 1)$. Inhibitory input current $I_e \sim \mathcal{N}(-.1, 1)$. Correlation coefficient $\rho(I_e(t), I_i(t)) = -0.99$ for $t < 0.5\,\text{s}$ and $\rho(I_e(t), I_i(t)) = -0.1$ for $t \geq 0.5\,\text{s}$ **(a)**: membrane potential. **(b)** inputs $I_e$ (blue) and $I_i$ (green). **(c)** effective input.

For the second part of the argument, consider a neuron which learns to balance excitatory and inhibitory inputs for a certain predictable input. If the degree of balance for an unpredictable input remains higher than for a predictable input, the resulting spike train after introducing a threshold will encode how surprising an input is–according to the previous paragraph. Figure B.8 (a) shows the sub-threshold membrane potential for a neuron which receives inputs from two neuronal populations. One input population provides excitatory inputs and the other one provides inhibitory inputs. The connections to all input neurons have random, none-negative synaptic weights. The two lines correspond to two different input spike patterns. Figure B.8 (b) shows the membrane potential after the synaptic weights were optimised to balancing excitation and inhibition for one input pattern. The fluctuations for the second pattern are actually increased. This result is robust over repeated simulations, for different parameters and optimisation methods.

Taken together, it follows from very basic neuronal principles that neurons which learn to balance excitation and inhibition become highly sensitive detectors to stimuli that are unpredictable from what the neuron has learnt. In a neuronal population, the "surprise" should be encoded in the population variance. Encoding signals in a population variance facilitates a much better response at high frequencies than encoding the signal in the population mean [SBM+04]. Therefore, fast predictive codes should robustly emerge in the brain. Even though the above arguments belong to a rather common ground in neuroscience, it appears as if this whole line of reasoning had not made as clearly and concisely before. For some further discussion, see chapter 16.

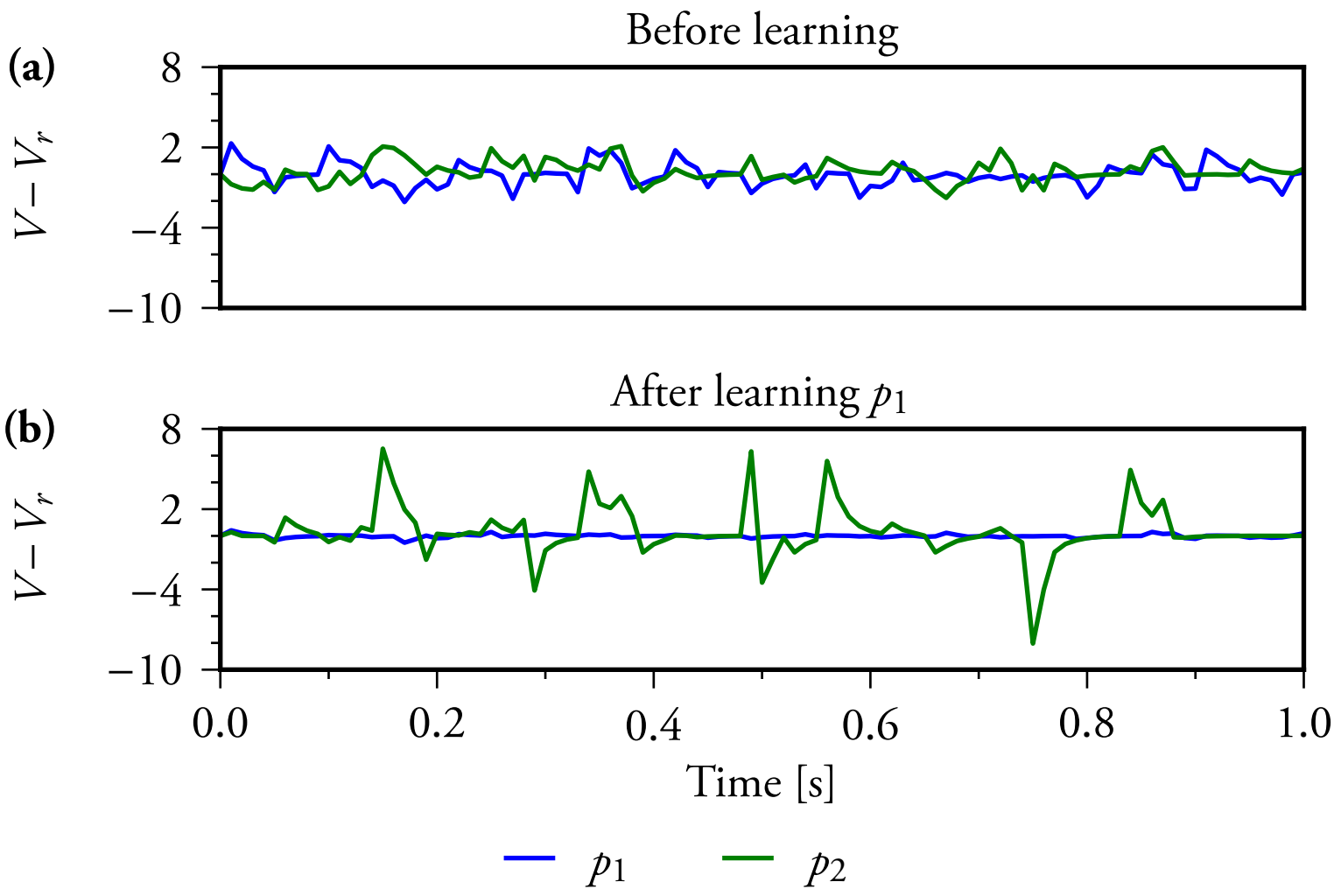

Figure B.8.: A leaky neuron without a threshold, which recieves inputs from 50 excitatory and 50 inhibitory neurons. We compare two input patterns $p_1$ and $p_2$. For each pattern, each input neuron provides a frozen Poissonian spike train with a firing rate of 1Hz. Output neuron parameters (in normalised units): membrane resistance $R = 10$, membrane time constant $\tau = 20\,\mathrm{ms}$. **(a)**: the relative membrane potential after initialising the synaptic weights for the inputs from a uniform distribution between 0 and 0.03. **(b)** the membrane potential after learning to suppress membrane potential fluctuations for $p_1$. Starting from the initialised weight vector, the MSE was minimised using the downhill simplex method, while keeping all weights non-negative. Qualitative results do not depend on the particular optimisation method.

# Part IV.

# Conclusions

# 15. Summary

> "Though this be madness, yet there is method in 't."
>
> WILLIAM SHAKESPEARE

In this work, we investigated apparently unreasonable outliers in the variation of two types of human behaviour. The largest outliers, however, are not singular, isolated events. Instead, they are the most visible extensions of self-similar event distributions spanning several scales. The underlying causes were investigated experimentally and theoretically using models that can explain the empirical event distributions, as well as important characteristics of their temporal structures. These models were constructed to account for the domain-specific requirements, yet share a novel overreaching principle which can be intuitively understood in terms of the behaviour investigated in part II.

Consider the task of balancing a stick on a finger tip. It is only possible to predict where the stick is going to fall once it starts falling. In other words, the more successfully a system is stabilised, the less its visible dynamics reveal about its structure. This Information Annihilation Instability (IAI) can induce self-similar fluctuations in an adaptive control system.

A model based on this principle with a realistic reaction time and smooth movements can quantitatively reproduce many details of the spatio-temporal structure of the movements of real subjects in a Virtual Stick Balancing (VSB) task. Furthermore, specific empirical measures can be directly linked to model properties. According to this theory, subjects predict how the stick is going to fall based on the most recently observed trend. They then move their hand slightly faster towards the anticipated location.

The model further reveals a trade-off between typical balancing errors and rare, extreme ones. This prediction was tested in a newly developed VSB task featuring a high score to impose different cost functions onto the subjects. Results are consistent with the model predictions while several alternative explanations have to be rejected: Subjects neither add multiplicative noise to stabilise an otherwise unstable system, nor is most multiplicative noise inherent to movement execution. It was furthermore demonstrated for the first time that subjects in this task can optimise their behaviour for distinct cost functions, leading to pronounced qualitative differences in behaviour.

In part III, financial markets were investigated. Here, individual and collective behaviour has to be distinguished. In chapter 11, a parsimonious market model was introduced where the impacts of trading strategies naturally adapt according to their success. This implements a learning rule for the whole market that, as we proved, minimises predictable price changes. Large price changes therefore reflect surprising information that reaches the market. When the trading strategies are based on past price patterns, however, equilibrium states become destabilised after the market has absorbed all predictable information. In fact, self-referential markets were found to generate surprising information by changing between attractors such that the consequent price changes quantitatively match the distribution and temporal structure of price changes in real markets very well.

In chapter 12, we introduced the seesaw game as a paradigm for group experiments. It was found that subjects were nearly information-efficient with respect to a simple type of information: the preceding price at each time step. This elimination of predictable trends was found to induce long-range correlated dynamics that resemble bubbles. This finding could be reproduced in a very simple and analytically tractable model. The difference between this type of instability and IAI of a fixed point can be understood intuitively in terms of stick balancing.

A model that adapts to complex return patterns evolves towards equilibrium prices. A dynamical instability arises when all observable information has been absorbed. This can be thought of as a high-dimensional analogue to the situation during stick balancing. When using absolute prices as information, however, the market cannot evolve towards an equilibrium price because mean reversion is eliminated. This corresponds to a stick-balancing controller that would eliminate predictable velocities and never decisively move the stick back towards the upright position. The stick's angle, after a sufficient amount of time, would then diffuse towards arbitrary values where the system behaves nonlinearly. Eventually, the stick will fall unless some additional control mechanism steps in. In a real market, both types of information are likely to be used by traders. The emergence of a combination of bubbles, patterns, and realistic log-return distributions was observed in an extended seesaw game presented in chapter 13. It was further demonstrated that the emergence of large-scale collective phenomena can be studied in group experiments even with moderate group sizes.

# 16. Discussion and outlook

We presented a spectrum of results ranging from contributions to technical and field-specific discussions to potentially far-reaching and very general findings.

## 16.1. Balancing

The main findings on motor control provide a new view on the origin of movement variability and on internal models used by the Central Nervous System (CNS). The results indicate that the latter are highly adaptive and reconfigure on short time-scales to reflect task-relevant information, which is extracted from observations only as immediately needed.

Similar strategies were found in very different tasks before (secs. 9.6.3 and 9.6.4), but the concept was apparently never discussed as a dynamical source of movement variability. The reason why this didn't happen may be that motor control experiments often involve tasks such as pointing or throwing, where the subjects is not interacting with an unstable object. While the temporal structure of postural sway shares a number of features with stick balancing, it involves reflexes that actively stabilise joints. Humans may also prioritise minimising effort over minimising movements. While humans may use similarly adaptive strategies in many different tasks, this might not always manifest in very pronounced power-law-distributed fluctuations.

Additional findings include explanations why the movement of a balanced stick appears highly damped without a plausible physical mechanism (sec. 7.8) and why distributions in difficult balancing tasks are less heavy-tailed than in easy ones (sec. A.6), as well as results on significance tests for power laws in time series where subsequent events are not independent (sec. A.1).

A possible long-term goal for future research is a complete theory for learning and performing motor control which explains experimental findings across different tasks. IAI may contribute pieces to this puzzle on different levels, which will be discussed further below.

## 16.2. Markets

The main findings on speculative trading are consistent with the hypothesis that information efficiency is at least a good first-order approximation of the dynamics of financial markets. It was, however, also found that information efficiency can be a double-edged sword, depending on which information a market adapts to. Three main types of information were investigated. For extrinsic information about states of nature that do not depend on the market, heterogeneous multi-agent models based on common and reasonable assumptions on speculative trading were found to self-organise towards efficient equilibrium states, as predicted by classical economic theories. These markets may react sensitively to arriving information, but only if the information is surprising (sec. 11.7). However, markets cannot be fully efficient with respect to their own past if no one exploits patterns that arise as epiphenomena of the interaction of dynamics on different time scales.

When markets adapt to self-generated information, equilibria can become destabilised. Dynamical instabilities were found in markets that adapt to local patterns in recent price changes (sec. 11.5). The absorption of mean reversion, which was observed in real markets on short to intermediate time scales, was found to facilitate the formation of bubbles (sec. 12.3). This principle was implemented in a model which might turn out to be the simplest incarnation of a basic principle behind many more complicated models that link market imbalance to bubbles and crashes (sec. 12.4). However, even though the models in chapters 11 and 12 generate similar log return distributions, there are some conceptual differences.

It was found that an invariant manifold in the space of asset distributions is a characteristic feature of trading (sec. 11.4.1). This finding gives insight into why real markets might be in a state of high excitability even if they appear calm. Consider that in some markets, the number of traded assets can exceed the number of existing assets after mere days, yet prices on average change less then one percent per day. Therefore, market forces are apparently in a delicate balance which can be disturbed suddenly by unexpected, possibly minor and market-internal events.

In contrast, the nonlinearity in the minimal bubbles model explicitly depends on the pricing rule. In reality, however, this rule arises from microscopic principles, the investigation of which may clarify whether there are any hidden connections between the mechanisms. After all, both mechanisms exhibit an increased volatility for imbalanced markets. It may also be possible that the "stylised facts" can arise in different ways and that more empirically measured quantities have to be included in future analyses to distinguish between them. If so, the next question would be whether the "stylised facts" in real markets can be attributed to one mechanism, or whether markets exhibit many different instabilities that somehow lead to similar price fluctuations. Note that if IAI is indeed a "universal" principle, it could even occur on different levels of the same system.

Additional findings include a clarification of the properties of different means of adaptation in multi-agent systems (sec. 14.1). It was shown that IAI can be found in suitably designed Minority Games (MGs). Nevertheless, we identified some differences between models where one versus two resources are adapted. More traditional MGs, however, where strategies are exchanged faster than their impact is adjusted, or where the impact is not adjusted at all, were found to behave completely differently because they cannot reach market equilibria at all.

Future research might concentrate more on identifying common phenomena among different models. A related question is: what can a market "learn" under which conditions? Another possible topic, that is particularly important for open (e.g. grand canonical) markets, is the influence of dividends and consumption (see chap. 14 for an explanation). Finally, identifying relevant information states and imbalances in real markets should become a main focus of future work. Besides the analysis of large data sets from real markets, behavioural experiments may serve as a semi-controlled testing ground. As we have shown, the emergence of extreme collective phenomena can be studied even in moderately sized groups. Such experiments can furthermore serve a double purpose as an educational tool.

## 16.3. Predictability

This work focused on the mechanism behind crises, but not on their prediction. Nevertheless, the presented models exhibit certain regularities that in principle allow for predictions. In section A.13, it is shown that the minimal balancing model can be predicted nearly perfectly by an external observer. All other models, however, are more complicated. While the trading model is completely deterministic, the naive approach to prediction requires the knowledge of all strategies and the distribution of assets. Furthermore, a prediction method most likely becomes less useful if it becomes widely used in a real market.

Nevertheless, informed statistical risk prediction based on knowledge of fundamental system dynamics might be a worthwhile topic for future research. Many systems exhibit "warning signs" near tipping points. One such sign is critical slowing down where a system recovers more slowly from perturbations while autocorrelation and variance increase. This phenomenon was reported for systems as diverse as lasers and neurons [SBB+09]. Similar effects are observed in bubble states in the model introduced in chapter 12, and also for the two-dimensional spin model briefly discussed in section 10.9 [KBB12].

## 16.4. Optimality and universality

Whether a controller is "truly optimal" is ill-posed question. Optimality can only be defined with respect to certain assumptions and limitations. The minimal balancing model is optimal given the restriction to a memory of two time steps. In the continuous balancing model, the reason why an infinite memory is not optimal lies in the restriction to smooth and cautious movements. The latter were found to be optimal, for example, under certain assumptions on movement execution noise (sec. 5.3). However, there may be other real-world advantages to fast adaptation. Furthermore, there is probably little evolutionary pressure to perform optimally in stick balancing. In more complex tasks, high performance with little effort and behavioural flexibility might favour parsimonious and adaptive internal representations of task-relevant features even more.

Nevertheless, the balancing model suggests that error distributions between the Lévy- and Gaussian regimes may reflect a nearly optimal compromise between the elimination of random local trends and rare large errors (sec. 7.6). A similar effect was reported in portfolio optimisation where "minimizing 'small' risks can lead to an increase of 'large' risks" [AS01]. In some sense, the trade-off is also analogous to the bias-variance dilemma that is well known in machine learning: models that reach a very low error in some training data are often bad predictors for unseen data due to overfitting (i.e., the model "learned the noise").

This raises the question whether the observed range of power-law exponents can be justified from some general optimisation. A slightly less exciting explanation, at least for the lower bound of observed exponents, may be that power laws outside of the Lévy-regime ($\xi > 2$) still feature a finite variance. If the variance were infinite, certain types of systems might not survive very long.

## 16.5. IAI and the brain

If IAI is indeed a fundamental instability in complex adaptive systems, similar phenomena should be observed in different contexts. One may ask in particular how the results relate to neuronal activity. After all, we likened resource redistribution among trading strategies with learning in neuronal networks. Furthermore, returns in the trading model effectively represent a predictive code. As it turns out, there are very self-evident arguments for the hypothesis that the brain might use IAI to its advantage. As shown in section B.4, it is even possible to construct neuronal models for predictive coding based on this principle. Instead of buy and sell orders, neurons balance inputs that increase or decrease their membrane potential. The output spikes of these neurons encode inputs that have not yet been learned, in other words: surprise. This furthermore raises the question whether similar mechanisms might be generally useful for rapid data processing with limited resources.

## 16.6. Information and the critical point

A thermodynamic system at a critical point is in equilibrium, and an external parameter is tuned to critical value. Self-organised critical systems, where a critical point is an attractor, are typically neither closed nor in equilibrium: they are slowly driven from the outside (sec. 4.2). Here, we investigated systems where internal interactions that annihilate predictable information create attractors that are critical points. In most presented models, equilibria are destabilised by IAI. While this might be considered a new type of Self-Organised Criticality (SOC), the underlying principle is distinct from previous instantiations since it also occurs in low-dimensional systems.

The phase transitions that occur in heterogeneous multi-agent models usually differ from those in thermodynamic systems. The latter mark a change in the spatial order of a system, while most market models have no spatial dimensions. In common MGs, the phase tran-sition marks the boundary between phases of independent and linearly dependent agent strategies. Non-Gaussian returns and long-range temporal correlations only occur in combination with secondary mechanisms that are not directly related to the phase transition. In models with resource redistribution, the phase transition separates the phases where trading strategies form an either incomplete or (over-)complete basis. In the trading model, the critical point at the phase transition is important because there the market "forgets in the right way". There are, however, many possible reasons why a real market might adapt locally even though it is not exactly at such a phase transition. The critical point at the phase transition is furthermore different from the dynamical critical point where equilibria become unstable.

The impact of locally minimising fluctuations in a SOC sand pile was investigated in a recent study [NBD13]. A trade-off was found between the control of common, small avalanches and inevitable rare, extreme ones. Preventing small avalanches can even make a subcritical sand pile critical.

These findings raise the question whether the description of (self-organised) criticality as information processing may lead to a unifying framework for some phenomena that currently appear unrelated. After all, information (entropy) is already fundamental to several physical theories and not exclusive to brains or computers. A major problem for such an endeavour, however, would be to first find a suitable general formalisation of information. IAI can be expressed for any system that can be linearised around a fixed point using Fisher information to naturally measure locally observable information (sec. 7.2). Shannon information (i.e. entropy) is a global measure of the absence of predictable structures in a signal. It is closely related to surprising information which, however, in market models is best characterised by a local measure (sec. 11.7). A more general theory would probably be based on a very general information measure. Nevertheless, if one is searching for a comprehensive information-based principle for criticality, then this work offers several examples for limiting cases that should be included.

# Acknowledgements

Many thanks to everyone who supported me over the last years and thereby allowed me to complete this work. My apologies to those I forgot to name below.

I am particularly grateful for having Klaus Pawelzik as a supervisor. He provided me with the most inspiring starting point for my research, offered outstanding support as I searched for my own way, and motivated me to aim higher. Our discussions never failed to open new avenues, often straying off the beaten path.

I would also like to thank my second referee Stefan Bornholdt who followed my research and provided great advice since I was an undergraduate student.

My very great appreciation goes to the proofreaders Julie Comparini, Philipp Heyken, and Christian Albers (sec. B.4). Andreas Kreiter greatly helped improving the general understandability of [PP13] which forms the basis for most of chapter 11.

Furthermore, I want to thank all past and current members of the Theoretical Neurophysics Lab who made my past years so much more enjoyable. I could not have wished for better personal and scientific interactions in a more sincere, cordial environment. In particular Agnes Janssen provided outstanding organisational support. David Rotermund kept the electronic infrastructure running, most importantly the computing cluster. Also Udo Ernst provided substantial advice and technical support. In addition, I want to mention my collaegues (in semi-random order, starting with current members of the group) Maren Westkott, Christian Albers, Daniel Harnack, Nergis Tömen, Axel Grzymisch, Philipp Heyken, Joscha-Tapani Schmiedt, Orlando Arevalo Acosta, Ingo Bathmann, Josephine Mielke, Marco Linke,

Joana Vieira, Sven Eberhardt, Niels Treiber, Nadja Schinkel-Bielefeld, Onno Böhler, Markus Riegel, as well as other (current and former) members of the center for cognitive sciences Cathleen Grimsen, Detlef Wegener, Dieter Gauck, Marc Schipper, Iris Grothe, Maren Prass, Andreas Kreiter, Manfred Fahle, Manfred Herrmann, Helmut Schwegler, Manuela Jagemann, Sabiene Melchert, Charlotte Herzmann, Margarete Korsch, Torsten Stemmler, and Sirko Straube. I furthermore thank Sebastian M. Krause, past member of the Complex Systems Lab, for the discussions over the past months. Many thanks also to Dennis Bredemeier, Daniel Henrichs, and Katharina Morosov who tested the experimental setup I had developed (chap. 8) during their internships.

I thank the Volkswagen foundation for funding significant parts of the work presented in part III.

Last, not least, I want to thank my family, in particular my parents for their endless support.

# Curriculum vitae

| | |
|---|---|
| Name | Felix Patzelt |
| Date of birth | 14.04.1982 |

EDUCATION

| | |
|---|---|
| 2008 - 2014 | Ph.D. student at the Theoretical Neurophysics lab of Prof. Dr. Klaus Pawelzik (University of Bremen) |
| 2008 | Diploma in Physics (University of Bremen). Thesis title: "Self-organised critical control in human behaviour". |
| 2002 - 2008 | Undergraduate studies in Physics |
| 2001 | Abitur (Kippenberg Gymnasium Bremen) |

WORK

| | |
|---|---|
| 2008 - 2013 | Research associate at the the Theoretical Neurophysics lab, University of Bremen. |
| 2002 - 2008 | Freelance work in the network "nachtlicht-media.de": print and screen design, programming and more. |
| 2005 | Internship: MeVis Medical Solutions AG. |
| 2001-2002 | Civil Service: Bremer Umweltinstitut für die Analyse und Bewertung von Schadstoffen. |
| 1998 | Internship: Kossann Professional Media GmbH. |

# Select publications